\documentclass{cernrep}
\usepackage{graphicx,here}
\usepackage{amssymb}
\usepackage{epsf}

\newcommand{\ar}{\rightarrow}

\def\as{\alpha_{\mbox{\tiny S}}}
\def\cM{{\cal M}}
\def\deg{^{\mbox{\scriptsize o}}}
\def\ee{e^+e^-}
\newcommand{\vect}[1] {\mbox{\bf #1}}
\newcommand\yc{y_{\rm c}}
\newcommand{\permil}{\raisebox{0.5ex}{\tiny $0$}$\!/$\raisebox{-0.3ex}{\tiny $\! 00$}{\normalsize}}
\def\ZP{Z.\ Phys.\ {\bf C}}
\def\PL{Phys.\ Lett.\ {\bf B}}

\def\CPC{Comp.\ Phys.\ Comm.\ }

\newcommand{\mz}{M_{\mathrm Z}}

\newcommand{\fig}{Figure~\ref}
\newcommand{\tab}{Table~\ref}

\newcommand{\gev}{\mbox{\,Ge\kern-0.15exV}}
\newcommand{\GeV}{\mbox{\,Ge\kern-0.15exV}}
\newcommand{\mev}{\mbox{\,Me\kern-0.15exV}}

\newcommand{\beq}{\begin{equation}}
\newcommand{\eeq}{\end{equation}}
\def\jetset{{\sc Jetset}}
\def\pythia{{\sc Pythia}}
\def\herwig{{\sc Herwig}}
\def\ariadne{{\sc Ariadne}}
\def\apacic{{\sc Apacic++}}
\def\amegic{{\sc Amegic++}}

\def\aleph{{\sc Aleph}}
\def\delphi{{\sc Delphi}}
\def\ldrei{{\sc L3}}
\def\opal{{\sc Opal}}

\def\sld{{\sc Sld}}
\def\lep{{\sc Lep}}
\def\lepone{{\sc Lep}1}
\def\leptwo{{\sc Lep}2}

\newcommand{\cambridge}    {{\sc Cambridge}}

\newcommand{\durham}    {{\sc Durham}}
\newcommand{\jade}      {{\sc Jade}}

\newcommand{\eqn}{Eq.~\ref}

\newcommand{\sect}{Sec.~\ref}
\newcommand{\asmz}{$\as(\mz^2)$}

\newcommand{\debrecen}{{\sc Debrecen}}
\newcommand{\fourjphact}{{\sc Fourjphact}}

\begin{document}

\title{REPORT OF THE QCD WORKING GROUP}
 
\author{
 A.~Ballestrero~$^a$,
 P.~Bambade~$^{b,d}$,
 S.~Bravo~$^c$,
 M.~Cacciari~$^d$,
 M.~Costa~$^e$,
 W.~deBoer~$^f$,
 G.~Dissertori~$^d$,
 U.~Flagmeyer~$^g$,
 J.~Fuster~$^e$,
 K.~Hamacher~$^g$,
 F.~Krauss~$^h$,
 R.~Kuhn~$^i$,
 L.~Lonnblad~$^j$,
 S.~Marti~$^{d,k}$,
 J.~Rehn~$^f$,
 G.~Rodrigo~$^{f,l}$,
 M.H.~Seymour~$^m$,
 T.~Sjostrand~$^j$,
 Z.~Trocsanyi~$^{d,n}$,
 B.R.~Webber~$^{d,o}$
 }
 
\institute{$^a$~INFN Torino, Italy, $^b$~LAL, ~Univ.~Paris-Sud, ~IN2P3/CNRS, 
France, $^c$~IFAE Barcelona, Spain, $^d$~CERN, Switzerland,
$^e$~IFIC, Univ.~Valencia - CSIC, Spain, $^f $Univ.~Karlsruhe, Germany,
$^g$~Univ.~Wuppertal, Germany, $^h$~Technion, Israel,
$^i$~T.U.~Dresden, Germany, $^j$~Univ.~Lund, Sweden, 
$^k$~Univ.~Liverpool, UK, $^l$~INFN Florence, Italy,
$^m$~Rutherford Appleton Laboratory, UK, $^n$~Univ.~Debrecen, Hungary,
$^o$~Univ.~Cambridge, UK}
%
%
%

\maketitle 
 
\begin{abstract}
The activities of the QCD working group concentrated on improving the
understanding and Monte Carlo simulation of multi-jet final states due to hard
QCD processes at \lep, i.e.\ quark-antiquark plus multi-gluon and/or secondary
quark production, with particular emphasis on four-jet final states and
$b$-quark mass effects. Specific topics covered are: relevant developments in
the main event generators \pythia, \herwig\ and \ariadne; the new multi-jet
generator \apacic; description and tuning of inclusive (all-flavour) jet
rates; quark mass effects in the three- and four-jet rates; mass, higher-order
and hadronization effects in four-jet angular and shape distributions;
$b$-quark fragmentation and gluon splitting into $b$-quarks.
\end{abstract}

\tableofcontents
\section{INTRODUCTION}

\subsection{Objectives of the working group}
\label{intro_object}

Fully hadronic multi-jet topologies play an important role at \leptwo, in
the contexts both of physics measurements and of searches for new phenomena.
For example four and more hadronic jet topologies dominate the statistics
both in the measurements of $W$ boson pairs and in the searches for Higgs 
bosons, because of the large hadronic decay branching ratios of all heavy 
bosons involved. Improving our understanding of the physics of QCD processes 
and of the modelling provided by our main generators is relevant at
\leptwo\ for two main reasons:

\begin{itemize}
\item
In contrast to the other two main decay topologies occurring in boson pair
production and studied at \leptwo\ (the semi- or fully leptonic ones), 
four-quark production processes leading to fully hadronic topologies must be 
analysed in the presence of large backgrounds from two-quark production, which 
can lead to similar multi-jet topologies via {\it hard QCD} processes. 
\item
The reconstruction of basic event observables such as for instance 
boson masses is intrinsically more difficult in fully hadronic channels 
because of {\it soft QCD} processes, which broaden the jets, create 
ambiguities in assigning the jets, and can also result in cross-talk 
between the produced bosons (if they are short-lived) which may be 
large enough to 
be noticeable in precision measurements such as that of the $W$ mass.
\end{itemize}

This working group on QCD generators has focussed its activity on the first of 
the two items above, dealing mainly with {\it hard QCD} processes. The 
second item (physics and modeling of {\it soft QCD}), has been and still 
is pursued in the framework of the WWMM-2000 (previously called Crete) 
workshop \cite{WWMM}.

The work described here was originally motivated by the desire to assess 
the performance of the various QCD generators used to model QCD backgrounds 
at \leptwo, as well as the expected corresponding theoretical uncertainties.
The point of view taken was that final publications at \leptwo\ should be
based on the best possible Monte Carlo programs, and that we should 
be able to specify corrections when needed, and to
quote uncertainties, in a reliable way, particularly when fully 
satisfactory treatments are not yet available.
 
In addition to serving the \leptwo\ community, the improvements of the 
programs and of the basic understanding also benefits a number of 
other genuine QCD studies.

In the following section the programs available and investigated by the
working group are described by their authors. In the case of standard programs
commonly used in the community, only those aspects relevant to the topics
studied, and the related improvements stimulated by the working group, are
covered. Also several new approaches and options are described.

Then follow five sections where the investigations of the main physics features
considered are reported :

\begin{itemize}
\item
{\it Inclusive (all flavour) jet rates} are not extremely well modelled 
and can result in significant discrepancies, even at \leptwo, when 
four-jet events are selected. The different Monte-Carlo approaches available, 
and the tuning strategies adopted by the different collaborations, are 
compared, and a procedure to extrapolate the uncertainty to \leptwo\
energies, based on the quality of the description achieved at \lepone, 
is outlined.
\item 
{\it Mass effects in 3- and 4-jet rates} were not previously considered 
in detail by the modellers, but are relevant to analyses in 
which $b$-tagging is used as a tool, such as the Higgs searches 
at \leptwo. In addition several features of the modelling result
in uncertainties in basic QCD measurements at \lepone, such as that of the
$b$-quark mass. A consistent method to quantify the theoretical
uncertainty is presented, and the performance of the different Monte-Carlo
programs available, including recent improvements, described.
Additional uncertainties from gluon splitting processes into 
$b {\bar b}$ (see below as well) in the case of the 4-jet rate are
also considered.
\item
Genuine {\it four-jet observables}, particularly angular distributions, are 
not well described by Monte-Carlo programs based on parton 
shower approaches matched to matrix elements at the level of three partons. 
This can result in biases when methods based on topological information
are used to select (or anti-select) the events.
An additional basic motivation for improving the description in this respect 
lies in the use of four-jet events to measure the strong interaction 
coupling constant $\alpha_{\mathrm{S}}$. The emphasis of the work
was to estimate uncertainties, and to evaluate 
new Monte-Carlo programs in which matching of the parton shower approach with
matrix elements is attempted beyond three partons.  
\item
The {\it $b$-quark fragmentation function} is relevant to a number of
topics involving
$b$ quarks, at both \lepone\ and \leptwo\ energies, as it affects for instance 
the lifetime of $B$-hadrons and selection efficiencies of $b$-tagging 
algorithms. Although this topic was not a central one in this working group,
it was felt important to report as much as possible the present status
and recent results on this topic.
\item
Processes involving 
{\it gluon splitting into $b {\bar b}$} are poorly known, both 
theoretically and experimentally, and become more important at \leptwo\ 
energies. Several new options exist in the different Monte-Carlo programs,
which enable one to alter the rate and kinematics of the production.
These are considered in the light both of analytical results and of
measurements at \lepone.
\end{itemize}

The evaluations were based on comparisons of the different Monte-Carlo 
programs, with analytical results when available, and with data at \lepone. 
An effort was made to define dedicated observables enabling meaningful 
comparisons, and to estimate the theoretical uncertainties quantitatively. 
In several cases the calculations, the Monte-Carlo simulations and the 
evaluations of systematic uncertainties were extrapolated to \leptwo\ 
energies as well.
In some cases discrepancies were found between the theoretical 
expectations, the data, and Monte-Carlo results. An attempt to quantify 
such discrepancies was then made, and the results served to stimulate 
improvements by the model builders. Several such improvements were 
actually achieved in the course of the workshop, and evaluations of the 
resulting new Monte-Carlo versions was carried out as well.

In the final section, overall conclusions are presented. Although in some
instances real progress was achieved thanks to this working group,
clearly in many cases still more work and checks are needed.
Such additional investigations and developments are mentioned, 
based on the present knowledge. General recommendations on the use of
the present programs are formulated in each of the relevant contexts.

\subsection{Jet clustering algorithms}
\label{intro_jetalgo}

The jet clustering algorithms used in this report are
those in most common use in $\ee$ experiments: the \jade\ \cite{jade},
\durham\ \cite{durham,CDOTW,BS} and \cambridge\ \cite{camjet} algorithms.
They are used to define the jets at parton level in the theoretical
calculations, and for grouping the selected charged and neutral particles
into jets at the experimental level.

The \jade\ algorithm was the earliest of these and established the method
of successive binary clustering that has been adopted in later algorithms.
For all pairs of final-state particles $(i,j)$, a test variable $y_{ij}$
is defined as indicated in \tab{tab:alg}.
The minimum of all $y_{ij}$ is compared with the
so-called jet resolution parameter, $y_c$ (often called $y_{cut}$).
If it is smaller, the two particles
are recombined into a new pseudo-particle with four-momentum
$p_k=p_i+p_j$.\footnote{Other possible recombination schemes are discussed
in \cite{jade}} The algorithm can be applied again to the new group of
pseudo-particles until all pairs satisfy $y_{ij} > y_c$. The number of jets
in the event is then the number of pseudo-particles one has at the end. In
perturbative theoretical calculations, this procedure leads to
infrared-finite quantities because one excludes the
regions of phase-space that cause trouble. For the same reason,
sensitivity to non-perturbative physics is limited and hadronization
corrections can be estimated from Monte-Carlo models. 

The \jade\ algorithm was nevertheless found to have some unpleasant
theoretical and experimental features, which arise from the fact that its
resolution criterion is approximately one of invariant mass,
$M_{ij}^2\simeq 2E_iE_j(1-\cos\theta_{ij}) > y_cE_{vis}^2$.
This means that particles at widely different angles can be combined
into the same jet, leading to theoretical predictions with large
higher-order corrections that cannot be resummed, and to the
possibility of ``ghost jets'' (jets in directions where no
particles are observed) at the experimental level.

The problems of the \jade\ algorithm are largely alleviated
by replacing the test variable by one that measures the relative
transverse momentum of pairs of particles rather than their
invariant mass. This led to the formulation of the \durham\ algorithm,
the most widely used for \lep\ physics, in which $\min(E_i^2,E_j^2)$
simply replaces $E_iE_j$ in the \jade\ formula (see \tab{tab:alg}).
The resolution criterion then becomes $k_{Ti}^2 > y_cE_{vis}^2$ at small
angles, where $k_{Ti}$ is the transverse momentum of a particle/jet
relative to the direction of any other in the event. 

The \cambridge\ algorithm has been introduced to cure some remaining
defects of the \durham\ algorithm at low values of the jet resolution $y_c$,
with a better understanding of the
processes involving soft gluon radiation, allowing one to explore regions of
smaller $y_c$, where furthermore the experimental error of three-jet ratios is
expected to be smaller. It uses the same recombination procedure and test
variable as \durham\ but with the new ingredients of angular ordering and
{\it soft freezing}. 

The selection of the first pair of particles to be compared with the
resolution parameter is now made according to the ordering variable
$v_{ij}=2(1-\cos\theta_{ij})$ (see \tab{tab:alg}). Then, for the pair of
particles with the smallest $v_{ij}$, one computes $y_{ij}$ and if
$y_{ij}<y_c$ the two particles are recombined. If not, the {\it soft
freezing} mechanism comes into the game by considering the softer particle as
a resolved jet and by bringing back the other one into the binary procedure.
The net effect of the new definition is that NLO corrections to the three-jet
fraction become smaller \cite{nosaltres}. 

In the \durham\ algorithm one can always define a transition value of
$y_c$, $y^{n{\leftarrow}n+1}$, in which an $(n+1)$-jet configuration event
becomes one with $n$ (or fewer) jets. Furthermore, the number of jets is
monotonically decreasing for increasing $y_c$. However, in \cambridge,
this property is lost due to the fact that the sequence of clustering depends
on the external $y_c$ and in some circumstances certain jet topologies are not
present for a specific event. In the case of three jets this affects
${\sim}1\%$ of the  events in the range $y_{c}~{\ge}~0.01$. 

For a more thorough discussion of these and other $\ee$ jet algorithms in
current use, see \cite{Moretti:1998qx}.

\begin{table}[hbt]
\centering
\vspace{7mm}
\begin{tabular}{||cccc||}
\hline\hline
 & & & \\
{\rm Algorithm} & {\rm Resolution} & {\rm Ordering}   & Recombination \\
 & & & \\
\hline
 & & & \\
\jade\ \cite{jade}  &   
$y_{ij}$ = $ {2 \cdot E_i E_j \cdot (1-\cos\theta_{ij})\over 
E_{vis}\sp{2} } $  
& $v_{ij}$ = $y_{ij}$  & $ p_k = p_i + p_j $  \\
 & & & \\
\durham\ \cite{durham,CDOTW,BS}  &   
$y_{ij}$ = $ {2 \cdot {\rm min}(E_i\sp{2},E_j\sp{2}) \cdot (1-\cos\theta_{ij})\over 
E_{vis}\sp{2} } $  
& $v_{ij}$ = $y_{ij}$  & $ p_k = p_i + p_j $  \\
 & & & \\ 
\cambridge\ \cite{camjet}  &   
$y_{ij}$ = $ {2 \cdot {\rm min}(E_i\sp{2},E_j\sp{2}) \cdot (1-\cos\theta_{ij})\over 
E_{vis}\sp{2} } $  

& $v_{ij}$ = $2 \cdot (1-\cos\theta_{ij})$ 
& $p_k = p_i + p_j$  \\
 & & & \\
\hline\hline
\end{tabular}
\vspace{0.5cm}
\caption{Definition of the jet resolution variable $y_{ij}$,
          ordering variable and recombination procedure of the
          \jade, \durham\ and \cambridge\ jet finders. 
          $E_{vis}$ is the total visible energy of the event,
          $p_{i} \equiv (E_{i},\vec{p}_{i})$ denotes a 4-vector
          and $\theta_{ij}$ is the angle between $\vec{p}_{i}$ and
          $\vec{p}_{j}$.}
          \label{tab:alg}
\end{table}

\subsubsection{Jet rates}
\label{intro_jetrates}
Having chosen a jet algorithm one may define the {\em n-jet rate},
$R_n$, by the fraction of hadronic final states that are clustered into
precisely $n$ jets at jet resolution $y_c$: 
\begin{equation}\label{eq:Rnj}
R_n(y_c)  =   \frac{\sigma_n(y_c)}{\sigma_{had}}
\end{equation}
where $\sigma_n$ and $\sigma_{had}$ are the $n$-jet and the total
hadronic cross sections, respectively.  Here we assume that all processes
other than the direct QCD one, $\ee\to Z^0/\gamma^*\to q\bar q\to$ hadrons,
have been eliminated by suitable cuts. For some purposes it will be useful
to define jet rates for a particular primary quark flavour:
\begin{equation}\label{eq:Rqnj}
R^q_n(y_c)  = 
\frac{\sigma_{q\bar q\to n\,jets}(y_c)}{\sigma_{q\bar q\to had}}
\end{equation}
where $q=\ell$, $c$ or $b$, with $\ell$ representing a light ($u,d,s$) quark.

\section{MONTE CARLO GENERATORS}
This Section gives brief descriptions of the main QCD event generators
for two-fermion processes at \leptwo, with emphasis on the features relevant
to multi-jet and $b$-jet fragmentation.

\subsection{PYTHIA}

\pythia\ is a general-purpose generator \cite{ts:pythia}. The current 
version, \pythia\ 6.1, combines and extends the previous generation
of programs, \pythia\ 5.7, \jetset\ 7.4 and {\sc Spythia} \cite{ts:spythia}.
Here we concentrate on those aspects of the program that 
have been modified as a consequence of the current workshop,
or are of specific interest to this working group. Program code, 
manuals and sample main programs are obtainable from
\texttt{http://www.thep.lu.se/}$\sim$\texttt{torbjorn/Pythia.html}.

\subsubsection{Gluon radiation off heavy quarks}\label{sec:pyglhvy}

The \pythia\ final-state shower \cite{ts:finshow} consists of an evolution
in the squared mass $m^2$ of a parton. That is, emissions are ordered
in decreasing mass of the radiating parton, and the Sudakov form
factor is defined as the no-emission rate in the relevant mass range.
Such a choice is not as sophisticated as the angular one in \herwig\ or 
the transverse momentum one in \ariadne, but usually the three tend to 
give similar results. (An exception, where small but significant 
differences were found, is the emission of photons in the shower 
\cite{ts:photonemission}.) One of the advantages is that a mapping 
between the parton-shower and matrix-element variables is rather 
straightforward to $\mathcal{O}(\alpha_{\mathrm{S}})$ for massless 
quarks, and that already the basic shower populates the full phase 
space region very closely the same way as the matrix element. It is 
therefore possible to introduce a simple correction to the shower 
to bring the two into agreement.

The other main variable in the shower is $z$, as used in the 
splitting kernels. It is
defined as the energy fraction in the CM frame of the event. That is,
in a branching $a \to b+c$, $E_b = z E_a$ and $E_c = (1-z) E_a$.
In the original choice of $z$, which is done at the same time as
$m_a$ is selected, the $b$ and $c$ masses are not yet known. A
cut-off scale $Q_0 \approx 1$~GeV is used to constrain the allowed 
phase space, by assigning fictitious $b$ and $c$ masses $\simeq Q_0/2$
so that $a$ can only branch if $m_a > Q_0$, but kinematics is 
constructed as if $b$ and $c$ were massless. At a later stage, when 
$m_b$ and $m_c$ are being selected, possibly well above $Q_0$, the 
previously found $z$ may be incompatible with these. The solution is 
to take into account mass effects by reducing the magnitude of the 
three-momenta $\mathbf{p}_b = - \mathbf{p}_c$ in the rest frame of 
$a$. Expressed in four-momenta in an arbitrary frame, this is 
equivalent to
\begin{eqnarray}
p_b & = & (1 - k_b) p_b^{(0)} + k_c p_c^{(0)} ~, \nonumber\\
p_c & = & (1 - k_c) p_c^{(0)} + k_b p_b^{(0)} ~,
\label{ts:eq:shuffle}
\end{eqnarray}
where $p_b^{(0)}$ and $p_c^{(0)}$ are the original massless momenta 
and $p_b$ and $p_c$ the modified massive ones. The parameters $k_b$ 
and $k_c$ are found from the constraints $p_b^2 = m_b^2$ and 
$p_c^2 = m_c^2$. 

Angular ordering is not automatic, but is implemented by vetoing 
emissions that don't correspond to decreasing opening angles. The 
opening angle of a branching $a \to b+c$ is calculated approximately 
as
\begin{equation}
 \theta \approx \frac{p_{\perp b}}{E_b} + \frac{p_{\perp c}}{E_c}
 \approx \sqrt{z(1-z)} m_a \left( \frac{1}{z E_a} + 
 \frac{1}{(1-z) E_a} \right) 
 = \frac{1}{\sqrt{z(1-z)}} \frac{m_a}{E_a} ~.
\label{ts:eq:angle}
\end{equation}

The procedure thus is the following. In the $\gamma^*/Z^0$ decay,
the two original partons 1 and 2 are produced, back-to-back in
the rest frame of the pair. In a first step, they are evolved 
downwards from a maximal mass equal to the CM energy, with the 
restriction that the two masses together should be below this CM 
energy. When the two branchings are found, they define $m_1$ and 
$m_2$ and the $z$ values of $1 \to 3+4$ and $2 \to 5+6$. These 
latter branchings obviously have smaller opening angles than the 
$180^{\circ}$ one between 1 and 2, so no angular-ordering constraints
appear here. The matching procedure to the matrix element is used
to correct the branchings, however, as will be described below.
In subsequent steps, a pair of partons like 3 and 4 are evolved in
parallel, from maximum masses given by the smaller of the mother (1)
mass and the respective daughter (3 or 4) energy. Here angular ordering 
restricts the region of allowed $z$ values in their branchings,
but there are no matrix-element corrections. Once $m_3$ and $m_4$ are 
fixed, the kinematics of the $1 \to 3+4$ branching needs to be 
modified according to eq.~(\ref{ts:eq:shuffle}).

Let us now compare the parton-shower (PS) population of three-jet 
phase space with the matrix-element (ME) one. With the conventional 
numbering $q(1) \bar q(2) g(3)$, and $x_j = 2 E_j / E_{CM}$, the 
matrix element is of the form
\begin{equation}
\frac{1}{\sigma_0} \frac{\mathrm{d}\sigma_{\mathrm{ME}}}%
{\mathrm{d}x_1\mathrm{d}x_2} = \frac{\alpha_{\mathrm{S}}}{2\pi} \, 
\frac{4}{3} \, \frac{M(x_1,x_2,r_q)}{(1-x_1)(1-x_2)} ~.
\end{equation}
For massless quarks
\begin{equation}
M(x_1,x_2,0) = x_1^2 + x_2^2 ~,
\end{equation}
while for massive ones
\begin{equation}
M \left( x_1,x_2,r_q = \frac{m_q^2}{E_{CM}^2} \right) = 
x_1^2 + x_2^2 - 4 r_q x_3 - 8 r_q^2 - (2 r_q + 4 r_q^2) 
\left( \frac{1-x_2}{1-x_1} + \frac{1-x_1}{1-x_2} \right) ~. 
\end{equation}

There are two shower histories that could give a three-jet
event. One is $\gamma^*/Z^0(0) \to q(i) \bar q(2) \to
q(1) \bar q(2) g(3)$, i.e. with an intermediate ($i$) quark 
branching $q(i) \to q(1) g(3)$. For massless quarks this gives
\begin{eqnarray}
Q^2 & = & m_i^2 = (p_0 - p_2)^2 = (1-x_2) E_{CM}^2  ~,  \\
z & = & \frac{p_0 p_1}{p_0 p_i} = \frac{E_1}{E_i} = 
\frac{x_1}{x_1 + x_3} = \frac{x_1}{2 - x_2}  ~, \\
& \Rightarrow & \frac{\mathrm{d}Q^2}{Q^2} \, \mathrm{d}z = 
\frac{\mathrm{d}x_2}{1-x_2} \, \frac{\mathrm{d}x_1}{2-x_2} ~.
\label{ts:eq:jacobian}
\end{eqnarray}
The parton-shower probability for such a branching is
\begin{equation}
\frac{\alpha_{\mathrm{S}}}{2\pi} \, \frac{4}{3} \, \frac{1+z^2}{1-z} 
\mathrm{d}z \frac{\mathrm{d}Q^2}{Q^2} =
\frac{\alpha_{\mathrm{S}}}{2\pi} \, \frac{4}{3} \,
\frac{1-x_1}{x_3} \left[ 1 + \left( \frac{x_1}{2-x_2} \right)^2 
\right] \frac{\mathrm{d}x_1\mathrm{d}x_2}{(1-x_1)(1-x_2)} ~.
\end{equation}
There also is a second history, where the r\^oles of $q$ and
$\bar q$ are interchanged, i.e. $x_1 \leftrightarrow x_2$.
(On the Feynman diagram level, this is the same set as for 
the matrix element, except that the shower does not include any
interference between the two diagrams.) Adding the two,
one arrives at a form
\begin{equation}
\frac{1}{\sigma_0} \frac{\mathrm{d}\sigma_{\mathrm{PS}}}%
{\mathrm{d}x_1\mathrm{d}x_2} = \frac{\alpha_{\mathrm{S}}}{2\pi} \, 
\frac{4}{3} \, \frac{S(x_1,x_2,r_q)}{(1-x_1)(1-x_2)} ~,
\end{equation}
with 
\begin{equation}
S(x_1,x_2,0) = 1 + 
\frac{1-x_1}{x_3} \left( \frac{x_1}{2-x_2} \right)^2 +
\frac{1-x_2}{x_3} \left( \frac{x_2}{2-x_1} \right)^2 ~.
\end{equation}

In spite of the apparent complexity of $S(x_1,x_2,0)$ relative to
$M(x_1,x_2,0)$, it turns out that $S(x_1,x_2,0) \approx M(x_1,x_2,0)$
everywhere but also that $S(x_1,x_2,0) > M(x_1,x_2,0)$. It is 
therefore straightforward and efficient to use the ratio 
\begin{equation}
\frac{\mathrm{d}\sigma_{\mathrm{ME}}}{\mathrm{d}\sigma_{\mathrm{PS}}}
= \frac{M(x_1,x_2,0)}{S(x_1,x_2,0)}
\end{equation}
as an acceptance factor inside the shower evolution, in order 
to correct the first emission of the quark and antiquark to
give a sum in agreement with the matrix element. 

Clearly, the shower will contain further branchings that 
modify the simple result, e.g. by the emission both from 
the $q$ and the $\bar q$, but these effects are formally of 
$\mathcal{O}(\alpha_{\mathrm{S}}^2)$ and thus beyond the accuracy 
we strive to match. One should also note that the shower modifies
the distribution in three-jet phase space by the appearance
of Sudakov form factors, and by using a running 
$\alpha_{\mathrm{S}} (p_{\perp}^2)$ rather than a fixed one. In 
both these respects, however, the shower should be an improvement 
over the fixed-order result.

The prescription of correcting the first branchings by a factor
$M(x_1,x_2,0)/S(x_1,x_2,0)$ was the original one, used up
until \jetset\ 7.3. In 7.4 an intermediate ``improvement'' was
introduced, in that masses were used in the matrix-element
numerator, i.e. an acceptance factor $M(x_1,x_2,r_q)/S(x_1,x_2,0)$.
(The older behaviour remained as an option.)
The experimental problems found with this procedure has
prompted new studies as part of this workshop. Starting with
\pythia\ 6.130, therefore also masses have been introduced in 
the shower expression, i.e. an acceptance factor
$M(x_1,x_2,r_q)/S(x_1,x_2,r_q)$ is now used.

In the derivation $S(x_1,x_2,r_q)$, one can start from the ansatz
\begin{eqnarray}
x_2 & = & 1 - \frac{m_i^2 - m_q^2}{E_{CM}^2} ~,\nonumber\\
x_1 & = & \left( 1 + \frac{m_i^2 - m_q^2}{E_{CM}^2} \right)
 \left( (1-k_1) z + k_3(1-z) \right) ~, \\
x_3 & = & \left( 1 + \frac{m_i^2 - m_q^2}{E_{CM}^2} \right)
 \left( (1-k_3) (1-z) + k_1 z \right) ~. \nonumber
\end{eqnarray}
The quark mass enters both in the energy splitting
between the intermediate quark $i$ and the antiquark 2,
and in the correction procedure of eq.~(\ref{ts:eq:shuffle})
for the sharing of energy in the branching $q(i) \to q(1) g(3)$.
The constraints $p_1^2 = m_q^2$ and $p_3^2 =0$ give
$k_1 = 0$ and $k_3 = m_q^2/m_i^2$. One then obtains
\begin{eqnarray}
Q^2 & = & m_i^2 = (1-x_2 + r_q) E_{CM}^2  ~,
\nonumber\\
z & = & \frac{1}{2 - x_2} \left( x_1 - r_q 
\frac{2 -x_1 -x_2}{1 - x_2} \right) ~.
\end{eqnarray}
By a fortuitous cancellation of mass terms, 
$\mathrm{d}Q^2/Q^2 \, \mathrm{d}z$ is the same as in
eq.~(\ref{ts:eq:jacobian}), but the $(1+z^2)/(1-z)$ factor 
is no longer simple. Therefore one obtains
\begin{equation}
S(x_1,x_2,r_q) = \frac{1-x_1}{x_3} \frac{1-x_2}{1 - x_2 + r_q} 
\left[ 1 + \frac{1}{(2-x_2)^2} \left( x_1 - r_q \frac{x_3}{1-x_2}
\right)^2 \right] + \left\{ x_1 \leftrightarrow x_2 \right\} ~,
\end{equation}
where the second term comes from the graph where the antiquark 
radiates. 

\begin{figure}
\begin{center}
\includegraphics[width=12cm]{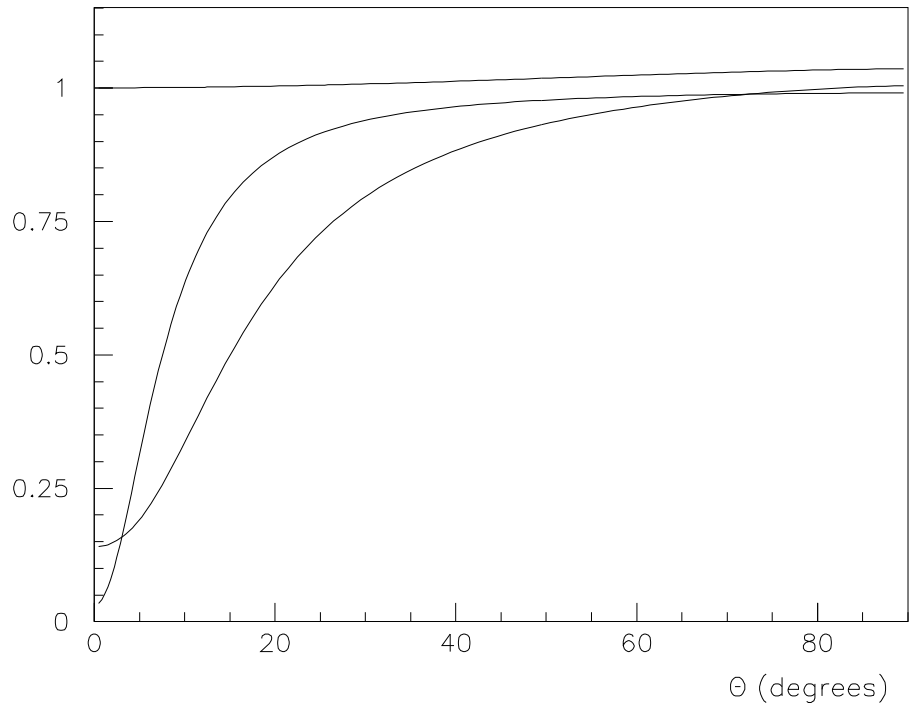}
\caption{The gluon emission rate as a function of emission angle,
for a 10~GeV gluon energy at $E_{CM} = 91$~GeV, and with 
$m_b = 4.8$~GeV. All curves are normalized to the massless 
matrix-element expression. Dashed (upper): massless parton shower, i.e. 
$S(x_1,x_2,0)/M(x_1,x_2,0)$. Dash-dotted (middle): massive matrix element, 
i.e. $M(x_1,x_2,r_q)/M(x_1,x_2,0)$. Full (lower): massive parton shower, 
i.e. $S(x_1,x_2,r_q)/M(x_1,x_2,0)$.}
\label{ts:fig:suprmass}
\end{center}
\end{figure}

The mass effects go in the ``right'' direction,
$S(x_1,x_2,r_q) < S(x_1,x_2,0)$, but actually so much so that
$S(x_1,x_2,r_q) < M(x_1,x_2,r_q)$ in major regions of phase space.
This is illustrated in \fig{ts:fig:suprmass}. The dashed curve 
here shows how well the PS and ME expressions agree in the
massless case. The dash-dotted one is the well-known ``dead cone effect''
in the matrix element \cite{deadcone}, and the full the corresponding 
suppression in the shower. Very crudely, one could say that the
massive shower exaggerates the angle of the dead cone by about 
a factor of two (in this rather typical example).

Thus the amount of gluon emission off massive quarks is 
underestimated already in the original prescription, where masses 
entered in the kinematics but not in the ME/PS correction factor. 
If instead the ratio $M(x_1,x_2,r_q)/S(x_1,x_2,0)$ is applied, 
the net result is a distribution even more off from the correct
one, by a factor $S(x_1,x_2,r_q)/S(x_1,x_2,0)$. Thus it would
have been better not to introduce the mass correction in \jetset\ 7.4.

Armed with our new knowledge, we can now instead use the correct 
factor, namely the ratio
$M(x_1,x_2,r_q)/S(x_1,x_2,r_q)$. A technical problem is that
this ratio can exceed unity, in the example of 
\fig{ts:fig:suprmass} by up to almost a factor of two. This could
be solved e.g. by enhancing the raw rate of emissions by this 
factor. However, another trick was applied, based on the fact that
the accessible $z$ range is smaller for a massive quark than a 
massless one. Therefore, without any loss of phase space, $z$ can be 
rescaled to a $z'$ according to
\begin{equation}
(1-z') = (1-z)^k ~, ~~~\mathrm{with}~~~ 
k = \frac{\ln(m_q^2/E_{CM}^2)}{\ln(Q_0^2/E_{CM}^2)} < 1 ~.
\end{equation}
The ME/PS correction factor then has to be compensated by $k$,
and thereby comes below unity almost everywhere --- the remaining
weighting errors are too small to be relevant.

In \sect{sec:R34comp} of this report it is shown that the corrected procedure 
now does a good job of describing mass effects in the amount of
three-jet events. Problems still remain in the four-jet sector,
however, where the emission off heavy quarks is reduced more in
\pythia\ than in the data. These four-jets come in several categories
in the Monte Carlo simulation. If one resolved gluon is emitted 
from the quark and another from the antiquark, or if a gluon branches
into two resolved partons, the mass effects should now be included.
If the quark emits both resolved gluons, however, the second
emission involves no correction procedure. Instead the dead
cone effect is exaggerated, similarly to what was shown in 
\fig{ts:fig:suprmass}. That might then explain the 
discrepancies noted above.

The intention is to find an alternative algorithm that better can
take into account mass effects at all steps of the shower. 
For instance, if the evolution is performed in terms of the
variable $Q^2 = m^2 - m_q^2$ rather than $Q^2 = m^2$, then the
dead-cone effect is underestimated rather than overestimated. 
A suppression factor could therefore be implemented to correct 
down to the desired level. The technical details have yet to be 
worked out.

\subsubsection{The total four-jet rate}

The above modifications partly address the four-jet rate off heavy 
quarks relative to light quarks, but not the shortfall in the 
overall four-jet rate in \pythia\ relative to the data. Currently the 
matrix-element correction procedure is used in the first branching
of both sides of the event, i.e. both the quark and the antiquark
ones. Thus not only the three-jet but also the four-jet rate is
affected. If the correction procedure is only used on the side with
the harder emission, here defined as the one occuring at the 
largest mass, one might hope to increase the four-jet rate
relative to the three-jet one. This possibility was studied, 
for simplicity only for massless quarks. The result was 
disappointing, however. To the extent that the four-jet rate is
at all changed, it is below the 1\% level. In retrospect, this is
maybe not so surprising, considering how close the matrix-element 
correction factor is to unity, cf. \fig{ts:fig:suprmass}.
A solution to the four-jet rate problem therefore remains to be found.

\subsubsection{Gluon splitting to heavy quarks}

A few new options have been included in \pythia, that allow
studies of the gluon splitting rate under varying assumptions.
These developments are described in \sect{ts:sec:gsplit}.

\subsubsection{Fragmentation of low-mass strings}
\label{ts:sec:stringfrag}

The Lund string fragmentation algorithm \cite{ts:stringfrag} has
remained essentially unchanged over the years, and generally
does a good job of describing data. Some improvements have 
recently been made (in \pythia\ 6.135 onwards) in the description
of low-mass strings \cite{ts:clusfrag}, however. 

Whereas gluon emission only adds kinks on the string stretched 
between a quark end and an antiquark one, a gluon splitting 
$g \to q \bar q$ splits an existing string into two. In this
process, one of the new strings can obtain a small invariant mass,
so that it can only produce one or two primary hadrons. Such 
a low-mass system is called a cluster, and is handled separately
from ordinary strings. If only one hadron is produced, 
``cluster collapse'', its flavour is completely specified by 
the string endpoints. 

In fixed-target $\pi p$ collisions, strings are often stretched 
between a produced central charm quark and a beam remnant
antiquark or diquark. Thus the cluster collapse mechanism
favours the production of charm hadrons that share a valence
flavour content with the incoming beam particles. This was 
predicted in \pythia, but the measurements have shown that
production asymmetries are smaller in data than in the model. 
The new data have therefore been used to tune some aspects of 
the cluster treatment, and some other improvements were included 
at the same time. The ones relevant for $e^+e^-$ physics are 
summarized below.

The quark masses assigned to ``on-shell'' quarks, e.g. in the
event listing, have been changed to $m_u = m_d = 0.33$~GeV,
$m_s = 0.5$~GeV, $m_c = 1.5$~GeV and $m_b = 4.8$~GeV. In previous
program versions, lower ``current-algebra'' masses were used to comply
with requirements e.g. for Higgs physics, but these latter needs
are now covered by the new running-mass function \texttt{PYMRUN}.
The change in masses has consequences in several places, e.g. 
for the rate of $g \to q \bar q$ branchings. In this Section, 
the main point is the change in the string mass spectrum, 
and thereby in the fate of strings. For a string $q_1 \bar q_2$, 
the cluster treatment is applied whenever 
$m(q_1 \bar q_2) < m(q_1) + m(\bar q_2) + 1$~GeV,
while the normal string routine is used above that.

A cluster can produce either one or two primary hadrons. The
choice is made dynamically, as follows. The cluster is assumed 
to break into two hadrons $h_1 = q_1 \bar q_3$ and 
$h_2 = q_3 \bar q_2$ by the production of a new $q_3 \bar q_3$ 
pair. The composition of the new flavour and the spin multiplet
assignment of the hadrons is determined by standard string 
fragmentation parameters. If $m(h_1) + m(h_2) < m(q_1 \bar q_2)$,
an allowed two-body decay of the cluster has been found. Even in
case of failure, a subsequent new try might succeed, with another
$q_3$ or another spin assignment. Therefore a very large number
of tries would make each cluster decay to two hadrons if at all 
possible, while only one try gives a more gradual transition 
between one and two hadrons as the various two-body thresholds are
passed. As a compromize between the extremes, up to two tries are 
made. If neither succeeds, the cluster collapses to one hadron  

In a cluster collapse, it is not possible to conserve
energy and momentum within the cluster. Instead other parts of the
events have to receive or donate energy to put the hadron on mass 
shell. The algorithm handling this has now been made more physically 
appealing, by performing the shuffling to/from the parts of the
event that are most closely moving in the same general direction
as the collapsing cluster. The technical details \cite{ts:clusfrag}
are not described here, but one may note that differences are small
relative to the previous simpler algorithm (still available as an 
option and as a last resort, should the more sophisticated one fail
to find a sensible solution).

The treatment of a two-body cluster decay has been improved to provide
a smoother match to the string description in the overlapping
mass region. At a first step, the cluster decay is isotropic. 
The decay is accepted with a weight $\exp(-p_{\perp}^2/2\sigma^2)$,
where the $p_{\perp}$ is defined relative to the $q_1 \bar q_2$
axis in the cluster rest frame. This agrees with the standard 
Gaussian string fragmentation $p_{\perp}$ spectrum well above 
threshold, but reverts to isotropic decay near the threshold. Even 
with $p_{\perp}$ fixed, two ``mirror'' solutions exist for the 
longitudinal momenta of the hadrons. The relative probabilities 
are well-defined in the string model, and are here used to make the
choice. Near threshold both are equally likely, while further above 
threshold the  $q_1 \bar q_3$ hadron is preferentially moving in the 
$q_1$ direction and vice versa. 

\subsubsection{A shower interface to four-jet events (massless ME)}
\label{sec:py4j}
A few years ago, an algorithm was developed to allow the \pythia\ shower
to start from a given four-jet configuration, $q \bar q g g$ or 
$q \bar q q' \bar q'$ \cite{ts:showmatch}. This was intended to 
allow comparisons e.g. of four-jet topologies between matrix-element
calculations and data, with a realistic account of showering
and hadronization effects not covered by the matrix-element
calculations. The standard \pythia\ shower does not do this well,
since it does not include any matching procedure to four-jet
matrix elements and therefore does not do e.g. the azimuthal angles
in branchings fully correctly. 

A problem is that the standard shower routine is really set up
only to handle systems of two showering partons, not three or more.
(Actually an option does exist for three, but it is primitive and
hardly used by anybody.) The trick \cite{ts:showmatch} therefore 
is to try to guess the ``prehistory'' of shower branchings that 
gave the specified four-parton configuration, and thereafter to 
run a normal shower starting from two partons. Here two of the 
subsequent branchings already have their kinematics defined, 
while the rest are chosen freely as in a normal shower. Benefits 
of having a prehistory include (\textit{i}) the availability of the
standard machinery to take into account recoils when masses are
assigned to partons massless in the matrix elements,  (\textit{ii})
a knowledge  of angular-ordering constraints on subsequent emissions 
and azimuthal anisotropies in them, and  (\textit{iii}) information 
on the colour flow as required for the subsequent string description.

The choice among possible shower histories is based on a weight
obtained from the mass poles and splitting kernels. As an example,
consider a $q(1) \bar q(2) g(3) g(4)$ configuration, which could
come e.g. from an initial $q(i) \bar q(2)$ configuration followed
by branchings $q(i) \to q(1) g(j)$ and $g(j) \to g(3) g(4)$.
The relative weight is then
\begin{equation}
\mathcal{P} = \mathcal{P}_{i \to 1j}\mathcal{P}_{j \to 34} =
\frac{1}{m_i^2} \, \frac{4}{3} \, 
\frac{1 + z_{i \to 1j}^2}{1 -z_{i \to 1j}} \cdot 
\frac{1}{m_j^2} \, 3 \, \frac{(1 - z_{j \to 34}%
(1 - z_{j \to 34}))^2}{z_{j \to 34}(1 - z_{j \to 34})} ~.
\end{equation}
Of course, one could imagine including further information,
e.g. on azimuthal angles or on a scale-dependent 
$\as$.

The original routines were not set up to handle massive quarks,
e.g. to correct the $z$ definition for the rescaling of 
eq.~(\ref{ts:eq:shuffle}). This has now been included, and also 
the interface has been simplified. The re-implementation
originally contained a bug, that was fixed in \pythia\ 6.137. 

Users can now \texttt{CALL PY4JET(PMAX,IRAD,ICOM)} to shower and
fragment a four-parton configuration. If \texttt{ICOM} is 0 or 1
the configuration is picked up either from the \texttt{HEPEVT}
or the \texttt{PYJETS} commonblock. The partons have to be stored 
in the order $q \bar q g g$ or $q \bar q q' \bar q'$, where 
$q' \bar q'$ is assumed to be the secondary quark pair. 
(Interference terms make the primary/secondary pair distinction
nontrivial in a matrix element, but pragmatic recipes should 
work well.) Initial-state photons can be interspersed anywhere
in the given initial state, and final-state photon radiation in 
the shower is off or on for \texttt{IRAD} 0 or 1. \texttt{PMAX}
sets the maximum mass scale allowed in the shower. In an exclusive
description, i.e. where one wants four-jet only and not five or
more jets, the logical choice would be to put \texttt{PMAX}
equal to the mass cutoff applied to the matrix elements. An 
inclusive picture, where all emissions are allowed below the 
lowest mass scale of the reconstructed shower, is obtained for
\texttt{PMAX}$ = 0$ (or, more precisely, \texttt{PMAX}$ < Q_0$).  

\subsubsection{Interfacing 4 parton LO massive ME: \fourjphact.}\label{sec:4jphact}


As already explained in the preceding Sections, complete matrix elements 
calculations are expected to give a good description of multijet events
when large separations among jets are involved and in particular when
 angular variables are considered. On the other hand, pure ME differential
cross sections lack PS and hadronization and cannot 
reproduce collinear and soft
radiation. It is therefore important to
have the possibility to start with pure ME calculations and complement them
with these additional features. The results obtained in this way 
(ME + PS + hadronization) can be compared with 
pure parton level ones  as well as with those from dedicated QCD MC's.

If one takes for example topologies with four or more jets, one expects that
a reasonable description for not too small values of the 
jet resolution $y_{cut}$ may be obtained 
starting with four jet ME at a much lower $y_{cut}$ and adding to it PS and 
hadronization.
One must however be aware of the fact that when starting with four parton
ME, all events  described by two or three parton ME + PS + hadronization are 
not taken into  account.
In this respect QCD MC's, like \herwig\ or \pythia, surely give a more complete 
description, as they  start PS from two parton
ME and match 3 parton production with the respective ME results.
 The above mentioned approach
of starting from 4 parton ME can however be considered
 as a complementary approach for some studies and a way to check MC 
results when for instance  angular variables or mass effects are involved.  

\fourjphact\ is a Fortran code which has been written to provide a tool for this 
kind of studies and comparisons. 
It computes exact LO {\em massive} ME for all 
{$e^+e^-\ar q\bar q q'\bar q'$} and {$e^+e^-\ar q\bar q  g g$} final states and
it interfaces them with the \pythia\ routine {\tt PY4JET} described in the 
preceding Section. It can   therefore be used to compute  total or 
differential four jet cross sections at parton level or  
to study fully hadronic events initiated by 4 partons. 

The program, together with instructions and examples, can be found in\\
{\tt http://www.to.infn.it/}$\sim${\tt ballestr/qcd/} .

Here we limit ourselves to a brief description of the main features of the 
program.
 
\fourjphact\ computes all $ee\ar 4q$ ME's with the method of  ref.~\cite{noi}
while  for $ee\ar 2q2g$ it makes use of the routine of ref~\cite{bmm}.
Numerical  integration over phase space is performed with VEGAS~\cite{veg}.
Unweighted event generation and  distributions at parton level
are implemented as in the four fermion program WPHACT \cite{wph}. 
Initial State Radiation is included, when requested,   via the  structure
function approach~\cite{sfa}.

When using the program, one starts by computing some cross section.
Unweighted events may be generated during this step, or in a second run
in order to obtain a predetermined  number of events.
These may be passed to \pythia\ which provides PS and hadronization.

In the cross section computation one may choose between fixed or running
$\as(M)$. In the second case, the scale $M$ for $4q$ diagrams is chosen
to be the invariant mass of the gluon propagator, while for $2q2g$ the
invariant mass of the two gluons is used.

An inventory of cuts at parton level are already defined in \fourjphact:
to implement them one has only to specify the numerical 
values for  minima and maxima of   energies, transverse momenta,
 angles among partons and invariant masses.
\jade\ or \durham\ or \cambridge\ $y_{cut}$ at parton level can be requested
in a similar way. Any other cut  can be easily defined 
 in an include file.  It must be noticed in this connection that massive LO ME
for $2q2g$ cannot be computed without any cut or $y_{cut}$. 
$4q$ final states can  in principle be computed   without any cut, as 
quark masses are exactly accounted for. It is however wiser  to 
use also in this case  realistic cuts, in order to avoid 
regions which are computationally demanding and of dubious physical 
interpretation at this level of approximation

Parton level distributions can be easily defined in the include file. 
Corresponding values for each bin will be 
given after cross section computation in output .dat file.
This feature might be useful when one wishes  to compare 
 partonic  distributions with hadron level ones obtained after 
 the call to {\tt PY4JET}.

\fourjphact\ can compute or generate events for one final state at a time
( eg. $u\bar u g g$ or $b\bar b c \bar c$), or for all 20 final states with 
quarks (not top) and gluons  at the same time. In this last case, the 
corresponding probability of every channel is determined or read from a file,
and the generated events  will have the correct fraction of all final states.   
This ``one shot'' option is often used when hadronization is required.

In the call to  {\tt PY4JET(PMAX,IRAD,ICOM)} the parameters {\tt PMAX, IRAD,
ICOM} are set respectively to 0.d0, 0, 0 in a data statement. Their meaning
is explained in the previous Section and they can  of course be changed if 
needed. 

The partons have to be stored in the proper order before the call to  
{\tt PY4JET}: this is unambiguous for $q\bar q gg$ while for 
$4q$ one has somehow to  decide which of the two $q\bar q$ pair 
corresponds to the secondary emission. Such a distinction between first
and second pair is not well defined in the case
of ME.  As the highest
contribution comes, event by event, from the diagrams which have the lower 
$q\bar q$ invariant mass as secondary emission, we choose this configuration
for giving the proper order to quarks. This we do also in the case of two 
identical  flavours (e.g. $u\bar u u \bar u$).

Examples of results obtained with \fourjphact+\pythia\ and comparisons
with other methods can be found in this report in \sect{sec:4jMZ},
\sect{subsec:FK_masscorr} and \sect{subsec:GD_pscomp}.

\subsection{HERWIG}

Like \pythia, \herwig\ \cite{Marchesini:1992ch} is a general-purpose event
generator which uses parton showering to simulate higher-order QCD effects.
The main differences are the variables used in the parton showers,
which are chosen to simplify the treatment of soft gluon coherence,
and the hadronization model, which is based on cluster rather than string
fragmentation. The current version, described here, is \herwig\ 6.1
\cite{Corcella:1999qn}. The program and documentation are available at\\
{\tt http://hepwww.rl.ac.uk/theory/seymour/herwig/}

\subsubsection{Parton showers}\label{sec_hwshower}

The \herwig\ parton shower evolution is done in terms of the parton energy
fraction $z$ and an angular variable $\xi$.  In the parton splitting $i\to jk$,
$z_j = E_j/E_i$  and $\xi_{jk} = 2 (p_j\cdot p_k)/(E_j E_k)$.  Thus
$\xi_{jk}\simeq\theta_{jk}^2$ for massless partons at small angles.

The values of $z$ are chosen according to the relevant DGLAP splitting
functions and the distribution of $\xi$'s is determined by the Sudakov form
factors. See e.g.\ \cite{Ellis:1996qj} for
technical details. Coherence of soft gluon emission is simulated by
angular ordering: each $\xi$ value must be smaller than the one for the
previous branching of the parent parton.

The initial conditions for each parton cascade are determined by the
configuration and colour structure of the primary hard process.  The
initial value of $\xi$ for the showering of parton $j$ is $\xi_{jk}$
where $k$ is
the parton that is colour-connected to $j$. For example, in
$\ee\to q\bar q g$ the gluon has a colour that is connected to
the antiquark and an anticolour
connected to the quark.  Therefore the initial angle for the quark
jet is the angle between the quark and the gluon.  For the gluon jet,
the initial angle is either the gluon-quark or gluon-antiquark angle,
with equal probability.

In general, the hard process may involve several possible colour flows,
which are unique and distinct only in the limit of an infinite number of
colours, $N_c\to\infty$.  For example in $\ee\to q\bar q g_1 g_2$ either
gluon 1 or gluon 2 may be connected to the quark.  In the limit
$N_c\to\infty$ these colour flows have distinct matrix elements-squared,
$|\cM_1|^2$ and $|\cM_2|^2$.  In \herwig\ colour flow 1 is chosen with probability
$|\cM_1|^2/(|\cM_1|^2+|\cM_2|^2)$ and flow 2 with probability
$|\cM_2|^2/(|\cM_1|^2+|\cM_2|^2)$, after using the full ($N_c=3$)
matrix element to generate the momentum configuration.
In this approximation, each final state has a unique colour flow
which tells us how to limit the angles in each parton shower.

The parton showers are terminated as follows. For partons of mass $m_i$
there is a cutoff of the form $Q_i = m_i + Q_0$, and showering from any
parton stops when a value of $\xi$ below $Q_i^2/E_i^2$ is selected for the next
branching. The condition $\xi > Q_i^2/E_i^2$ corresponds to the ``dead cone"
for heavy quarks \cite{deadcone}.
Then the parton is put on mass-shell, or given a small
non-zero effective mass in the case of gluons. Working backwards from these
on-shell partons, one can now construct the virtual masses of all the internal
lines of the shower, and the overall jet mass, from the energies and opening
angles of the branchings. Finally one can assign the azimuthal angles of the
branchings, including EPR-type correlations, and deduce all the 4-momenta
in the shower.

Next the parton showers are used to replace the (on mass-shell) partons
that were generated in the original hard process. This is done in such a
way that the jet 3-momenta have the same directions as the original partons in
the c.m.\ frame of the hard process, but they are boosted to conserve
4-momentum taking into account their extra masses.

We see that combining any tree-level hard process matrix element
with parton showers is quite straightforward in \herwig.  Double-counting
is avoided, or at least suppressed, by angular ordering, which limits the
showers to cones defined by the hard process and its colour structure.
The price for this simplicity is that one must know both the overall
($N_c=3$) matrix element-squared and the separate ones ($|\cM_1|^2$ etc) for all
the possible colour flows in the limit $N_c\to\infty$.

One must bear in mind that results from combined matrix elements and parton
showers are only likely to make sense if all the energy scales in the hard
process being modelled by the matrix element are bigger, or at least not much
smaller, than those in the parton showers.  Otherwise, the structure of the
final state
will be determined mainly by the showers and the details of the matrix element
become irrelevant.  This is ensured in \herwig\ by a variable {\tt EMSCA}, set by
the hard process subroutine, which acts as an upper limit on the relative
transverse momentum of any branching in the associated parton showers.  For
example, in the $\ee\to$ 4-jets matrix element option, discussed below
in \sect{sec_hw4jet},
{\tt EMSCA} is (the square root of) the smallest of the invariant
quantities $s_{ij}=2p_i\cdot p_j$ for the 4 partons generated
in the hard process.

While the above procedure of attaching parton showers to a hard process
generated by a tree-level matrix element may be straightforward, the
problem of matching matrix elements and showers beyond tree level is
certainly not. So far, this has only been done up to order $\as$
in \herwig\ (as in \jetset), for a limited class of processes including
$\ee\to q\bar q (g)$.  In \herwig\ the problem separates into two
parts.  First (``hard" matrix element corrections) there is a region of
phase space that $\ee\to q\bar q$ + parton showers does not populate
at all to order $\as$.  That region can easily be filled by generating
a gluon according to the matrix element. Second, there are the (``soft")
matrix element corrections that have to be applied inside the parton showers.
As shown in Ref.~\cite{Seymour:1995df},  the right way
to do this is to apply a correction not only to the first branching in
each shower but also to every branching that is the ``hardest so far". This
is especially important in \herwig\ where the evolution in $\xi$ means that
several relatively soft (i.e.\ low $p_t$) wide-angle branchings can precede
a harder one with a smaller angle.

To provide full matrix-element matching for 2-, 3- and 4-jets would mean
extending the above procedure to next-to-next-to-leading order. There will
be unpopulated regions of 4-parton phase space to be filled using the
hard 4-jet matrix element, and ``semi-hard" regions in which the 3-jet
matrix element should be used in combination with order $\as$ ``soft"
corrections within a shower.  However the bulk of the cross section
will be in regions where order $\as^2$ corrections within the showers
must be computed and applied -- a daunting prospect.

One may, however, implement a less ambitious procedure for `combining'
2, 3 and 4-jets so as to describe multijet distributions to leading
order, which is discussed in \sect{sec:hw234jet}.

\subsubsection{Hadronization}

Hadronization in \herwig\ is done using a cluster model. First of all,
any ``on mass-shell" gluons at the ends of the parton showers are split
into light quark-antiquark pairs.  As mentioned above, a unique colour
flow is generated for each final state, so that each final-state quark
is uniquely colour-connected to an antiquark and vice-versa.  These
connected pairs can therefore form colour-singlet clusters carrying the
combined flavour and 4-momentum of the pair.  In the simplest case these
clusters decay directly into pairs of hadrons according to the density of
states for possible pairs of the right flavour.  The transverse momentum
$\sim 300$ MeV generated in hadronization is a reflection of the typical
momentum release in cluster decay, which is determined by the cutoff
$Q_0$, the quark masses and the QCD intrinsic scale $\Lambda$. 

If a cluster is too light to decay into two hadrons, it is converted into
a single hadron of that flavour by donating some 4-momentum to a
neighbouring cluster.  If its mass is above a flavour-dependent
value set by the parameter {\tt CLMAX} (default value 3.35 GeV),
$$M_{jk} > [{\tt CLMAX}^p+(Q_j+Q_k)^p]^{1/p}$$
where the power $p$ is given by a parameter {\tt CLPOW} (default 2.0),
it is split collinearly into two lighter clusters.  A further parameter
{\tt PSPLT} (default 1.0) specifies the mass distribution of the resulting
lighter clusters, which is taken to be proportional to $M^{\tt PSPLT}$.

The cluster mass spectrum falls rapidly at high masses and its peak lies
below the threshold for cluster splitting. One can show that these features
are asymptotically independent of the energy scale of the hard process.
However, there is always a finite probability of producing a very massive
cluster.  In this case sequential collinear splitting is invoked,
leading to string-like hadronization.

\subsubsection{$b$-jet fragmentation}\label{sec:hwbfrag}

We concentrate here on primary $b$-quark showering and
hadronization, leaving discussion of gluon $\to b \bar b$ to
\sect{sec:hwgbbar}

The main point to note in connection with $b$-quark
showering is the treatment of quark masses in \herwig\
parton showers.  In the basic algorithm, the quantity $m_i$ appears only in
the shower cutoff $Q_i=m_i+Q_0$, but this affects the distributions of $\xi$
and $z$ throughout the shower via the constraint
$$Q_j/(E_i\sqrt{\xi_{jk}}) < z_j < 1-Q_k/(E_i\sqrt{\xi_{jk}})$$
at each branching $i\to jk$.  Since this is always a low-energy cutoff it
seems clear that the relevant value of $m_i$ is the pole or constituent mass.
On the other hand a running mass might well be more appropriate in evaluating
the hard process matrix element and the corresponding matrix element
corrections.

In the process of $b$-quark hadronization, the input value of $m_b$ clearly
affects the fraction of $b$-flavoured clusters that become a single B
meson, the fractions that decay into a B meson and another meson, or into
a B baryon and an antibaryon, and the fraction that are split into
more clusters. Thus the properties of $b$-jets depend on the parameters
$m_b$, {\tt CLMAX}, {\tt CLPOW} and {\tt PSPLT} in a rather complicated way.

In practice the parameters {\tt CLMAX}, {\tt CLPOW} and {\tt PSPLT}
are tuned to global
final-state properties and one needs extra parameters to describe $b$-jets.
A parameter {\tt B1LIM} has been introduced to allow clusters somewhat 
above the B$\pi$ threshold mass $M_{th}$ to form a single B meson if
$$M < M_{lim} = (1+{\tt B1LIM}) M_{th}\;.$$
The probability of such single-meson clustering is assumed to decrease
linearly for $M_{th} < M < M_{lim}$.
This has the effect of hardening the B spectrum if {\tt B1LIM} is increased
from the default value of zero.

Finally one should note that the properties of $b$-jets in \herwig\ are also
affected by the parameters {\tt CLDIR} and {\tt CLSMR}, which control the
decay angular distribution of clusters containing a perturbative quark
(as opposed to
the quark-antiquark pairs produced by the non-perturbative gluon splitting
at the end of the parton showers -- see above). If {\tt CLDIR}=0, the
decay of such a cluster is taken to be isotropic in its rest frame, as
for other clusters. But if {\tt CLDIR}=1 (the default value), the decay
hadron carrying the flavour of the perturbative quark is assumed to
continue in the same direction as that quark in the cluster rest-frame.
This is suggested by the observation that the leading hadron in a quark
jet preferentially carries the quark flavour. The value of {\tt CLSMR}
determines the amount of smearing [exponential in $(1-\cos\theta)$] of this
angular correlation.  The default value of zero corresponds to perfect
correlation. Thus increasing {\tt CLSMR} tends to soften and broaden the
B-hadron distribution in $b$-jets. In practice, the predicted spectrum
tends to be too soft and {\tt CLSMR}=0 is preferred.

In \herwig\ version 6.1, the parameters {\tt PSPLT}, {\tt CLDIR} and
{\tt CLSMR} have been converted into two-dimensional arrays, with the first
element controlling clusters that do not contain a $b$-quark and
the second those that do. Thus tuning of $b$-fragmentation can now
be performed separately from other flavours, by setting
{\tt CLDIR(2)}=1 and varying {\tt PSPLT(2)} and {\tt CLSMR(2)}.
By reducing the value of {\tt PSPLT(2)}, a harder B-hadron spectrum
can be achieved.

\subsubsection{4-jet matrix element + parton shower 
option (massless ME)}\label{sec_hw4jet}

A new option available in \herwig\ version 6.1 is to
generate events starting from the 4-parton processes $\ee\to q\bar q gg$
and  $\ee\to q\bar q q\bar q$.   The relevant process code is
{\tt IPROC} = 600 +{\tt IQ} for primary quark flavour {\tt IQ} or 600 for
a sum over all flavours.  The matrix elements used are those
of Ellis Ross and Terrano \cite{Ellis:1981wv} and Catani and Seymour
\cite{Catani:1997vz}, which include the relative orientation of initial
and final states but not quark masses. As explained
in \sect{sec_hwshower}, the kinematic effects of quark masses
are taken into account in the subsequent parton showers and in matching
the showers to the momentum configurations generated according to the
matrix elements.  As also explained there, the variable
${\tt EMSCA} = \min\{\sqrt{s_{ij}}\}$ sets a limit on the transverse
momenta in the showers and is also used as the scale for $\as$.
The latter feature has the effect of enhancing the regions of small
$s_{ij}$ relative to  matrix element calculations with $\as$ fixed.

To avoid soft and collinear divergences in the matrix elements,
an internal parton resolution parameter {\tt Y4JT} (default value 0.01)
must be set. The interparton distance is calculated using either the
\durham\ or \jade\ metric. This choice is governed by the logical parameter
{\tt DURHAM} (default {\tt .TRUE.}). For reliability of the results,
one should use the same metric for parton and final-state jet resolution,
with a value of {\tt Y4JT} smaller than the $y_{cut}$ value to be
used for jet resolution.

\subsubsection{Combined 2,3 and 4-jet matrix element + parton shower option}
\label{sec:hw234jet}

As a result of discussions in the working group, a preliminary version of
a combined 2,3 and 4-jet option based on \herwig\ 6.1 was developed.
The strategy for combining matrix elements and parton showers
follows that of \cite{Webber:2000xx,Catani:2000xx}, with some
simplifications, as follows.

The program first generates conventional \herwig\ $\ee$ hadronic events
starting from matched $q\bar q$ and $q\bar q g$ matrix elements
(process code {\tt IPROC}=100).  After parton showering, the \durham\ clustering
algorithm is applied, and those events with precisely four jets at resolution
scale $y_1\equiv{\tt Y4JT}$ (default value 0.008) are {\em replaced} by
events generated using the (massless LO) 4-parton matrix element
({\tt IPROC}=600), with \durham\ cutoff $y_{ij}>{\tt Y4JT}$. The 4 parton
momenta are distributed according to the matrix element
{\em multiplied by a weight factor}, which for $q\bar q g g$ is
\begin{equation}\label{eq:wt4j}
{\cal W}(y_1,y_3,y_4) = \frac{\as(y_3s)}{\as(y_1s)}\frac{\as(y_4s)}{\as(y_1s)}
\Delta_g(y_1s,y_3s)\Delta_g(y_1s,y_4s)
\end{equation}
where $y_{3,4}$ are the jet resolution values at which the partons are
just resolved into 3,4 jets, and $\Delta_g$ is the Sudakov form factor of the
gluon (see e.g.\ \cite{Ellis:1996qj}).

As explained in \cite{Webber:2000xx,Catani:2000xx}, the extra weight
factor (\ref{eq:wt4j}) is necessary to ensure smooth
matching to the parton showers at small values of $y_1$ --- more specifically,
to cancel leading and next-to-leading logarithms of $y_1$. Since this
factor is always
less than unity, reweighting is simply achieved by rejecting configurations
with ${\cal W}(y_1,y_3,y_4)<{\cal R}$ where ${\cal R}$ is a random number.

After a 4-parton configuration has been generated, parton showers are
generated in the usual way except that (for the 4-parton events only)
parton branchings that would lead to sub-jets resolvable at resolution
$y_1$ are {\em vetoed}. This means they are not allowed, but the
evolution scale for subsequent branching is reduced as if they had occurred.
Again, this is  necessary to cancel LL and NLL $y_1$-dependence between
ME and PS.  In \herwig\ it is simply ensured by resetting
${\tt EMSCA} =\sqrt{y_1 s}$ after the 4-parton hard process.

Combining 2,3 and 4-jet events in \herwig\ 6.1 according to the above
``replacement" algorithm is done by the (Fortran) program {\tt hw234jet.f}.
A prerelease version and some further discussion
can be found at {\tt http://home.cern.ch/webber/} .
To run the program one must link the slightly revised \herwig\ version 6.103,
also available there.

\subsection{ARIADNE} 

The \ariadne\ program is based on the Colour Dipole model \cite{coldip}
where the QCD cascade is described in terms of gluon emissions from
independent colour-dipoles between colour-connected partons. The
program is described in detail elsewhere \cite{ariman,Knowles:1995kj}, and
the following will mainly discuss issues related to gluon radiation
off heavy quarks.  Gluon splitting into heavy quarks in \ariadne\ is
discussed in \sect{ll:gbbplit}.

One of the main advantages of the dipole model is that, since gluons
are emitted by the dipoles between partons, the interference between
diagrams where a gluon is emitted by either of two partons is
automatically taken into account, and there is no need to introduce
explicit angular ordering. Another related advantage is that, since
the first gluon emitted in an e$^+$e$^-\rightarrow q\bar{q}$ event,
again is emitted coherently by the $q$ and $\bar{q}$, the full leading
order matrix element can be used explicitly in this emission, and
correction procedures necessary in conventional parton shower models
are not needed.

\subsubsection{Gluon radiation off heavy quarks}\label{sec:arihvy}
For heavy quarks, the default current version of the program uses an
approximate extra suppression to suppress gluon radiation close to the
direction of the quark to account for the dead-cone
effect\cite{deadcone}. This extra suppression can be switched
off\footnote{By setting the switch \texttt{MSTA(19)=0} in the
  \texttt{/ARDAT1/} common block.} and,
as discussed in \sect{sec:3jMZ}, it seems that this actually
improves the description of the heavy-to-light jet-rate
measurement somewhat. Recently, the full massive
leading order matrix element was implemented in \ariadne\footnote{Not
  yet released. A prerelease can be obtained on request to
  {\tt leif@thep.lu.se}} for the first gluon emission, and it seems that
this also describe jet rates a bit better than the approximate
dead-cone suppression, although excluding mass effects still seems to
give the best desctiption. This needs to be studied further.

\subsection{APACIC++}

Paradigm of the program:
\begin{center}
Employ {\em matrix elements} to describe the {\em production of jets},\\
model the {\em evolution of jets} with the {\em parton shower}.
\end{center}

\subsubsection{Introduction}
As stated already in the introduction, due to various reasons, 
the modelling of 
multijet events in high--energy reactions becomes increasingly important with 
rising energies. 

With emphasis on this modelling of multijet events, the program package 
\apacic/\amegic\ has been developed only recently. 
The philosophy 
of the new approach presented here is to use matrix elements (ME) and parton 
showers (PS) in the corresponding regimes of their 
reliability \cite{APP,jpg} : {\em matrix elements} are employed 
to describe the production of jets, and {\em parton showers} to
model their evolution.
A general algorithm to match them \cite{Long} has been proposed 
and implemented in 
\apacic\ \cite{APA}, the PS part of the package. The algorithm is based on the 
paradigm above, namely to restrict the validity of the ME's for the description of 
particle emission to the regions of jet--production, i.e. to regions of comparably large 
angles and energies -- or to large $y_{\rm cut}$ of the corresponding jet--clustering scheme. 
In contrast, the PS is restricted to the disjunct region of jet--evolution, i.e. small angles and 
low energies -- or low $y_{\rm cut}$, respectively. 

However, in its current state, the package is capable to deal with multijet 
production in $e^+e^-$--annihilations only, where the jet--configurations available 
are determined by the ME generator. In addition to the generic ME part of the package, 
\amegic\ \cite{AME}, interfaces to \debrecen\ \cite{zoltan2}
and {\sc Excalibur} \cite{Exc} are provided as well. 
The hadronization of the partons is left to well--established schemes. At the moment, 
an interface only to the hadronization in the Lund--string picture as implemented in
\pythia\ \cite{ts:pythia} is supplied. 

The short description of the package follows closely the steps of event generation, namely
\begin{enumerate}
\item Initialization of matrix elements and jet rates,
\item Choice of jet structure of the single event,
\item Evolution of the jets with the parton shower, and
\item Hadronization.
\end{enumerate}

\subsubsection{Initialization of matrix elements and jet rates}
The use of matrix elements for the determination of the large--scale jet structure of 
the single events enforces their initialization and the calculation of the corresponding 
jet rates before the generation of single events. Since the description of the two other
ME--generators can be found elsewhere, only the ME--part \amegic\ of the
package will be discussed here. At the present stage, it is capable to deal with the
following processes 
\begin{eqnarray}
e^+e^-&\to& \gamma\,,\;\;Z\to\, (\le 5)\,\mbox{\rm QCD--jets}\nonumber\\
e^+e^-&\to& (\le 4)\,\mbox{\rm fermions}
\end{eqnarray}
at tree--level in the Standard Model. All particles can be taken massless or massive,
which allows for the inclusion of Higgs interactions. Effects due to photonic initial state radiation 
off the incoming electron pair can be included in the structure function approach.

\amegic\ constructs and integrates the matrix elements fully
automatically. It proceeds in the following steps,
\begin{enumerate}
\item{Building of topologies with unspecified internal lines and specified external legs in 
all combinations. Mapping of predefined Feynman--rules onto the topologies.}
\item{Construction of helicity amplitudes corresponding to the Feynman--amplitudes.
Gauge test and transformation into a word--string, which is stored in a library.}
\item{Integration over the phasespace of the outgoing particles. Here, a significant 
acceleration is gained by using the compiled and linked word--strings out of the library.} 
\end{enumerate}
As a result of this procedure, \amegic's source code of roughly 13 000 lines
grows considerably to up to 200 000 lines when libraries for all possible processes are 
added.

However, as a well--known fact, the integration over the phasespace is plagued with real 
divergencies related to the soft and collinear emission of massless particles. To handle them, 
usually the phasespace is cut to avoid the dangerous regions. Then, outgoing particles are
identified with jets, which are well--separated in phasespace with a measure $y_{\rm cut}$
depending on the jet--scheme. The package provides different jet--clustering schemes. 
Consequently, the cross--sections for $n_j\ge3$ jets depend sensitively on the choice of the 
scheme and the corresponding $y_{\rm cut}$. Concentrating for the moment on pure 
QCD events and defining 
\begin{eqnarray}
\sigma_{\rm QCD} = \sum _q\sigma_{ee\to q\bar q}
\end{eqnarray}
the package provides three different schemes to determine jet rates. With ${\tilde\sigma}$ the various 
cross sections with the appropriate powers of $\as$ pulled out, the jet rates in the ``direct'' 
scheme read
\begin{eqnarray}\label{dirsch}
{\cal R}_{n_j}^{\rm dir.} = \as^{n_j-2}(\kappa_{n_j} s_{ee}) \,
 \frac{\tilde\sigma_{n_j}}{\sigma_{\rm QCD}}\;\;\;\mbox{\rm and}\;\;\;
{\cal R}_{2}^{\rm dir.} = 1-\sum\limits_{n_j=3} {\cal R}_{n_j}^{\rm dir.}\,,
\end{eqnarray}
whereas in the two ``rescaled'' schemes they are defined by
\begin{eqnarray}\label{ressch}
{\cal R}_{n_j}^{\rm res.1} &=& {\cal R}_{n_j}^{\rm dir.} - {\cal R}_{n_j+1}^{\rm dir.}\nonumber\\ 
{\cal R}_{n_j}^{\rm res.2} &=& {\cal R}_{n_j}^{\rm dir.} \cdot
\prod\limits_{m>n_j}\left(1-{\cal R}_m^{\rm res.2}\right)\;\;\;\mbox{\rm and}\;\;\;
{\cal R}_{2}^{\rm res.2} = 1-\sum\limits_{n_j=3} {\cal R}_{n_j}^{\rm res.2}\,.
\end{eqnarray}
To account for the effect of higher order corrections, the package supplies scale factors 
$\kappa_S^{n_j}$ for the corresponding $n_j$--jet rate. They enter in the form of 
$\as=\as(\kappa_S s_{ee})$.

Going beyond pure QCD--events, the final states are divided into two ensembles, namely
an electroweak one and the QCD ensemble. The former consists of all events with at least 
four fermions in the final state, where the normalization is given by the appropriate sum of 
the cross sections taken into account. This division obviously assigns a small amount of QCD events, 
namely the ones containing four quarks, to the electroweak ensemble. However, it should be noted, 
that so far this issue of electroweak events is still under further investigation.

\subsubsection{Choice of jet structure of the single event}
The jet structure of the single events is now determined following the steps below,
\begin{enumerate}
\item{The number of jets and their flavour structure are chosen according to the rates given
above, Eqs.,\ref{dirsch}, \ref{ressch}.}
\item{The kinematical configuration is chosen. An appropriate number of equally distributed
four vectors for the outgoing on--shell partons is produced. Their minimal $y_{\rm cut}$,
$y_{\rm min}$, is forced to be larger than the $y_{\rm cut}=y_0$ used for the initialization of the
jet rates. These fourvectors are then reweighted to reproduce the kinematical configurations 
as determined by the matrix element, potentially including the effect of higher order corrections. 
Defining $|\tilde{\cal M}|^2({\rm max})$ the largest matrix element squared which can be obtained,
\apacic\ again offers three choices for the corresponding weights, namely
\begin{eqnarray}\label{kinweight1}
{\cal W}^{\rm L.O.} &=& \frac{|\tilde{\cal M}|^2(p_i)}{|\tilde{\cal M}|^2({\rm max})} \nonumber\\
{\cal W}^{\alpha} &=& \left[\frac{\as(y_{\rm min} s_{ee})}{\as(y_0 s_{ee})}\right]^{n_j-2}
                                          \cdot\frac{|\tilde{\cal M}|^2(p_i)}{|\tilde{\cal M}|^2({\rm max})} 
\end{eqnarray}
for leading order--  and $\as$--corrected weights and a more involved one, 
which follows closely the reasoning of resummed jet rates. For example, in this scheme \cite{Catani:2000xx},
the weight for three--jet events reads
\begin{eqnarray}\label{kinweight2}
{\cal W}^{\rm NLL} = \frac{\as(y'_{\rm min} s_{ee})}{\as(y_0 s_{ee})}
                                          \cdot\frac{|\tilde{\cal M}|^2(p_i)}{|\tilde{\cal M}|^2({\rm max})}
                                          \cdot\frac{\Delta_g(y'_{\rm min} s_{ee})} {\Delta_g(y_0 s_{ee})}\,, 
\end{eqnarray}
where $y'_{\rm min} = {\rm min}\{y_{qg},y_{g\bar q}\}$ the minimal $y_{\rm cut}$ related to the
gluon of the $q\bar qg$--configuration of the three--jet event and the Sudakov form factors
for the gluon, $\Delta_g$ to be found in \cite{CDOTW}. 

Note, that at this stage, the outgoing momenta are still on their mass--shell.}
\item{The colour configuration is determined. This is achieved by constructing relative
probabilities ${\cal P}_i$ of the different parton histories. Here, \apacic\ provides three
schemes, two of them based on the corresponding amplitudes ${\cal M}_i$ related to the single 
diagrams $i$ in the form
\begin{eqnarray}\label{prob1}
{\cal P}_i &=& \frac{1}{\sum_j |{\cal M}_j|^2}\cdot |{\cal M}_i|^2\nonumber\;\;\;\;\mbox{\rm or}\\
{\cal P}_i &=& \frac{1}{\sum\limits_j|{\cal M}_j\sum\limits_l{\cal M}_l^*|}
                          \cdot |{\cal M}_i \sum_k{\cal M}_k^*|\,.
\end{eqnarray} 
The third scheme relies on a shower oriented picture and was proposed in a similar fashion already 
in \cite{ts:showmatch} and discussed in \sect{sec:py4j}. To illustrate this scheme,
consider the diagram displayed in \fig{exgraph}. Its 
relative probability reads up to a suitable normalization
\begin{figure}
\begin{center}
\includegraphics[height=4cm]{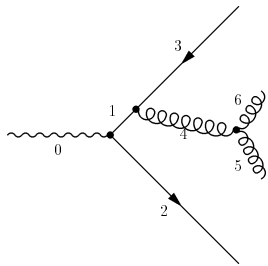}\\
\parbox{12cm}{\caption{\label{exgraph} 
Example diagram for the production of a $q\bar qgg$ final state.}}
\end{center}
\end{figure}
\begin{eqnarray}
{\cal P}_i = \frac{1}{t_1}\cdot P_{q\to qg}(z_{1\to 34})\cdot\frac{1}{t_4}
                                         \cdot P_{g\to gg}(z_{4\to 56})\,,
\end{eqnarray}
where the $t_{ij} = (p_i+p_j)^2$ and the $z$ are the energy fractions related to the various emissions
encountered. The parton history and equivalently the colour configuration of the parton ensemble are 
then chosen according to their relative probabilities.}
\item{The task left now, is to use the parton history to supply the outgoing on--shell particles with
virtual masses to allow them to experience a jet evolution via multiple emission of secondary partons.
This is achieved by means of the appropriate Sudakov form factors for each of the outgoing legs, where
the starting scale for the form factor is given by the internal $t_i$, like $t_1$ for the virtual mass
of leg $3$ and  $t_4$ for $5$ and $6$, in the exemplary diagram.
To ensure, that no additional jet under the jurisdiction of the initial $y_0$ is produced by
means of the parton shower, an appropriate veto is introduced into the subsequent shower algorithm.
To account for local four momentum conservation during the change from on--shell to 
off--shell particles the kinematics are slightly rearranged, resulting in slight changes in the energy 
fractions and the opening angles of the outgoing partons. }
\end{enumerate}

\subsubsection{Evolution of the jets}
The evolution of the jets proceeds in the standard way employing the Sudakov form factors
\cite{Ellis:1996qj}.
In addition to the usual switches allowing for the optional inclusion of prompt photons 
and azimuthal contributions, \apacic\ provides the possibility to use either the ordering 
by virtualities (LLA) or the ordering by angles (MLLA). However, in contrast to the pure 
MLLA--parton shower as performed in \herwig\ \cite{Marchesini:1992ch}, 
\apacic\ uses a hybrid solution when
switching to ordering by angles. Anticipating, that the MLLA--scheme is valid only in the domain 
of small angles, the first branching in each jet is done using the LLA--prescription, i.e. using the 
proper virtual mass as evolution parameter in the Sudakov form factor. After that, \apacic\
continues with the scaled angles as evolution parameters. In both cases, the Sudakov form factor
yields the probability for no observable branch between scales $t_+$ and $t_-$ and has the 
following form
\begin{eqnarray}\label{Sudform}
\Delta(t_+,t_-) = \exp\left\{-\int\limits_{t_-}^{t_+}\,\frac{dt}{t}
\int\limits_{z_-(t)}^{z_+(t)}
\,dz\,\as\!\left[p_\perp^2(z,t)\right] P(z)\right\}\,.
\end{eqnarray}
The boundaries for the $z$ integration are given by
\begin{eqnarray}\label{zbound}
z_{\pm}^{\rm LLA}(t) &=& \frac12\pm\frac12\sqrt{1-\frac{4t_0}{t}}\;\;\mbox{\rm and}\nonumber\\
\sqrt{\frac{t_0}{t}} < &z^{\rm MLLA}& < 1-\sqrt{\frac{t_0}{t}}\,,
\end{eqnarray}
where $t_0$ is the minimal virtual mass allowed in the parton shower. Within \apacic,
however, all partons leaving the jet--evolution have virtual mass $t_{\rm fin}$
\begin{eqnarray}\label{minmass}
t_{\rm fin} = \mbox{\rm min}\{t_0,m_f^2\}\,.
\end{eqnarray}
The transversal momentum squared for the decays is 
\begin{eqnarray}
p_\perp^2 \stackrel{\rm LLA}{\longrightarrow} z(1-z) t 
\stackrel{\rm MLLA}{\longrightarrow} z^2(1-z)^2 t
\end{eqnarray}  
reflecting the interpretation of the $t$ in LLA and MLLA as the virtual mass of the decaying particle
and the scaled opening angle of the branching, respectively.
\subsubsection{Treatment of mass effects}\label{sec:apamas}
Within the framework of LLA parton showers, \apacic\ treats non--vanishing quark 
masses in the following way:
\begin{enumerate}
\item{For the Sudakov form factors of \eqn{Sudform}, the minimal virtual mass $t_0^f$ is 
flavour dependent, see \eqn{minmass}. Consequently, the minimal virtual mass for 
decaying quarks and gluons are changed, too.}
\item{The boundaries of the $z$--integration are determined by \eqn{zbound} but with the
corresponding replacement 
\begin{eqnarray}
4t_0 \to \left(\sqrt{t_0^b}+\sqrt{t_0^c}\right)^2
\end{eqnarray}
for branchings of the form $a\to bc$. For example, this results in
$4t_0\to \left(m_b+\sqrt{t_0}\right)^2$ for $b\to bg$ branchings and
$4t_0\to 4m_b^2$ for $g\to b\bar b$ branchings, respectively.}
\item{Accordingly, for $g\to q\bar q$ splittings individual Sudakov form factors are summed.
The $t-z$--pair are chosen according to the sum, when picking the resulting quark flavour 
forbidden flavours are respected. For example, if $t<4m_b^2$, and equivalently, if the $z$ value 
results in quark energies $E_q<m_q$, $b$ quarks are not picked any more.}
\end{enumerate}

\subsubsection{Hadronization}
At its present stage, \apacic\ performs the hadronization of the outgoing particles
with the help of the Lund--string as provided by \pythia. For this purpose, an appropriate interface 
has been written and included. A similar interface for the cluster--hadronization of \herwig\ is
planned. However, at this place, it should again be
noted, that all particles leaving the parton shower of \apacic\ have a non--vanishing mass
as given by \eqn{minmass}. Therefore, before entering the Lund--string all particles
have to be set on their mass--shell resulting in a small rescaling of their four--vectors.

\subsubsection{Summary : physics and computer features}
\underline{Physics features :}
\begin{enumerate}
\item{The program package \apacic/\amegic\ is designed for the modelling of 
multijet events. It is capable to produce and evaluate matrix elements for the production of up to five 
massive partons in QCD and at least all electroweak processes of the type $e^+e^-\to$ four fermions
allowed in the Standard Model. Additional interfaces to various different M.E. generators describing 
the production of multijet topologies are available, too.}
\item{The MEs are matched to the parton shower (PS) via a generically new matching algorithm. This 
algorithm is capable to deal with -- in principle -- any number of jets produced via 
the strong, weak or electromagnetic interaction on equal footing.}
\item{The hadronization is modelled with the LUND--string approach as provided by \jetset, the 
corresponding interface is provided, an similar interface to the cluster--hadronization of
\herwig\ is planned.}
\end{enumerate}
\underline{Computer features: }
\begin{enumerate}
\item{The programming language is C++, allowing for a transparent and 
user--friendly programming style.}
\item{The package \apacic/\amegic\ has been developed
under Linux with the GNU--compilers. In addition, it has been tested
under AIX, Digital Unix and IRIX.}
\item{Size of the package is :\\
\begin{tabular}{lccc}
Source code :& \apacic\ &$\sim$ &7 000 lines\\
                        & \amegic\ &$\sim$ &13 000 lines\\
Own libs : & \amegic\ & up to & 200 000 lines
\end{tabular}}
\end{enumerate}

\subsection{Tuning and tests of \apacic\ to reproduce event shape data}

\subsubsection{Introduction}
High precision measurements of event shape distributions and inclusive 
particle 
spectra, based on \lepone\ data taken with the \delphi\ detector at \lep, are
used to determine \apacic\ parameters. Extensive studies are performed to
compare predictions of \apacic\ with \delphi\ data.

Definitions of used observables
and a description of their measurement, together with a short description of
the \delphi\ detector, can be found in \cite{Abreu:1996na,WUB-DIS95-11}

\subsubsection{The tuning procedure}
The tuning procedure is based on a simultaneous fit of Monte Carlo
parameters to
physical observables, taking correlations between parameters into account.
The fit is based on the minimisation of the variable
\begin{equation}
  \chi^{2}(\vec{p}) := 
       \sum_{\mbox{observables}} \sum_{\mbox{bins}} \left(
       \frac{X_{meas.} - X_{MC}(\vec{p})}{\sigma_{meas.}}  
       \right)^{2}
  \label{chi2}
\end{equation}
The sum extends over all bins of all physical observables included in the fit,
$\sigma_{meas.}$ being the total (statistical and systematic) error on 
the measured value $X_{meas.}$, $X_{MC}(\vec{p})$ being the Monte Carlo
prediction of bin $X$ for the parameter setting $\vec{p}$.
To perform the fit a fast prediction of $X_{MC}$ for 
any parameter setting $\vec{p}$ is needed. 
This is approximated by a Taylor expansion:
\begin{equation}
  X_{MC}(p_{1},p_{2},\cdots,p_{n}) = 
     A_{0} + \sum_{i=1}^{n}B_{i}p_{i} + \sum_{i=1}^{n}C_{i}p_{i}^{2}
       + \sum_{i=1}^{n-1} \sum_{j=i+1}^{n}D_{ij}p_{i}p_{j} + \cdots
    \label{taylor-app}
\end{equation}
The coefficients $A_{0},B_{i},C_{i}$ and $D_{ij}$ are extracted from a
systematic parameter variation by applying a singular value decomposition.

For a detailed description of the tuning procedure, 
see \cite{Abreu:1996na,WUB-DIS95-11}.

\subsubsection{Tuning of \apacic}
The generator \apacic\ together with \amegic, restricted to 
at most five massless jets from matrix element calculation, 
followed by a LLA parton shower, is chosen to be tuned. The 
initial jet-finder is the \durham\ algorithm, and fragmentation 
is achieved by the Lund string model.

The tuning of a new Monte Carlo generator is an iterative process. 
Each tuning triggers a learning process, resulting in improvements 
in the program, followed by a re-tuning. This process has not yet 
finally converged, but the quality of the program has reached a 
competitive state.

Each iteration starts with the selection of parameters to be tuned
and the definition of their variation ranges. Since the fit result 
is difficult to predict, the variation ranges have to be chosen
generously, degrading the precision of the fit. A second tuning
around the optimal values provides further improvements to the
generator. 

The fit is performed to a sample of observables, sensitive to the 
varied parameters. Exchanges in the composition of the observables
give hints to systematic uncertainties of the tuning result. 

\subsubsection{\apacic\ parameters}
Within \apacic\ there are parameters describing the matrix 
element calculation, the parton shower evolution and the Lund fragmentation.

\begin{itemize}
  \item{Matrix element \\
    \begin{itemize}
    \item[$\circ$]{$y_{cut}^{ini}$\\
    Emissions of colour charged partons are restricted to resolution 
    parameters    $y_{cut} > y_{cut}^{ini}$. In previous tunings of \apacic\
    an adequate agreement to the reference data could only be achieved
    for large values of $y_{cut}^{ini}$ $(\simeq 0.05)$, 
    resulting in predictions for hard QCD processes dominated by the
    parton shower. For that reason $y_{cut}^{ini}$ was fixed to some
    sensible value ($y_{cut}^{ini} = 0.005$). Improvements 
    by re-weighting the kinematic distribution of jets
    cured this problem. One of the future projects will
    be to restore $y_{cut}^{ini}$ to the list of tuning parameters.
    }
    \item[$\circ$]{$\kappa_s^{3,4,5}$\\
    Due to the truncation of the perturbative expansion, matrix element
    calculations show a significant dependence on the QCD renormalisation 
    scale. \apacic\ accounts
    for these dependences by a scale parameter $\kappa_s^{3,4,5}$ for 
    each $n$-jet configuration: $\as = \as(\kappa_s^n \cdot s)$\\
    }
    \end{itemize}
  }
  \item{parton shower\\
    \begin{itemize}
    \item[$\circ$]{\asmz\\
    The strong coupling constant \asmz\ is responsible for
    the parton shower evolution
    }
    \item[$\circ$]{cutoff PS\\
    The parton shower ends at a given energy scale, where fragmentation 
starts\footnote{The parameter ``cutoff PS'' in \apacic\ is different 
from the cutoff parameter $q_0^2$ in \pythia: $4\cdot \mbox{cutoff} =
q_0^2$}.\\
    }
    \end{itemize}
  }
  \item{fragmentation\\
    \begin{itemize}
    \item[$\circ$]{Lund A,B\\
    Lund A and B enter the Lund fragmentation function. Due to the strong
    anticorrelation between A and B it is sufficient to tune one and
    keep the other fixed.
    }
    \item[$\circ$]{$\sigma_q$\\
    The width of the Gaussian distribution of transverse momentum for
    fragmentation quarks is given by $\sigma_q$.
    }
    \end{itemize}
  }
\end{itemize}
\tab{tunpara} summarises the parameters considered, their 
variation ranges and an illustrative tuning result.

\begin{table}[hbtp]
\begin{center}
\renewcommand{\arraystretch}{1.3}
\begin{tabular}{||c|c|c||}
 \hline\hline
  parameter       &   variation range              & fit result   \\ \hline \hline
 $y_{cut}^{ini}$  &    \multicolumn{2}{|c||}{0.005 (fixed)   } \\ \hline
 $\kappa_s^3$     &    $10^{-0.5}$ -- $10^{-1.3}$  & $10^{-1.10}$ \\ \hline  
 $\kappa_s^4$     &    $10^{-2.0}$ -- $10^{-2.8}$  & $10^{-2.65}$ \\ \hline  
 $\kappa_s^5$     &    $10^{-1.5}$ -- $10^{-2.3}$  & $10^{-1.62}$ \\ \hline  
 \asmz            &    0.107 -- 0.113  & 0.108 \\ \hline            
 cutoff PS        &    0.80 -- 1.40  & 1.267 \\ \hline            
 Lund A           &    0.75 -- 0.90 & 0.905 \\ \hline            
 Lund B           &    \multicolumn{2}{|c||}{0.85 (fixed)}    \\ \hline            
 $\sigma_q$       &    0.38 -- 0.46 & 0.422 \\ \hline  \hline          
\end{tabular}
\caption{Tuned \apacic\ parameters.}
    \label{tunpara}
\renewcommand{\arraystretch}{1.0}
\end{center}
\end{table}
\subsubsection{Data distributions}
Measurements of event shape distributions and inclusive particle spectra 
are taken from \cite{Abreu:1996na,WUB-DIS95-11}; for definitions of the
observables used see there. 

\apacic\ parameters from \tab{tunpara} are simultaneously fitted to 
a set of data distributions. For the fit result depending on the composition
of the data set some systematic checks are performed to estimate the
stability of the fit. The strategy followed within the composition 
is to include at least one distribution that is sensitive 
to the parameters being fitted. Within this constraint systematic 
exchanges in the composition are performed to study uncertainties in the fit 
result. \tab{datasets} gives a summary of 15 different compositions used.

\begin{table}[htb]
 \begin{center}
 \begin{small}
 \begin{tabular}{||c|c|c|c|c|c|c|c|c|c|c|c|c|c|c|c||}
\hline\hline
                       & \multicolumn{15}{c||}{fit to data sample}\\
\cline{2-16} 
\raisebox{1.3ex}[-1.3ex]{observable}
                       &     1   &     2   &    3    &    4    
                       &     5   &     6   &    7    &    8 
                       &     9   &     10  &    11   &    12 
                       &     13  &     14  &    15   \\
 \hline
 \hline
                        
$1-$thrust             & $\surd$ &         & $\surd$ & $\surd$                  
                       & $\surd$ & $\surd$ & $\surd$ & $\surd$                 
                       & $\surd$ & $\surd$ & $\surd$ & $\surd$
                       & $\surd$ & $\surd$ & $\surd$ \\
\hline
$D_{32}^{Jade}$        &         &         &         & $\surd$                        
                       &         &         &         &                         
                       &         &         &         &        
                       & $\surd$ &         & $\surd$        \\
\hline
$D_{43}^{Jade}$        &         &         &         & $\surd$                 
                       &         &         &         &                         
                       &         &         &         &        
                       & $\surd$ &         &                \\
\hline
$D_{54}^{Jade}$        &         &         &         & $\surd$                 
                       &         &         &         &                         
                       &         &         &         &        
                       & $\surd$ &         &                \\
\hline
$D_{32}^{Durham}$      & $\surd$ & $\surd$ & $\surd$ &                         
                       & $\surd$ & $\surd$ & $\surd$ & $\surd$                 
                       & $\surd$ & $\surd$ & $\surd$ & $\surd$
                       & $\surd$ & $\surd$ & $\surd$        \\
\hline
$D_{43}^{Durham}$      & $\surd$ & $\surd$ & $\surd$ &                         
                       & $\surd$ &         & $\surd$ & $\surd$                 
                       & $\surd$ & $\surd$ & $\surd$ & $\surd$
                       & $\surd$ & $\surd$ & $\surd$        \\
\hline
$D_{54}^{Durham}$      & $\surd$ & $\surd$ & $\surd$ &                         
                       & $\surd$ & $\surd$ &         & $\surd$                 
                       & $\surd$ & $\surd$ & $\surd$ & $\surd$
                       & $\surd$ & $\surd$ & $\surd$        \\
\hline
sphericity             &         & $\surd$ &         &                         
                       &         &         &         &                         
                       &         &         &         &        
                       & $\surd$ &         &                \\
\hline
aplanarity             &         &         &         &                         
                       &         &         &         &                         
                       &         &         & $\surd$ &         
                       & $\surd$ & $\surd$ & $\surd$        \\
\hline
planarity              & $\surd$ & $\surd$ & $\surd$ & $\surd$                 
                       & $\surd$ & $\surd$ & $\surd$ & $\surd$                 
                       & $\surd$ &         & $\surd$ & $\surd$
                       & $\surd$ & $\surd$ & $\surd$        \\
\hline
major                  & $\surd$ & $\surd$ & $\surd$ & $\surd$                 
                       & $\surd$ & $\surd$ & $\surd$ & $\surd$                 
                       &         & $\surd$ & $\surd$ & $\surd$
                       & $\surd$ & $\surd$ & $\surd$        \\
\hline
minor                  & $\surd$ & $\surd$ & $\surd$ & $\surd$                 
                       & $\surd$ & $\surd$ & $\surd$ & $\surd$                 
                       &         & $\surd$ & $\surd$ & $\surd$
                       & $\surd$ & $\surd$ & $\surd$        \\
\hline
eec                    &         &         &         &                         
                       &         &         &         &                         
                       &         &         &         & $\surd$
                       & $\surd$ & $\surd$ & $\surd$        \\
\hline
\hline
$N_{ch}$               & $\surd$ & $\surd$ & $\surd$ & $\surd$                 
                       &         & $\surd$ & $\surd$ & $\surd$                 
                       & $\surd$ & $\surd$ & $\surd$ & $\surd$
                       & $\surd$ & $\surd$ & $\surd$        \\
\hline
$p_t^{in}(t)$          & $\surd$ & $\surd$ & $\surd$ & $\surd$                 
                       & $\surd$ & $\surd$ & $\surd$ & $\surd$                 
                       & $\surd$ & $\surd$ & $\surd$ & $\surd$
                       & $\surd$ & $\surd$ & $\surd$        \\
\hline
$p_t^{out}(t)$         & $\surd$ & $\surd$ & $\surd$ & $\surd$                 
                       & $\surd$ & $\surd$ & $\surd$ & $\surd$                 
                       & $\surd$ & $\surd$ & $\surd$ & $\surd$
                       & $\surd$ & $\surd$ & $\surd$        \\
\hline
$p_t^{in}(s)$          &         &         &         &                         
                       &         &         &         &                         
                       &         &         &         &        
                       & $\surd$ &         &                \\
\hline
$p_t^{out}(s)$         &         &         &         &                         
                       &         &         &         &                         
                       &         &         &         &        
                       & $\surd$ &         &                \\
\hline
$y(t)$                 &         &         & $\surd$ &                         
                       &         &         &         &                         
                       &         &         &         &        
                       & $\surd$ &         &                \\
\hline
$y(s)$                 &         &         &         &                         
                       &         &         &         &                         
                       &         &         &         &        
                       & $\surd$ &         &                \\
\hline
$x_p$                  & $\surd$ & $\surd$ & $\surd$ & $\surd$                 
                       & $\surd$ & $\surd$ & $\surd$ & $\surd$                 
                       & $\surd$ & $\surd$ & $\surd$ & $\surd$
                       & $\surd$ & $\surd$ & $\surd$        \\
\hline\hline

\end{tabular}
\end{small}
\caption{Composition of different data sets used to fit the 
         \apacic\ parameters of \tab{tunpara}. $\surd$ means
         that the observable is included in the fit.}
        \label{datasets}
\end{center}
\end{table}
\subsubsection{Results}
\begin{figure}[bt]
  \begin{center}
    \includegraphics[height=10cm]{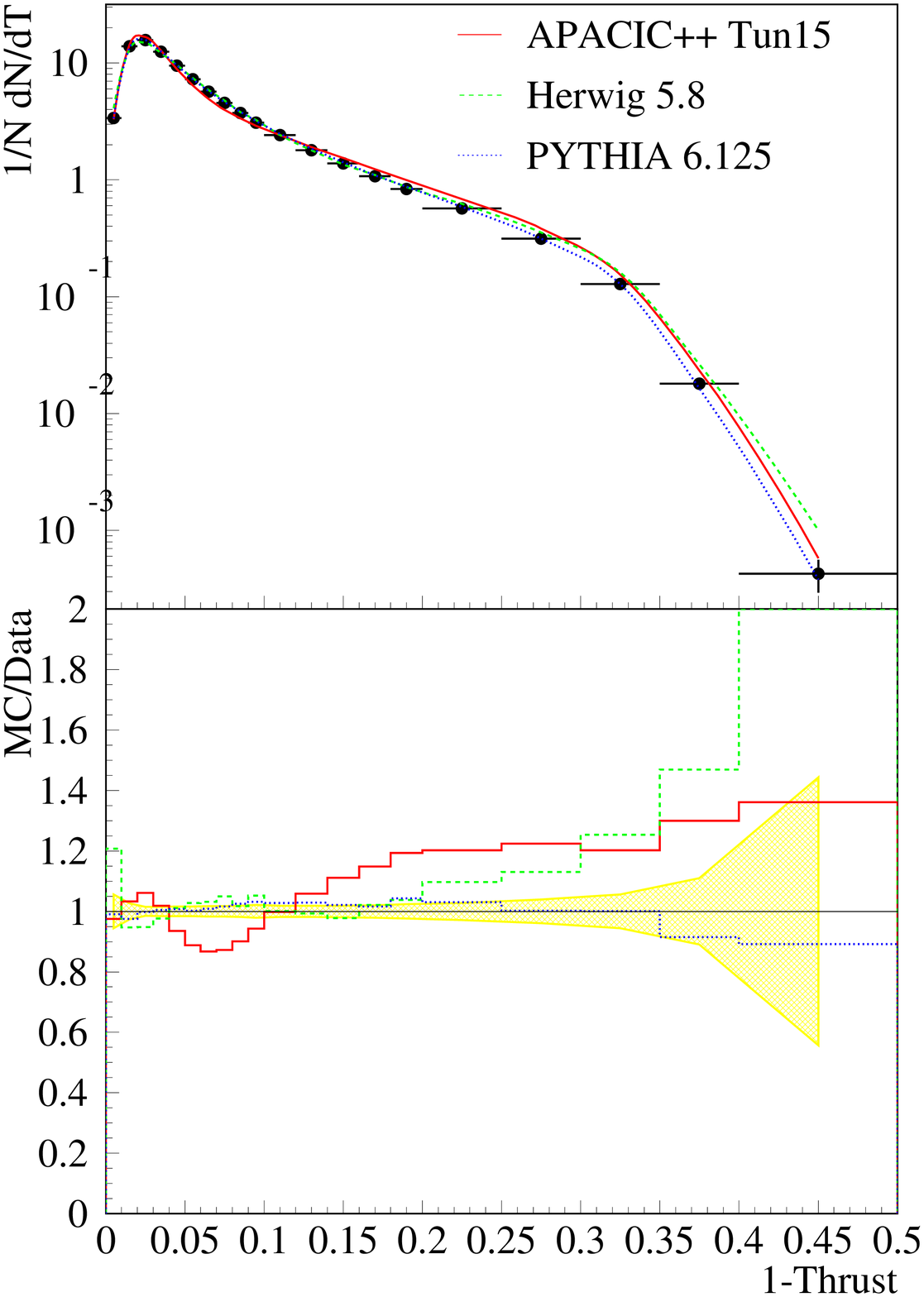}
    \includegraphics[height=10cm]{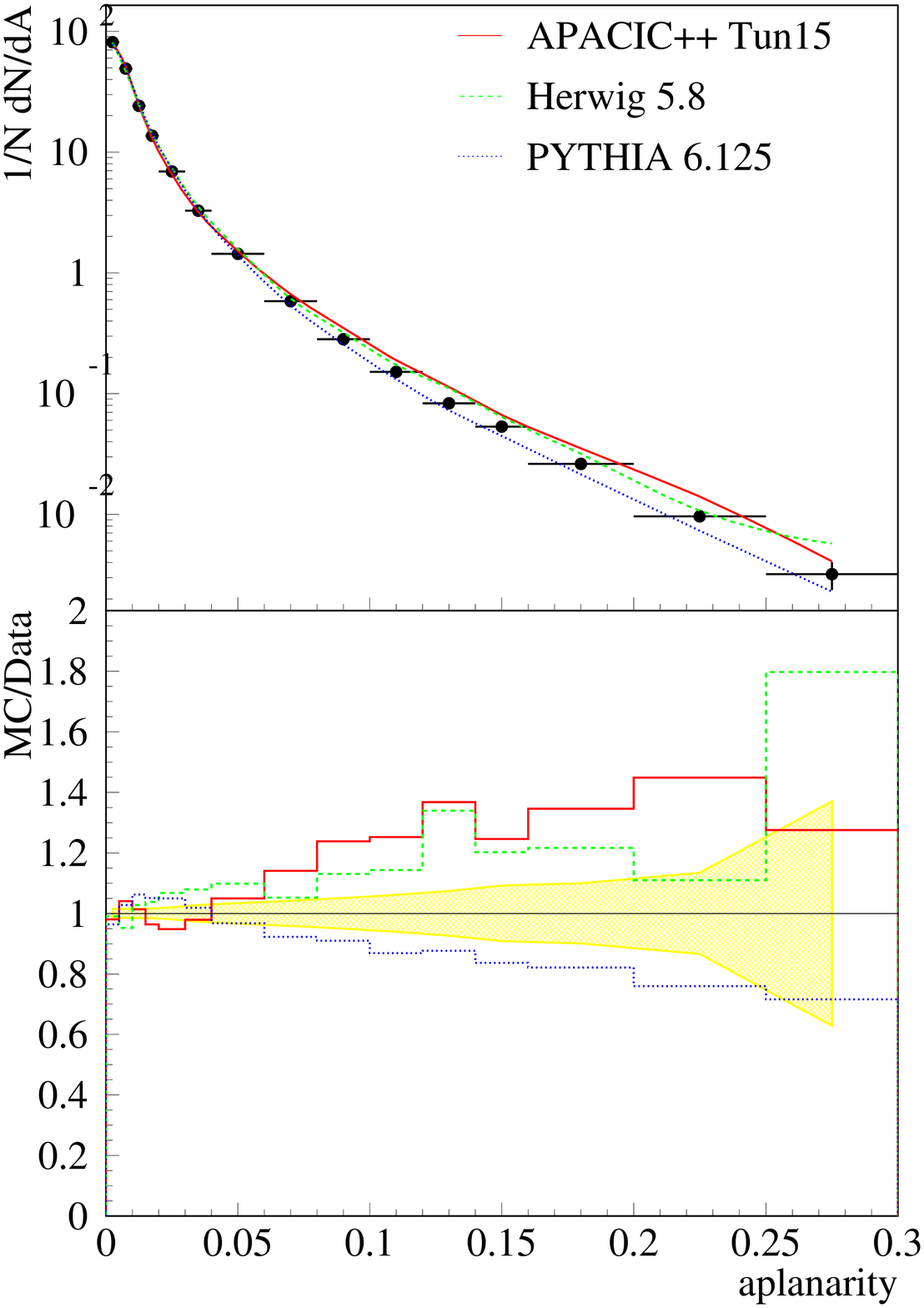}
  \end{center}
  \caption {Agreement between various Monte Carlo generators 
            and \delphi\ event shape distributions.
            The upper part of the plot shows the observable,
            the lower one the ratio of Monte Carlo and data; the
            shaded band represents the error of the data.}
  \label{evshapes1}
\end{figure}

\begin{figure}[bt]
  \begin{center}
    \includegraphics[height=10cm]{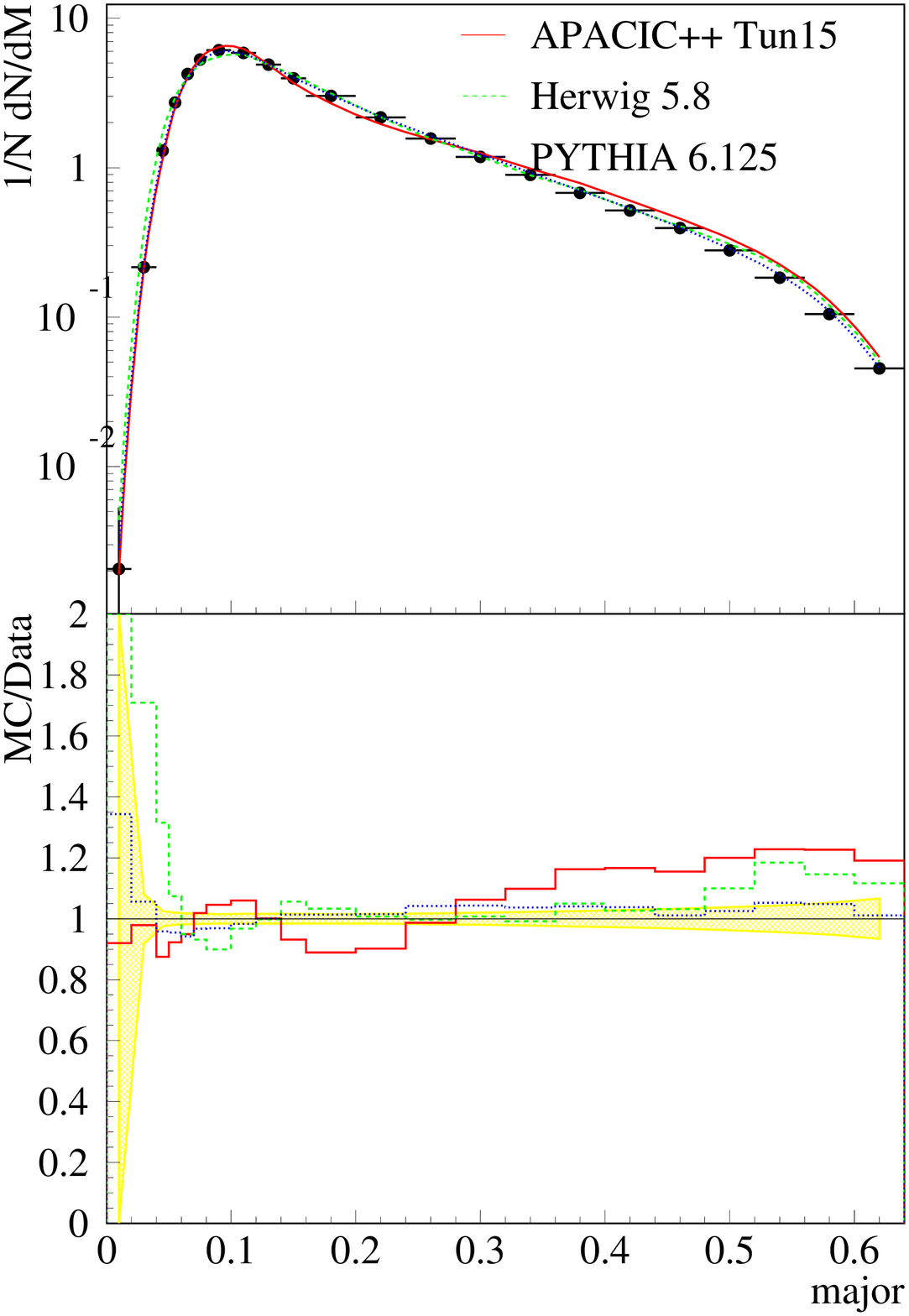}
    \includegraphics[height=10cm]{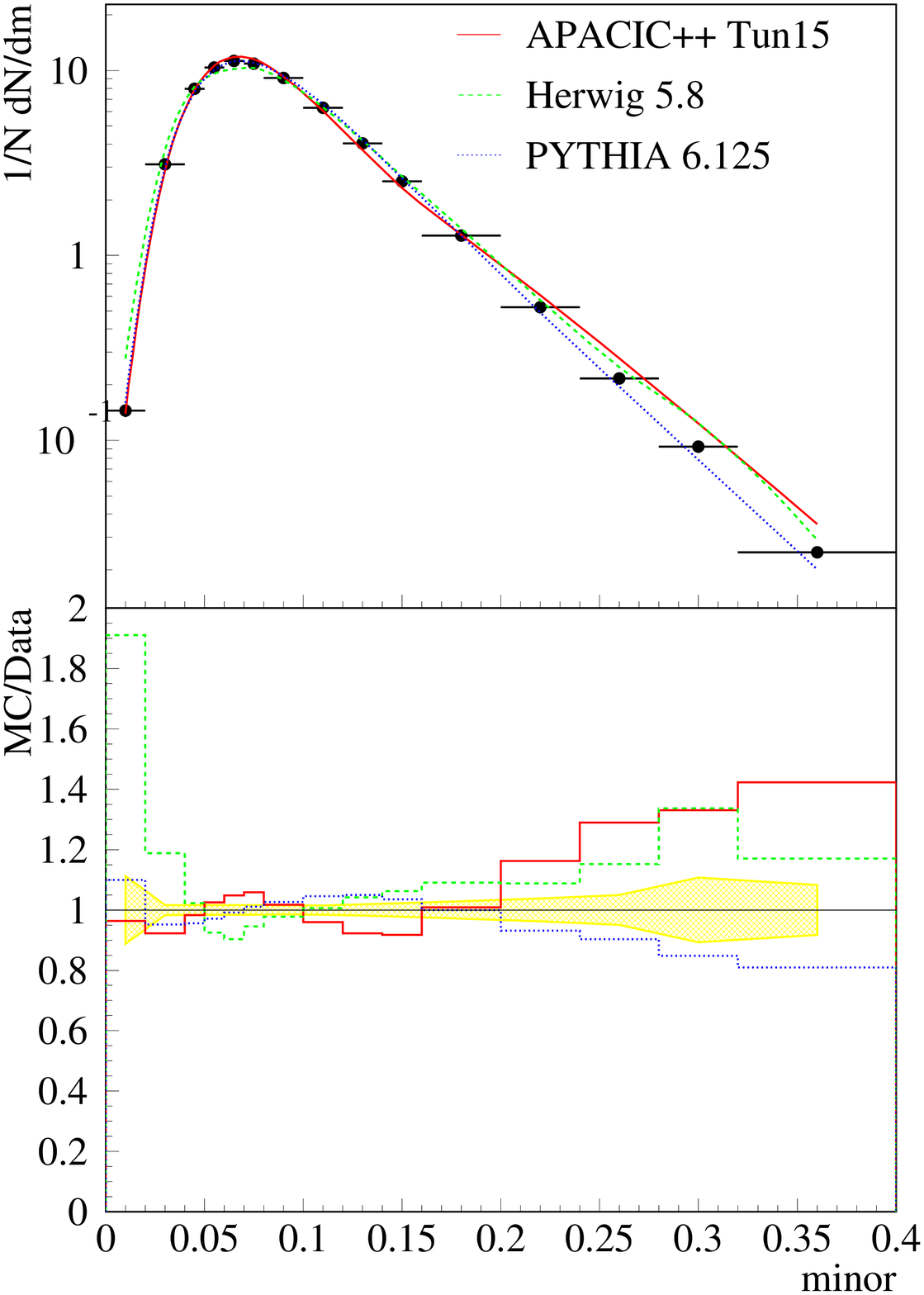}
  \end{center}
  \caption {Agreement between various Monte Carlo generators 
            and \delphi\ event shape distributions.
            The upper part of the plot shows the observable,
            the lower one the ratio of Monte Carlo and data; the
            shaded band represents the error of the data.}
  \label{evshapes2}
\end{figure}

\begin{figure}[bt]
  \begin{center}
    \includegraphics[height=10cm]{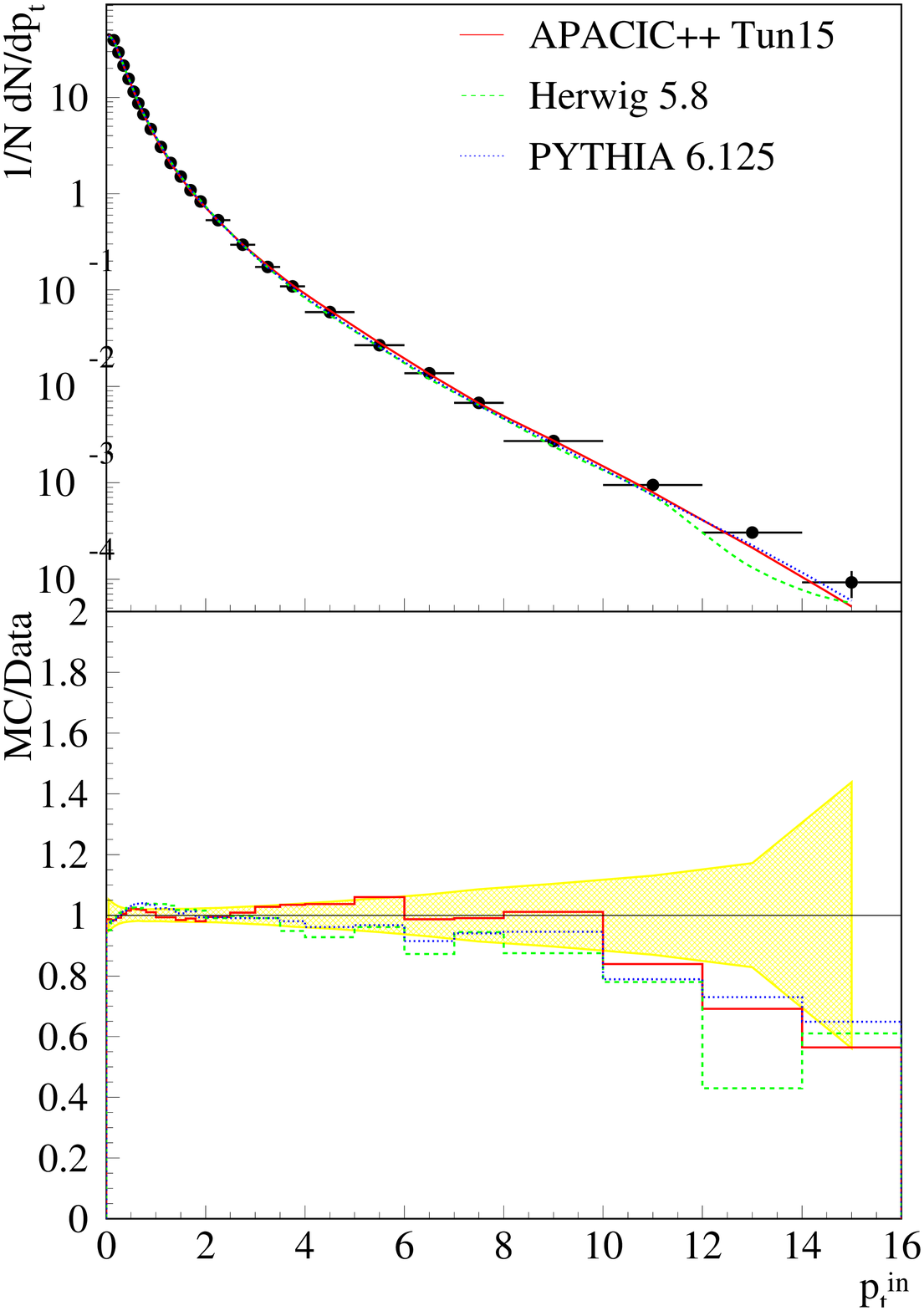}
    \includegraphics[height=10cm]{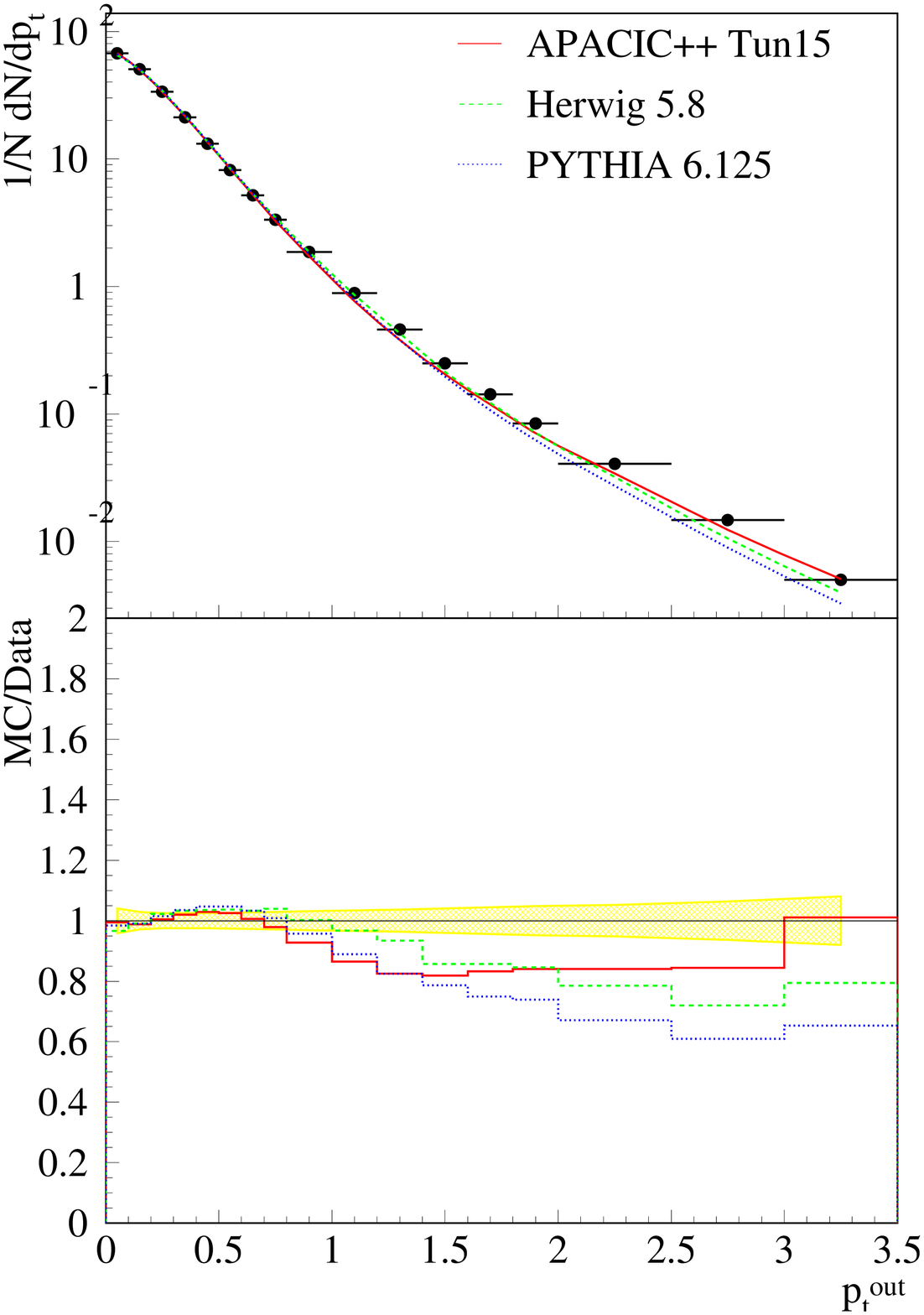}
  \end{center}
  \caption {Agreement between various Monte Carlo generators 
            and \delphi\ inclusive particle spectra.
            The upper part of the plot shows the observable,
            the lower one the ratio of Monte Carlo and data; the
            shaded band represents the error of the data.}
  \label{incpart1}
\end{figure}

\begin{figure}[bt]
  \begin{center}
    \includegraphics[height=10cm]{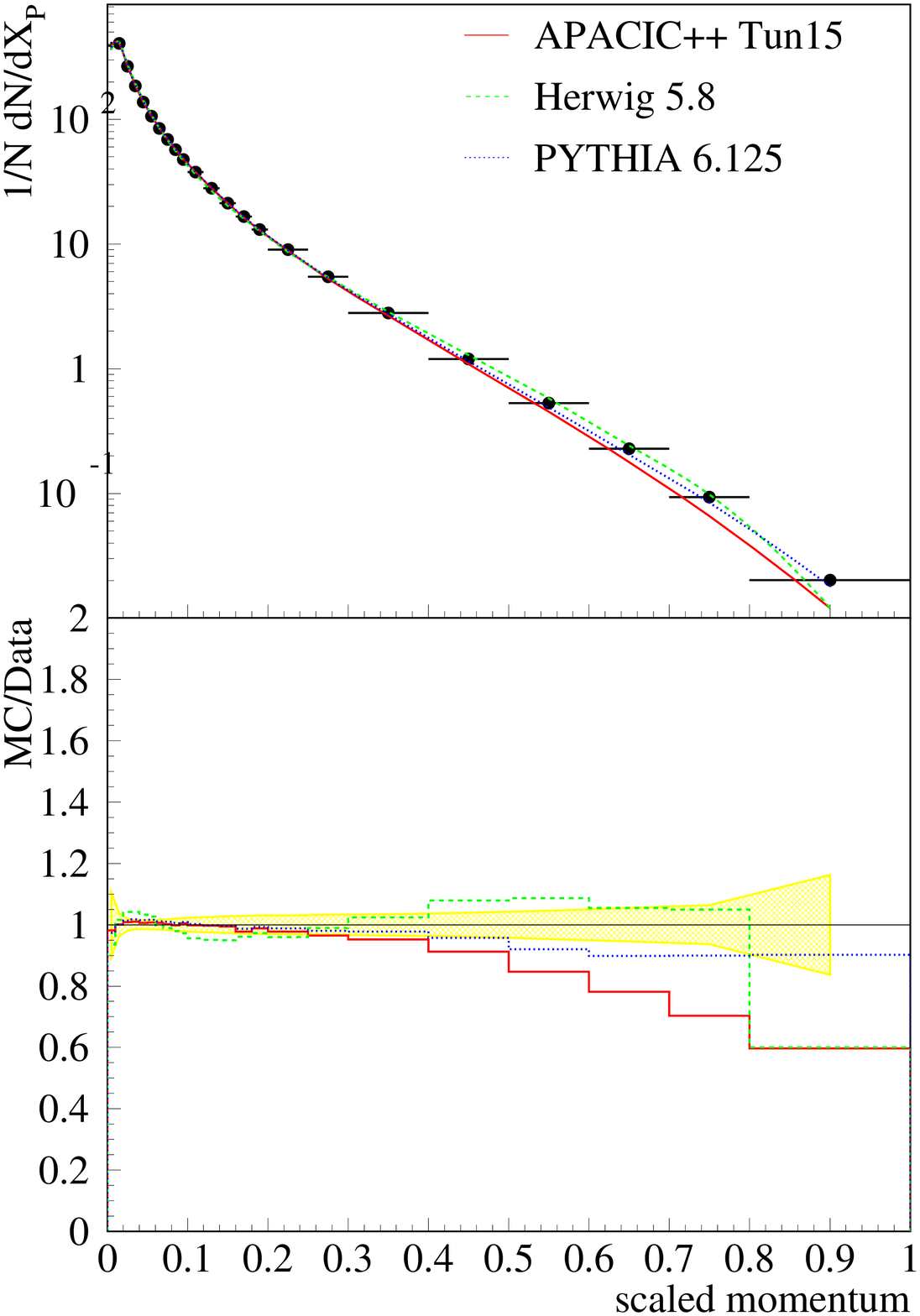}
    \includegraphics[height=10cm]{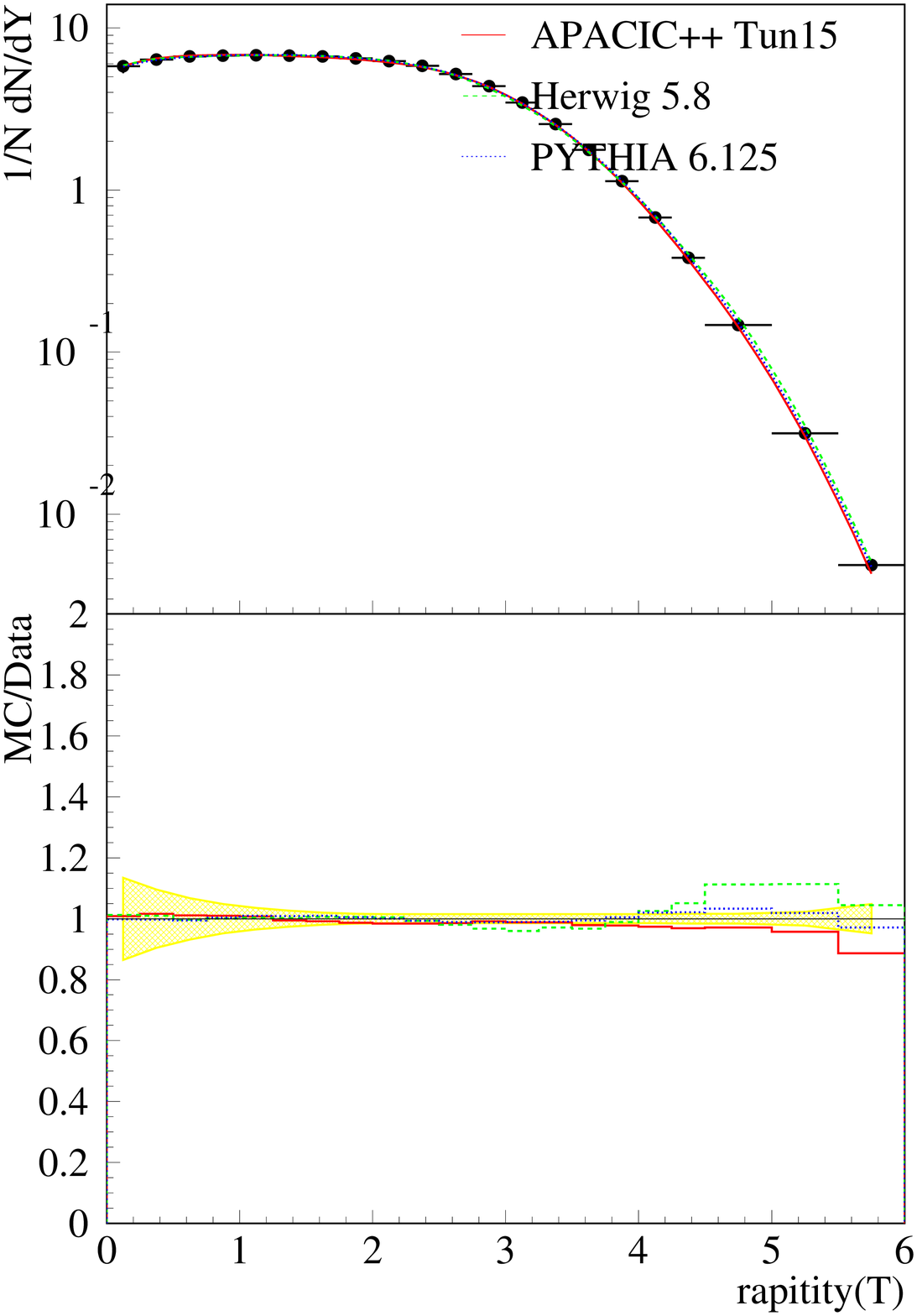}
  \end{center}
  \caption {Agreement between various Monte Carlo generators 
            and \delphi\ inclusive particle spectra.
            The upper part of the plot shows the observable,
            the lower one the ratio of Monte Carlo and data; the
            shaded band represents the error of the data.}
  \label{incpart2}
\end{figure}

\begin{figure}[bt]
  \begin{center}
    \includegraphics[height=7cm]{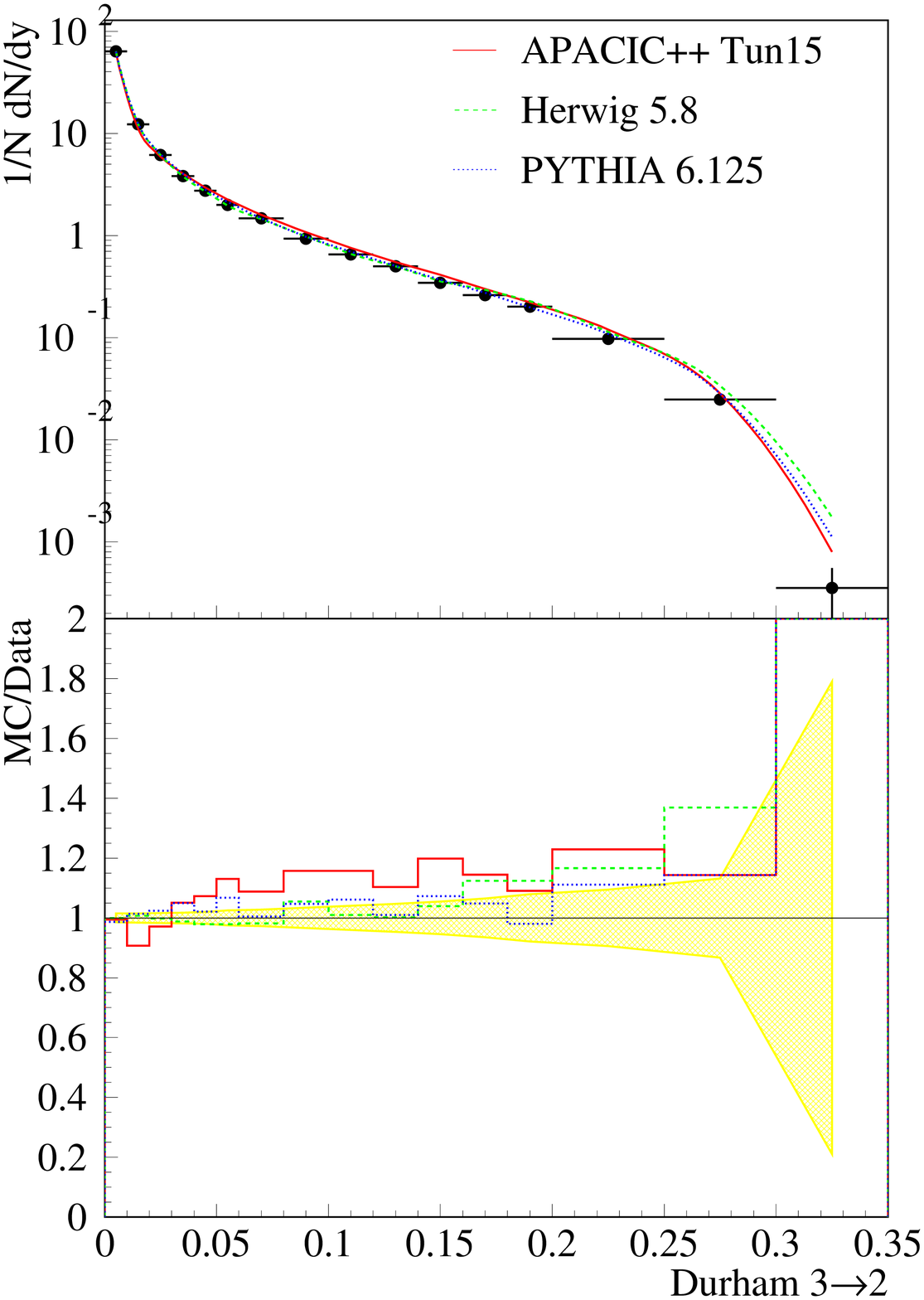}
    \includegraphics[height=7cm]{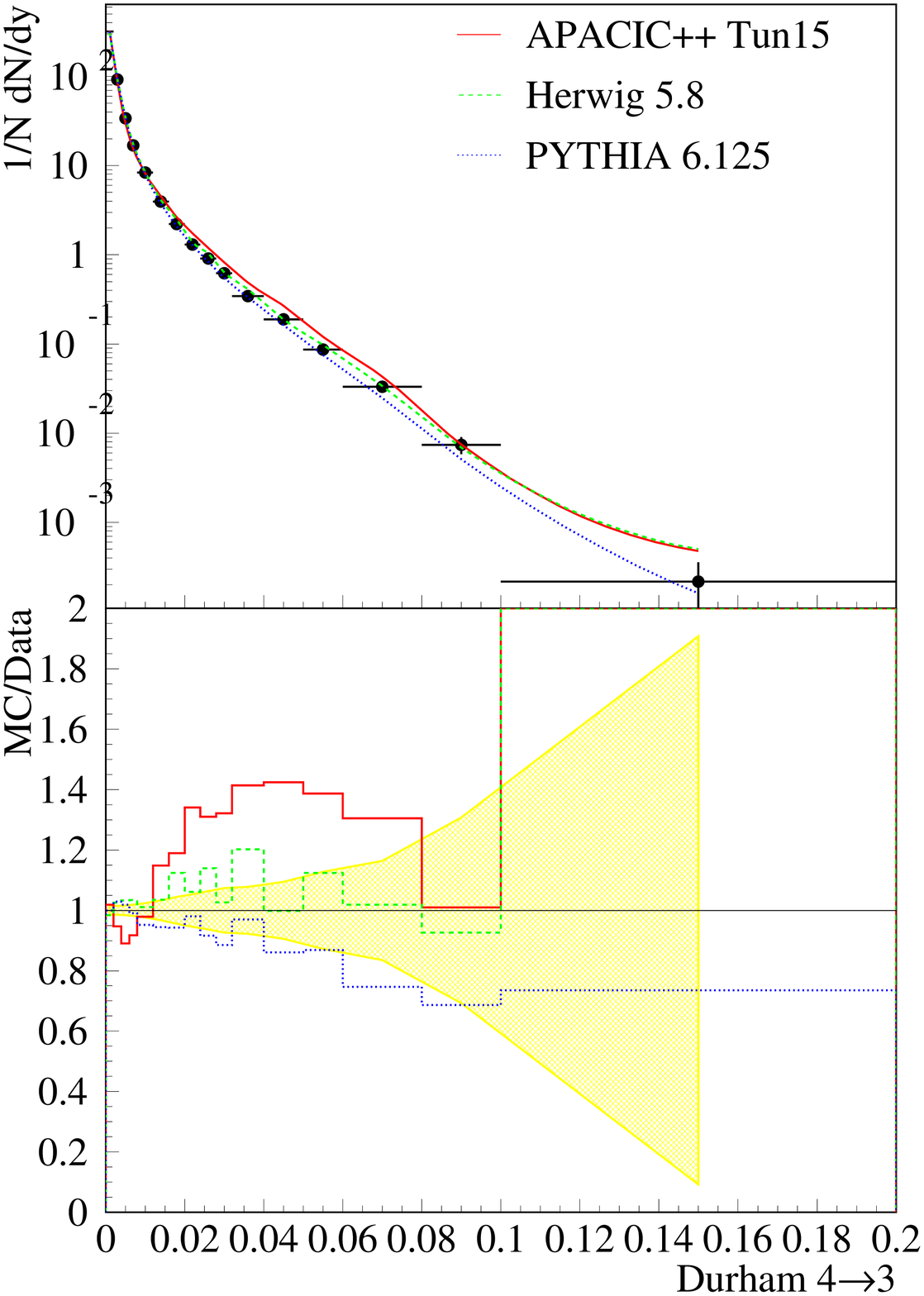}
    \includegraphics[height=7cm]{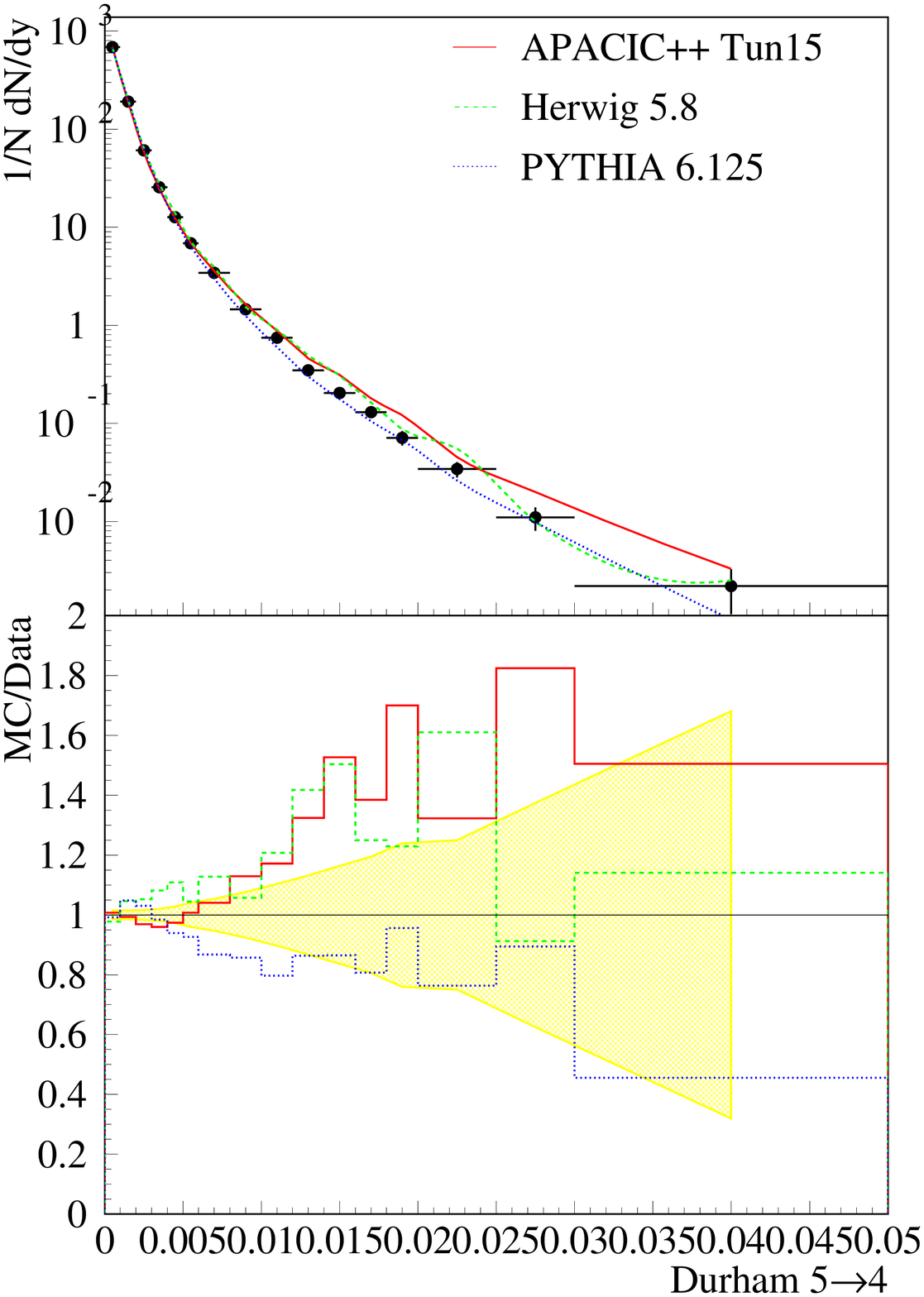}
  \end{center}
  \caption {Agreement between various Monte Carlo generators 
            and \delphi\ jet rates.
            The upper part of the plot shows the observable,
            the lower one the ratio of Monte Carlo and data; the
            shaded band represents the error of the data.}
  \label{jr}
\end{figure}

Predictions of \apacic\ event shape distributions, jet rates and 
inclusive particle spectra are compared to established Monte Carlo
generators (like \pythia, \herwig, \ariadne) and to \delphi\ data.
Figures 
\ref{evshapes1},\ref{evshapes2},\ref{incpart1},\ref{incpart2} and \ref{jr}
give an
overview of the behaviour and the relative (dis-)advantages of 
\apacic. 

%
%
%

%
%
%

%
%
%

%
%
%

%
%

\subsubsection{Conclusion and outlook}
\apacic\  parameters have been tuned to various sets of \delphi\
event shape distributions, jet rates and inclusive particle 
spectra.

The fits converged, the tuned parameters came out to be
basically reasonable: The parameter $y_{cut}^{ini}$ has been 
fixed in the latest tuning.
The parameter for the cutoff of the
parton shower is high, giving large weight to the fragmentation 
and minor to the parton shower. This has to be investigated. 

\apacic\ is able to predict all examined observables reasonably well.
Still none of the examined Monte Carlo generators is able to
predict the tail of the $p_t^{out}$ distribution (see however
\sect{sec:modelperf}).

\section{INCLUSIVE (ALL FLAVOUR) JET RATES}


\subsection{Tuning issues}

During the \lepone\ phase a qualitative improvement of the description of the
hadronic final state by parton shower fragmentation models has been reached,
mainly due to the possibility to precisely tune the models to a vast amount of
high quality data \cite{Knowles:1995kj}.
For this task flexible tuning procedures were used allowing interpolation
between model responses generated with different parameter settings
\cite{Buskulic:1992hq,Abreu:1996na}.

The effects on the model response of the individual parameters of the two major
aspects of the models -- the parton shower and the actual hadronisation
phase -- turn out to be strongly correlated. 
This requires one to determine the  most important model parameters in
global fits
to high statistic event shape and inclusive charged particle spectra and to
identified particle data.
A recent example for such a fit is discussed in \cite{Rudolph2000}.

\subsection{Model performance and multi-jet rates}\label{sec:modelperf}

It turns out that the string as well as the cluster hadronisation model are
able to represent the major features of particle production, especially the
identified particle rates, reasonably well. 
More detailed discussions can be found in
\cite{Knowles:1995kj,Knowles:1997dk,Bohrer:1997pr}.

Distributions depending mainly on the parton shower phase of the models are
in general very well represented. 
Especially for most of the event shape distributions, data and models agree 
within a few percent.
There are two important exceptions to this rule:

\begin{figure}
\begin{center}
\includegraphics[width=15cm]{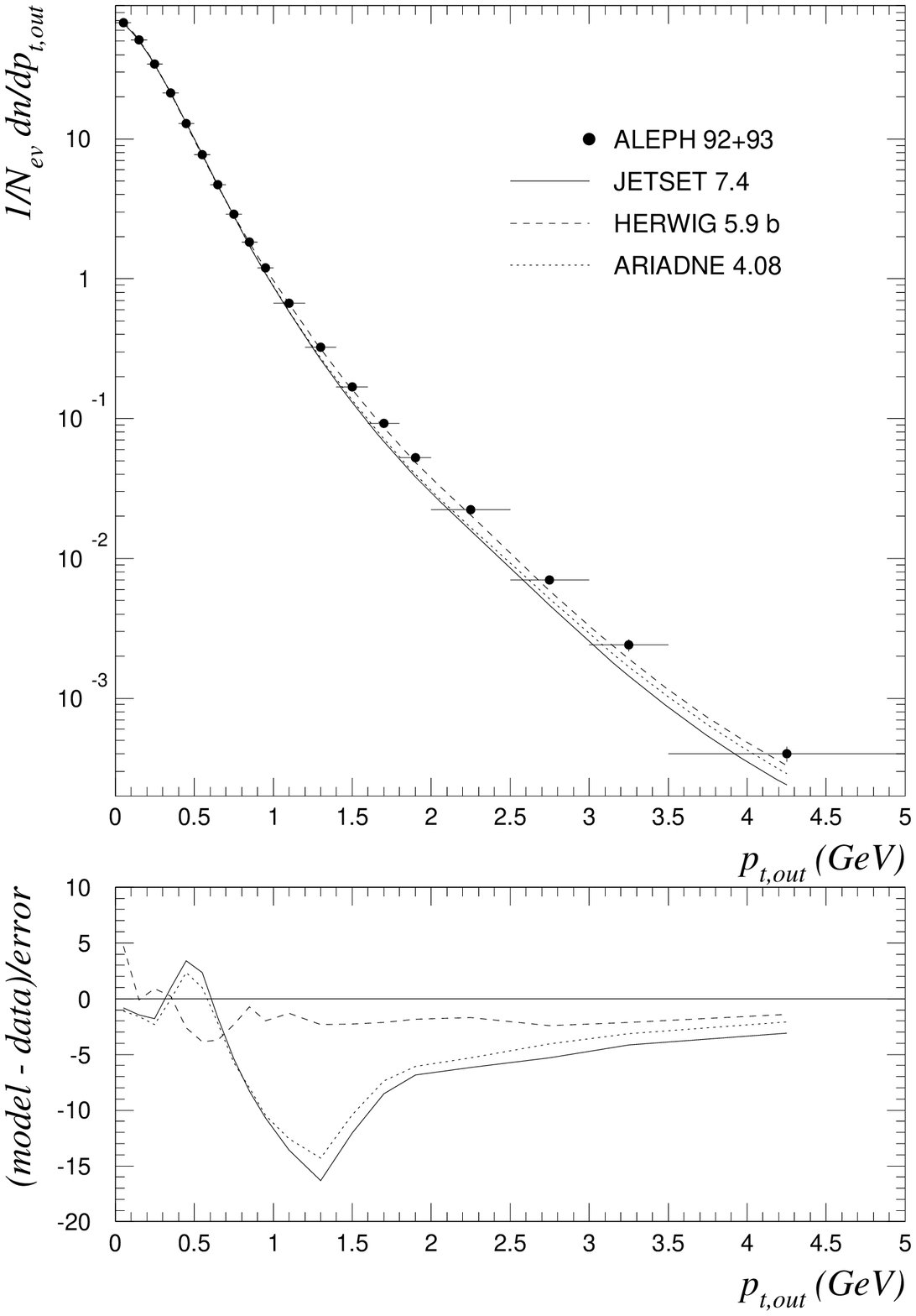}
\caption{$p_t^{out}$--distribution as measured by 
\aleph\ compared to predictions by
\ariadne, \jetset\ and \herwig\ \cite{Rudolph2000}}
\label{fig:ptos}
\end{center}
\end{figure}

Firstly the tail of the transverse momentum distribution of particles out of
the event plane is underestimated by about 30\% by most models 
\cite{Knowles:1995kj,Abreu:1996na}.
A possible explanation for this deficiency is that
the parton shower models account for part of the angular
structure of multi-jet events by tracing the polarization of the emitted 
gluons (see e.g. \cite{Ellis:1996qj}) to further splittings. 
This approximation cannot account for interference effects like a full
matrix-element calculation.
It should be emphasized, however, that the most recent tuning
\cite{Rudolph2000}  of the latest
version of \herwig\ shows a remarkable improvement of the
$p_t^{out}$-description.
This distribution (see \fig{fig:ptos}) now seems almost perfectly 
reproduced.

\begin{figure}
\begin{center}
\includegraphics[width=15cm]{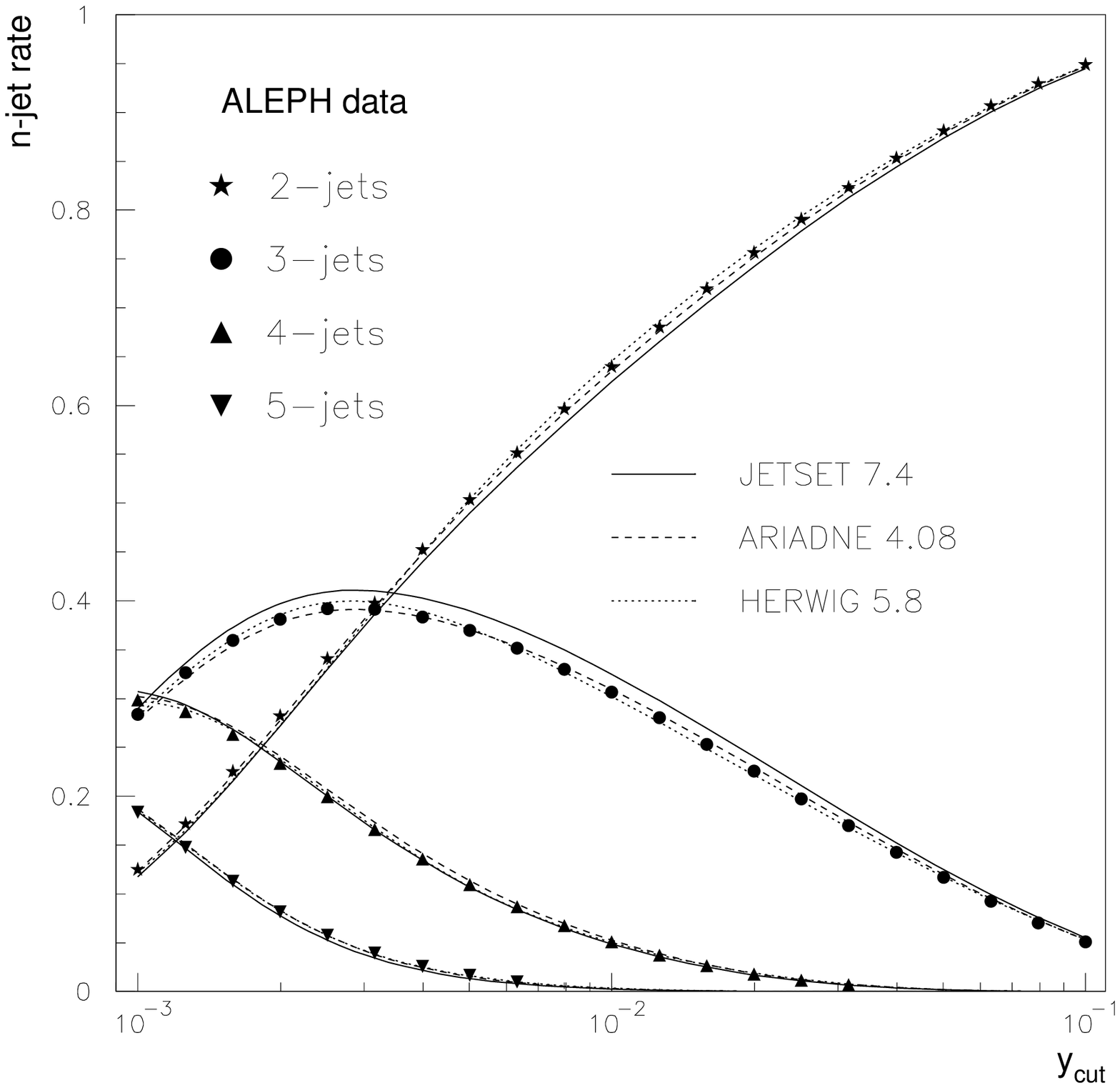}
\caption{Jet-rates for the \durham\ algorithm as measured by 
\aleph\ \cite{Barate:1998fi}}
\label{fig:r_aleph}
\end{center}
\end{figure}

\begin{figure}
\begin{center}
\includegraphics[width=15cm]{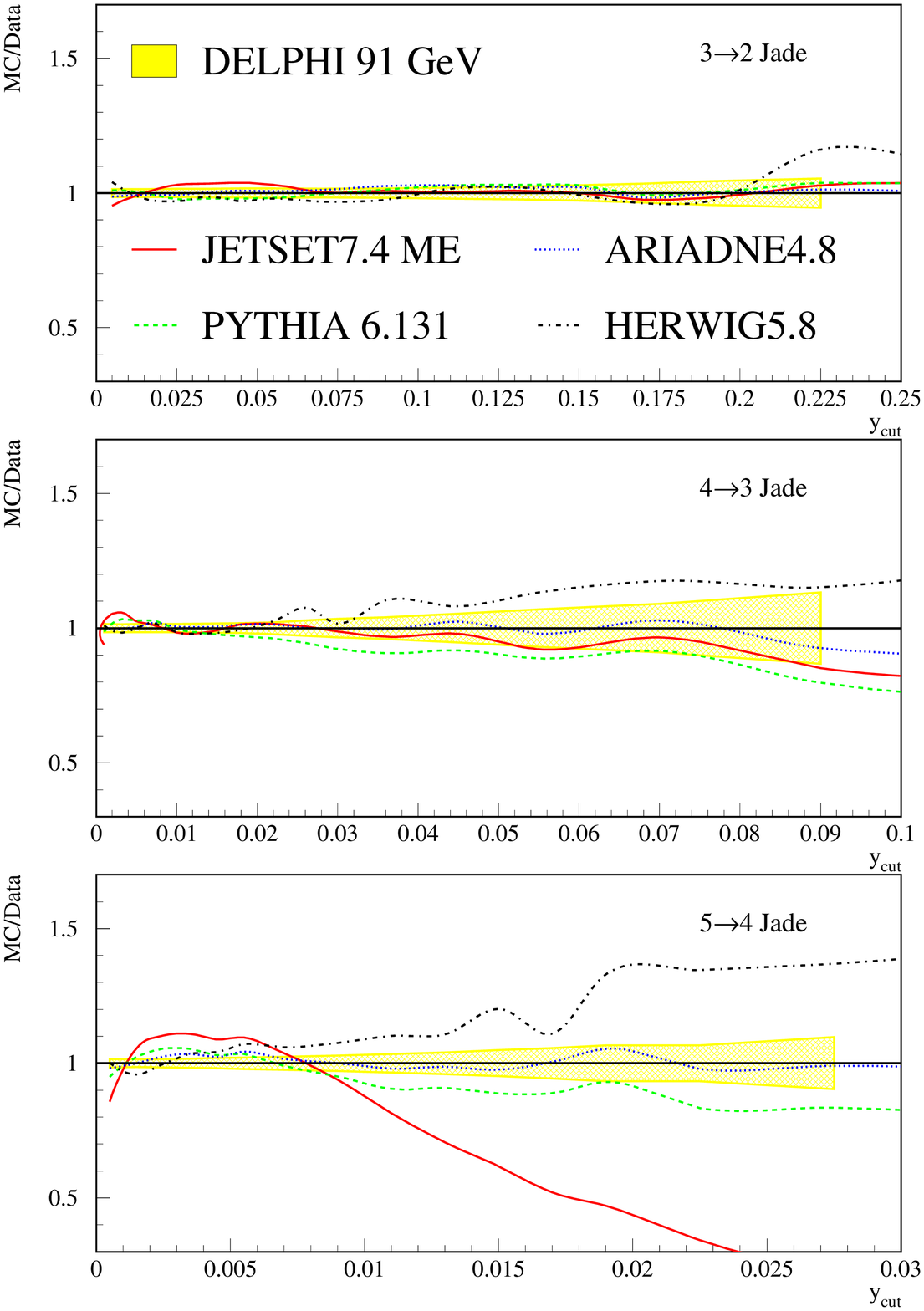}
\caption{Ratio data/Monte Carlo for the differential 2- (3$\rightarrow$2), 
3- (3$\rightarrow$2) and 4 jet (3$\rightarrow$2) rates (\jade\ algorithm).
Data measured by \delphi. The bands represent the statistical errors.
Model tunings as in \cite{Abreu:1996na}. 
\jetset\ ${\cal O}(\as^2)$ matrix element option for comparison.}
\label{fig:rj_delphi}
\end{center}
\end{figure}

\begin{figure}
\begin{center}
\includegraphics[width=15cm]{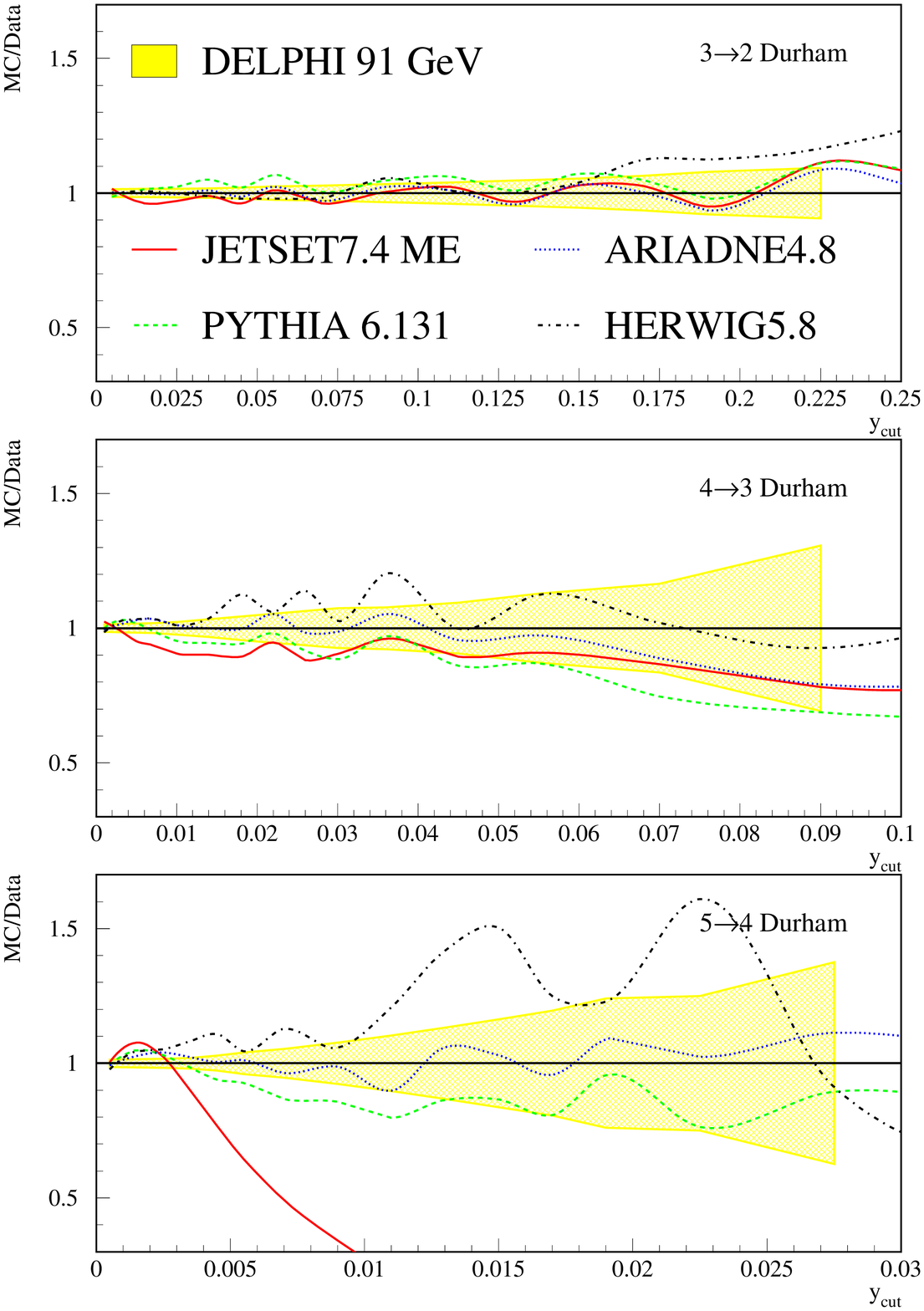}
\caption{Ratio data/Monte Carlo for the differential 2- (3$\rightarrow$2), 
3- (3$\rightarrow$2) and 4 jet (3$\rightarrow$2) rates (\durham\ algorithm).
Data measured by \delphi. The bands represent the statistical errors.
Model tunings as in \cite{Abreu:1996na}. 
\jetset\ ${\cal O}(\as^2)$ matrix element option for comparison.}
\label{fig:rd_delphi}
\end{center}
\end{figure}

\begin{figure}
\begin{center}
\includegraphics[width=12cm]{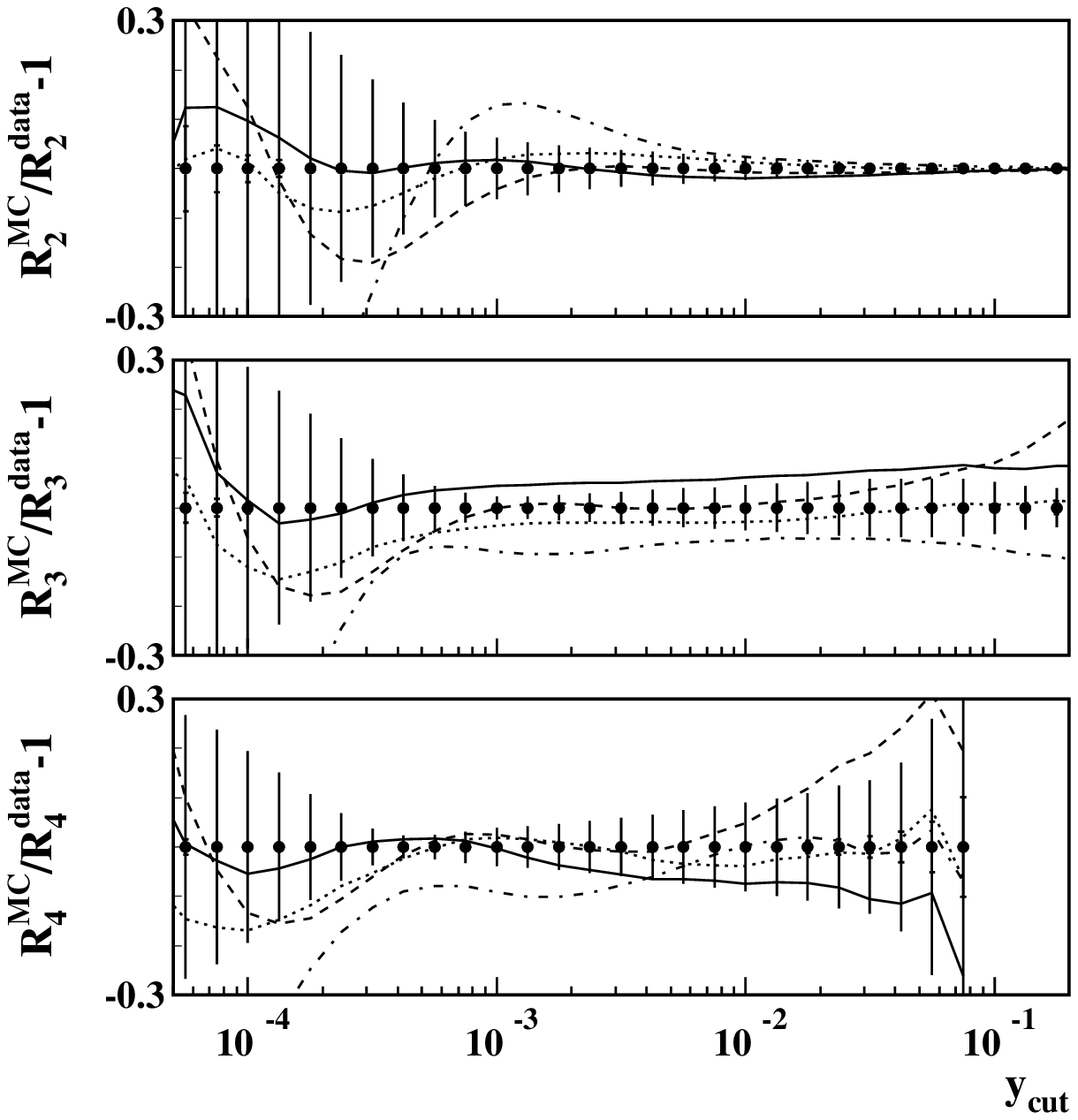}
\caption{Residuals (data/Monte Carlo -1) for the \cambridge\ jet algorithm. Data measured by \opal\
\cite{Abbiendi:1999rz}. Error bars depict the experimental errors. Full line:
\pythia. Dashed: \herwig. Dotted: \ariadne. Dot-dashed: COJETS \cite{cojets}.}
\label{fig:r_opal}
\end{center}
\end{figure}

The second exception concerns the inability of \herwig\ and \pythia/\jetset\ to
simultaneously describe different multi-jet rates with the precision desired by
the experiments.
This can already be 
seen from the $y_{cut}$ dependence of the multi-jet rates 
shown in \fig{fig:r_aleph} but is more clearly evident from the direct
data/model comparisons in Figures  \ref{fig:rj_delphi}, \ref{fig:rd_delphi}
and \ref{fig:r_opal}.
For a well represented three-jet rate, as was perhaps required in the tunings
of the \delphi\ Collaboration \cite{Abreu:1996na}, the (differential) four and
more jet-rates are systematically overestimated (underestimated) by \herwig\ or
\pythia/\jetset, respectively. This is already observed for the four-jet rate,
which is of special importance at \leptwo, but is even more so for the five-jet
rate. 
This general trend remains valid even for the aforementioned latest
version/tune of \herwig\ \cite{rudolphpriv}.
Depending on the strategy followed by the experiments, this misrepresentation 
can be distributed differently among the individual jet-rates. For
example, the \opal\ tuning mediates between the rates (see \fig{fig:r_opal}).
The general discrepancy between multi-jet-rate data and the corresponding 
model predictions has been
reported during the workshop by all experiments. 

Only \ariadne\ so far is able to well represent all jet-rates simultaneously
(see Figures \ref{fig:rj_delphi}, \ref{fig:rd_delphi} and \ref{fig:r_opal}).
The likely explanation for this difference between \ariadne\ and the other
models lies in the matching between the parton shower and the first order
$\rm q\bar{q}g$ matrix element simulations performed in \pythia\ and \herwig\ in
order to well represent the initial hard gluon radiation.
This matching is not needed in the dipole model implemented in \ariadne\ as here
the splitting probability for all splittings is given by the lowest order
matrix element expression. 
The ``opposite'' behavior observed for \pythia\ and \herwig\ may indicate,
however, that a better agreement may be reachable by suitably improving the 
matching procedure.

\subsection{Residual uncertainties}

Residual uncertainties due to the imperfect description of multi-jet rates are
difficult to review globally as they will depend critically on the individual 
analyses as well as on the tuning chosen by the different experiments. 

An incorrect 4-jet rate at high energies may require a 
reweighting of the Monte Carlo
to properly account for the QCD background in W or Higgs analyses.
Due to the correlation of the number of jets of an event 
with other properties, e.g.\ the
charged multiplicity or the momentum spectrum,
this is likely to have unwanted side effects.
An incorrect value of the strong coupling (which in the tunings is often
fixed by the 3-jet rate)
may cause an incorrect energy extrapolation of the models which is
hard to control at high energy because of the limited data statistics. 

A possible strategy for a determination of systematic error for a QCD type
observable such as the four jet rate at \leptwo\ energies may be the following:
The quality of the description of the observable is checked at the Z$^0$.
A possible misrepresentation at the Z$^0$ and at high energy is corrected using
the same correction factor. 
A large fraction of the deviation of the correction factor from unity
has to be taken as systematic uncertainty of the correction factor
at high energy, since the reason for the bad data description, and
consequently a possible energy dependence, is unknown.

The additional error for the uncertainty of the energy evolution of the model
will in general be small. In the case of the four-jet cross section at high
$y_c$, it will be dominated by the uncertainty of the strong coupling.
From the expected QCD evolution of the four-jet rate \cite{webeq} 
this uncertainty is at $\sqrt{s}=200$ GeV:
\begin{equation}
\nonumber
\frac{\delta R_4(\sqrt{s})}{R_4(\mz)} = 4 b
\ln\left(\frac{\sqrt{s}}{\mz}\right)
\delta\as(\mz) \simeq 1.9 \,\delta\as(\mz)  
\end{equation}
Here $b=(33-2n_f)/12\pi\simeq0.61$.
For an optimistically reachable error of $\as$ in the models of
$\delta\as(\mz)=0.003$ this yields $0.6\%$.

Employing alternative models and alternative model tunings will provide an
important cross check of the above error estimate. 
A model which correctly represents the four-jet cross section
at the Z$^0$, but overestimates the three-jet cross section 
($\propto \as$) by about 10\% 
(compare Figure \ref{fig:r_aleph}
at $y_{cut}\sim 0.01 ~{\rm to}~ 0.02$)
may
in fact lead to a more optimistic error estimate. The error for the correction
factor will vanish in this case at the expense of an increased error of the
model extrapolation. This error, however, is still small 
($1.9 \,\delta\as = 1.9 \times 0.012 \sim 2\%$).

For some analyses already today the abovementioned deficiency of the 
multi-jet description of the models leads to 
important contributions to the systematic error. An example is the \delphi\
measurement of the W pair production cross section with a fully hadronic final
state. A systematic error of 5\% 
(including a possible misrepresentation of the jet angular distributions)
is here assigned to the major background of QCD events.
This error was estimated by comparing different (uncorrected) models
and dominates the overall systematics. 
With increasing statistics this systematic error will be of similar size 
to the statistical error. \delphi\ therefore starts to 
employ \ariadne\, which certainly in terms of the jet rates
provides the best description of the data, as an 
alternative model for the full simulation.

\section{STUDY OF MASS EFFECTS IN 3- AND 4-JET RATES}

\subsection{Introduction}

The aim of this Section is to study the theoretical precision in the 
modeling of the rate of QCD processes leading to 3 and 4 final state jets, 
at \lepone\ and \leptwo, and involving $b$ quarks. This is important both to 
help understanding how to treat mass effects in our phenomenological QCD 
models, and to ensure precise enough control of backgrounds to new particle 
searches with $b$ quarks in the final state, such as for instance the Higgs 
search. With this aim in mind we compare mass effects on jet rates 
in the different MC approaches both to data and to analytic calculations.

It is natural to consider that because of their higher mass, $b$ quarks 
must from kinematics radiate fewer gluons than light quarks. More 
generally such a suppression enters in what is often refered to as 
the dead cone effect. What we want to know is how well the magnitude 
of this suppression is modeled in our Monte-Carlo approaches, and what is
the related theoretical uncertainty. In some sense this question can also 
be formulated as that of specifying the appropriate mass which should be
used for the $b$ quark.

From basic kinematics arguments, it can be shown that the magnitude of the 
suppression of gluon radiation from mass effects should scale as
$m_b^2/(s.y)$, where $m_b,s$ and $y$ are the $b$ quark mass, the collision 
energy squared and jet resolution parameter, respectively.
From this scaling law can be anticipated that the effects on jet rates  
are reduced at \leptwo\ energies as compared to \lepone, but that they can still 
remain substantial if jets are defined with a very small $y$ parameter.

In practice, from the results obtained, we find that the above scaling law 
is very approximate, and can really only be used to give a rough indication 
of the magnitude of effects. In addition to the purely kinematic effects,
resulting from the more limited phase-space, significant corrections - 
often referred to as dynamic mass effects - arise when taking into 
account properly the $b$ quark mass in the matrix elements for 3 or 4 
final state partons. Moreover, when comparing the data to the behaviour
of the analytic calculations or to that of the MC, one must be careful
to appropriately take into account in a consistent way effects resulting 
in final state jets with $b$ quarks other than the radiation of gluons off
$b$ quarks, such as processes with a gluon splitting into $b {\bar b}$.

In what follows, we first describe the method used for the evaluation.
This involves describing briefly the experimental analysis and the procedure
enabling meaningful comparisons with analytic calculations and with
predictions from Monte-Carlo generators. We then describe the analytic 
calculations which are available, show how these can be parametrised as
a function of $m_b,s$ and $y$ to enable the above mentioned comparisons,
as well as extrapolations. A proposed definition for the theoretical 
uncertainty to be quoted is also presented and discussed. Then follows 
a Section presenting the results of the comparisons of the different
Monte-Carlo approaches studied (\ariadne, \pythia, \herwig\ and \fourjphact) 
with the analytic calculations and with the \delphi\ data (some of which 
is still preliminary).  The comparisons with data are performed using the high 
statistics data available from \delphi\ at \lepone. Several of the used 
Monte-Carlo programs 
provide different options for the treatment of mass effects. Also, some 
have recently benefited from improvements. The main features of the 
different results and behaviours are described. Finally, remaining 
issues are discussed. One of these concerns the contribution 
to the observables used in the evaluation from processes involving 
gluon splitting into $b {\bar b}$. A quantitative measure of the precision 
of the description provided by the different programs for the 3 and 4 
jet cases as a function of $y$, at both \lepone\ and \leptwo\ energies, is given.

\subsection{Procedure used for evaluation}

\subsubsection{Appropriate choice of experimental observable}\label{sec:varcho}

Any observable with an explicit mass dependence can be computed 
either using the quark pole mass ($M_q$) or the running mass ($m_q(\mu)$). 
Both methods are equally valid, and the results have to be the same
if computed to all orders in perturbation theory.
However at any fixed finite order the predictions using 
both schemes do not necessarily agree. At a fixed order, calculations 
have a dependence on the scale at which the observable is computed. 
This dependence reveals the size of the higher order terms, as the 
sum of all the series must have no dependence on the scale.

\aleph\ \cite{alephEPS99} has performed a study of the $b$ mass
effects on several event shape observables comparing their Leading 
Order (LO) and NLO terms and hadronization effects. Some observables 
have quite large mass dependence as shown in that study. However as
the raw measurements have to be corrected accounting for the hadronization 
effects, it so happens that some observables have a correction even 
larger than the size of the effect to be measured.

Therefore the ideal observable to study quark mass effects will  
exhibit a large mass dependence, low higher-order corrections and
small hadronization effects.

\subsubsection{$b$ quark mass effects on the 3-jet ratio}

In order to study the $b$ mass effects, both \aleph\ \cite{alephEPS99} and 
\delphi\ \cite{mbatmz,delphiEPS99} 
have chosen an observable which fully complies with the above requirements.
The observable is the following:
\begin{equation}\label{eq:Rbl3}
R_{3}^{b{\ell}}(y_c)  =  \frac{R^b_3(y_c)}{R^\ell_3(y_c)}
\end{equation}
where $R^b_3(y_c)$ and $R^\ell_3(y_c)$ are the $b$-quark and light-quark
3-jet rates as defined in \eqn{eq:Rqnj}.

\subsubsection{Data analysis}

The jets of the hadronic events  can be reconstructed by means  of a jet
finding algorithm  (e.g. \durham\ or \cambridge\ -- see
\sect{intro_jetalgo}). By choosing the appropriate jet resolution
parameter $y$, it is possible to force the reconstruction of just three jets
in every event. Then, a set of quality cuts are applied over each jet 
(e.g. minimum charged multiplicity, enough visible energy, \ldots).

\subsubsection{Flavour definition}

The criterion adopted in \cite{mbatmz} and \cite{Rodrigo:1999qg} is such that
the flavour of the event is defined as that of the quark coupled
to the $Z$ in the  $Z\rightarrow q\bar{q}$ vertex. By convention, the 
production of $b$ quarks via the splitting of a bremsstrahlung gluon 
is ignored in this definition. It was shown that for this particular 
3-jet observable, the corresponding effect is practically negligible. 
As will be seen in \sect{sec:gbbimpact}, this is no longer 
the case in the case of 4-jets.

\subsubsection{Flavour tagging technique \label{sec:proba}}

The sample of hadronic events is split in two categories, with both
events strongly enriched in $b\bar{b}$, by using $b$-tagging, and events
strongly enriched in light quarks, by using $anti-b$-tagging. 

The tagging procedure normally uses the impact parameter information of all 
charged particles in the event. \lep\ experiments are equipped with 
silicon vertex detectors allowing accurate track reconstruction.
$b\bar{b}$ events can be selected by considering the presence of particles 
with large impact parameter significance (impact parameter over its error), 
while $\ell\bar{\ell}$ events can be tagged just by the lack of those tracks.
In addition the presence or absence of a reconstructed secondary vertex 
can be used in the $b$-tagging criterion.

\subsubsection{Measuring and correcting $R_3^{b\ell}$}

The measured value of our  observable can be computed from the
ratio of the reconstructed three-jet rates for $b$ and light quark tagged
flavours ($R^{\rm  obs}_{3q}(\yc), q=b,\ell$):

\[R^{b\ell\rm  -obs}_3 (y_{\rm c}) =
R^{\rm  obs}_{3b}(\yc)/R^{\rm obs}_{3\ell}(\yc)\]
 
This raw value of the observable must be converted into a parton level
one which may be compared with the LO and NLO calculations
\cite{Rodrigo:1999qg,Brandenburg:1997pu}.  The correction method accounts for the detector
effects, biases introduced in the flavour tagging, and hadronization effects.

The contribution   of    each flavour, $R^i_{3q}$,    to  the observed
three-jet cross section is given by:
\[ 
R^{\rm obs}_{3q} = c^b_q \cdot R^b_{3q} + c^c_q \cdot R^c_{3q} + 
c^{\ell}_q \cdot R^\ell_{3q}
\] 
where $c^i_q$ represents the flavour content for $i=b,~c,~\ell$ in each
of the experimentally tagged samples.  

The reconstruction level and parton level three-jet rates for each flavour
and defined sample ($R^i_{3q}$ and $R^{\rm par}_{3q}$, respectively) are
related by:
\[
R^i_{3q}(y_{\rm c}) = d^i_{3q}(\yc)\cdot h_{3i}(\yc)\cdot 
R^{\rm par}_{3i}(\yc) 
\]
where $d^i_{3q}$ and $h_{3i}$ are correction factors, accounting, respectively, 
for detector acceptance and tagging effects (deduced from the modeling of the 
detectors response to hadronic events), and for hadronization effects (estimated 
by comparing the hadron and parton level distributions obtained from MC programs).

By taking the jet rates from $c$ quarks equal to those from light quarks 
($R^{\rm par}_{3c} \equiv  R^{\rm par}_{3\ell}$), the measured jet rates can be 
expressed as:
\[ 
R^{\rm obs}_{3b}(\yc) = A_b(\yc) \cdot R_{3b}^{\rm par}(\yc) + 
B_b(\yc) \cdot R_{3\ell}^{\rm par}(\yc)  
\] 
\[ 
R^{\rm obs}_{3\ell}(\yc) = A_\ell(\yc) \cdot R_{3b}^{\rm par}(\yc) + 
B_\ell(\yc) \cdot R_{3\ell}^{\rm par}(\yc)  
\] 
where $A_q$ and   $B_q$   are a redefinition  of   the
original set  of  parameters:  $c^i_q,~d^i_{3q}$ and  $h_{3i}$.  This
parametrization allows expressing the corrected observable as:
\[
R_3^{b\ell}(\yc) = \frac{R_{3b}^{\rm par}}{R_{3\ell}^{\rm par}} =
\frac{B_b - B_{\ell}\cdot R_3^{b\ell{\rm -obs}}}
{A_{\ell}\cdot R_3^{b\ell{\rm -obs}} - A_b}
\]


The results are described in \sect{sec:R34comp}. At $\sqrt{s}=\mz$ the total 
correction to the raw $R_3^{b\ell}$ is typically about 10\%, the bulk 
of which corresponds to the detector and tagging effects, while hadronization 
corrections are of the order of 1\%. 

Uncertainties in the detector modeling were studied.
The main uncertainty results from the limited statistics of fully simulated 
events with detector response, resulting in limited knowledge 
of the factors $c^i_q$ describing the flavour content of each of the tagged 
samples. This error was estimated to be roughly 0.3\%, but has a strong 
dependence on the jet resolution parameter $y_c$ as it is directly related to 
the statistics of the three-jet simulated sample. 
Uncertainties from the modeling of the hadronization were also studied,
and are described in \sect{sec:hadunc}.

\subsubsection{$b$ quark mass effects on the 4-jet ratio}\label{sec:Rbl4}

The study of the $b$ quark mass in the 4-jet rate was performed by \delphi\ in 
\cite{delphiEPS99}. The analysis uses an observable defined in a similar 
manner to $R_3^{b\ell}$: 
\begin{equation}\label{eq:Rbl4}
R_{4}^{b{\ell}}(y_c)  =  \frac{R^b_4(y_c)}{R^\ell_4(y_c)}
\end{equation}
where $R^b_4(y_c)$ and $R^\ell_4(y_c)$ are the $b$-quark and light-quark
4-jet rates as defined in \eqn{eq:Rqnj}.

The analysis was similar to the $R_3^{b\ell}$ and also reveals a clear 
dependence of the 4-jet rate on the quark mass. However $R_4^{b\ell}$ has 
only been computed at LO level (see \sect{sec:anacal}).
The result of the analysis are shown in \sect{sec:R34comp}.




\subsection{Analytical calculations}\label{sec:anacal}

The next-to-leading order (NLO) matrix element (ME)
calculation for the process $e^+e^- \rightarrow 3$ jets, with
complete quark mass effects, has been performed independently by three 
groups~\cite{Rodrigo:1999qg,Brandenburg:1997pu,Nason:1997nw}.
These predictions are in agreement with each other and were successfully 
used in the measurements of the bottom quark mass far above 
threshold~\cite{Abreu:1997ey,Brandenburg:1999nb,delphiEPS99,alephEPS99}
and in the precision tests of the universality of the strong
interaction~\cite{Abreu:1997ey,Abbiendi:1999fs,Abe:1998kr} at the $Z$-pole.
Instead, only leading order (LO)
predictions for 4-jet final states with heavy quarks are
available~\cite{noi,bmm} at present.

In this Section we discuss in detail the ME
calculation for the process $e^+e^- \rightarrow 4$ jets
with quark mass corrections.
A procedure to estimate the theoretical uncertainty
of the LO calculation, i.e. of the expected higher order
corrections, is also described.
First, we present and test this procedure in the
3-jet case, where the recent available NLO corrections can
be compared with the LO calculation.
Then, these results are extrapolated to the 4-jet case for which 
the higher order corrections are estimated.

Since quarks are {\it not free} particles it seems natural
to treat their mass like a coupling constant. In other words,
we can work with quark masses defined in several renormalization schemes.
As was already pointed out in \sect{sec:varcho}, 
the physical result cannot depend on which mass definition is used
in the ME calculation, but if at a fixed order in perturbation theory
the ME calculation gives different results for different quark mass
definitions, this difference should come from higher order corrections.
Therefore, at a given order, the spread of the results for different
mass definitions can be taken as an estimate of the theoretical 
uncertainty of the calculation, i.e.\ as an estimate of higher order 
corrections.

In the following we consider the ME calculation for
the production of bottom quarks through the processes 
$e^+e^- \rightarrow$ 3-jets, 4-jets in two different schemes:
the perturbative pole scheme and the running scheme.
In the former, the calculation is performed with the 
perturbative pole mass $M_b\sim 5$~GeV.
In the later, we use the $\overline{MS}$ scheme running
mass $m_b(\mu)$, normalized at the center of mass energy
of the collision, $m_b(\mz)\sim 3$~GeV at \lepone\ energies.
The two results define a band which can be taken as an
estimate of the higher order corrections.

\begin{figure}
\begin{center}
\includegraphics[width=15cm]{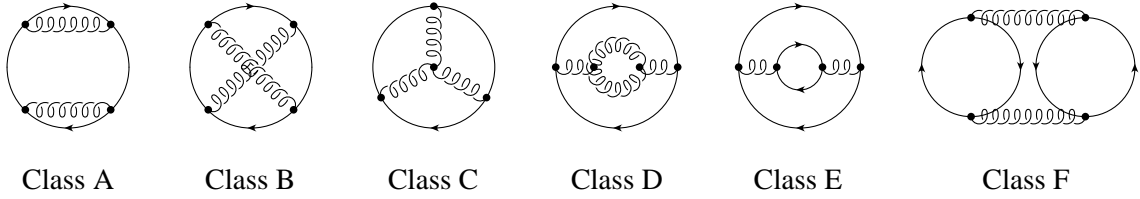} 
\caption{Feynman diagrams relevant for the processes 
$e^+e^- \rightarrow$ 3-jets, 4-jets.}
\label{fig:bubbles}
\end{center}
\end{figure}

At the NLO the bottom quark 3-jet cross section receives
contributions from one-loop corrected three parton final states,
$e^+e^- \rightarrow Z, \gamma^* \rightarrow b \bar{b} g$,
and tree level four parton final states,
$e^+e^- \rightarrow Z, \gamma^* \rightarrow b \bar{b} g g,
b \bar{b} b \bar{b}, b \bar{b} q \bar{q}$, with $q \ne b$. 
These contributions can be handled and classified through the
different cuts to the bubble diagrams of \fig{fig:bubbles}
giving rise to three and four parton final states.

We first evaluate the ratio of three-jet rates defined by \eqn{eq:Rbl3},
which can be interpreted as the measure of the suppression of gluon
radiation off $b$-quarks with respect to gluon radiation off light
quarks, $\ell=u,d,s$.
In this ratio, most of the electroweak corrections cancel out.

\begin{figure}
\begin{center}
\includegraphics[width=7.5 cm]{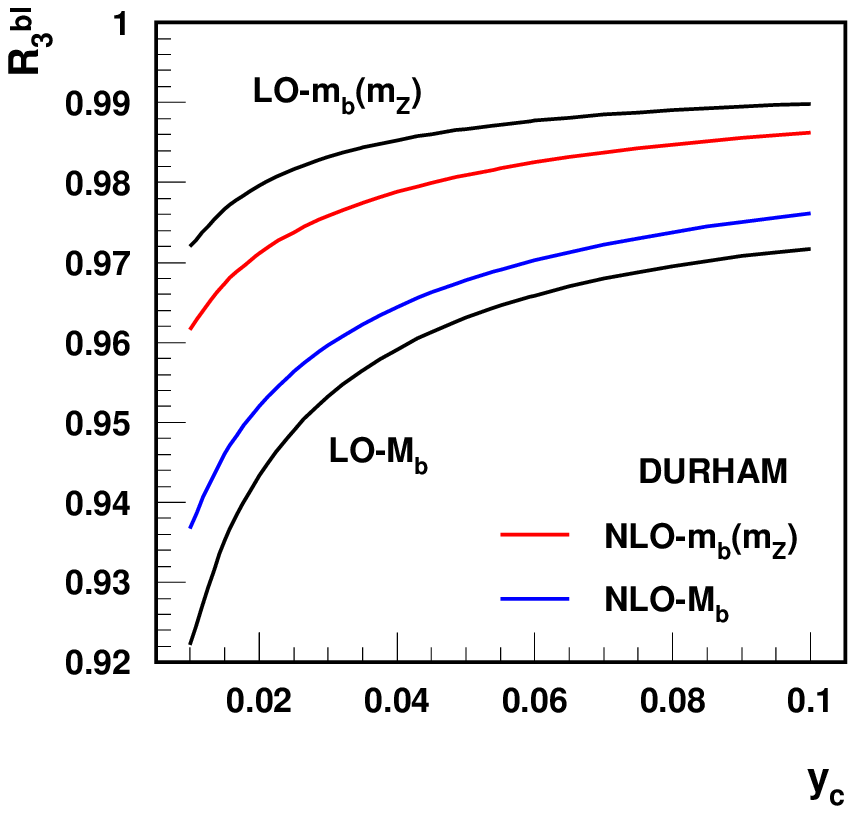}
\includegraphics[width=7.5 cm]{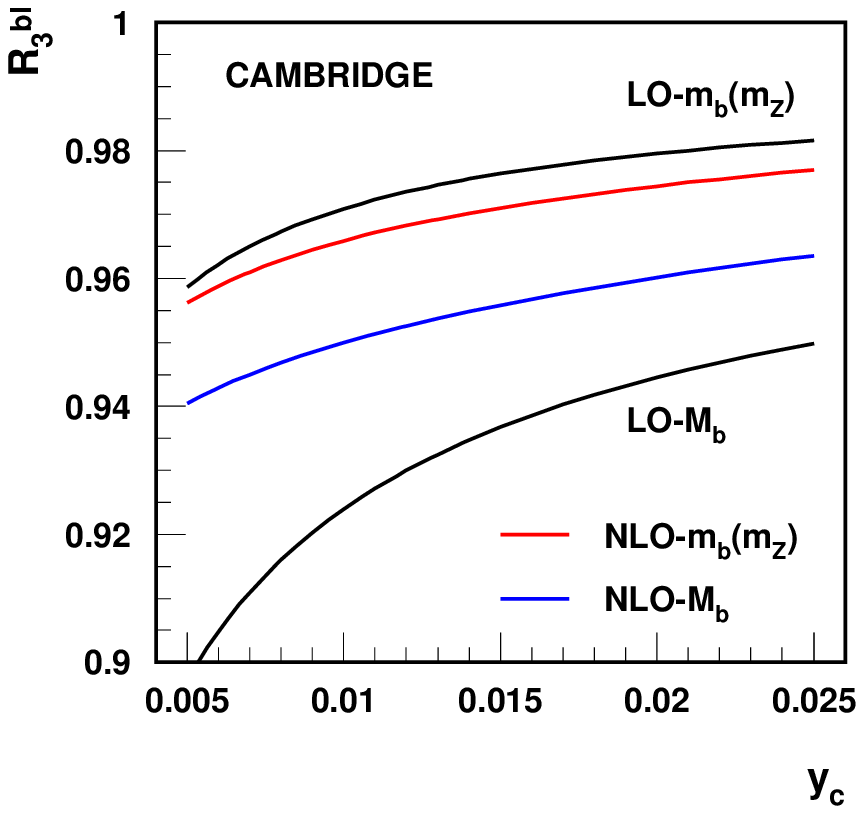}
\caption{The observable $R_3^{b\ell}$ as a function of $y_c$ at NLO for
the \durham\ and \cambridge\ jet-algorithms at the center of mass
energy $\sqrt{s}=\mz$. 
The blue (lower inner) lines give the observable computed at the NLO in 
terms of the pole mass $M_b=5$~GeV.
The red (upper inner) lines show the result for a 
running mass $m_b(\mz)=3$~GeV. In both cases the renormalization scale
is fixed to $\mu=\mz$, with $\as(\mz)=0.118$. For comparison we also
plot in solid lines the LO results for $M_b=5$~GeV (LO-$M_b$) and
$m_b(\mz)=3$~GeV (LO-$m_b(\mz)$).}
\label{fig:r3bl}
\end{center}
\end{figure}

In \fig{fig:r3bl}, the $R_3^{b\ell}$ observable is presented
at NLO for the \durham\ and the \cambridge\ algorithms at the $Z$-peak
energies, $\sqrt{s}=\mz$, in the running mass (NLO-$m_b(\mz)$) and
the pole mass (NLO-$M_b$) schemes. For comparison, the LO results
-- LO-$m_b(\mz)$ and LO-$M_b$ -- are also plotted. At a fixed order,
the band defined by the results in both schemes is taken as our
estimate of the theoretical uncertainty of the calculation at this order.
As one would naturally expect, the width of this band is 
reduced at the NLO with respect to the LO result,
roughly by a factor two in \durham\ or even more in \cambridge.
Furthermore, we found that the two LO predictions
bound the NLO results in both algorithms. This 
suggests that higher order contributions cannot be
too large and may be bounded by the lower order results.

Next we consider the process $e^+e^- \rightarrow 4$ jets.
The ratio of four-jet rates is defined in \eqn{eq:Rbl4}.
At LO, only four parton final state cuts to the
bubble diagrams of \fig{fig:bubbles} have to be
considered. Let's interpret for the moment, $R_4^{b\ell}$
as the suppression of gluon radiation from {\it primary} $b$-quarks
in four-jet events.
By primary quarks we mean that the flavour of an
event is defined by the flavour of the quark directly coupled
to the $Z$ or $\gamma^*$ bosons. This implies that events where a
bottom-antibottom pair is produced from gluon radiation off
a light quark pair (the so-called {\it gluon splitting into $b {\bar b}$}) are
considered as light events, even though they actually involve $b$ quarks,
and their contribution is added to the denominator of \eqn{eq:Rbl4}).
In $R_3^{b\ell}$, the contribution from gluon splitting into $b {\bar b}$ was
calculated to be small, resulting in effects of the order of a few permil.
This is not the case for four-jet events, where larger contributions 
can justify, 
as will be explained in \sect{sec:gbbimpact}, the consideration of a different 
convention for the jet rate ratio defined above, in which the 
numerator receives contributions from all events involving $b$-quarks.

At LO and for the definition with the primary quark convention, the $R_4^{b\ell}$ ratio can 
be parameterized in the following way 
\begin{equation}
R_4^{b\ell} = 1 + \frac{m_b^2}{s}
\left[\sigma_V(s) H_V\left(\frac{m_b^2}{s},y_c\right)
+ \sigma_A(s) H_A\left(\frac{m_b^2}{s},y_c\right)
\right]~, 
\end{equation}
were $m_b$ can be either the pole mass, $M_b$, or the running mass at some
renormalization scale, $m_b(\mu)$, typically $\mu=\sqrt{s}$, of the
bottom quark,
$\sigma_{V,A}$ is a function of the vector (axial-vector)
couplings of the quarks to the $Z$-boson and the photon, and
$H_{V,A}$ gives the behaviour as a function of the resolution parameter $y_c$,
and can also contain a small residual dependence on the ratio $m_b^2/s$.

\begin{figure}
\begin{center}
\includegraphics[width=15 cm]{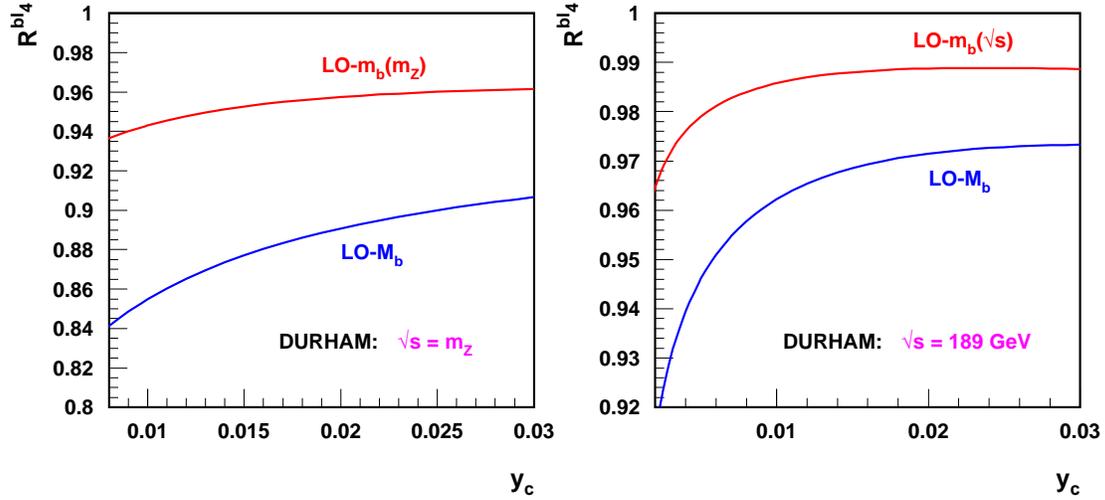}
\caption{The observable $R_4^{b\ell}$ as a function of $y_c$ at LO
in the \durham\ algorithm for $\sqrt{s}=\mz$ and $\sqrt{s}=189$~GeV.}
\label{fig:r4bl}
\end{center}
\end{figure}

In \fig{fig:r4bl}, the $R_4^{b\ell}$
ratio is shown at LO in the \durham\ algorithm at
the center of mass energies $\sqrt{s}=\mz$ and $\sqrt{s}=189$~GeV.
For the same center of mass energy,
the suppression of gluon radiation from $b$-quarks is a larger
effect in four-jet events than in three-jet events. 
At $\sqrt{s}=\mz$, it amounts to roughly $10\%$, which also corresponds
to the difference between the two LO predictions, LO-$m_b(\mz)$
and LO-$M_b$, taken as the theoretical uncertainty of the LO prediction. 
If the behaviour for three-jet events really could be extrapolated to the 
four-jet case, we would expect that this difference be reduced hopefully 
by half, if the still uncalculated NLO corrections were included. 
At $\sqrt{s}=189$~GeV, we get a plot that is roughly scaled by a 
factor $4$ with respect to the result at the \lepone\ energies. 
But even in this case, the theoretical uncertainty, i.e. the difference 
between the two LO predictions, can be as large as $5\%$ for small 
values of the jet resolution parameter $y_c$.

\subsection{Comparisons for $R_3^{b\ell}$ and $R_4^{b\ell}$ at 
LEP1 and LEP2}\label{sec:R34comp}

Following the procedure described above, we compare at both \lepone\ and
\leptwo\ energies, the double jet rate ratios $R_3^{b\ell}$ and $R_4^{b\ell}$
obtained with the 
different Monte-Carlo approaches studied (\ariadne, \pythia, 
\herwig  ~and \fourjphact) with the analytic calculations, and, in the 
case of \lepone, with the data. All these comparisons are done at the
level of partons (hadronization and non-perturbative effects are considered 
in \sect{sec:hadunc}). In 
Figures~\ref{3jetpt}, ~\ref{3jethw} and ~\ref{3jetar} are 
indicated both the LO matrix element results for  $m_b=$ 3 and 5 GeV and 
the NLO results for $m_b=$ 3 GeV, in the case of the 3-jet rates at \lepone. In
Figures~\ref{4jetpt}, ~\ref{4jethw}, ~\ref{4jetar} and 
~\ref{4jetfj} are
indicated only the LO results for $m_b=$ 3 and 5 GeV for the 4-jet rates at 
\lepone.  In Figures~\ref{4jetpt6125200}, ~\ref{4jetpt6131200} and 
~\ref{4jetar200} are 
indicated the LO results for $m_b=$ 3 and 5 GeV for the 4-jet rates at 
\leptwo.  The band defined by the pair of LO curves is in each case taken, 
conservatively, to represent the theoretical uncertainty in the 
ratios $R_3^{b\ell}$ and $R_4^{b\ell}$, as was explained in 
\sect{sec:anacal}.

\subsubsection{Three jet rates at $\sqrt{s}=\mz$}\label{sec:3jMZ}

\begin{figure}[htbp]
\begin{center}
  \begin{tabular}{ll}
  \includegraphics[width=14cm]{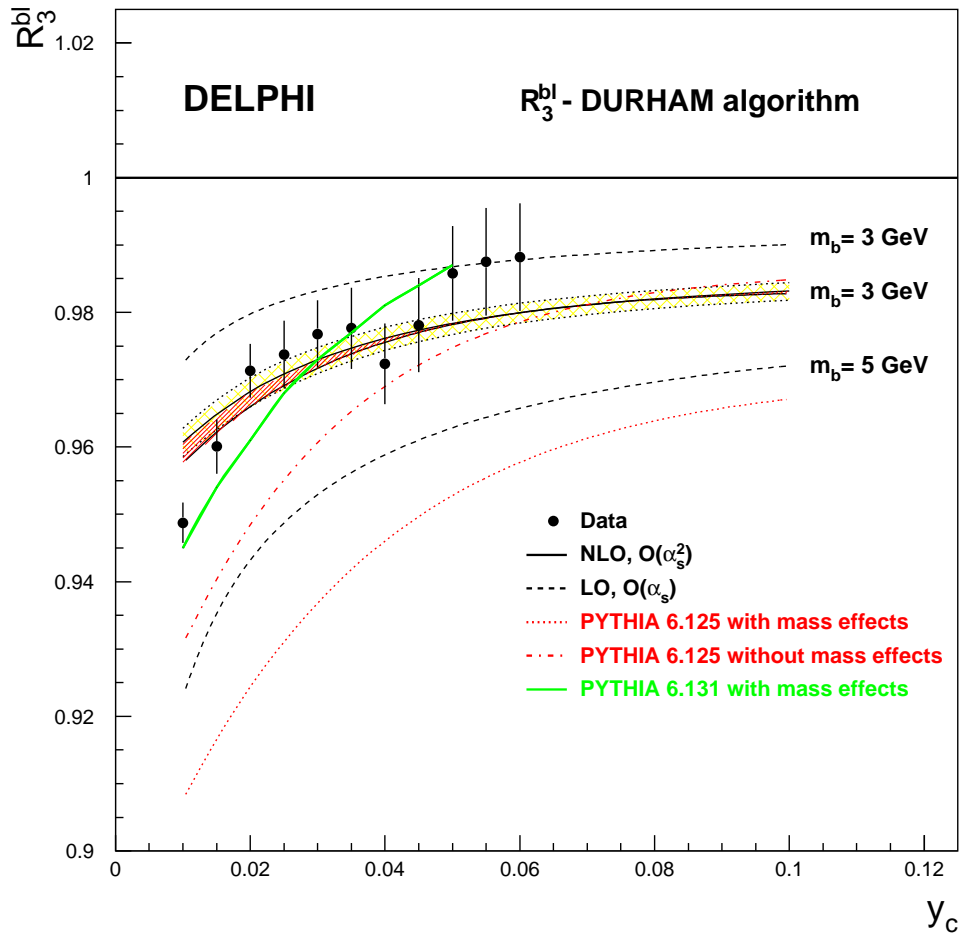} 
\end{tabular}
\caption{$R_3^{b\ell}$ double ratios at $\sqrt s = \mz$ for \pythia\ 6.125, with 
mass switch on and off, and for \pythia\ 6.131, with mass switch on} 
\label{3jetpt}
\end{center}
\end{figure}

\begin{figure}[htbp]
\begin{center}
  \begin{tabular}{ll}
  \includegraphics[width=14cm]{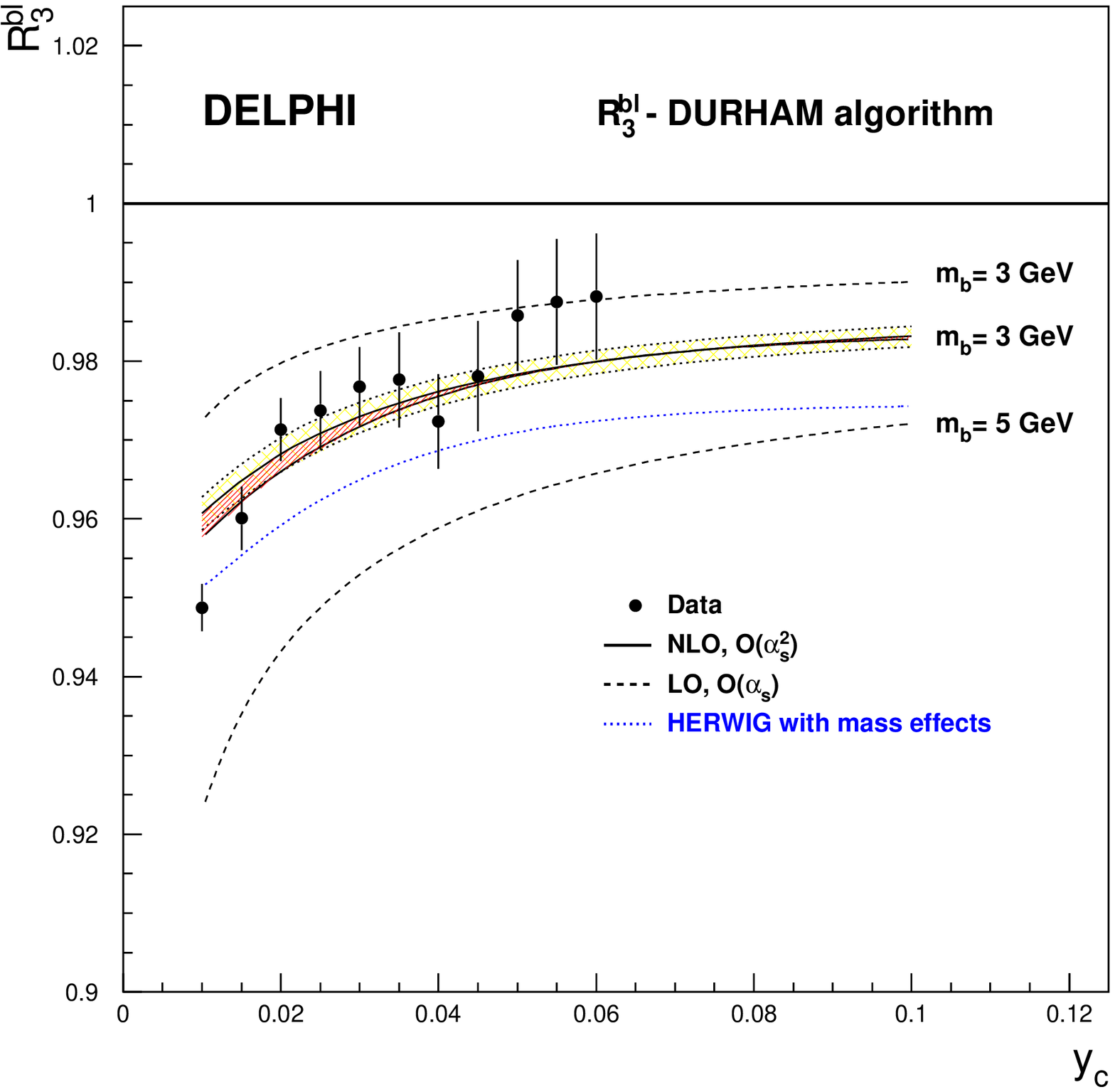} 
\end{tabular}
\caption{$R_3^{b\ell}$ double ratios at $\sqrt s = \mz$ for \herwig\ 5.8, with 
mass effects on.}
\label{3jethw}
\end{center}
\end{figure}

\begin{figure}[htbp]
\begin{center}
  \begin{tabular}{ll}
  \put(-200.0,-8.0){\includegraphics[width=14cm]{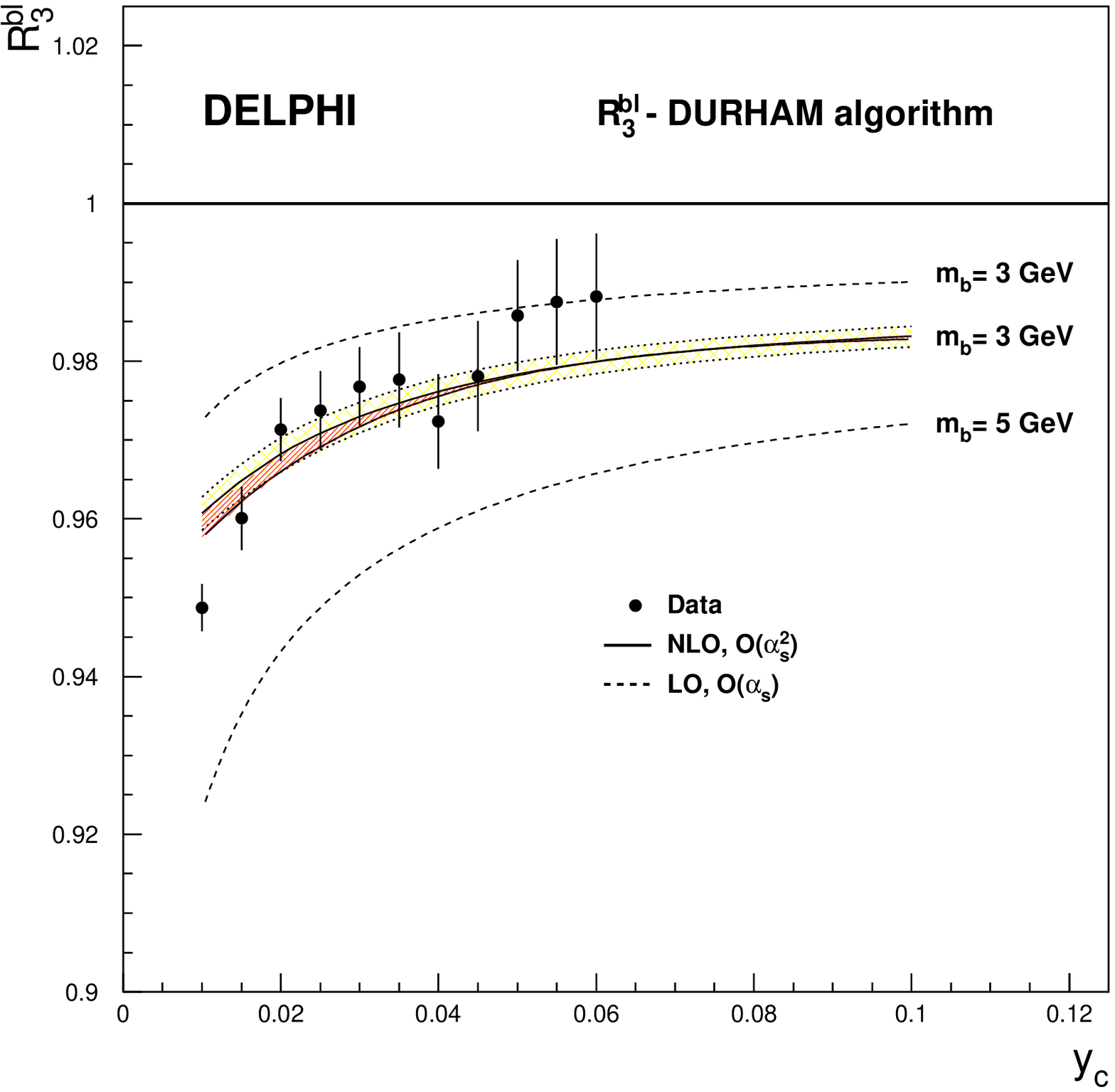}
} 
  \put(-200.0,-8.0){\includegraphics[width=14cm]{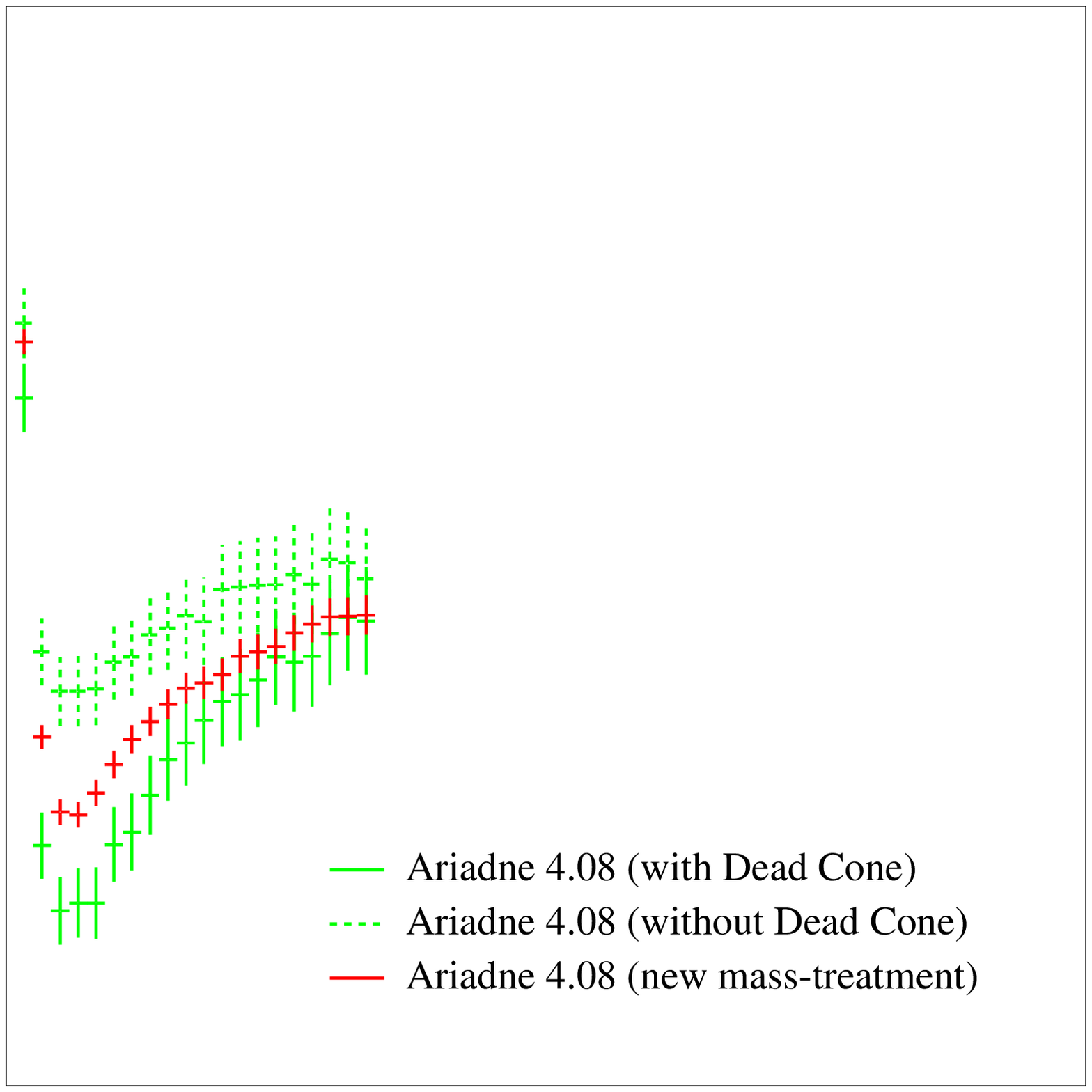}
} 
\end{tabular}
\caption{$R_3^{b\ell}$ double ratios at $\sqrt s = \mz$ for \ariadne\ 4.08,
with dead cone switch on and off, 
and with the new mass treatment described in \sect{sec:arihvy}.}
\label{3jetar}
\end{center}
\end{figure}

In Figures~\ref{3jetpt}, ~\ref{3jethw} and ~\ref{3jetar}   
are shown the comparisons
for $R_3^{b\ell}$ at $\sqrt{s}=\mz$ between the analytic calculations, the 
data\cite{mbatmz,delphiEPS99} and the Monte-Carlo generators 
\pythia, \herwig, \ariadne\ and \apacic, respectively. 

The \pythia\ comparison in
\fig{3jetpt} shows three curves corresponding to:
\begin{itemize}
\item
The initial treatment, with mass effects only present in the limitation of 
the phase-space available in $b \rightarrow bg$ branchings in the parton 
shower, but not in the kinematics of these branchings, and with 
massless matrix 
elements used in the matching procedure applied to the 3 jets generated
(\jetset\ versions $\leq$ 7.3, or any present version with the switch
{\tt MSTJ(47)} set to 1 to turn off subsequent additional mass effects).
\item
Mass effects present in the limitation of 
the phase-space available in $b \rightarrow bg$ branchings in the parton 
shower, but not in the kinematics of these branchings, and with
massive matrix elements used in the matching procedure applied to 
the 3 jets generated (\pythia\ versions 5.7-6.125).
\item
Mass effects present in the limitation of 
the phase-space available in $b \rightarrow bg$ branchings in the parton 
shower, in the kinematics of these branchings, and with
massive matrix elements used in the matching procedure applied to 
the 3 jets generated (\pythia\ versions $\geq$ 6.130).
\end{itemize}
As can be seen the recent changes consisting in introducing mass effects 
at all stages in the treatment (see \sect{sec:pyglhvy}) result in a very good
behaviour. In the prior versions which are presently still used by a 
majority of \lep\ experiments, a behaviour almost as good 
is obtained by using massless
expressions for the matching procedure of the generated 3 jets (switch 
{\tt MSTJ(47)} set to 1). 

The \herwig\ comparison in
\fig{3jethw} shows a single curves corresponding to
the massive options. A reasonable description is evident, even if the
predicted rate is slightly lower than the data and NLO result.

The \ariadne\ comparison in
\fig{3jetar} shows three curves corresponding to
using the optional extra dead cone suppression available ({\tt MSTA(19)}=1),
and to the new treatment of heavy masses described in
\sect{sec:arihvy}, in which the full leading order massive 
matrix element was introduced to describe the branching of the first 
gluon emitted in $q {\bar q}$ events.
As can be seen the rate is too low when no 
optional extra dead cone suppression is used, and then gets even worse
when it is used. The new treatment of heavy masses is a clear improvement
compared to the old treatment with the dead cone suppression option turned on.
But the best behaviour is still achieved when no mass corrections are
used.

\subsubsection{Four jet rates at $\sqrt{s}=\mz$}\label{sec:4jMZ}

\begin{figure}[htbp]
\begin{center}
  \begin{tabular}{ll}
  \put(-200.0,-8.0){\includegraphics[width=14cm]{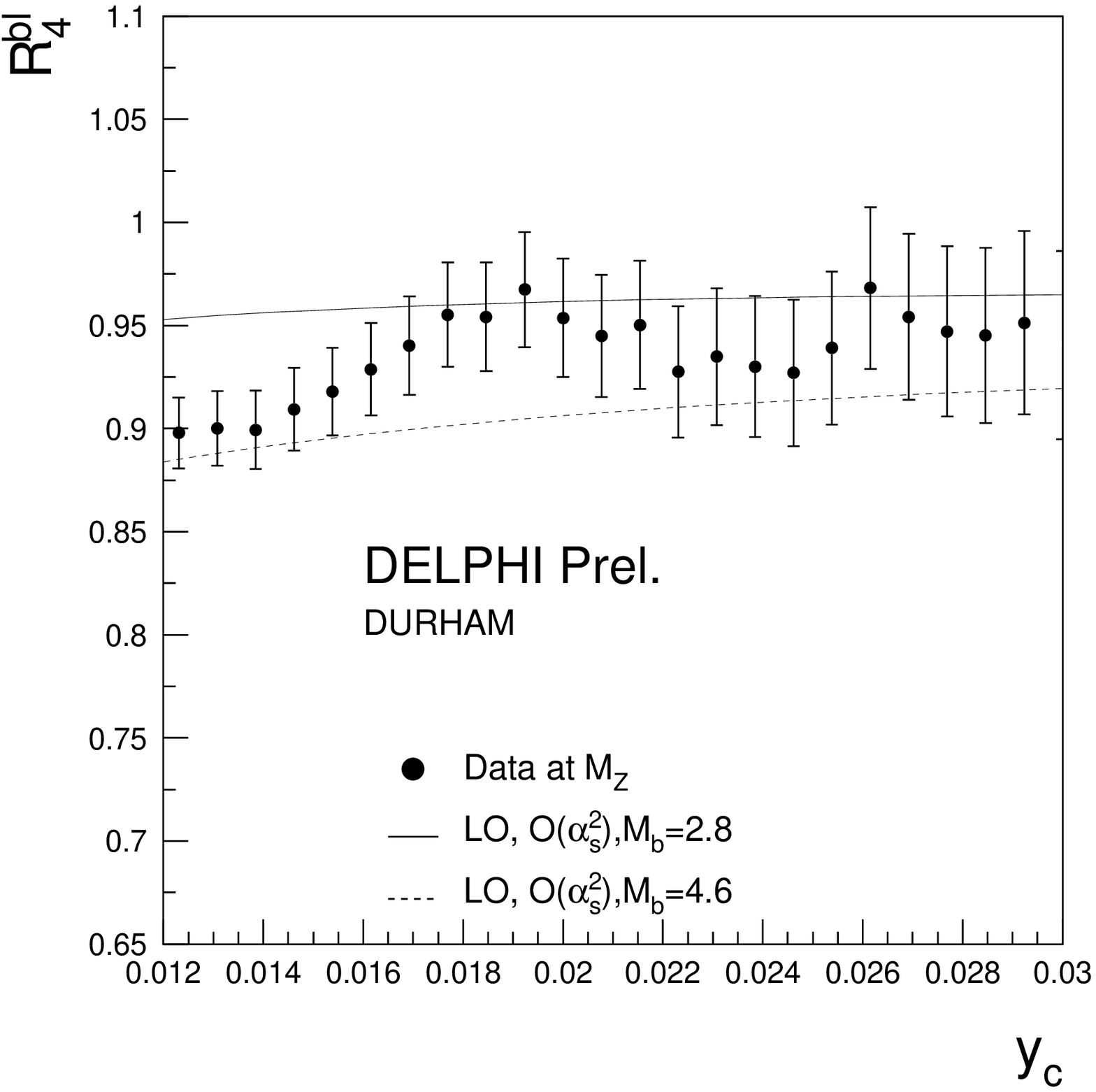}
} 
  \put(-200.0,-8.0){\includegraphics[width=14cm]{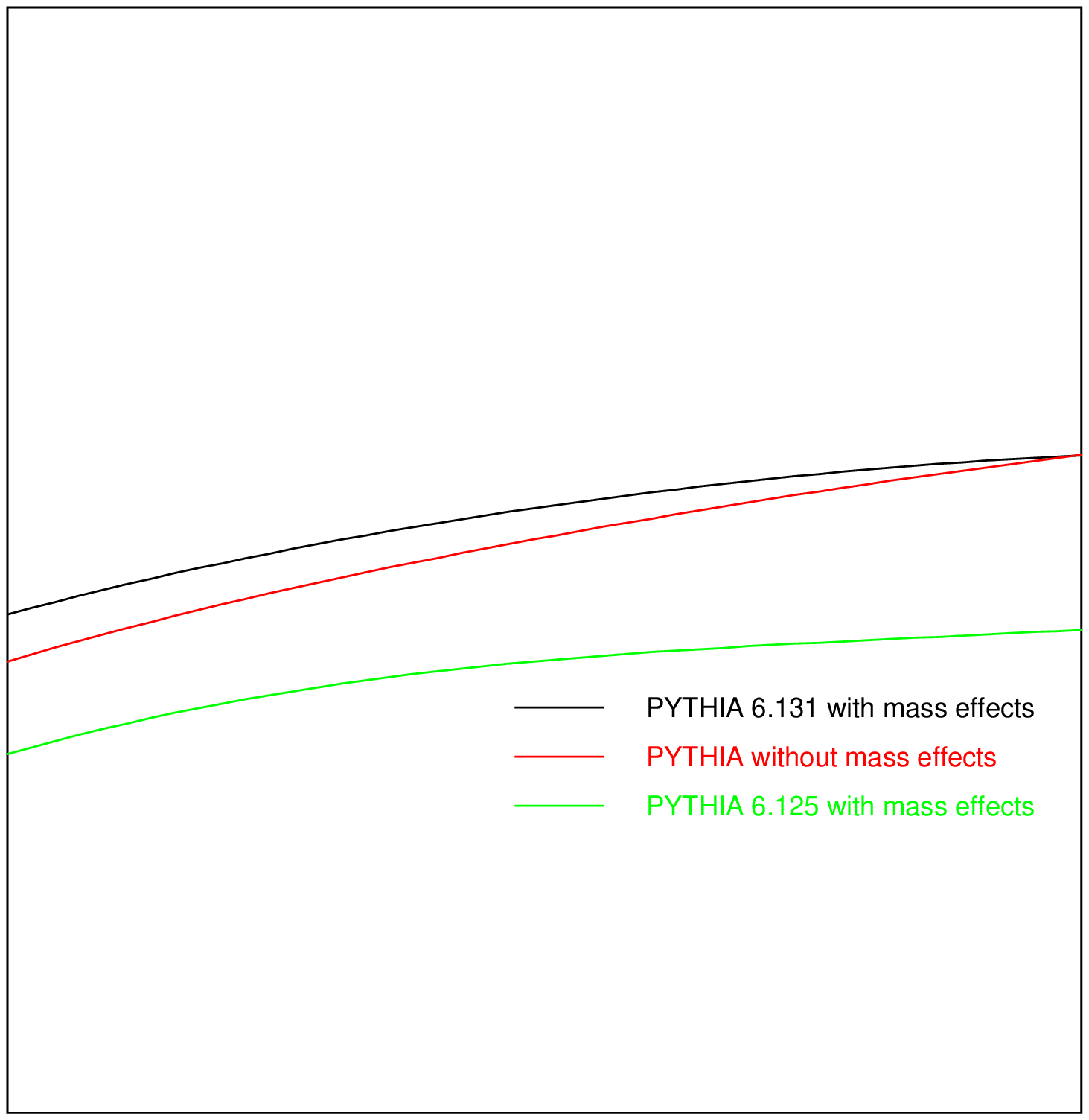}
} 
\end{tabular}
\caption{$R_4^{b\ell}$ double ratios at $\sqrt s = \mz$ for \pythia\ 6.125, with 
mass switch on and off, and for \pythia\ 6.131, with mass switch on.} 
\label{4jetpt}
\end{center}
\end{figure}

The \pythia\ comparison in
\fig{4jetpt} shows three curves corresponding to the three cases
described above.
As can be seen the recent changes consisting in introducing mass effects 
at all stages in the treatment also improve the description, as for 
the three jet case, but still results in a rate which is too low
by about 5-7\% with respect to data and to the LO 
matrix element predictions. The same trend is seen as for the three 
jet case that in the prior versions still presently used by a 
majority of \lep\ experiments, the best behaviour is obtained by using massless
expressions for the matching procedure of the generated 3 jets (switch 
{\tt MSTJ(47)} set to 1). 

\begin{figure}[htbp]
\begin{center}
  \begin{tabular}{ll}
  \put(-200.0,-8.0){\includegraphics[width=14cm]{delphi_r4bl.eps}
} 
  \put(-200.0,-8.0){\includegraphics[width=14cm]{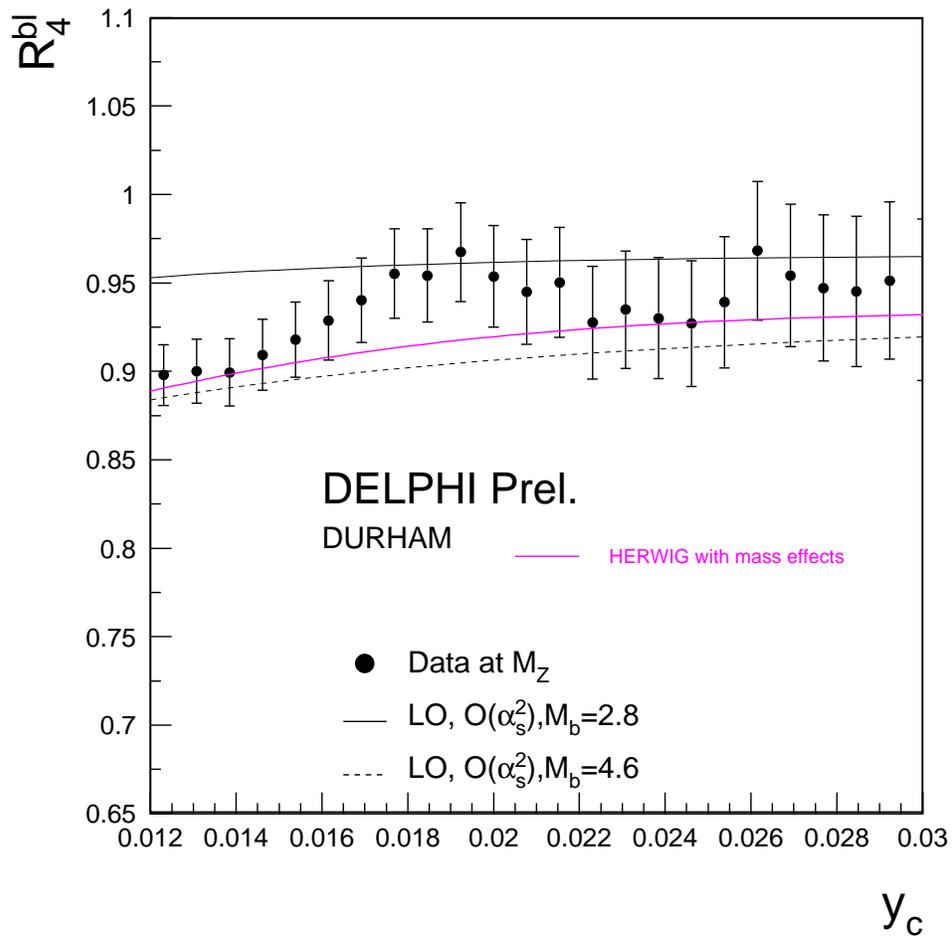}
} 
\end{tabular}
\caption{$R_4^{b\ell}$ double ratios at $\sqrt s = \mz$ for \herwig\ 5.8, with 
mass effects on.} 
\label{4jethw}
\end{center}
\end{figure}

The \herwig\ comparison is shown in \fig{4jethw}. A description compatible
with the data and with the LO matrix element prediction can be seen.

\begin{figure}[htbp]
\begin{center}
  \begin{tabular}{ll}
  \put(-200.0,-8.0){\includegraphics[width=14cm]{delphi_r4bl.eps}
} 
  \put(-200.0,-8.0){\includegraphics[width=14cm]{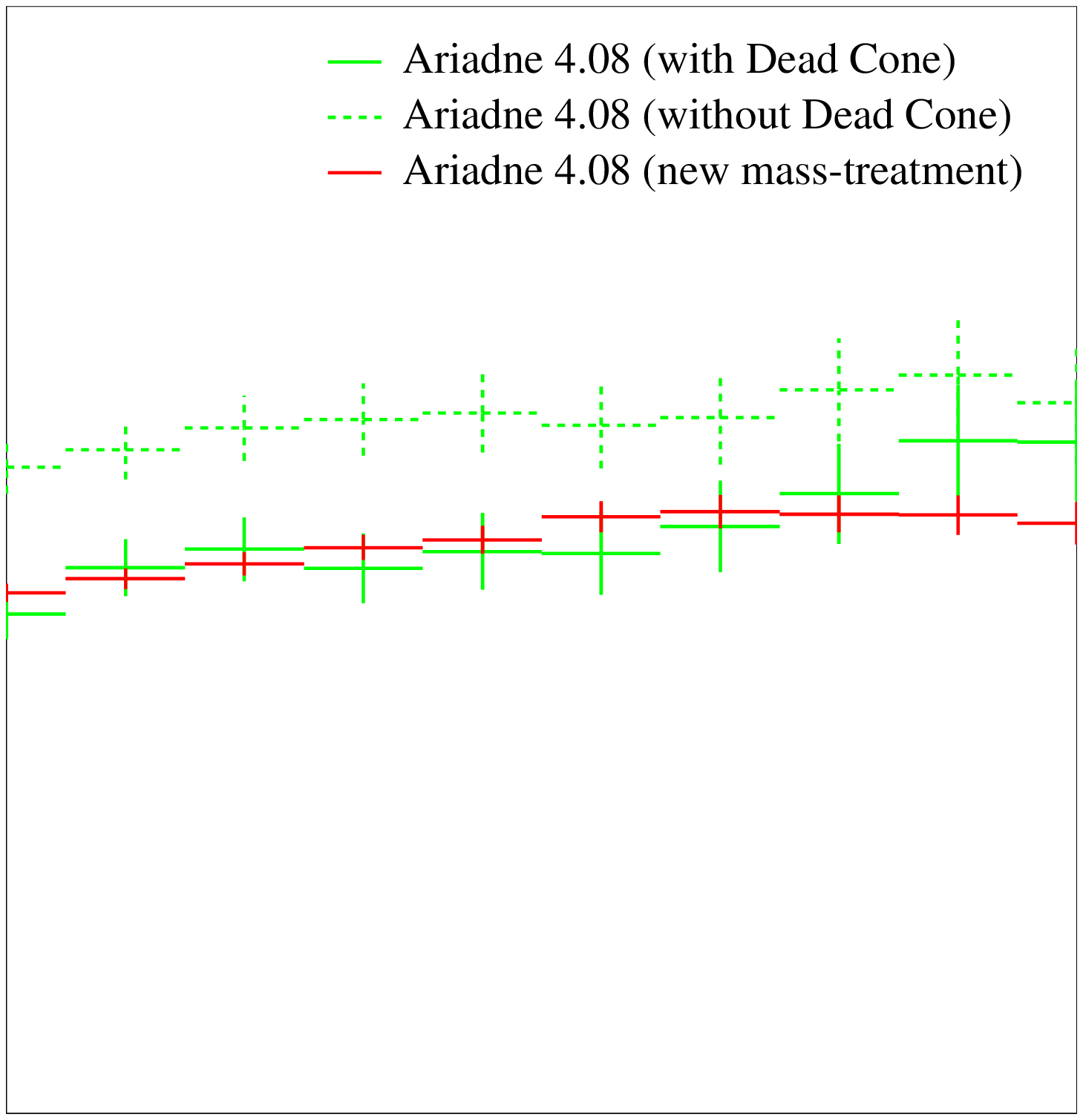}
} 
\end{tabular}
\caption{$R_4^{b\ell}$ double ratios at $\sqrt s = \mz$ for \ariadne\ 4.08,  
with dead cone switch on and off, and with the new mass treatment 
described in \sect{sec:arihvy}.}
 
\label{4jetar}
\end{center}
\end{figure}

The \ariadne\ comparison in
\fig{4jetar} shows three curves corresponding to
using the optional extra dead cone suppression available ({\tt MSTA(19)}=1), 
and to the new treatment of heavy masses described in
\sect{sec:arihvy}.
As can be seen, contrary to the three jet case, the overall behaviour 
is quite reasonable when no optional extra dead cone suppression is used, 
but somewhat too low when it is used.
The behaviour with the newly changed treatment of heavy masses
does not improve the situation significantly: the best behaviour is 
still achieved when no mass corrections are
used, as in the case of $R_3^{b\ell}$.

\begin{figure}[htbp]
\begin{center}
  \begin{tabular}{ll}
  \includegraphics[width=14cm]{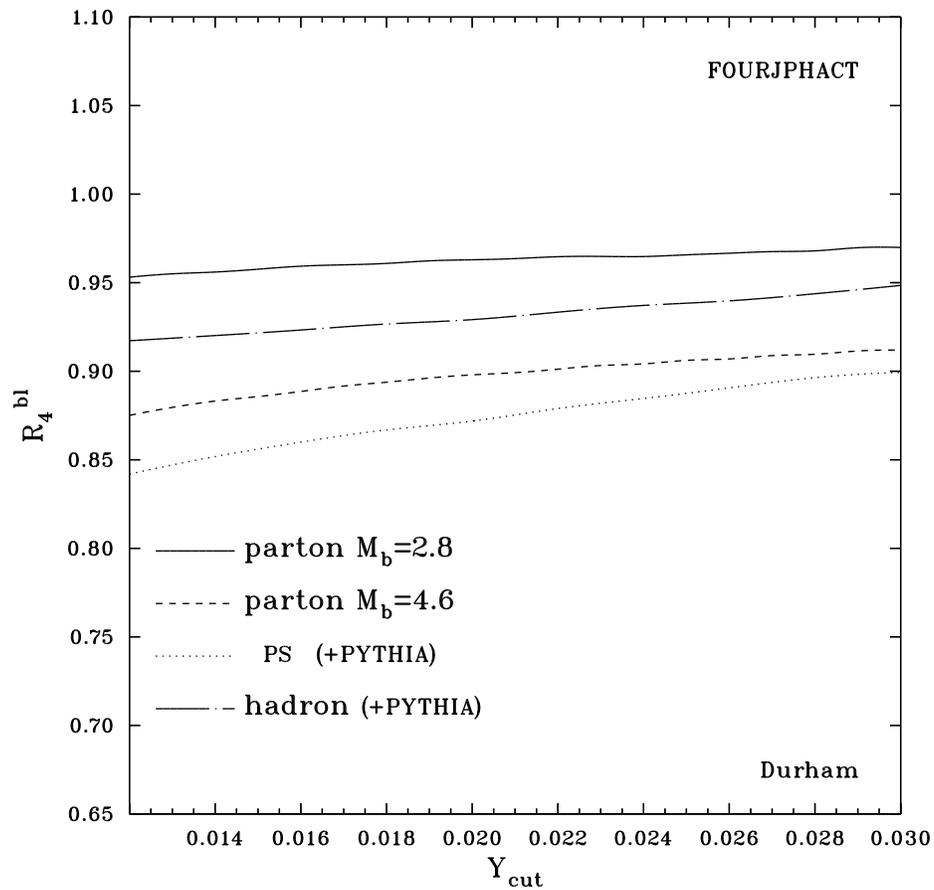} 
\end{tabular}
\caption{$R_4^{b\ell}$ double ratios at $\sqrt s = \mz$ for \fourjphact, with 
mass effects on. The full (dashed) line shows the result at parton level
only, before any subsequent parton showering (see \sect{sec:4jphact}),
for a $b$ quark mass of 2.8 (4.6) GeV. These curves are 
equivalent to the LO ME curves in \fig{4jetpt}. The dotted 
and dashed dotted lines 
shows the result after subsequent PS and hadronisation, and after 
only the subsequent PS, respectively.}
\label{4jetfj}
\end{center}
\end{figure}

Finally the result of an initial investigation of $R_4^{b\ell}$ with the new 
\fourjphact\ program is shown in \fig{4jetfj}. The LO ME curves calculated
within \fourjphact\ are identical to those shown in \fig{4jetpt}, \fig{4jethw}
and \fig{4jetar}. The results after only the subsequent parton shower are also
shown, as calculated starting from the LO ME with 4.6 GeV, and indicate as
expected an exageration of the suppression of gluon radiation from the 
$b$ quark mass. Presumably, if a way could be found in this program
to start the matching
procedure from the LO ME with 2.8 GeV (rather than 4.6 GeV) 
while preserving the corresponding 
jet angles, one could perhaps contemplate getting the LO ME + parton shower
only results to lie in the middle of the band of uncertainty defined in
\sect{sec:anacal}, where the NLO results are expected.
The result after both
parton shower and hadronisation are also shown, although
here it is fair to say that the large effect seen from the 
hadronisation is not understood and should be studied more.

\subsubsection{Four jet rates at $\sqrt{s}=189$ GeV}

\begin{figure}[htbp]
\begin{center}
  \begin{tabular}{ll}
  \includegraphics[width=14cm]{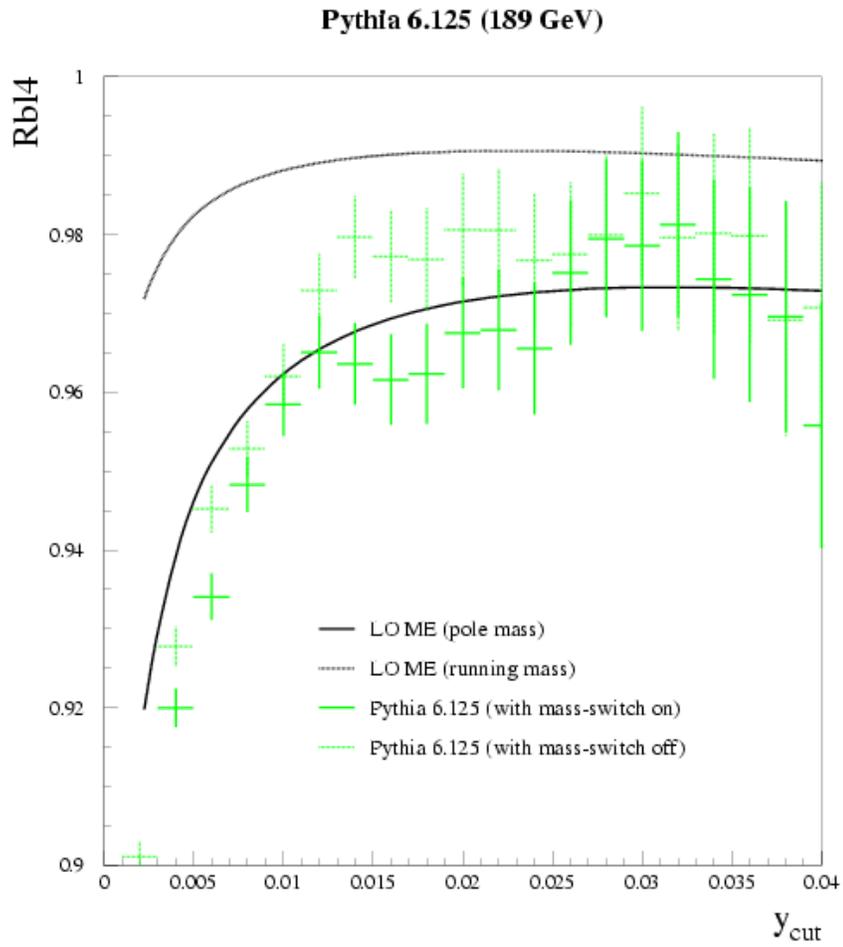} 
\end{tabular}
\caption{$R_4^{b\ell}$ double ratios at 189 GeV for \pythia\ 6.125, with mass switch on
and off} 
\label{4jetpt6125200}
\end{center}
\end{figure}

\begin{figure}[htbp]
\begin{center}
  \begin{tabular}{ll}
  \includegraphics[width=14cm]{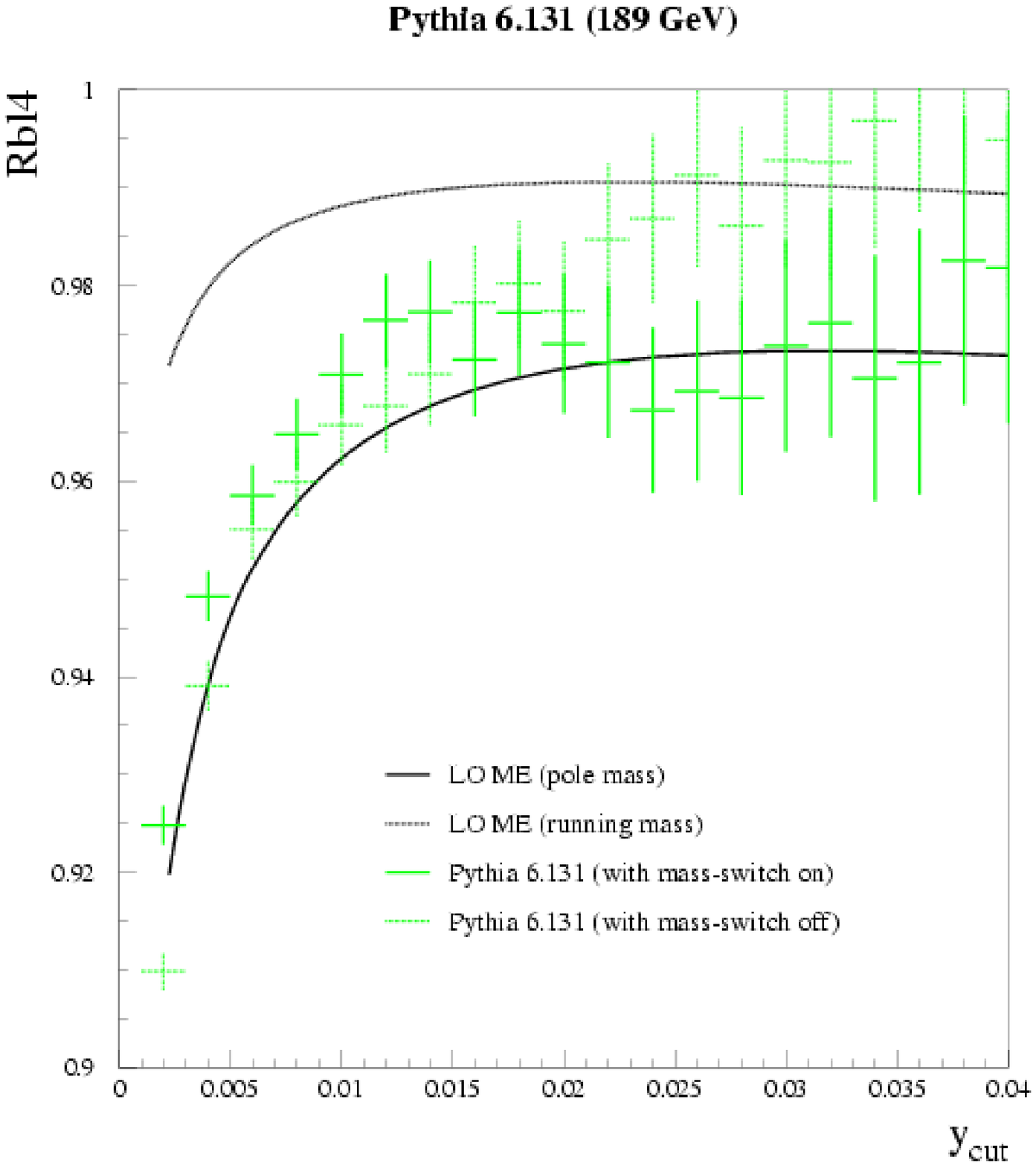} 
\end{tabular}
\caption{$R_4^{b\ell}$ double ratios at 189 GeV for \pythia\ 6.131, with mass switch on
and off} 
\label{4jetpt6131200}
\end{center}
\end{figure}

The \pythia\ 4 jet rate comparison was repeated at $\sqrt{s}=189$ GeV. This is
shown in Figures~\ref{4jetpt6125200} and \ref{4jetpt6131200}. 
A trend similar to that at $\sqrt{s}=\mz$
can be seen. For values of the jet resolution parameter in the range 
$y=0.002-0.008$, even with the most recent treatment consisting in 
introducing mass effects at all stages in the handling of the three jets,
a residual deficit of about 1-3\% with respect to the LO 
matrix element predictions. Also, in the prior versions 
still presently used by a majority of \lep\ experiments, 
the best behaviour is obtained by using massless
expressions for the matching procedure of the generated three jets (switch 
{\tt MSTJ(47)} set to 1). If the massive treatment is used in these prior 
versions is used, a deficit as large as about 2-6\% results in the range
$y=0.002-0.008$ (see \fig{4jetpt6125200}).

\begin{figure}[htbp]
\begin{center}
  \begin{tabular}{ll}
  \includegraphics[width=14cm]{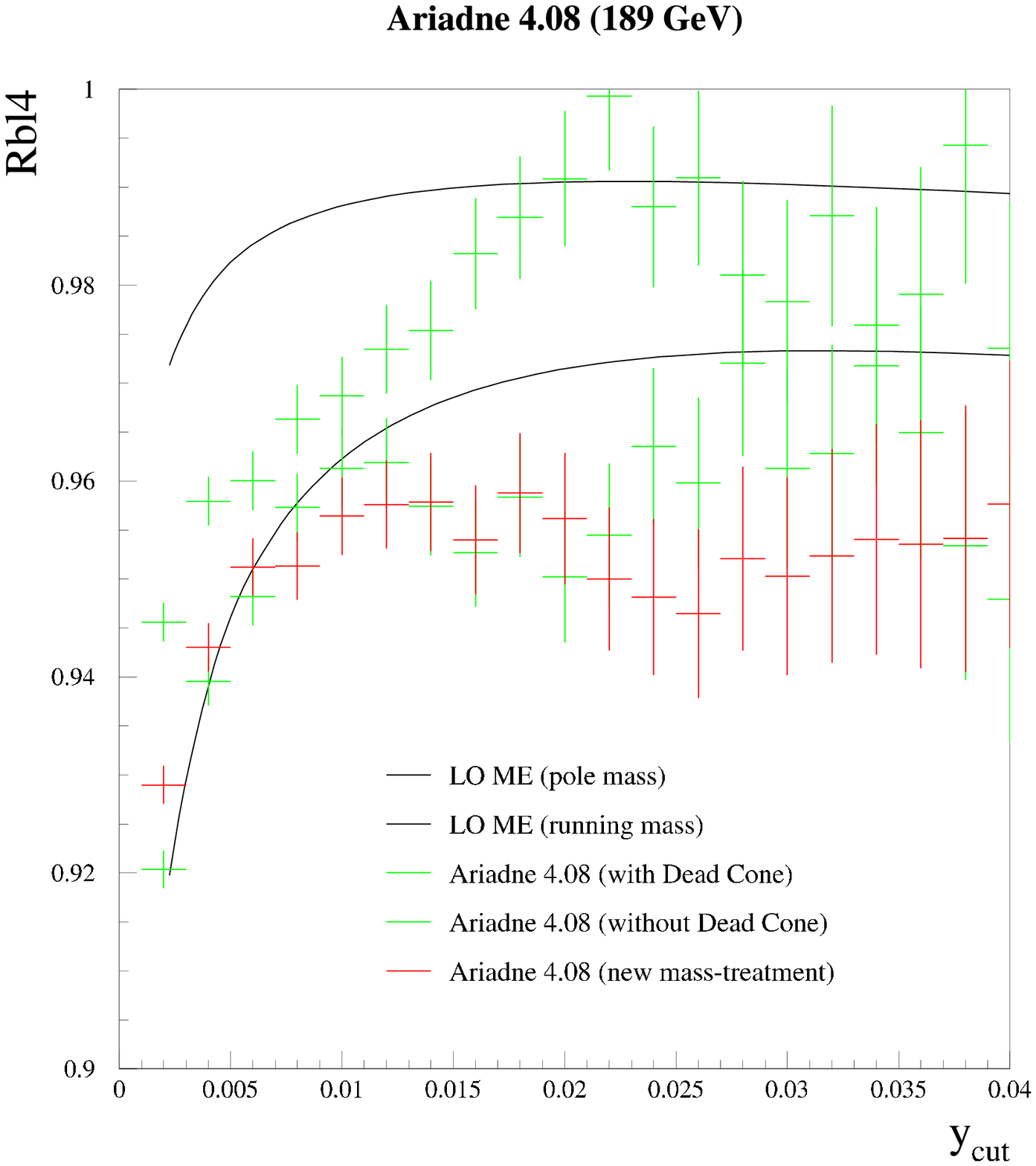} 
\end{tabular}
\caption{$R_4^{b\ell}$ double ratios at 189 
GeV for \ariadne\ 4.08, with dead cone
switch on and off, and with the new mass 
treatment described in \sect{sec:arihvy}.} 
\label{4jetar200}
\end{center}
\end{figure}

The \ariadne\ 4.08 4 jet-rate comparison was also repeated at 
$\sqrt{s}=189$ GeV. This is shown in \fig{4jetar200}. The same trend
is seen as for the 4-jet rate at \lepone\ energies: a quite reasonable behaviour
is observed when no optional extra dead cone suppression is used, 
but the rate is somewhat too low when it is used.
The behaviour with the recently changed treatment of heavy masses described 
\sect{sec:arihvy} is also shown. 
As can been the new treatment does not improve the description 
at \leptwo\ energies.

\subsubsection{Effects of gluon splitting on $R_4^{b\ell}$}
\label{sec:gbbimpact}

An additional issue was raised during the meetings of this working 
group concerning the impact on the 
evaluation of $R_4^{b\ell}$ arising from uncertainties in
processes involving gluon splittings into $b {\bar b}$. 
In the case of the three-jet rate ratio $R_3^{b\ell}$,
the effect was investigated and found to be small. On the contrary, for
four jets, because of the lower rate, and because the two $b$ quarks
emerging from gluon splittings are often resolved, effects are larger.

\begin{figure}[htbp]
\begin{center}
  \begin{tabular}{ll}
  \includegraphics[width=14cm]{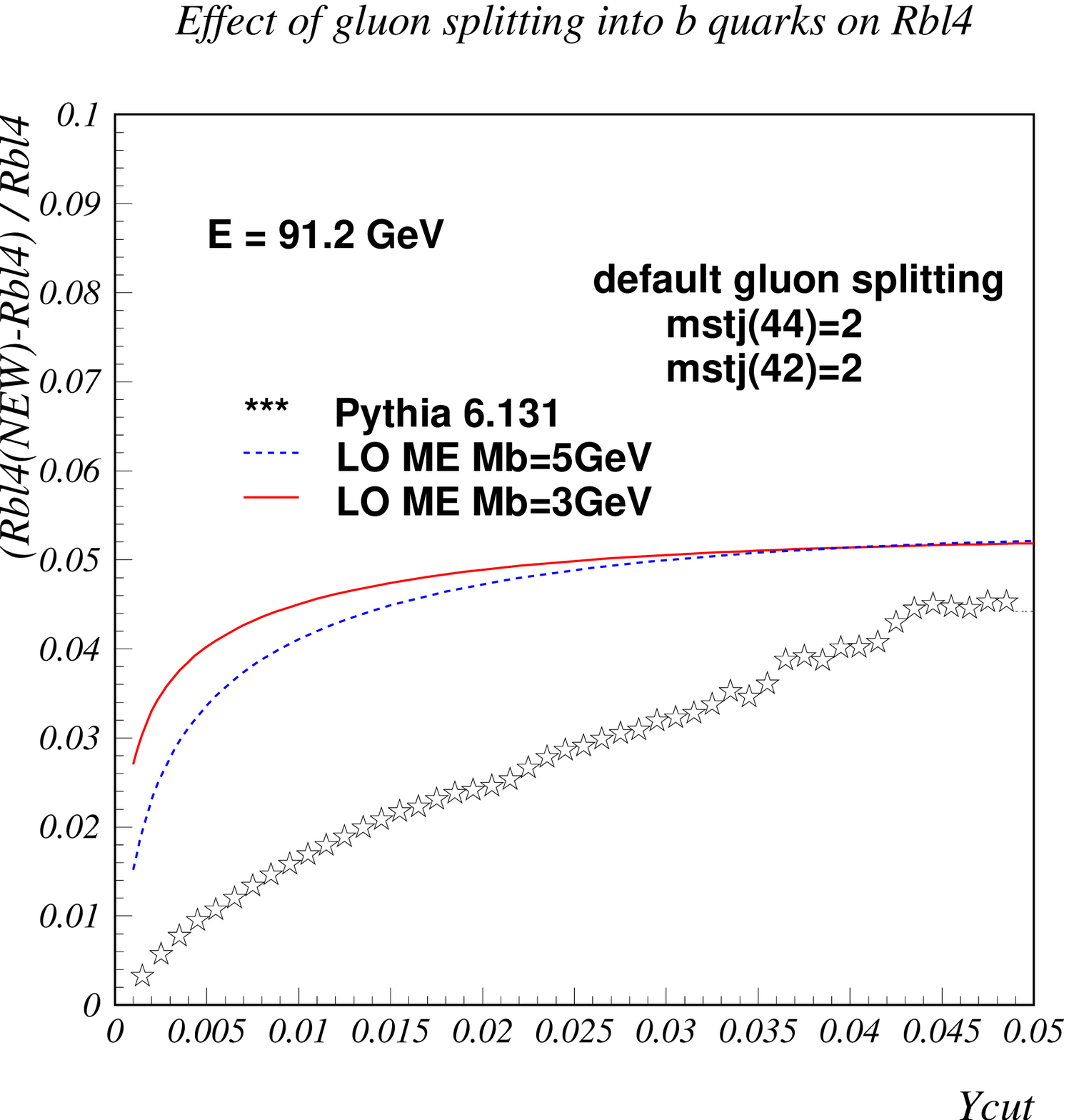} 
\end{tabular}
\caption{Difference of the $R_4^{b\ell}$ double jet ratio at 
$\sqrt s = \mz$ for \pythia\ 6.131 (shown with stars) 
in the NEW and standard definitions, for the default settings with
unmodified rate of gluon splitting into $b {\bar b}$, 
corresponding to {\tt MSTJ(44)}={\tt MSTJ(42)}=2, and in the LO matrix element 
calculation.}
\label{delta:rbl4:91:22}
\end{center}
\end{figure}

\begin{figure}[htbp]
\begin{center}
  \begin{tabular}{ll}
  \includegraphics[width=14cm]{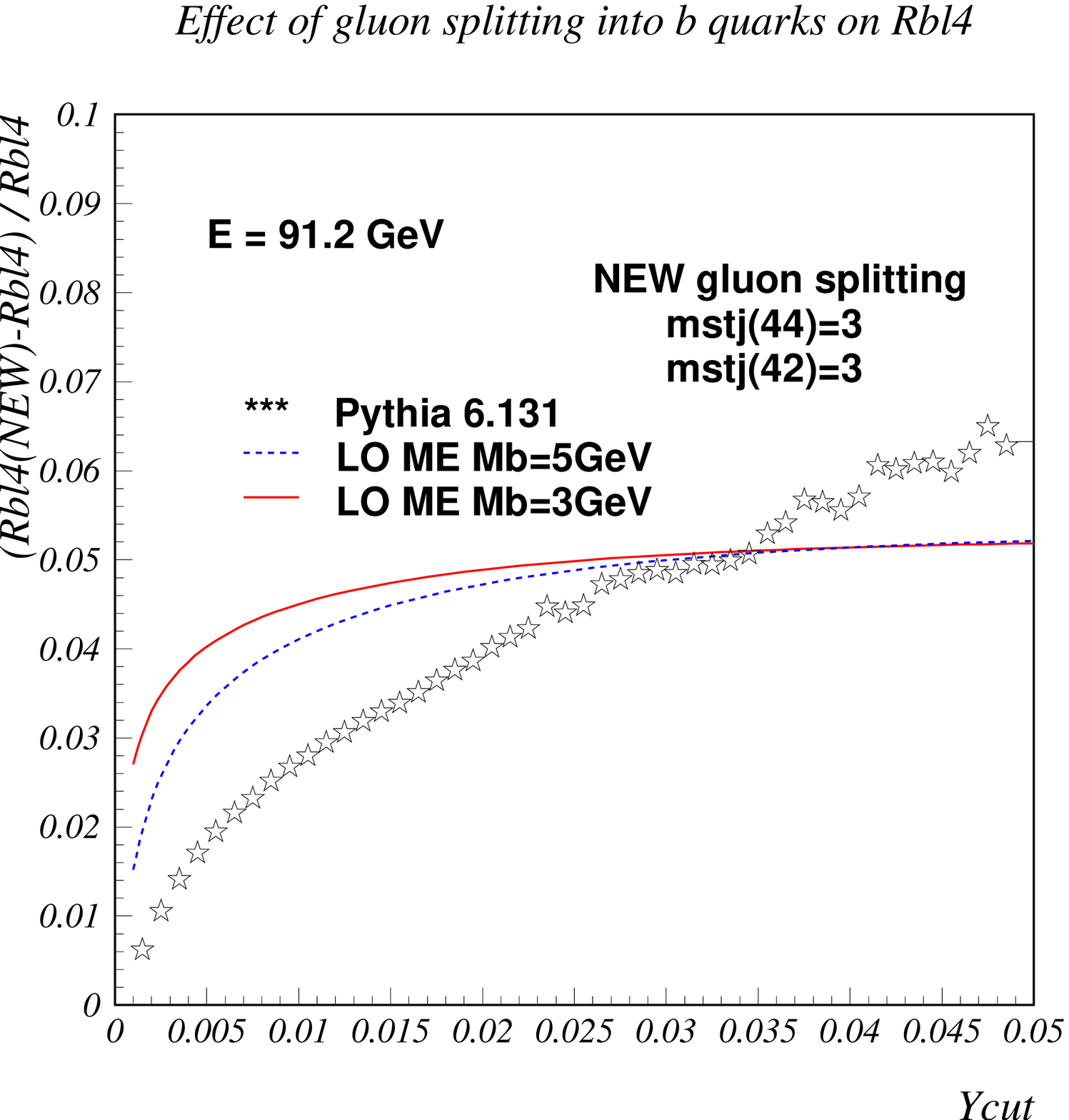} 
\end{tabular}
\caption{Difference of the $R_4^{b\ell}$ double jet ratio at 
$\sqrt s = \mz$ for \pythia\ 6.131 (shown with stars) 
in the NEW and standard definitions, for the settings with increased 
rate of gluon splitting into $b {\bar b}$ corresponding to
{\tt MSTJ(44)}={\tt MSTJ(42)}=3, and in the LO matrix element 
calculation.}
\label{delta:rbl4:91:33}
\end{center}
\end{figure}

\begin{figure}[htbp]
\begin{center}
  \begin{tabular}{ll}
  \includegraphics[width=14cm]{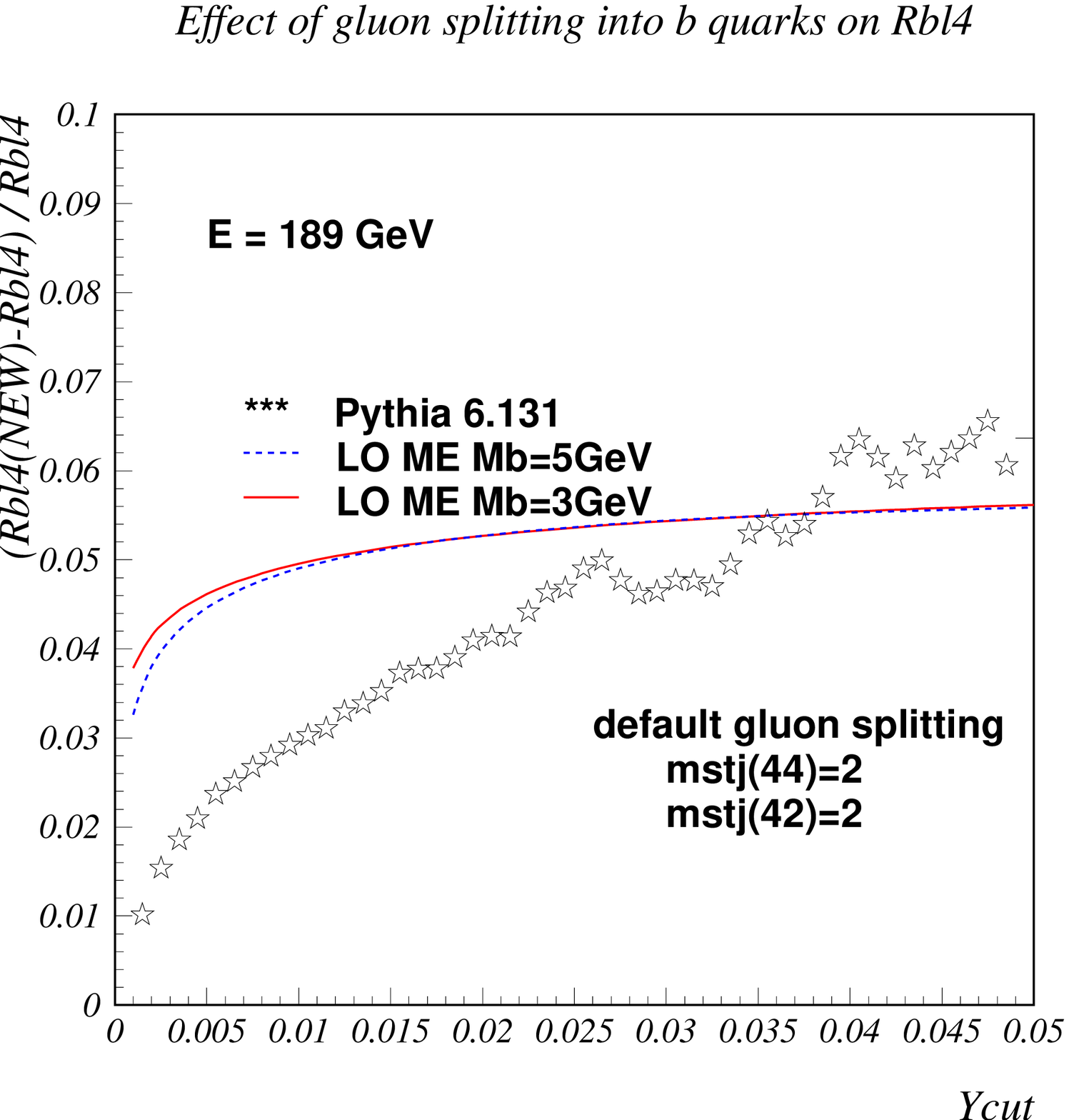} 
\end{tabular}
\caption{Difference of the $R_4^{b\ell}$ double jet ratio at 
$\sqrt s = 189$ GeV for \pythia\ 6.131 (shown with stars) 
in the NEW and standard definitions, for the default settings with
unmodified rate of gluon splitting into $b {\bar b}$, 
corresponding to {\tt MSTJ(44)}={\tt MSTJ(42)}=2, and in the LO matrix element 
calculation.}
\label{delta:rbl4:189:22}
\end{center}
\end{figure}

\begin{figure}[htbp]
\begin{center}
  \begin{tabular}{ll}
  \includegraphics[width=14cm]{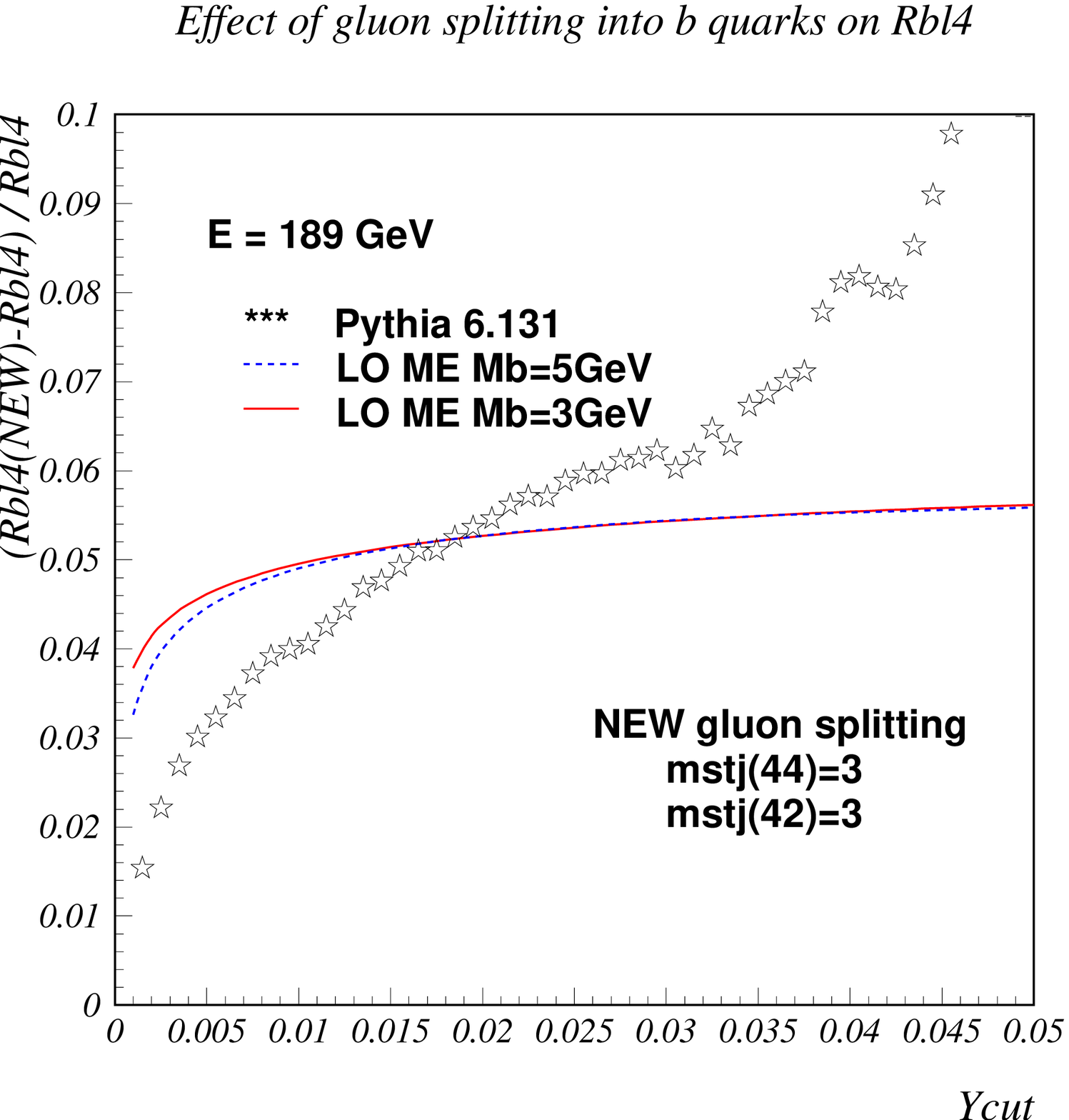} 
\end{tabular}
\caption{Difference of the $R_4^{b\ell}$ double jet ratio at 
$\sqrt s = 189$ GeV for \pythia\ 6.131 (shown with stars) 
in the NEW and standard definitions, for the settings with increased 
rate of gluon splitting into $b {\bar b}$ 
corresponding to {\tt MSTJ(44)}={\tt MSTJ(42)}=3, and in the LO matrix element 
calculation}
\label{delta:rbl4:189:33}
\end{center}
\end{figure}

The standard definition of $R_4^{b\ell}$ presented in \sect{sec:Rbl4} 
considers {\it primary} quarks, and is advantageous from the theoretical
point of view, but not from the experimental one, where such a
distinction is obviously not straightforward. Since the data are extrapolated
to the parton level using a Monte-Carlo to compare with the calculations,
a wrong assumption on the gluon splitting into $b {\bar b}$ translates
directly into a bias. In order to estimate this bias, a new definition
was proposed, labeled $R_4^{b\ell}$(NEW) in which are counted in the 
numerator any event containing $b$ quarks, irrespective whether they
originate from primary or secondary production, and at the denominator
only events with light quarks, excluding those with a gluon splitting 
into $b {\bar b}$. This new observable is closer to the experimental 
situation, but is known to carry larger NLO corrections, and is hence 
more uncertain theoretically. It was nonetheless evaluated both
at $\sqrt{s}=\mz$ and at at $\sqrt{s}=189$ GeV, analytically and
using \pythia\ 6.131, with the different settings corresponding to
different recent treatments of the gluon splitting process into $b {\bar b}$
developed in the framework of the working group 
(see \sect{ts:sec:gsplit}). As an example the normalised difference 
$(R_4^{b\ell}\mbox{(NEW)} - R_4^{b\ell})/R_4^{b\ell}$ is shown for
$\sqrt{s}=\mz$ in Figures~\ref{delta:rbl4:91:22} and \ref{delta:rbl4:91:33},
respectively for the present default settings 
({\tt MSTJ(44)}={\tt MSTJ(42)}=2), and for one of the
proposed set of new settings described in \sect{ts:sec:gsplit}
({\tt MSTJ(44))}={\tt MSTJ(42)}=3), corresponding to a $g \rightarrow b\bar{b}$
rate which is roughly doubled. The same comparison is shown for
$\sqrt{s}=189$ GeV in  Figures~\ref{delta:rbl4:189:22} and \ref{delta:rbl4:189:33}.

A general feature of these plots is that at small $y_c$ 
\pythia\ is always lower. This arises because in \pythia, contrary to the
analytic calculation, the four-jet
rates in both the denominator and numerator of $R_4^{b\ell}$ 
receive contributions also from two and three jets, which reduce the 
relative impact from the fraction of events containing a gluon splitting
into $b \bar{b}$. Considering firstly the results at 
$\sqrt{s}=\mz$, one can see that indeed 
doubling the gluon splitting rate into $b \bar{b}$ in 
\pythia\ results in a difference between the two definitions which
at large $y_c$ becomes similar to that obtained in the analytical calculation.
Since doubling the gluon splitting rate into $b \bar{b}$ tends to be
favoured by both experimental results and theoretical work 
(see \sect{ts:sec:gsplit}), one can in principle take this found
consistency as evidence confirming that it indeed needs to be doubled
in \pythia.
To check this further, the fraction of events containing a $g\to b \bar{b}$
splitting in which the two $b$ quarks are clustered into
separate jets (case of resolved gluon splittings) was evaluated
and found to be largely dominant. This indicated that large NLO corrections
in the new definition $R_4^{b\ell}$(NEW) are not expected to arise from 
gluon splittings into $b \bar{b}$, and that most of the effect does occur
at LO. From this can be concluded that a procedure consisting in correcting
the measured four-jet events in data to extrapolate to the parton level 
using a Monte-Carlo induces a sensitivity to the correct rate of 
gluon splittings into $b \bar{b}$, at the level of these discrepancies
between the differences. A second conclusion is that indeed at $\sqrt{s}=\mz$
doubling the rate of gluon splittings into $b \bar{b}$ would seem to be 
justified from the found 
consistency of the comparison with the analytical results.

The picture changes however at $\sqrt{s}=189$ GeV.
As can be seen from \fig{delta:rbl4:189:33},
the \pythia\ curve with doubled gluon splitting into 
$b \bar{b}$ rate now overshoots significantly at large $y_c$
To understand the origin of this behaviour,
the same study was performed as at $\sqrt{s}=\mz$
to evaluate the fraction of events 
containing a gluon splitting into $b \bar{b}$ in which the gluon splitting
is resolved, and it was found that at 189 GeV only about half are. 
Hence in this case the origin of the overshooting could be traced
to the fact that the new definition receives large 
contributions at NLO from gluon splittings into $b \bar{b}$.
Such a behaviour is in fact expected from the scaling with energy of this
last contribution, which grows like $\log(m_b^2/s)$.

None of these results prevent one from evaluating the performance
of Monte-Carlo programs using the standard definition for 
$R_4^{b\ell}$. However one must be careful as soon as one wants
to compare with data. Moreover, the results at 189 GeV should not be taken 
as evidence against a larger rate of gluon splittings 
into $b \bar{b}$ in \pythia, which may be needed as explained in
\sect{ts:sec:gsplit}, but just that for the case of the observable 
$R_4^{b\ell}$(NEW) the comparison is not meaningful, because of the
large NLO contributions affecting it.
A full calculation at NLO would be helpful to study this further.

\subsection{Discussion of hadronization corrections to $R_3^{b\ell}$}
\label{sec:hadunc}

In most of the experimental analyses that involve hadronic 
final states, the data
need to be corrected to the parton level in order to be 
compared  with the theoretical
predictions. In particular, this is true for the studies considered in this
report, for instance the determination of the $b$-quark mass, or the studies of
the flavour
independence of the strong coupling constant and the multi-jet production
rates. This procedure necessarily implies unfolding the data for detector
and acceptance effects as well as for the hadronization process. This
introduces
biases and uncertainties which need to be carefully studied and quantified to
extract reliable measured values within the quoted errors. The detector and
acceptance corrections depend on each particular experiment and consequently
will not be discussed here. Only the hadronization correction will be
the subject of this section. As shown in references
\cite{mbatmz,alephEPS99,Abe:1998kr,Brandenburg:1999nb,Abbiendi:1999fs}
the uncertainty arising from the lack of precise knowledge of how the
hadronization process takes place limits the experimental precision
of the experimental quantities and QCD tests. Any progress
leading to a better understanding of the transition from partons to hadrons
or finding new observables with better behaved properties will immediately
result in improving these measurements.

In particular, let us consider the hadronization corrections associated with
the $R_3^{b{\ell}}$ observable introduced in Eq. 34 (though on a qualitative
basis the same procedure can be easily applied and generalized to the
$R_4^{b{\ell}}$ observable of Eq. 35 or other event shape
variables). This observable summarizes most of the features
commented on in previous sections and at the same time has the advantage of
being calculated up to NLO (see section 4.3) and of having
relatively small fragmentation corrections (of roughly $\sim$1\%). 
The fragmentation
models considered in this exercise are \pythia\ and \herwig. The
analysis is also performed using
two jet clustering algorithms: \durham\ and \cambridge, and the
potential results of using one or the other are compared and discussed. The
determination of $b$-quark mass or the test of the flavour independence of the
strong coupling constant can be derived from this observable and the
implications are obvious: {\it a better understanding of $R_3^{b{\ell}}$ with
a smaller error leads to a more precise determination of the $b$-quark mass or
a more stringent test of the flavour independence of $\as$}.

In general there is no prescription to unambiguously define the fragmentation
correction factor to be applied to the $R_3^{b{\ell}}$ observable but the
procedure described below can be regarded as a reasonable approach. It is
based on the {\sc Delphi} procedure  \cite{mbatmz}, though others methods
could
also be envisaged for this purpose \cite{alephEPS99}. It seems appropiate then to
consider all models which give a good description of the data and calculate the
correction
factor corresponding to each model. The average of the correction factors
obtained is taken as the best estimate of the correction factor and the
distribution of these values
defines the uncertainty. This also means that the models or generators
considered in the analysis should be tuned in order to properly describe the
data. The overall fragmentation uncertainty can then be quantified by adding in
quadrature the two different source of errors: $\sigma_{mod}$, the uncertainty
due to the dependence of the hadronization correction factors on the two models
considered, \pythia\ and \herwig, and $\sigma_{tun}$, uncertainty due
to the possible variation of the main fragmentation parameters in
\pythia. Hence, the total uncertainty is expressed as:
\begin{equation}
\sigma_{had}(y_c) = \sqrt{\sigma_{mod}^2(y_c)+\sigma_{tun}^2(y_c)}\;.
\end{equation}

\begin{figure}[hbt]
\epsfverbosetrue
\begin{center}\mbox{\epsfxsize=10.cm\epsfysize=10.cm
\epsffile{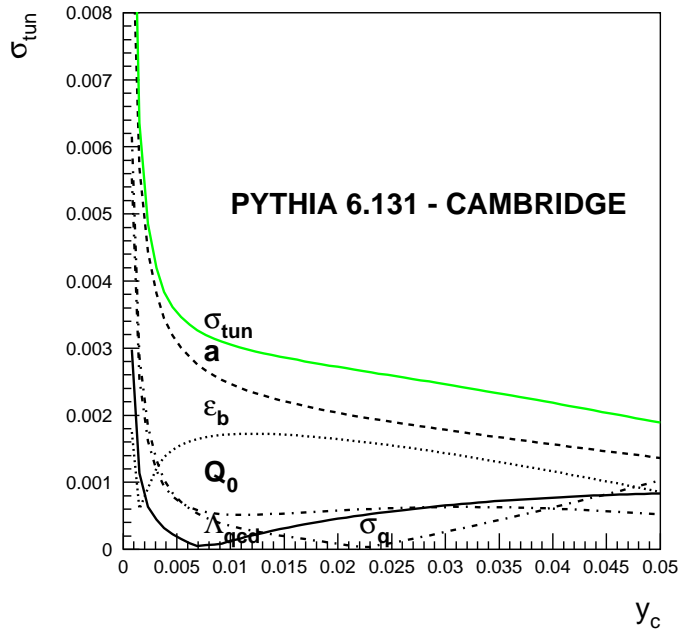}
}
\end{center}
\caption[]
{The $\sigma_{tun}$ uncertainty and the contributions of each individual
fragmentation parameter considered to the the total error, obtained with the
\cambridge\ algorithm. }
\label{fig:tun}
\end{figure}

Following the \delphi\ procedure the $\sigma_{tun}$ uncertainty
is obtained by varying the most relevant parameters of the string fragmentation
model incorporated in \pythia\ ($Q_0$, $\sigma_q$, $\epsilon_b$, a and b)
within
an interval of $\pm 2\sigma$ from their central tuned values and assuming that
the individual parameter errors are all independent. \fig{fig:tun}
shows, for the \cambridge\ algorithm, the $\sigma_{tun}$ uncertainty as a
function of the jet resolution parameter as well as the contribution of each
individual parameter error. For large enough $y_c$ values the overall
$\sigma_{tun}$ uncertainty is seen to be around 3\permil. The different tuned
versions of \jetset\ or \pythia\ have also been tested and
cross-checked
to give the same correction factors to the observable $R_3^{b{\ell}}$ within
1\permil.

In this exercise, the two fragmentation schemes considered are \pythia\
and \herwig, therefore the average of the two correction factors obtained
is considered as the fragmentation correction factor to $R_3^{b{\ell}}$ and the
uncertainty $\sigma_{mod}$ is taken to be half of their difference. The
generators
differ not only in the fragmentation process (cluster fragmentation in 
\herwig\
and Lund string fragmentation in \pythia) but also in the way the
particle decays are implemented. Therefore $\sigma_{mod}$ has two
contributions, one from the fragmentation scheme itself, $\sigma_{mod-frag}$,
and the other one from the decay tables used,  $\sigma_{mod-dec}$, so it can be
written as:

\begin{equation}
\sigma_{mod}(y_c) = \sqrt{\sigma_{mod-frag}^2(y_c)+\sigma_{mod-dec}^2(y_c)}\;.
\end{equation}

\begin{figure}[hbt]
\epsfverbosetrue
\begin{center}\mbox{\epsfxsize=10.cm\epsfysize=10.cm
\epsffile{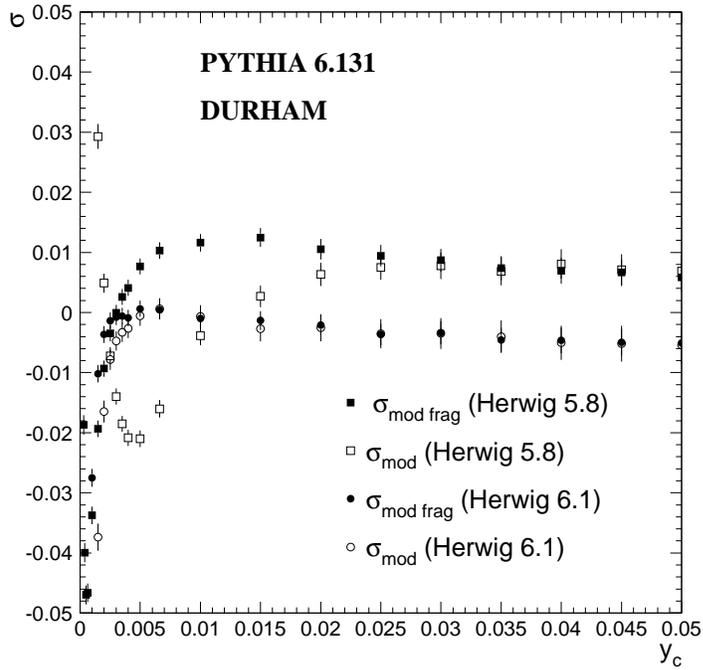}
}
\end{center}
\caption[]
{The $\sigma_{mod}$ and $\sigma_{mod-frag}$ uncertainties as a function of
the jet resolution paramenter, obtained with \durham. The error bars are
statistical.}
\label{fig:decays}
\end{figure}

We present here the results of the $\sigma_{mod}$ uncertainty obtained with
\herwig\ version 5.8 as tuned by \delphi\ and
version 6.1 as tuned by \aleph. For \pythia, the \delphi\ tuning
is used. Presently all \lep\  experiments are working on the tuning of new
versions, therefore new and better sets are expected soon.
\fig{fig:decays}
shows $\sigma_{mod}$ and $\sigma_{mod-frag}$ at different $y_c$ values as
calculated using the \durham\ jet clustering algorithm. For $y_c \ge 0.02$
the contribution of $\sigma_{mod-dec}$ to the total model uncertainty becomes
small. Also a better agreement between the latest generator versions of 
\herwig\ and \pythia\ is observed over the entire $y_c$ region. This
result probably is a consequence of the fact that in the latest versions of
both generators the $b$-fragmentation functions are now similar. How well these
models describe this
distribution in data is, however, discussed elsewhere (section 6) and new
analyses are
still emerging from the various LEP and SLC collaborations on this subject.

Apart from improving the performance of the generators by including more
precise
calculations and better modelling of the various processes taking part, the
search and use of observables having better theoretical or experimental
properties is also worthwhile to reduce the total uncertainty.
Still in the context of measuring the $b$-quark mass or of testing the flavour
independence of $\as$, the use of different algorithms to reconstruct jets
has extensively been studied and compared in references
\cite{Abe:1998kr,Brandenburg:1999nb,Rodrigo:1999qg,alephEPS99,delphiEPS99,nosaltres} and
for different event shape variables in references
\cite{Abbiendi:1999fs,alephEPS99}. The main
conclusion derived from these studies is that not all observables are
equally suited to make the above measurements because they are influenced by
different higher order corrections which in some cases can be
large. This can explain some of the spread the $b$-mass values measured by the
various experiments. Therefore studying each observable property in both its
theoretical 
and hadronization aspects is mandatory for making precise measurements.

\begin{figure}[hbt]
\epsfverbosetrue
\begin{center}\mbox{\epsfxsize=10.cm\epsfysize=10.cm
\epsffile{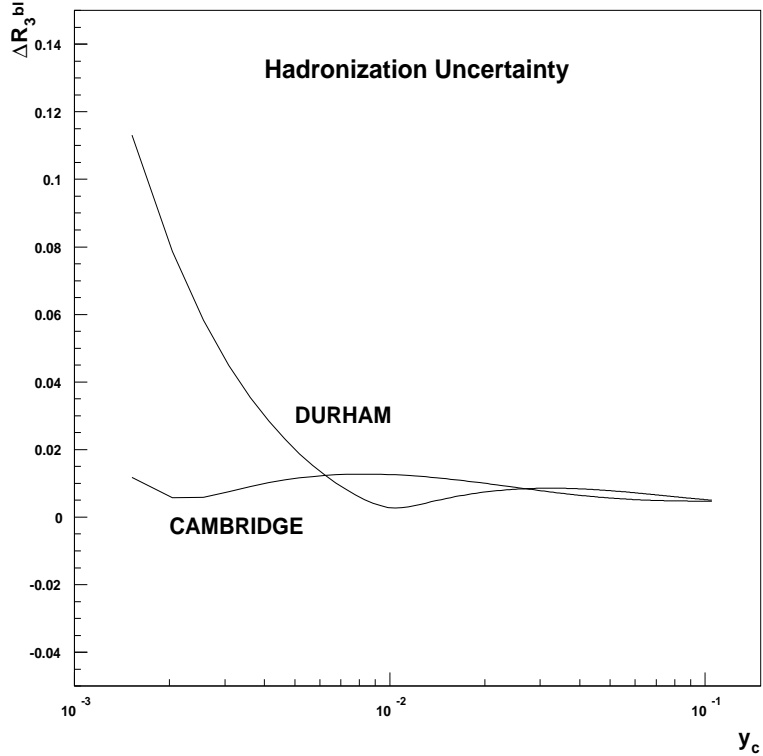}
}
\end{center}
\caption[]
{Evolution of the total hadronization error of $R_3^{b{\ell}}$ with the
resolution parameter $y_c$. It is presented in terms of
$\Delta R_3^{b{\ell}} = 2 \times |\sigma_{had}|$ with $\sigma_{had}$ being
defined
in the text.}
\label{fig:had}
\end{figure}

Following the spirit of this section, \fig{fig:had}
presents the size of the total hadronization correction uncertainty
($\sigma_{had}(y_c)$) for $R_3^{b{\ell}}$ using either the \durham\ or
\cambridge\ algorithms for the jet reconstruction. The use of the
\cambridge\ algorithm reduces the theoretical error on the $b$-quark mass
determination \cite{delphiEPS99,nosaltres} though the hadronization error
is about the same for $y_c \geq 0.02$. The use of \durham\ is however
limited to the $y_c$-region above 0.015 in order to keep the four jet rate
below 5\% and the hadronization correction small and flat with respect
to $y_c$. The same arguments applied to \cambridge\ allows the extension
of the  $y_c$ region down to 0.004, which, although it 
increases the sensitivity to the
$b$-quark mass marginally, does increase the sensitivity to the
difference
of the LO and the NLO predictions, enabling a better experimental distinction
between the
$\overline{MS}$ and {\em pole mass} schemes. The curves shown in
\fig{fig:had} have been taken from \cite{delphiEPS99,nosaltres} where
\pythia\ 6.131 and \herwig\ 5.8 were used and, therefore, promising
further reductions in those uncertainties can probably still be
obtained with the latest versions of these generators.

\subsection{Conclusions and remaining issues}\label{sec:R34conc}

In this section effects from the $b$ quark mass on three and four-jet
rates have been analysed, both in real data, and via analytic calculations
and Monte-Carlo approaches. A method for evaluating the relevant observables
and for estimating the theoretical uncertainty in the predictions in a
consistent way has been presented, and the predictions from several 
Monte-Carlo programs (\ariadne, \pythia, \herwig\ and \fourjphact) 
has been evaluated, including in some cases (\ariadne, \pythia)
recently improved versions. Moreover results from on-going studies aiming at 
controlling the additional uncertainties arising from the modeling of 
non-perturbative effects in the main Monte-Carlo programs used have been 
reported.

Here below we first quantify the theoretical 
uncertainty which is appropriate for each of the observables studied, and 
then summarize the performance of each Monte-Carlo program.
Finally we mention remaining issues relevant to the four-jet case, and 
improvements which are still needed. 

\subsubsection{Theoretical uncertainty affecting $R_3^{b\ell}$ and 
$R_4^{b\ell}$}

Following the prescription described in \sect{sec:anacal}, we
quote the theoretical uncertainty in the $R_3^{b\ell}$ and $R_4^{b\ell}$
ratios as $\pm$ half the 
difference between the LO ME results corresponding to the 
pole mass $M_b\sim 5$~GeV and to the running mass $m_b(\mu)$,
with $m_b(\mz)\sim 3$~GeV. This definition is also supported by the 
analysis of the \delphi\ data at $\sqrt s = \mz$, which for
both $R_3^{b\ell}$ and $R_4^{b\ell}$ lie within these two calculations.
As expected the uncertainties are found to be larger for $R_4^{b\ell}$
than for $R_3^{b\ell}$, and are reduced for large values of $y_c$ and of
$\sqrt s$.

At $\sqrt s = \mz$, for $R_3^{b\ell}$, from the 
difference between the LO ME results the error may be quoted as ranging
from $\pm$ 1 to 2\% for values of $y_c$ between 0.02 and 0.1 
(see \fig{3jetpt}). Arguably, this
estimate is conservative since in this case an NLO calculation exists. 
For $R_4^{b\ell}$, from the 
difference between the LO ME results the error may be quoted as ranging
from $\pm$ 2 to 4\% for values of $y_c$ between 0.012 and 0.03 
(see \fig{4jetpt}).

At $\sqrt s = 189$ GeV, for $R_4^{b\ell}$, from the 
difference between the LO ME results the error may be quoted as ranging
from $\pm$ 1 to 3\% for values of $y_c$ between 0.002 and 0.04 
(see \fig{4jetpt6125200}).

\subsubsection{Performance of the different Monte-Carlo programs}

\ariadne\ underestimates $R_3^{b\ell}$ at $\sqrt s = \mz$
in all of its versions (see \fig{3jetar}). The recently improved one
does however provide slightly better values than the version
with the dead cone suppression switched on. The best description is 
nonetheless achieved by switching off all mass treatments altogether.
In this case the result is within the band of uncertainty defined by the 
LO ME curves, but is about 1.5\% lower than the NLO curve and the data.
With the default setting of the present official version of the
program (dead cone suppression turned on) the shift with respect to the NLO
ME curve ranges from 2 to 4\%. On the other hand $R_4^{b\ell}$ at 
$\sqrt s = \mz$ is reasonably described in the version with all mass 
treatments switched off. For this observable both the new improved version
and the old mass treatment of the dead cone give similarly low results, 
A similar qualitative behaviour is observed at $\sqrt s = 189$ GeV
(see \fig{4jetar200}). For both of the versions of \ariadne\ with either
the default treatment of the $b$ mass via the dead cone suppression or with
the new improved treatment, the underestimation reaches about 3\%
for $y_c \simeq 0.002-0.3$.

\pythia\ results for $R_3^{b\ell}$ at $\sqrt s = \mz$
(see \fig{3jetpt}) show a strong underestimation for the old 
version (prior to 6.130) with mass effects turned on as per the default
of that version. The bias is about 1 to 4\% for $y_c$ ranging between
0.01 and 0.12. A better behaviour is obtained by turning off all mass effects.
The best description is however obtained thanks to the recently improved 
treatment of mass effects in versions following 6.131. In this case 
the MC prediction overlaps nicely with the NLO results and with data.
On the other hand for $R_4^{b\ell}$ at $\sqrt s = \mz$ all versions 
underestimate the rate (see Fig. ~\ref{4jetpt}). The recently 
improved one (following version 6.131) does nonetheless provide the
best description. The bias is in this case about 
2.5 to 7\% for $y_c$ ranging between 0.012 and 0.03. With the old default
mass treatment (prior to version 6.130) the bias becomes as large as
10 to 12\%. The discrepancies are somewhat less pronounced at 
$\sqrt s = 189$ GeV (see \fig{4jetpt6125200} and \ref{4jetpt6131200}). 
In this case also the newest version (following 6.131) is the best,
and appears to be about 1 to 3\% too low. The old default mass treatment
(versions prior to 6.130) is the worst, with the suppression exaggerated
by 1.5 to 4\% for $y_c$ ranging from 0.002 to 0.03. 

\herwig\ was studied for the $R_3^{b\ell}$ and 
$R_4^{b\ell}$ observables only at $\sqrt s = \mz$ (see \fig{3jethw}
and \ref{4jethw}). A fair agreement is seen in both cases although with
some slight underestimation.

The new \fourjphact\ program with massive four-parton matrix elements
matched to the parton shower algorithm of \pythia\ was investigated
as well, and an initial preliminary result was shown. 
More work is needed here to study whether this approach to matching
can provide a solution to the description of $R_4^{b\ell}$ once the
full parton shower and hadronisation treatements are implemented.


\subsubsection{Remaining issues and improvements needed}

As described above it has been 
found that all programs tend to exaggerate the suppression
which arises from the $b$ quark mass, by varying amounts, 
either in the three-jet rate, or in the four-jet rate, or in both.
The best global behaviour is seen for \herwig, although in the version 
5.8 of this program which was studied, hadronization corrections were 
quite a bit larger than for example in \pythia. In the most recent 
versions (version 5.9 and 6.1), hadronization corrections have become 
closer to those in \pythia. So from the particular point of view discussed in
this section, \herwig\ would seem to be best. More studies are of course 
needed to confirm that this is also the case at higher energies.
Moreover, additional work towards improving further 
the remaining discrepancies 
in \ariadne\ and \pythia\ is still on-going at the time of this 
writing, and will hopefully also bring these two programs in line in the near
future. Finally the new \fourjphact\ program looks quite
promising and needs to be investigated further in this context.

On the theoretical side work towards carrying out a massive four-parton
matrix element calculation at NLO would enable the estimate of the
uncertainty in the $R_4^{b\ell}$ observable described in this section 
to be checked explicitly and refined. It could then also be used
experimentally to probe the running of the $b$-quark mass, as has been done
with $R_3^{b\ell}$.

An additional issue -- which would also benefit from the availability of
an NLO  4-parton massive calculation -- concerns the impact on
$R_4^{b\ell}$ from uncertainties in
processes involving gluon splitting into $b {\bar b}$.
Monte Carlo studies indicate that enhancing the rate of
$g\to b {\bar b}$ splitting in \pythia, an option discussed
in \sect{ts:sec:gsplit}, gives consistency with the LO estimate of
the impact at $\sqrt{s}=\mz$ but overshoots it by up to a factor of two at
high $y_c$ at \leptwo\ energies. This could be taken as
an estimate of higher-order uncertainties resulting from  
$g\to b {\bar b}$ splitting processes at high energy.

\section{STUDY OF FOUR JET OBSERVABLES}

\subsection{Introduction}
  \label{sec:GD_intro}

The study of 4-jet final states is of great interest for
\lepone\ as well as \leptwo\ physics analyses, and reliable 
predictions of the properties of such final states by the
various Monte Carlo programs are mandatory.
From a QCD standpoint of view, 4-jet final states have their 
origin in the processes $\mathrm{Z}\rightarrow q\bar{q}gg$ 
and $\mathrm{Z}\rightarrow q\bar{q}q'\bar{q}'$, with the
secondary partons coming from double gluon Bremsstrahlung
and gluon splitting into gluon or quark pairs.

At \lepone\ these processes have been employed for  the tests of the
structure of the underlying gauge group (\cite{DG98} and references
therein), which is SU(3) in the case of QCD. In order to get
sensitivity to the gauge structure of the theory, a specific class of
observables has been employed, namely angular distributions of jets in
4-jet events. The perturbative expansion for the differential
distributions of these observables starts at ${\mathcal O}(\as^2)$,
and only leading-order (LO) predictions were available until recently.
However, now the next-to-leading order (NLO) corrections have been
computed \cite{signer1}-\cite{weinzierl}, which allows refined studies
of 4-jet observables, such as improved tests of the gauge structure or
measurements of the strong coupling constant with variables for which
the perturbative predictions start at ${\mathcal O}(\as^2)$, only.

At \leptwo\ the interest in 4-jet final states is more related to
background studies in analysis of fully hadronic W decays and searches,
such as the 4-jet channel in the search for the Higgs boson. As an
example for variables which enter the selection algorithms of those
analyses, the sum of the six interjet angles and the angle between the
second and third most energetic jets can be mentioned in case
of the W cross section measurements \cite{ALEPH99}.  The variable
$y_{34}$, which will be explained in the next section, enters in a
typical preselection of Higgs searches \cite{ALEPH00}, and a
further background rejection is obtained by looking at functions of
interjet angles.  Therefore it is clear that a good description of
4-jet observables by the Monte Carlo programs is necessary in order to
obtain reliable estimates of the QCD backgrounds.

In the following first the observables are defined which are used
for the studies of this section, then the predictions of the
various models are compared. These comparisons are first made
for the leading order matrix element predictions for four-jet
observables, then the effects of next-to-leading order contributions
and mass corrections are investigated, and finally the Monte Carlo models
are compared to each other and to the data for quantities
computed from hadrons instead of partons.

\subsection{Observables}
  \label{sec:GD_observ}

The observables which will be studied in detail, are described in the
following. For those events where exactly four jets are found by the
\durham\ jet clustering algorithm with the $E$
recombination scheme and a cut-off value of $y_{cut}=0.008$, the
energy-ordered jet momenta are used to compute the four-jet angular
variables listed below~:
        \begin{itemize}
            \item the Bengtsson-Zerwas angle \cite{BZ88,Ben89}~:
                     $\chi_{BZ} = \measuredangle [ (\vect{p}_1 \times \vect{p}_2),
                                                   (\vect{p}_3 \times \vect{p}_4) ]$
            \item the K\"orner-Schierholz-Willrodt angle \cite{JKW81}~:\\
                      $\Phi_{KSW} = 1/2, \{ \measuredangle [ (\vect{p}_1 \times \vect{p}_4)
                                   (\vect{p}_2 \times \vect{p}_3) ] +
                                   \measuredangle [ (\vect{p}_1 \times \vect{p}_3),
                                    (\vect{p}_2 \times \vect{p}_4) ] \}$
             \item the (modified) Nachtmann-Reiter angle \cite{NR82,SBZ91}~:
                       $\theta^*_{NR} = \measuredangle [ (\vect{p}_1 - \vect{p}_2),
                                                         (\vect{p}_3 - \vect{p}_4) ]$
             \item the angle between the two lowest energy jets \cite{Abr93}~:
                       $\alpha_{34} = \measuredangle [ \vect{p}_3, \vect{p}_4 ]$
        \end{itemize}
        These variables have already been used extensively for the measurements of the
        QCD colour factors \cite{DG98} because the shape of these
        distributions is sensitive to the group structure. 

For all hadronic events, the following event shape variables have been
considered~:
       \begin{itemize}
         \item D-parameter  $D$  \cite{Parisi78}, which is defined as the product
               $D = 27 \lambda_1 \lambda_2 \lambda_3$, with $\lambda_i$ being the
               three eigenvalues of the infrared safe momentum tensor
               \begin{equation}
                  \label{eq:GD_momtensor}
                  \Theta^{ij} = \sum_a \frac{p_a^i p_a^j}{|\vect{p}_a|} / \sum_a |\vect{p}_a| \quad.
                \end{equation}
               The sum on $a$ runs over all the final state particles 
               (partons or hadrons), and $p_a^i$ is the
               $i$th component of the three-momentum of the particle $a$ 
               in the centre-of-mass system.                
         \item $y_{34}$ (\durham, $E$ recombination scheme), 
                which is the jet resolution parameter when going from four to
                three jets, i.e., the event is clustered into jets 
                until only four jets are left, and then $y_{34} = \min y_{ij}$,
                where the minimum is taken over all distance measures (defined by
                the \durham\ metric) between the remaining jets.
       \end{itemize}

Since at the end of this section a comparison with corrected data will be given,
a short description of a typical data analysis is in place. Hadronic events 
are selected by requiring a minimum number of charged tracks and a minimum
charged energy per event. This reduces backgrounds from $\tau^+\tau^-$ and
two-photon events to negligible levels. 

Then the observables have to be corrected for detector effects such as
finite acceptance and resolution. This is done by computing
the observables from a Monte Carlo before and after the detector simulation
and imposing the same track and event selection cuts as 
for the data. Then bin-by-bin correction factors are computed
for every bin $i$ of the distribution,
\begin{equation}
 \label{eq:GD_detcorr}
           C^{det}_i = \frac{N^{had}_i}{N^{det}_i} \quad,
\end{equation}
where $N^{had}_i (N^{det}_i)$ denotes the number of entries in the
distribution at the hadron (detector) level. The hadron level distributions
are obtained by switching off any photon radiation in the initial and
final state (ISR, FSR), with all particles having mean lifetimes
less than $10^{-9}$ s required to decay, and all other particles being
treated as stable. So from a  measured distribution $D^{meas}_i$
a corrected distribution $D^{corr}_i$ is obtained according to
\begin{equation}
 \label{eq:GD_detcorr1}
           D^{corr}_i = C^{det}_i D^{meas}_i \quad.
\end{equation}
The detector correction factors are typically found within the 5-10\% range, 
increasing at the edges of the phase space.

These corrected distributions can be compared to the predictions from
perturbative QCD, which have to be corrected for hadronization effects,
i.e., long-distance non-perturbative effects. This is achieved by  
computing the relevant observables at parton and at hadron level, which 
allows to define bin-by-bin correction factors similarly to the
detector corrections,
\begin{equation}
 \label{eq:GD_hadcorr}
           C^{had}_i = \frac{N^{had}_i}{N^{part}_i} \quad.
\end{equation}
The superscript \textit{part} refers to the parton level. So 
from a purely perturbative prediction $D^{pert}_i$ a 
corrected QCD prediction $D^{QCD}_i$ is obtained according to
\begin{equation}
 \label{eq:GD_hadcorr1}
    D^{QCD}_i = C^{had}_i D^{pert}_i \quad,
\end{equation}
which is to  be compared to the corrected data $D^{corr}_i$.

The Monte Carlo simulations which most frequently are employed for the 
computation of these correction factors, as well as for the
simulations of QCD backgrounds to other physics channels, are the parton shower
models as implemented in \pythia\ \cite{ts:pythia} or 
\herwig\ \cite{Marchesini:1992ch},
together with the string fragmentation for the former and 
a cluster fragmentation in case of the latter.
It should be  considered that the basic idea of the parton shower is to
describe well the structure of jets in two-jet like events, since it is
based on a collinear approximation of the matrix elements for gluon
radiation off quarks.  Because of the
matching of the first parton branchings to the exact LO matrix elements, also
three-jet like quantities are described rather well. However, it can not
really be expected  that the parton shower approach gives a good description
of four-jet quantities. In fact, rather large discrepancies have been
observed in the past \cite{gluino}. 

A different approach can be tested by using the matrix element option in 
the \pythia\ program, where at the parton level two-, three- and four-parton
final states are generated according to the exact NLO matrix elements, and
then the hadronization step is performed via the string fragmentation scheme.
This model should give a better description of four-jet related quantities.
However, it is known not to describe well the
energy evolution of basic quantities such as the charged multiplicity
\cite{ALEPHqcd130,Acc98}.

Therefore new approaches are tried, based on the idea of matching matrix
element  calculations to parton shower evolutions, as described in the
previous sections. In the following these models will be discussed in
detail with respect to their description of 4-jet quantities.

\subsection{Comparison of model predictions}

\subsubsection{Leading order predictions (parton level)}
  \label{subsec:FK_lopred}

\begin{figure}[ht]
\begin{tabular}{cc}
\epsfxsize=8cm\epsffile{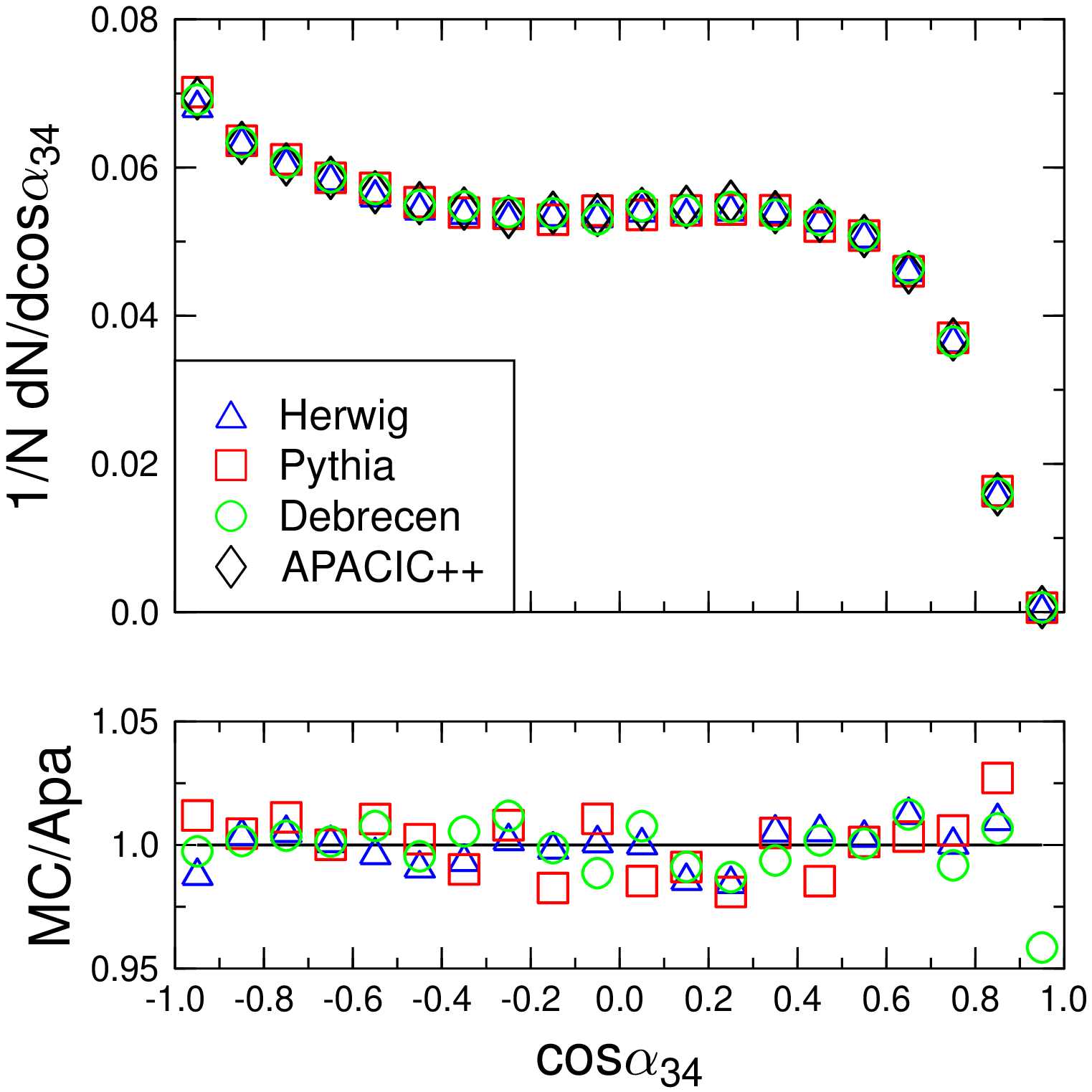} &
\epsfxsize=8cm\epsffile{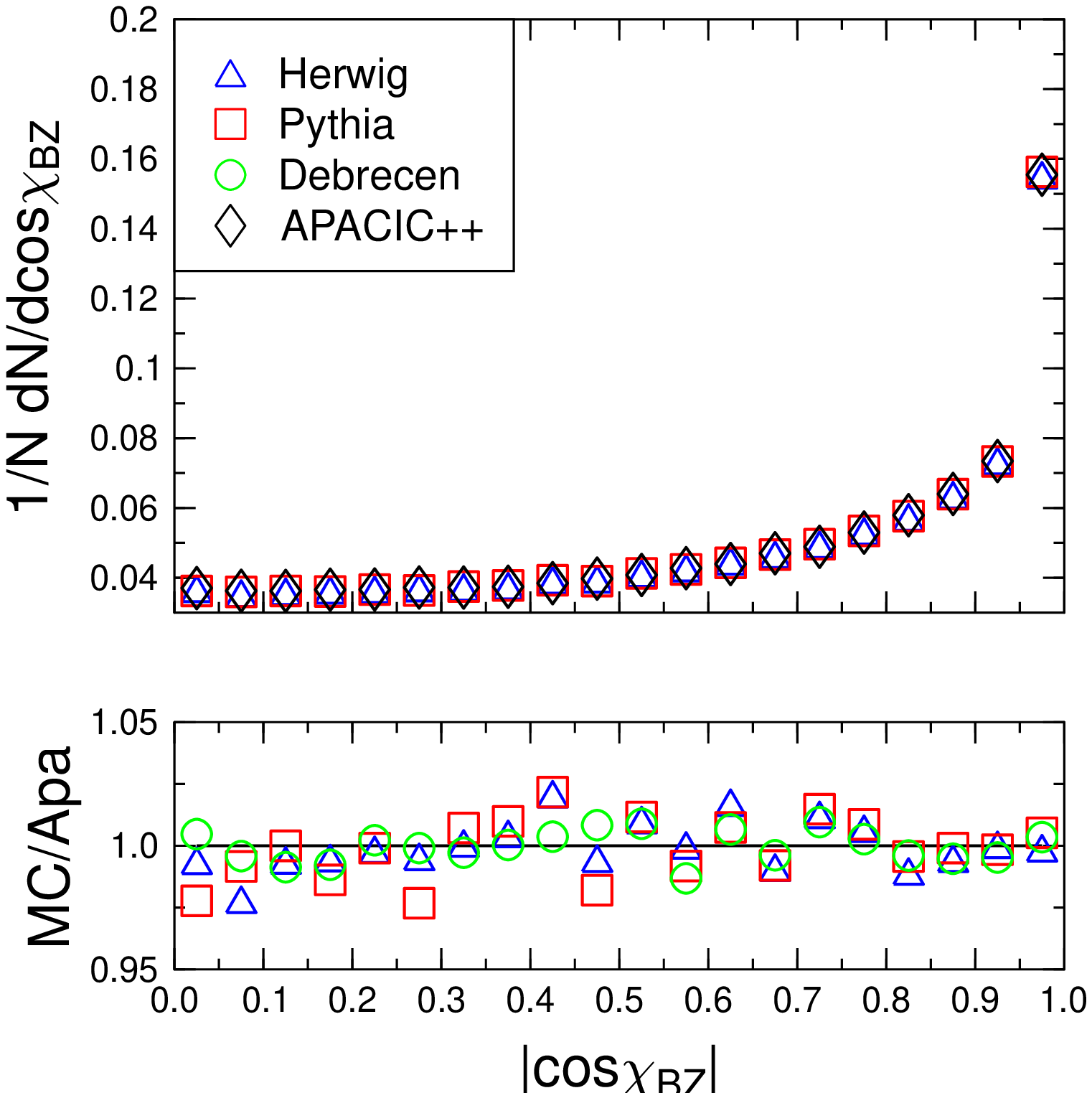}  \\
\epsfxsize=8cm\epsffile{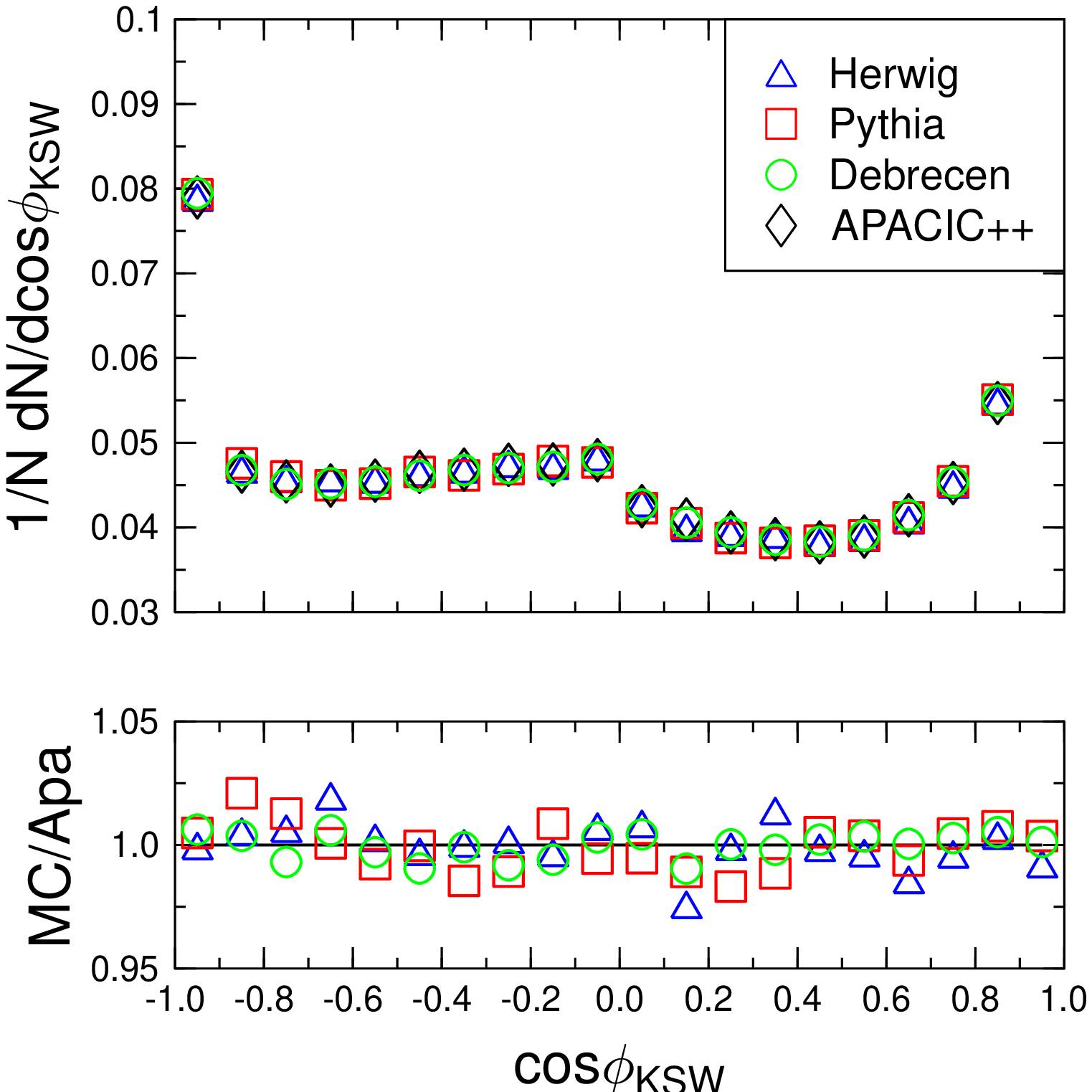} &
\epsfxsize=8cm\epsffile{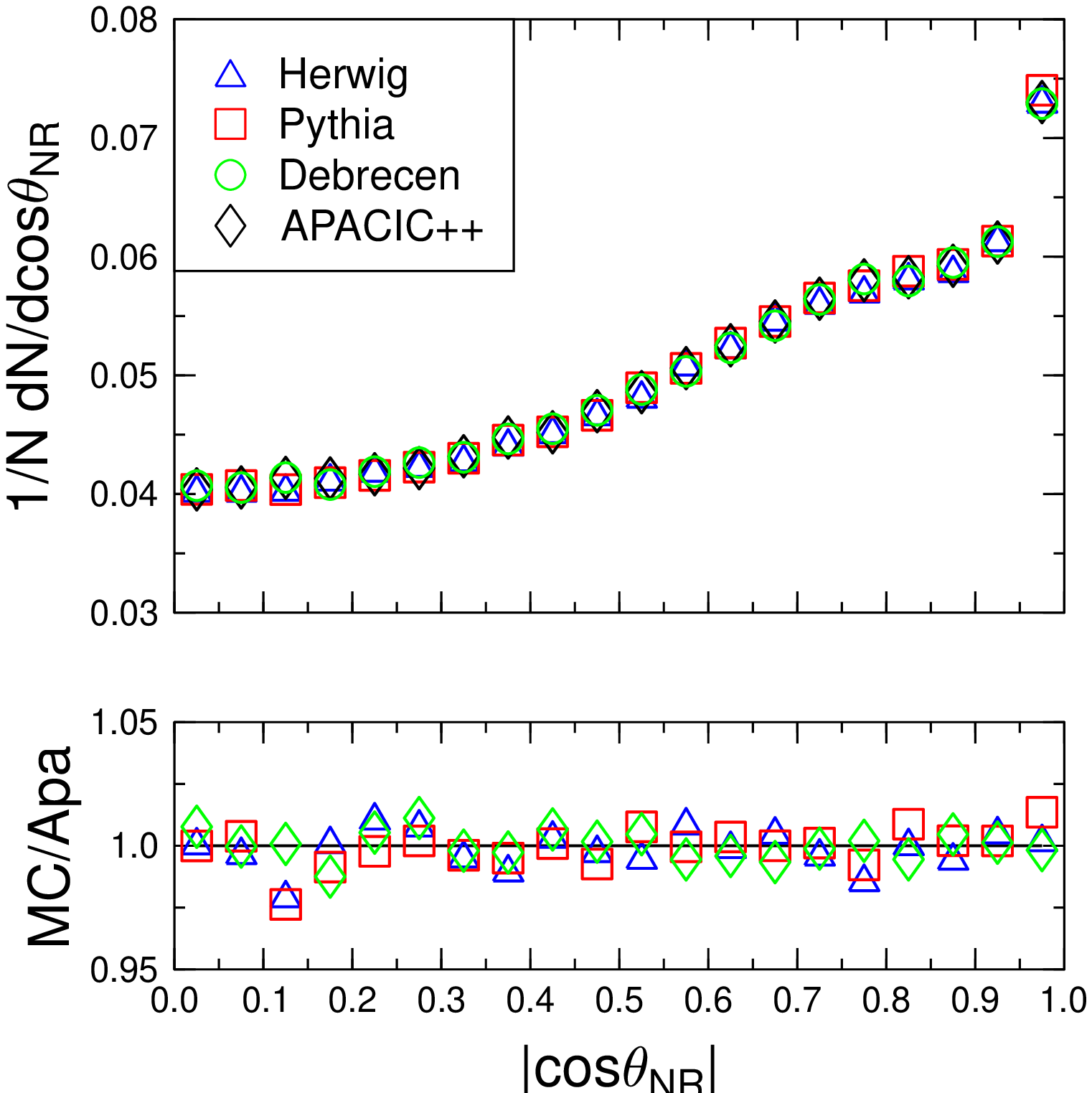} 
\end{tabular}
\caption{\label{ang0} Comparison of LO--results on the level of matrix elements for the
four jet angles with jets clustered at $y_{cut}^D=0.008$. 
The upper plots depict the normalized number of events per bin, the lower ones 
the ratios of \herwig, \pythia, \debrecen\ and \apacic.}
\end{figure}

\noindent There is quite a large variety of programs featuring the production
of four jets via QCD at the tree--level. Here, the performance of 
four of them, namely \herwig, \pythia, \debrecen\ and the 
package \apacic/\amegic\ (denoted as \apacic\
in the following), is compared. 

Note that according to the corresponding manuals, the four jet 
expressions within \pythia\ and \herwig\ are for massless partons 
(apart some mass effects which are built in for \pythia)
and they contain only the structures to be found for the exchange of 
virtual photons \cite{Ellis:1981wv}.
However, the claim is, that the additional terms 
related to intermediate $Z$--bosons have only a minor effect \cite{ts:pythia}.
At least for the observables studied here this claim has been verified. 

The focus is on the observables defined in \sect{sec:GD_observ},
specifically the four jet angles $\alpha_{34}$, $\chi_{BZ}$, $\phi_{KSW}$ 
and $\theta_{NR}$, and $y_{34}^D$, the $y_c$--value according to the 
\durham\--scheme, where four--jet events turn to three resolvable jets.
All results shown and discussed here are on the primary parton level, i.e.,
results obtained by the appropriate matrix elements squared, and at a
centre-of-mass energy of $91.2$ GeV, with the argument of $\as$ kept fixed.

\begin{figure}[thb]
\centerline{\epsfxsize=12cm\epsffile{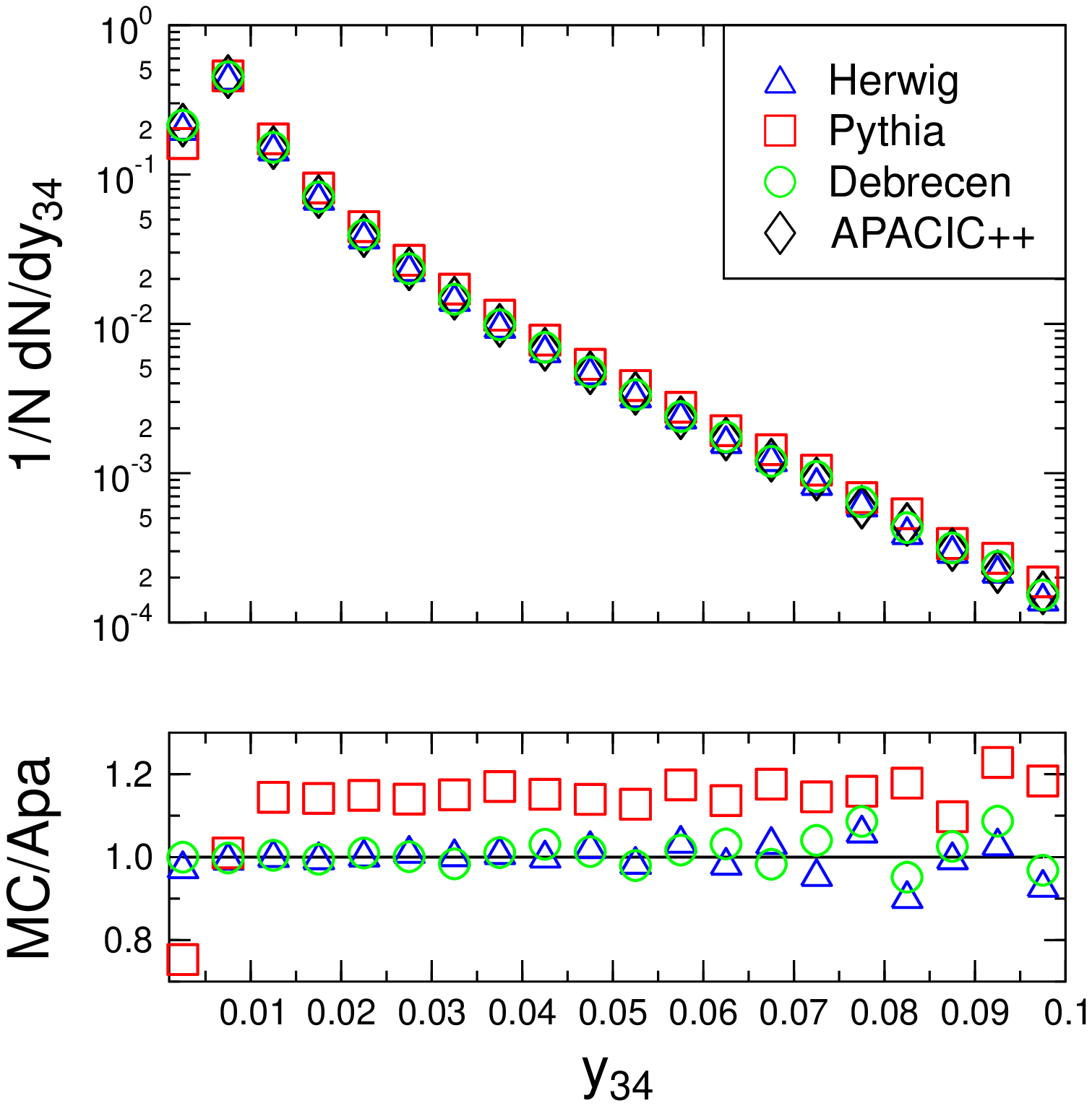}
} 
\caption{\label{y0} Comparison of LO--results for the $y_{34}^D$--distributions at the 
level of matrix elements. The upper plot exhibits the normalized number of 
events per bin, 
the lower one again the ratios of the other generators and \apacic.}
\end{figure}
\begin{figure}[bht]
\centerline{\epsfxsize=12cm\epsffile{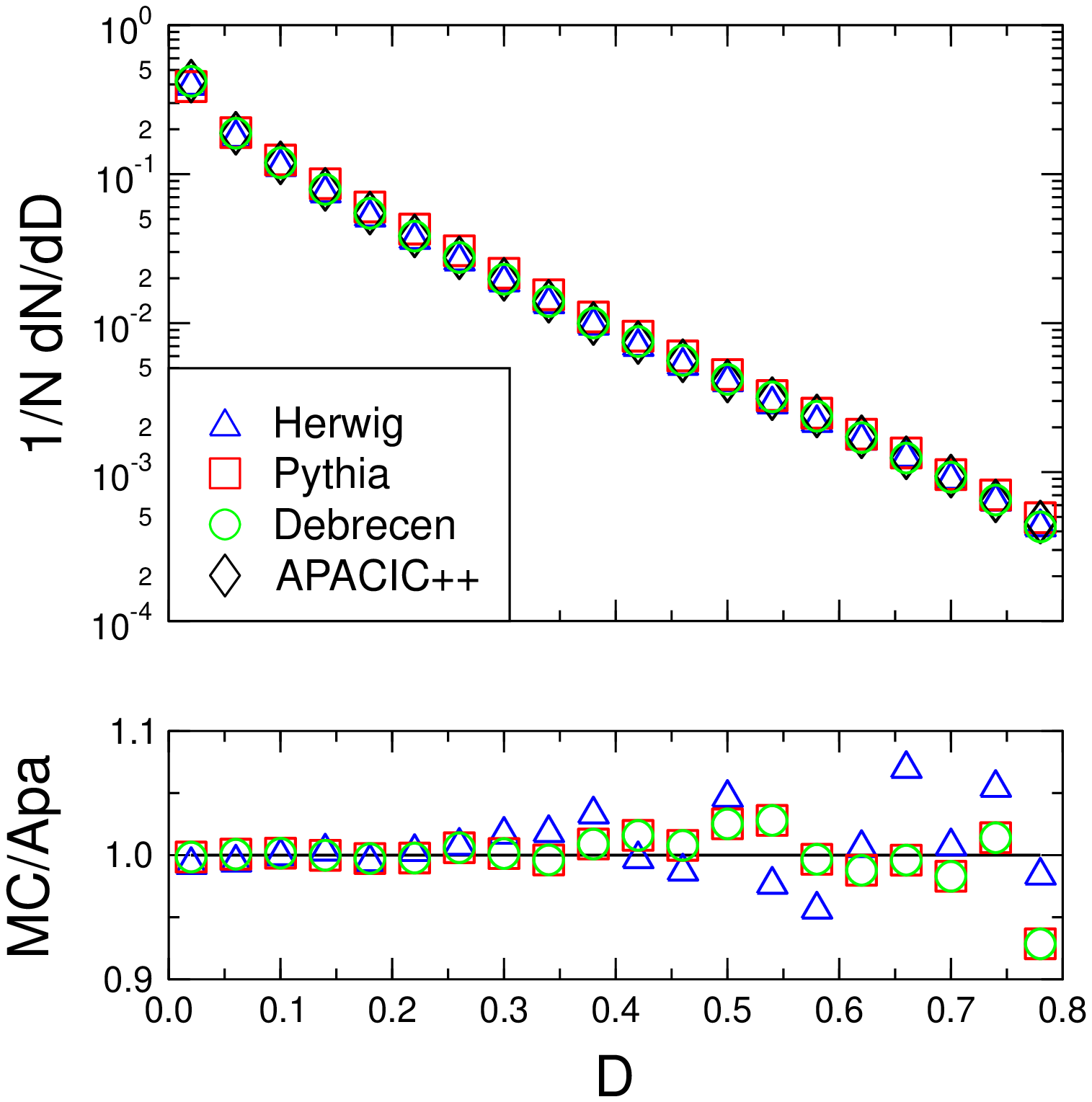}
} 
\caption{\label{d0} Comparison of LO--results for the $D$--parameter 
distributions at the 
level of matrix elements. The upper plot exhibits the normalized number 
of events per bin, 
the lower one again the ratios of the other generators and \apacic.}
\end{figure}

The Monte--Carlo points were produced adopting the following strategy :
\begin{enumerate}
\item{For \pythia\ a sample of four--jet events was generated
with $y_{34}^J \ge y_{cut}^J = 0.008$. 
This is due to the fact that in \pythia\
only the \jade\ scheme is available. Over a large region of phase space, as a
rule of thumb,  $y_{cut}^J\sim 4 y_{cut}^D$ for the same kinematical 
configurations.}
\item{Out of this first sample, only events with $y_{34}^D \ge y_{cut}^D = 0.004$
have been selected. For the other three generators, \herwig,
\debrecen\ and \apacic\ the events were directly generated
in the \durham\--scheme with $y_{cut}^D = 0.004$.}
\item{For the four jet angles, jets were defined according to
$y_{cut}^D = 0.008$, thus reducing the sample of step 2 by roughly
$50\%$. For the $y_{34}$--distribution no additional cuts have been
applied.\\}
\end{enumerate}

The resulting distributions of $\cos\alpha_{34}$, $|\cos\chi_{BZ}|$, 
$\cos\phi_{KSW}$ and $|\cos\theta_{NR}|$ can be found in \fig{ang0}. Here, 
the upper plots exhibit the total number of events per corresponding bin 
normalized to the total number of events with $y_{cut}^D = 0.008$, and 
in the lower plots the relative deviations from the \apacic\ results 
are displayed. With the exception of the last bin in $\cos\alpha_{34}$,
the relative (statistical) errors on each distribution are of the size of the symbols. 
Clearly, the results show a satisfying coincidence with no sizeable relative 
deviations.

This situation changes when considering the $y_{34}^D$--distribution, 
see \fig{y0}. Again, the upper plot shows the normalized number of events 
per bin, and the lower plots depicts the relative deviations from the 
\apacic\ results. Here, the spread of the statistical errors covers a region 
from barely visible in the left bins up to three times the size of the symbols in the 
right bins. However, the deviations of the generators from each other are larger than 
their individual relative errors and reach up to $15\%$. Seemingly, \herwig,
\debrecen\ and \apacic\ coincide. 
The results obtained by \pythia\ are somewhat
-- ${\cal O}(15\%)$  -- higher, with the first bin as the only significant exception.
Here, \pythia\ is well below ($\sim 25\% $) the other generators. However, it should be
noted here, that this is probably due to the way the \pythia\ sample was produced.
Since for the production of the \pythia\ sample in the first step the intrinsic
\jade\ scheme was employed, deviations can be expected especially in the regions 
where the phase space is cut, i.e., for low $y_{34}$. Normalising in the region
$y_{34}>0.01$, for example, would remove the discrepancy.
Turning to the $D$--parameter, the different generators agree very well with each other.
The relative errors reach roughly the size of the symbols for $D\approx 0.2$ and are
of the order of $10\%$ in the last bin. Note that for the $D$--parameter as well as for the
four jet angles, any difference seen in the $y_{34}$--distribution is washed out.

\begin{figure}[ht]
\centerline{\epsfxsize=8cm\epsffile{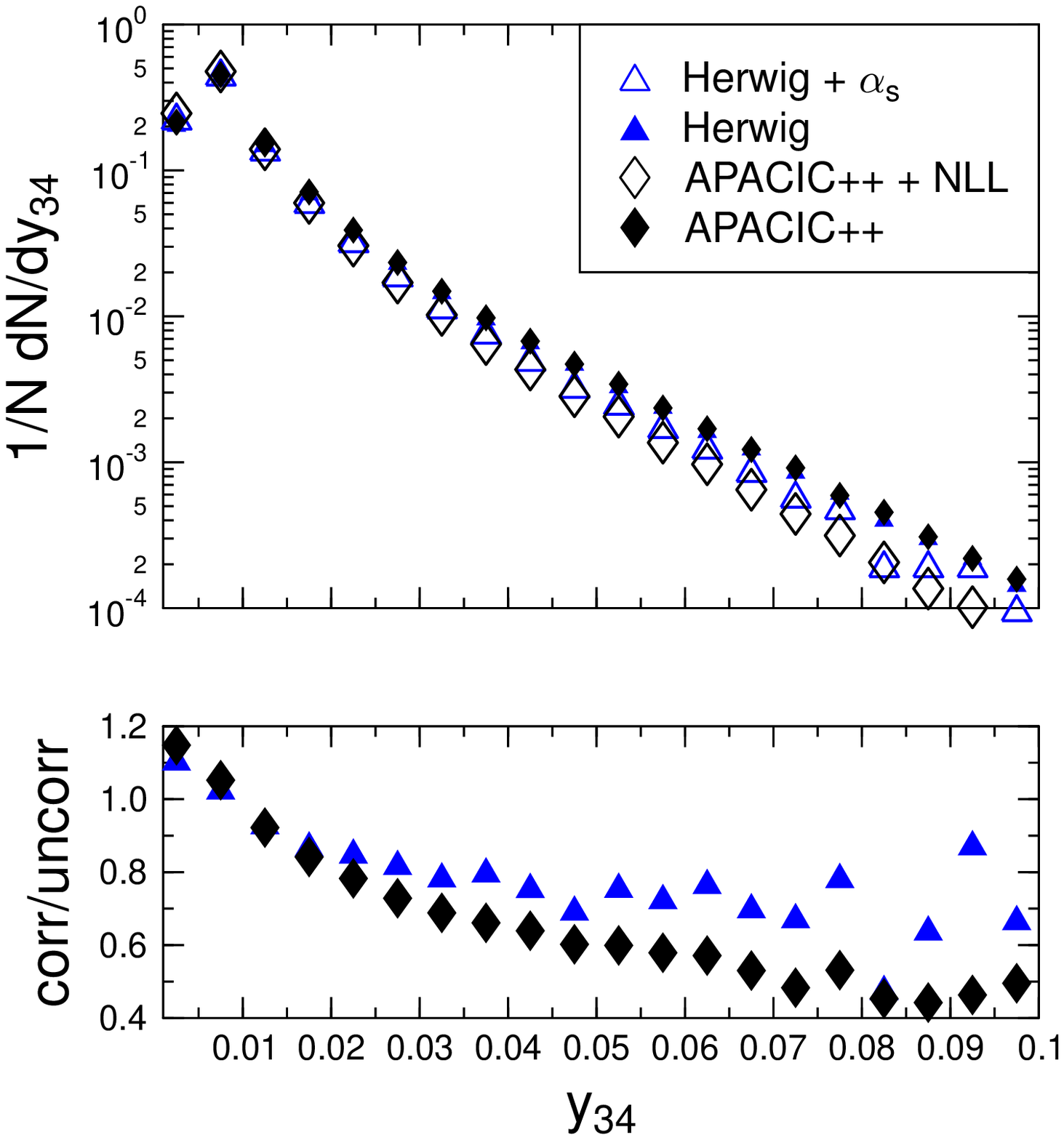}
} 
\caption{\label{y34_nll} The shift in the $y_{34}^D$--distributions induced via
the running of $\as$ in \herwig\ and the additional Sudakov weights in
\apacic. The upper plot shows the normalized number of events per bin, the lower
one shows the ratios of the corrected versus the uncorrected version of each
generator.}
\end{figure}
\begin{figure}[ht]
\begin{tabular}{cc}
\epsfxsize=8cm\epsffile{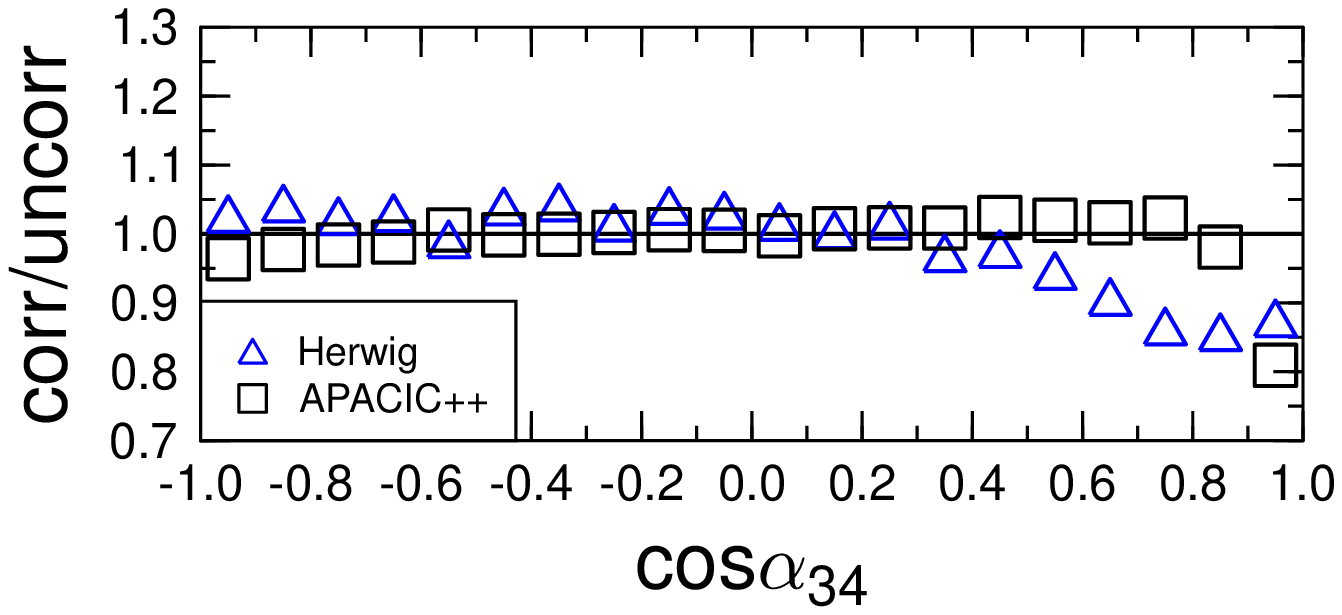} &
\epsfxsize=8cm\epsffile{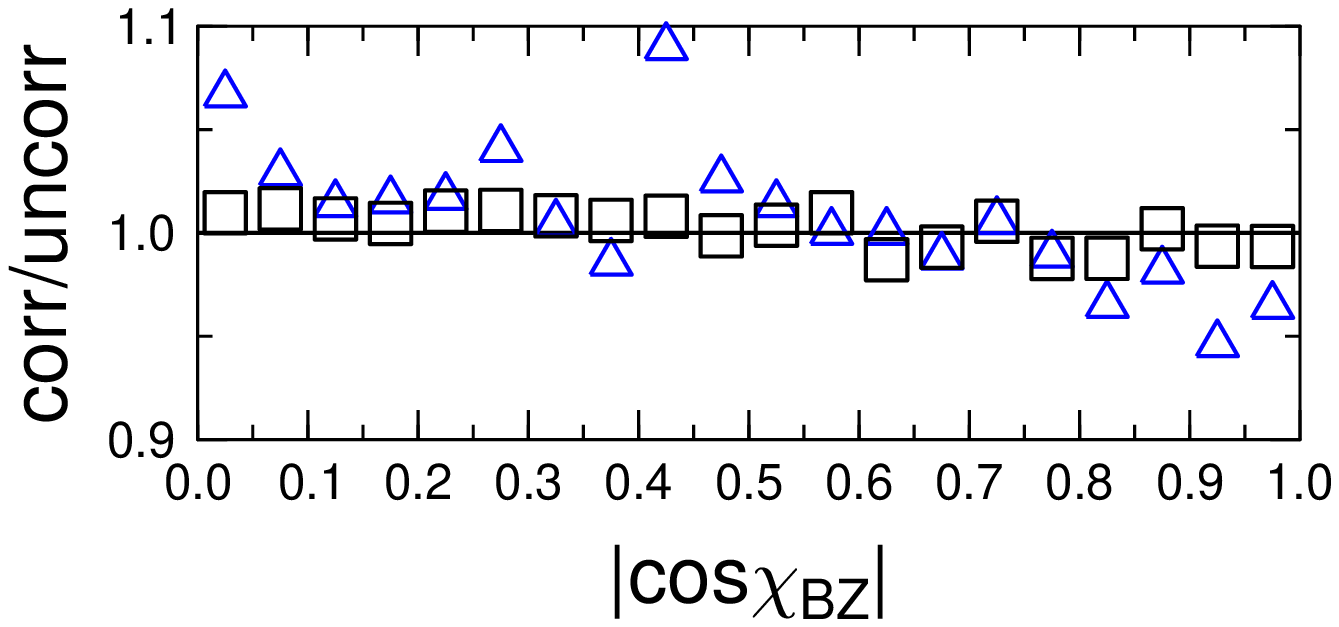} \\ 
\epsfxsize=8cm\epsffile{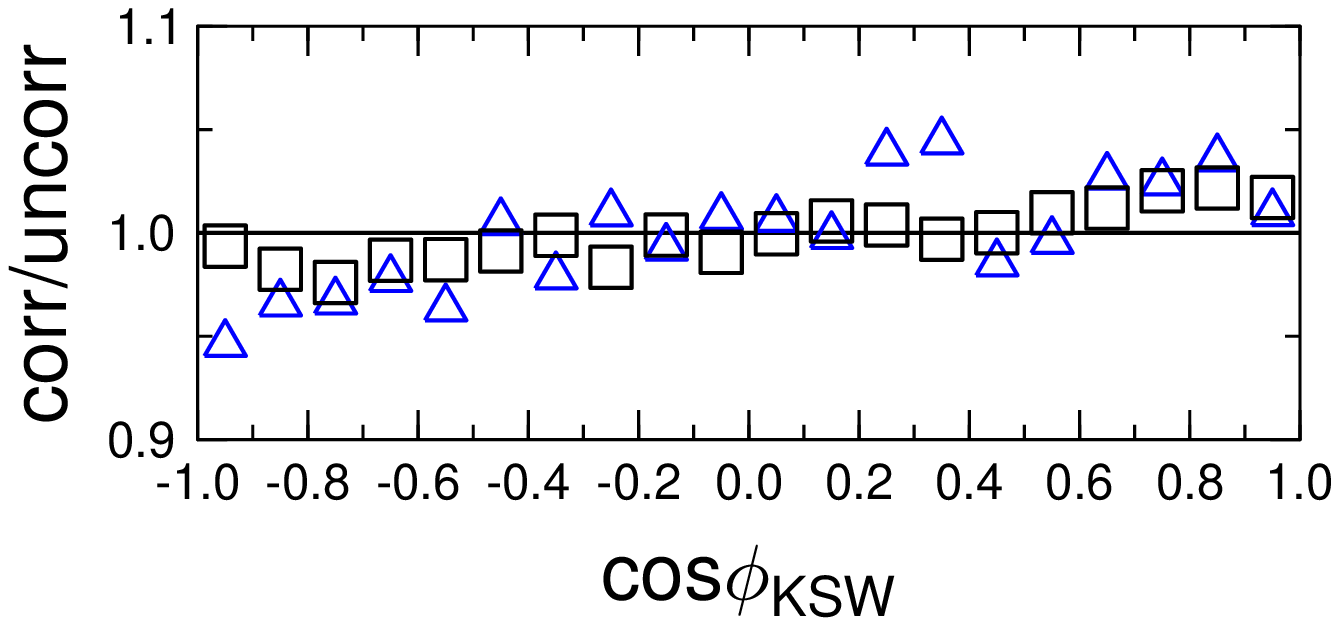} &
\epsfxsize=8cm\epsffile{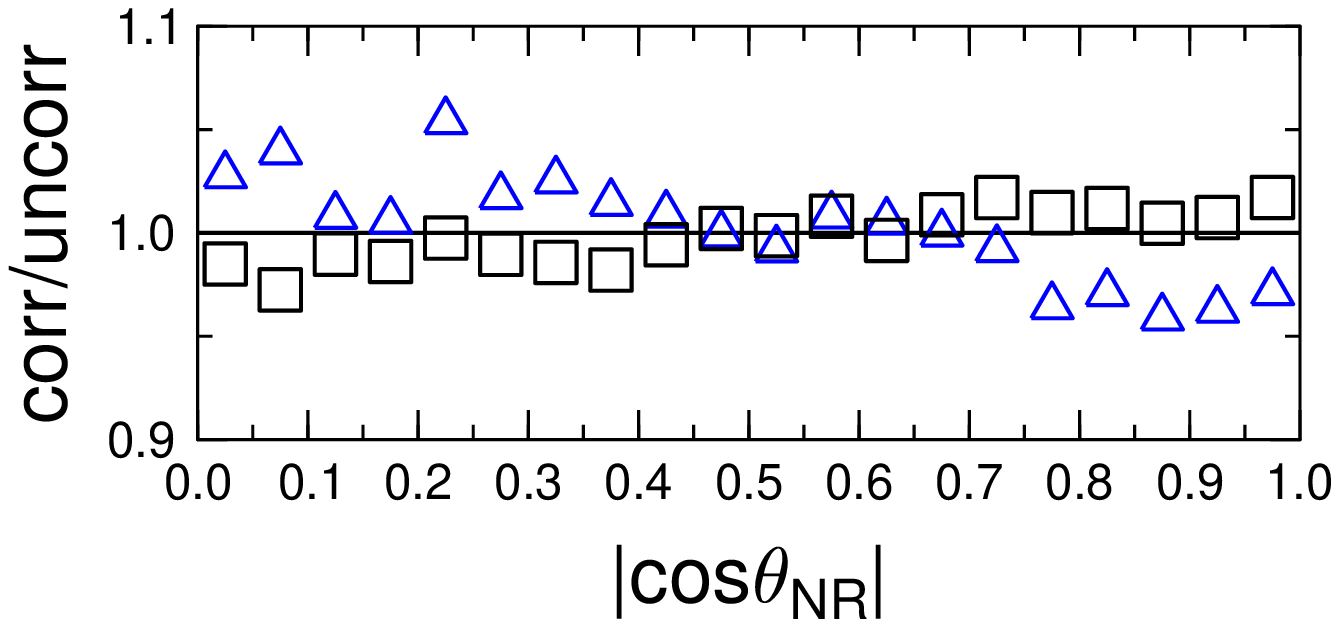} 
\end{tabular}
\caption{\label{angnll} The ratio of the corrected and the uncorrected versions of \herwig\
and \apacic\ for the four jet angles.}
\end{figure}

\herwig\ and \apacic\ provide additional options to supplement the pure
matrix elements with running $\as$ instead of the fixed one with a scale depending 
on the specific kinematical situation (\herwig\ and \apacic) or with some 
appropriate Sudakov weights (\apacic), which depend on the flavours and 
the kinematics of the individual event. These two options are meant to model
some aspects of higher order corrections to the pure LO matrix elements and result
basically in a shift of events from regions with large $y_{34}$ to region with
small $y_{34}$, see \fig{y34_nll}. Here, the upper plot shows the number of 
events per bin in the $y_{34}$--distribution normalized to the total number of events
and in the lower plot the ratio of the numbers per bin in the uncorrected and the corrected
versions of the generators is depicted. The full and empty triangles correspond to 
\herwig\ without and with the running $\as$ option, the diamonds refer to
\apacic\ without and with the Sudakov weights (``NLL''), respectively.
Obviously, these options ``soften'' the $y_{34}$--distribution of the samples. On the other hand, 
their effect on the angular distributions is only minor in most of the phase space, 
see 
\fig{angnll}. In the four plots the ratios of the corrected (corr) versus the uncorrected 
(uncorr) options for the four jet angles are displayed. It can be read off, that over the dominant 
region of phase space available, the inclusion of these corrections does not alter the angular 
distributions significantly. Rather, their effect is of the order of roughly $5\%$ with the only 
exception of the last bins for small $\alpha_{34}$, where the additional weights induce a 
drastical decrease of up to $15\%$. Note, however, that this region is strongly disfavoured,
see the corresponding plot in \fig{ang0}, thus, there are comparably large errors on the
results.
%

\subsubsection{Next-to-leading order corrections}
  \label{subsec:ZT_nlocorr}

\def\res#1#2#3 {$(#1\pm#2)\cdot10^{#3}$}

\newcommand{\ycut}    {\ensuremath{y_{\mathrm{cut}}}}
\newcommand{\Tr}      {{\rm Tr}}

\renewcommand{\d}     {{\rm d}}

For four-jet observables large differences were observed between
LO order perturbative predictions and corrected experimental data.
For instance, the LO perturbative result for the mean value of the D parameter
is $\langle D \rangle^{LO} = 0.0216$, while the experimental value is
$0.0618\pm 0.0024$ \cite{L3meanD}, and even after including the large
hadronization corrections, a significant discrepancy remains.  Such
a discrepancy and the large renormalization scale dependence of the LO
preturbative result \cite{zoltan1} shows clearly that if QCD is to work
for four-jet observables, then there have to be large higher-order or
non-perturbative
corrections.  On the other hand, the normalized angular distributions at
LO are independent of the strong coupling,  and therefore were expected
to be insensitive to the renormalization scale, indicating small higher
order corrections. For the clarification of the situation the NLO
calculations were indispensable. During the second phase of \lep, four
parton level NLO programs using different regularization methods were
published, {\sc Menlo parc} \cite{signer1}, \debrecen\
\cite{zoltan1}, {\sc Eerad2} \cite{campbell} and {\sc Mercutio}
\cite{weinzierl}.
The one-loop matrix
elements implemented in these codes were also different: {\sc Eerad2}
uses the one-loop matrix elements of Ref.~\cite{CGM}, while the other
three programs implemented the one-loop matrix elements published in
Ref.~\cite{BDK}. The results of the four programs
were compared for distributions of many four-jet observables and very 
good agreement was found as exemplified here in
\tab{table:jetcmp}, where the NLO four-jet fractions at three
different \ycut\ values for the \durham\ clustering algorithm are
compared. In the following, we shall present results obtained using the
\debrecen\ code. This program is the only one of the four that
gives the results in a colour decomposed form \cite{zoltan5}, which is
useful for colour charge measurements.
\begin{table}[ht]
\centering
\begin{tabular}{||c|c|c|c||} \hline\hline
Algorithm      & $y_{cut}$ & {\sc Menlo parc} & \debrecen\  \\ \hline
       & 0.005 & \res{1.04}{0.02}{-1} & \res{1.05}{0.01}{-1}  \\
\durham\ & 0.01  & \res{4.70}{0.06}{-2} & \res{4.66}{0.02}{-2}  \\
       & 0.03  & \res{6.82}{0.08}{-3} & \res{6.87}{0.04}{-3}  \\ \hline
               & $y_{cut}$ & {\sc Eerad2} &  {\sc Mercutio}     \\ \hline
       & 0.005 & \res{1.05}{0.01}{-1} & \res{1.06}{0.01}{-1}  \\
       & 0.01  & \res{4.65}{0.02}{-2} & \res{4.72}{0.01}{-2}  \\
       & 0.03  & \res{6.86}{0.03}{-3} & \res{6.96}{0.03}{-3}  \\
\hline\hline
\end{tabular}
\caption{\label{table:jetcmp}
The four-jet fraction as calculated by  {\sc Menlo parc},
\debrecen, {\sc Eerad2} and {\sc Mercutio}, 
for the \durham\ jet algorithm.}
\end{table}

The general form of the NLO differential cross section for a four-jet
observable $O_4$ (for instance, D parameter, $O_4 = D$, or 
Bengtsson-Zerwas angle, $O_4 = \chi_{BZ}$) is given by the following
equation:
\beq\label{O4}
\frac{1}{\sigma_0} \frac{d\sigma }{\d O_4}(O_4)=
 \eta(\mu)^2\,B_{O_4}(O_4)
+\eta(\mu)^3
\left[C_{O_4}(O_4) + B_{O_4}(O_4)\,\beta_0\,\ln(x_\mu^2) \right]\:,
\eeq
where $\sigma_0$ denotes the Born cross section for the process
$e^+e^-\to \bar{q} q$, $\eta(\mu) = \as(\mu)\,C_F/(2\pi)$, $x_\mu$ is the
ratio of the renormalization scale to the total centre-of-mass energy, and
$B_{O_4}(O_4)$, $C_{O_4}(O_4)$ are the perturbatively
calculable coefficient functions in the Born approximation and the
radiative correction, respectively, which are independent of the
renormalization scale.

\begin{figure}[thbp]
\begin{minipage}[t]{.47\linewidth}
\centerline{ \includegraphics[width=8.5cm]{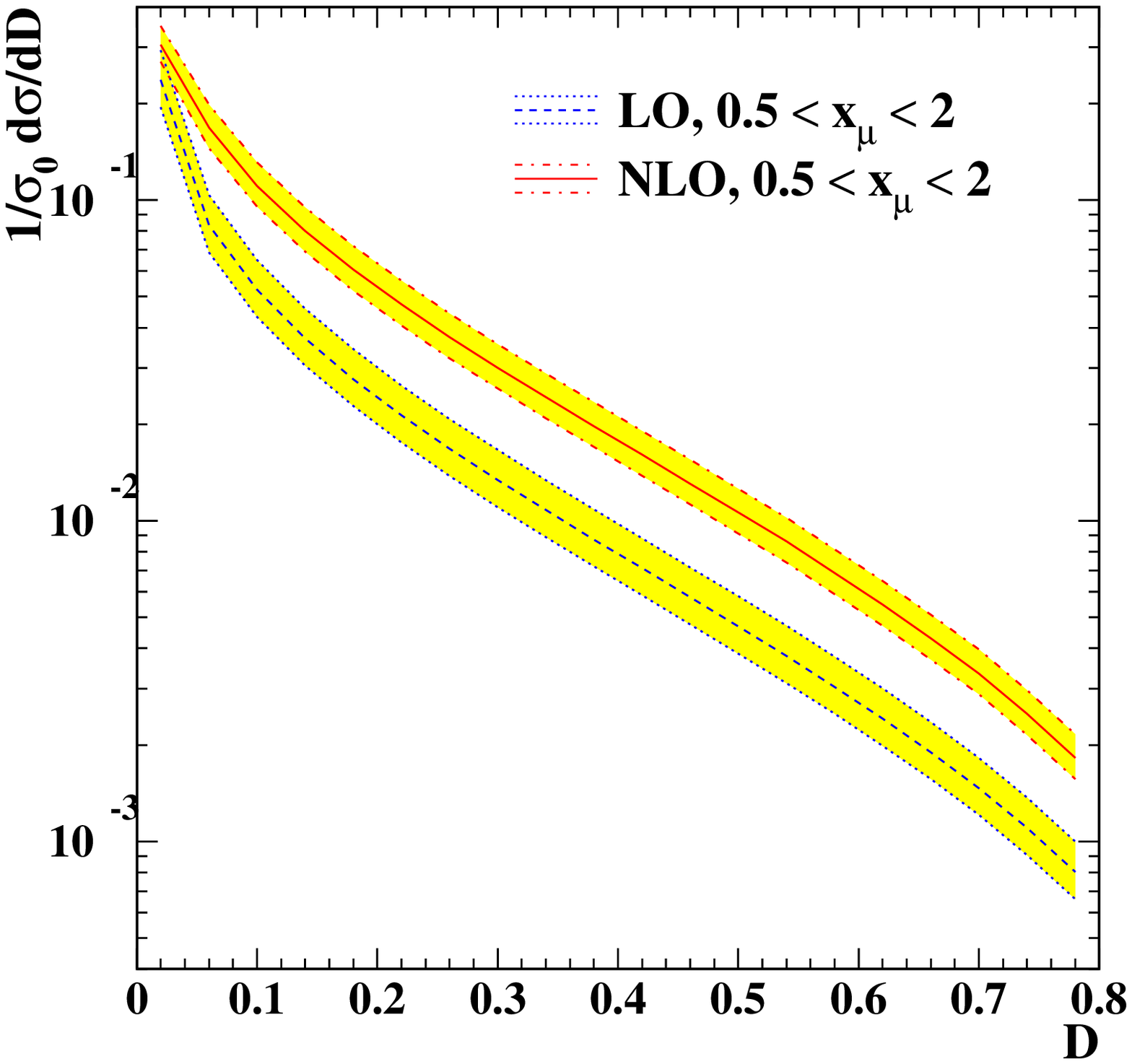}
 }
\end{minipage}
\hfill
\begin{minipage}[t]{.47\linewidth}
\centerline{ \includegraphics[width=8.5cm]{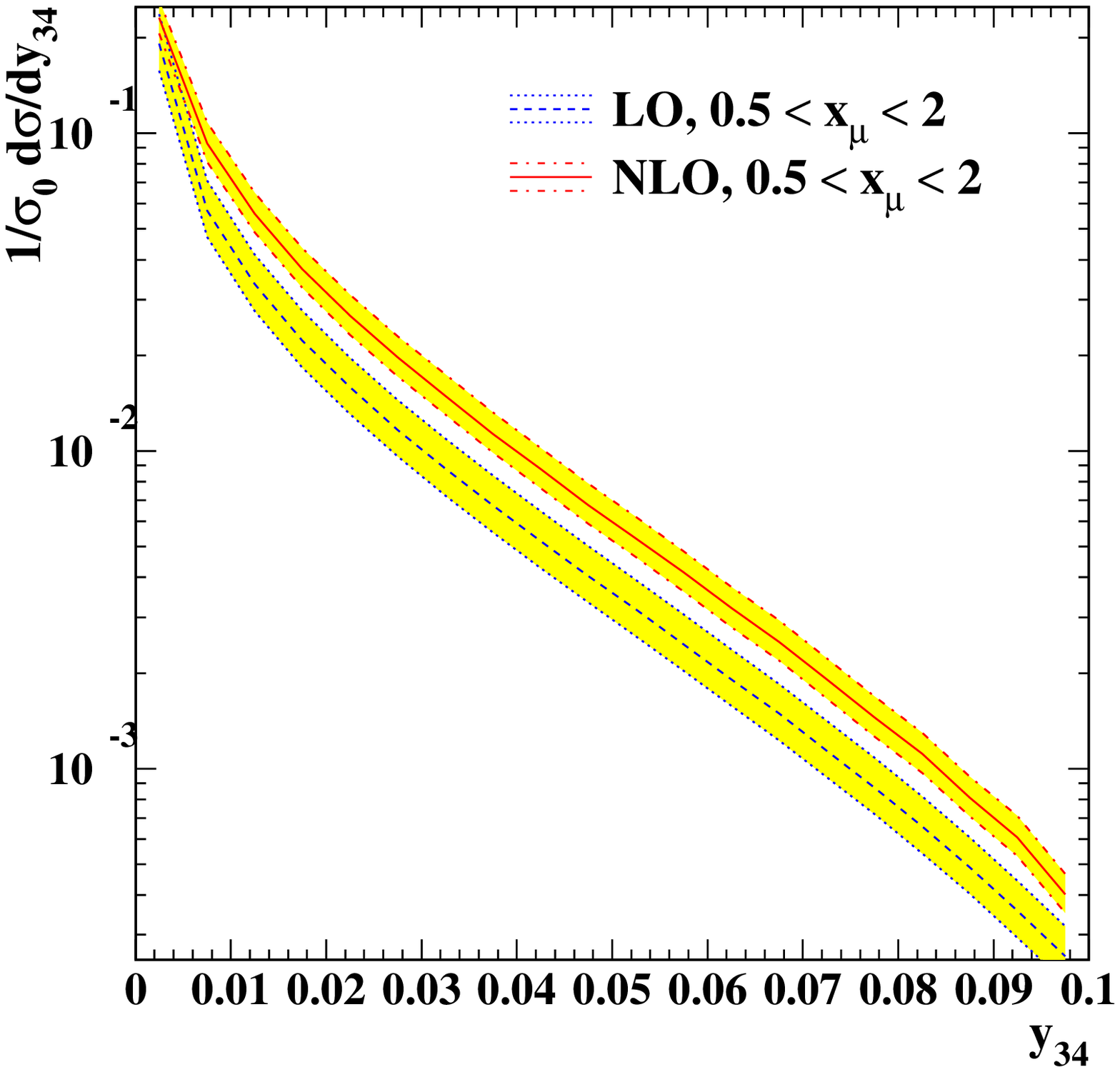}
 }
\end{minipage}
\caption{\label{fig:D_y34}
Comparison of the LO and NLO predictions for the D parameter (left) and the 
 $y_{34}$ distribution (right). The shaded regions indicate the renormalization
scale dependencies.}
\end{figure}
\begin{figure}[bthp]
\begin{minipage}[t]{.47\linewidth}
\centerline{ \includegraphics[width=8.5cm]{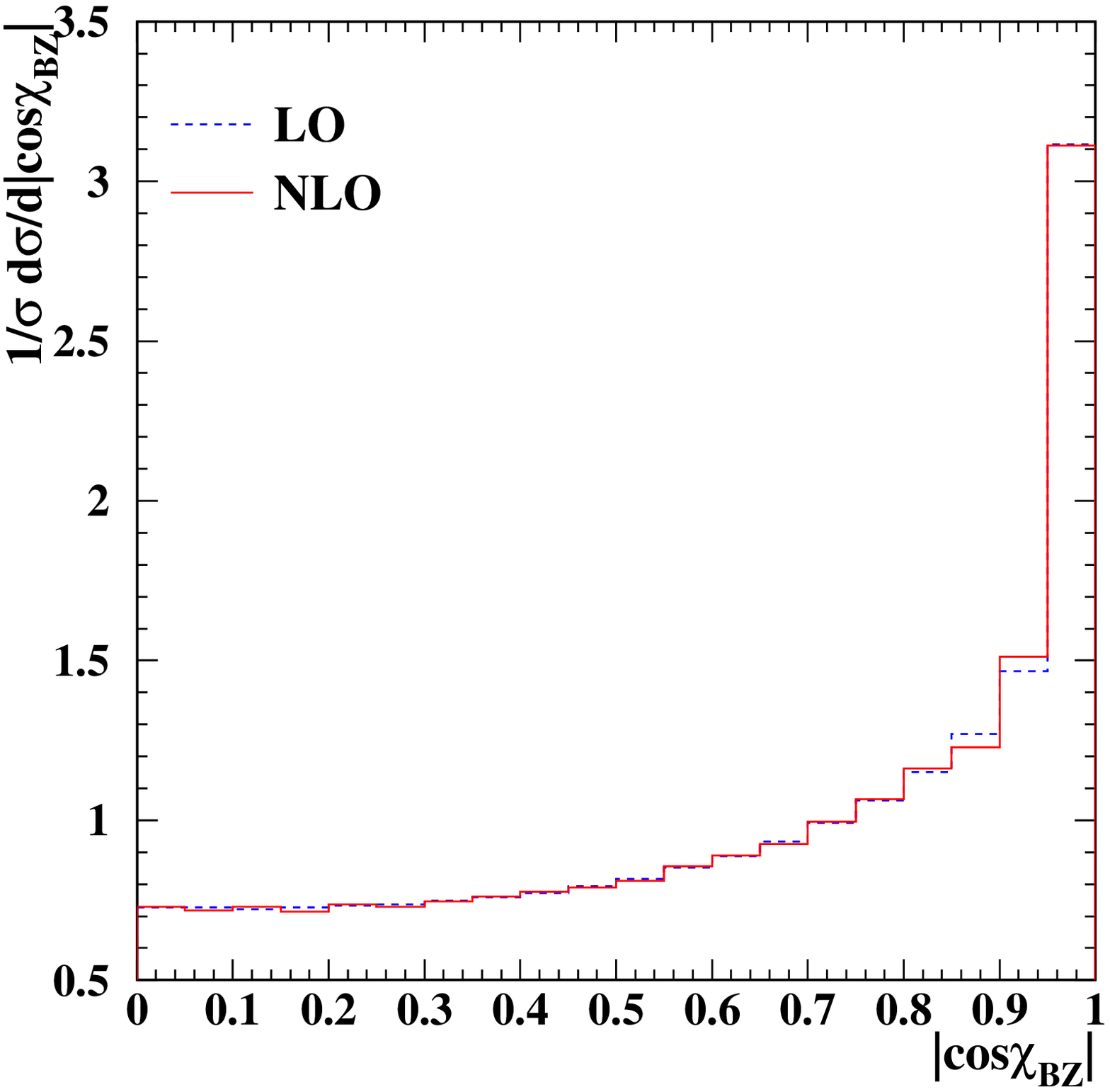}
 }
\end{minipage}
\hfill
\begin{minipage}[t]{.47\linewidth}
\centerline{ \includegraphics[width=8.5cm]{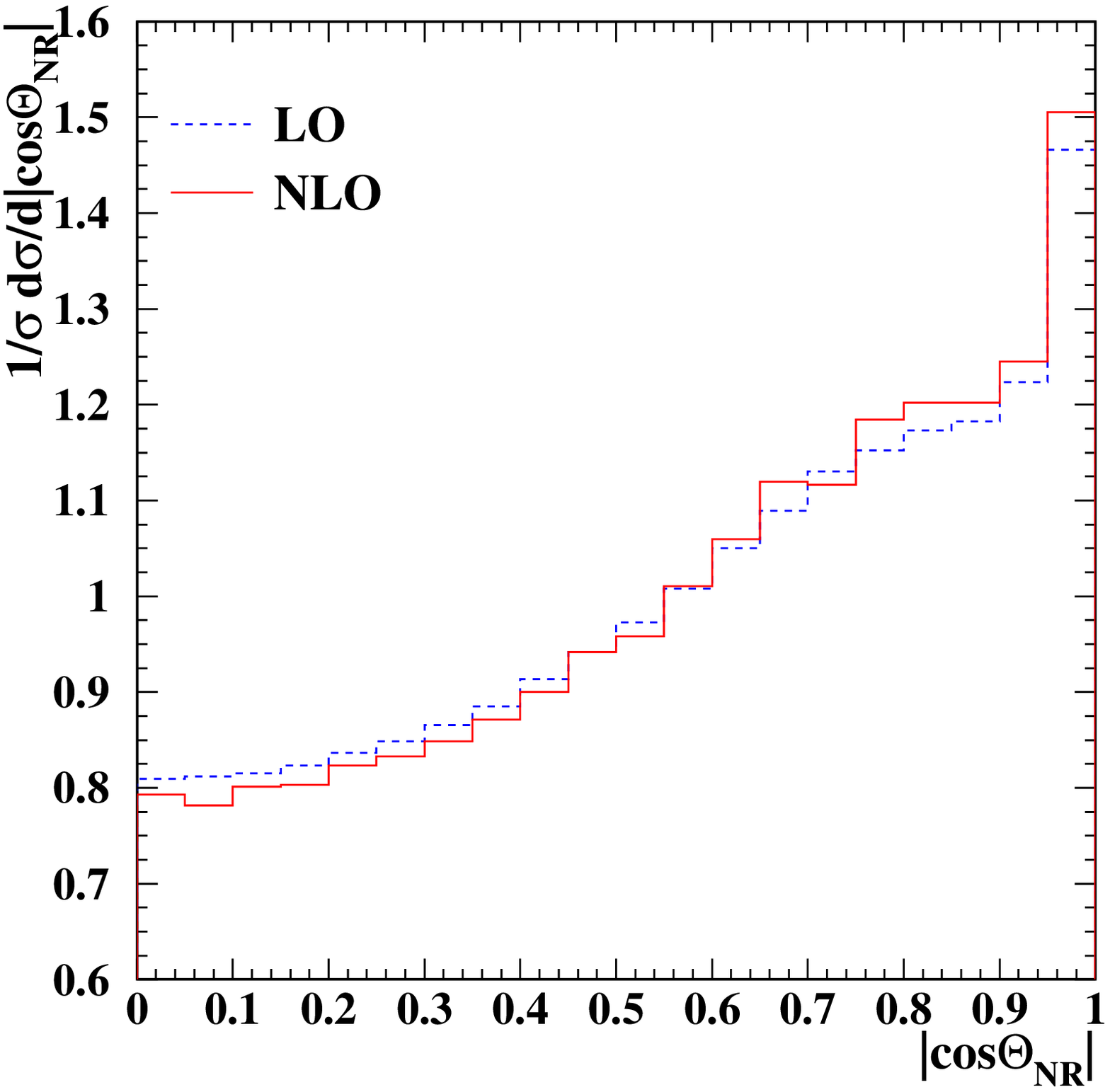}
 }
\end{minipage}
\caption{\label{fig:BZ_NR}
Comparison of the LO and NLO predictions for the Bengtsson-Zerwas (left) and the Nachtmann-Reiter distribution (right) for $x_\mu = 1$.}
\end{figure}

We have performed a high statistics calculation of the $B$ and $C$
functions for the two event shape variables and for the four angular
correlations defined in \sect{sec:GD_observ}.
\fig{fig:D_y34} shows the LO and
NLO perturbative cross sections for the two event shapes. We observe
that the inclusion of the radiative corrections increases the overall
event rate substantially (by 70--130\%). For instance, the mean value of
the D parameter at NLO is $\langle D \rangle^{NLO} = 0.0383(2)$, which is 
77\% larger than the LO value, but still far from the measured result. The
NLO predictions exhibit significant renormalization scale dependence
indicating the importance of even higher orders, which are not likely
to be known in the foreseeable future. Thus, for event shapes the
perturbative description remains unsatisfactory unless one attempts to
use an optimized scale choice \cite{campbell}. The only possible
exception is the four-jet rate for the \durham\ algorithm, where the
relative size of the NLO correction is around 60\%, and the resummation
of large logarithms exists \cite{CDOTW}. Indeed, after matching the
next-to-leading logarithmic and NLO results one finds a small
renormalization scale dependence and a remarkably good description of the
corrected experimental data \cite{zoltan4}.

The perturbative result is much more convincing in the case of the
angular correlations, although the NLO calculations have brought some
surprises, too. In \fig{fig:BZ_NR} we plot the
LO and NLO perturbative predictions for the distributions of the
Bengtsson-Zerwas and modified Nachtmann-Reiter angles. We see that the
shapes of the distributions change very little when going from LO to
NLO. However, when these predictions are used for measuring the QCD color
factors, then, at NLO, one uses a quadratic form of the color charge ratios
instead of the linear form used in a LO analysis \cite{zoltan5}. The
different functional form of the fitted function leads to different
fitted parameters, even if the shape of the distribution is only slightly
changed. In particular, the coefficient of the $T_R/C_F$ colour factor
ratio receives a very large negative contribution leading to a significant
shift in the measured value of this ratio if the NLO prediction is used instead
of the LO one \cite{zoltan3}.

\subsubsection{Mass corrections}
  \label{subsec:FK_masscorr}

For the investigation of mass effects \fourjphact\ and the 
package \apacic/\amegic\ were used with the mass parameters of
\tab{masses}.
Note that \apacic\ adds the value of a cutoff to the
mass parameters, see \sect{sec:apamas}.
\begin{table}[ht]
\begin{center}
\begin{tabular}{||l|ccccc||}
\hline\hline
 &      u         &      d        &      s       &      c        &      b           \\\hline
\fourjphact    & 0.35 GeV &0.35 GeV & 0.5 GeV &  1.5 GeV & 4.8 GeV    \\
\apacic\ & 0.01 GeV & 0.01 GeV& 0.2 GeV &  1.7 GeV & 4.7 GeV \\\hline\hline
\end{tabular}
\caption{\label{masses} Quark mass values used for the generators
                        \fourjphact\ and \apacic. }
\end{center}
\end{table}
\begin{figure}[ht]
\begin{tabular}{cc}
\epsfxsize=8cm\epsffile{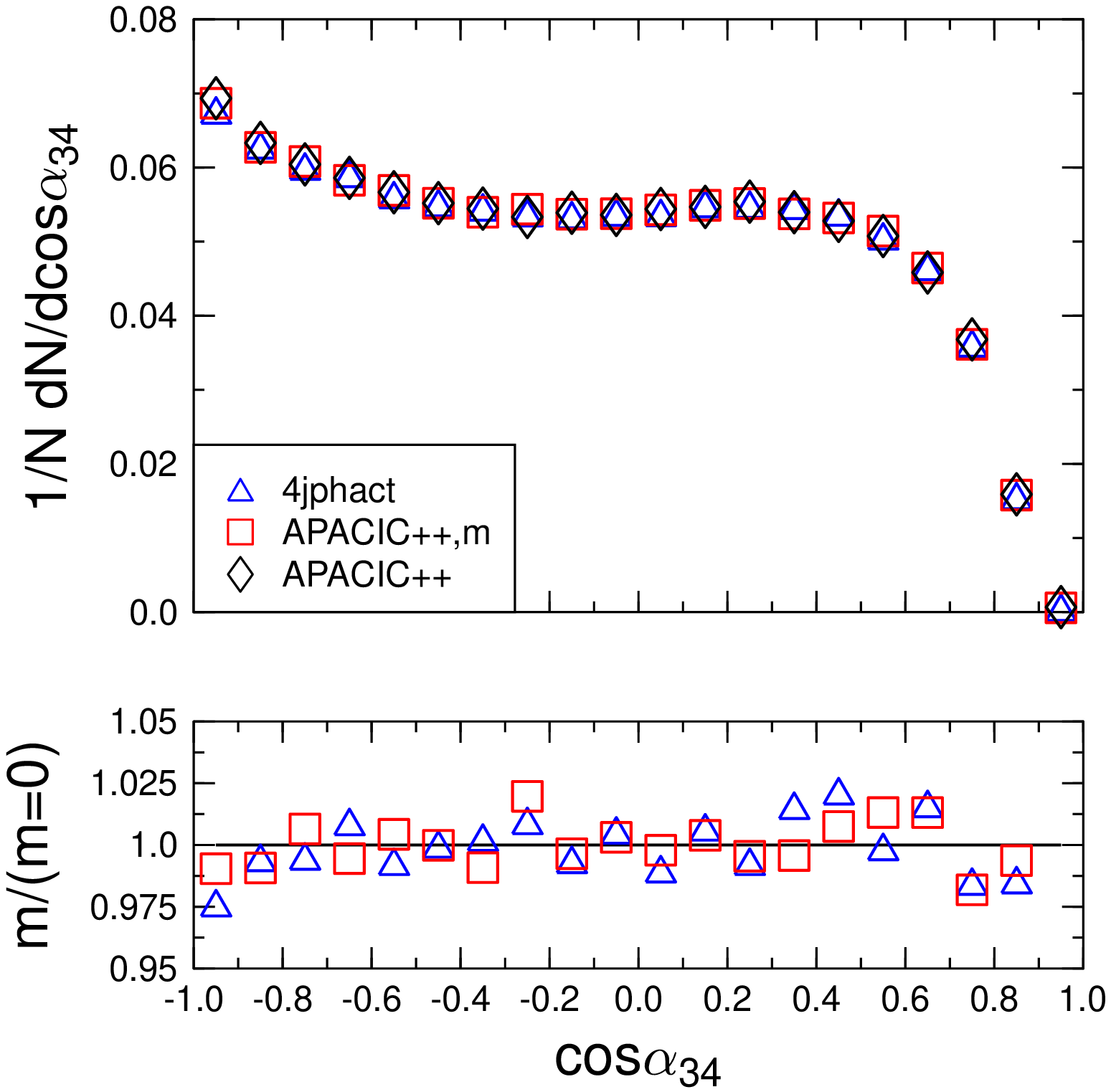} &
\epsfxsize=8cm\epsffile{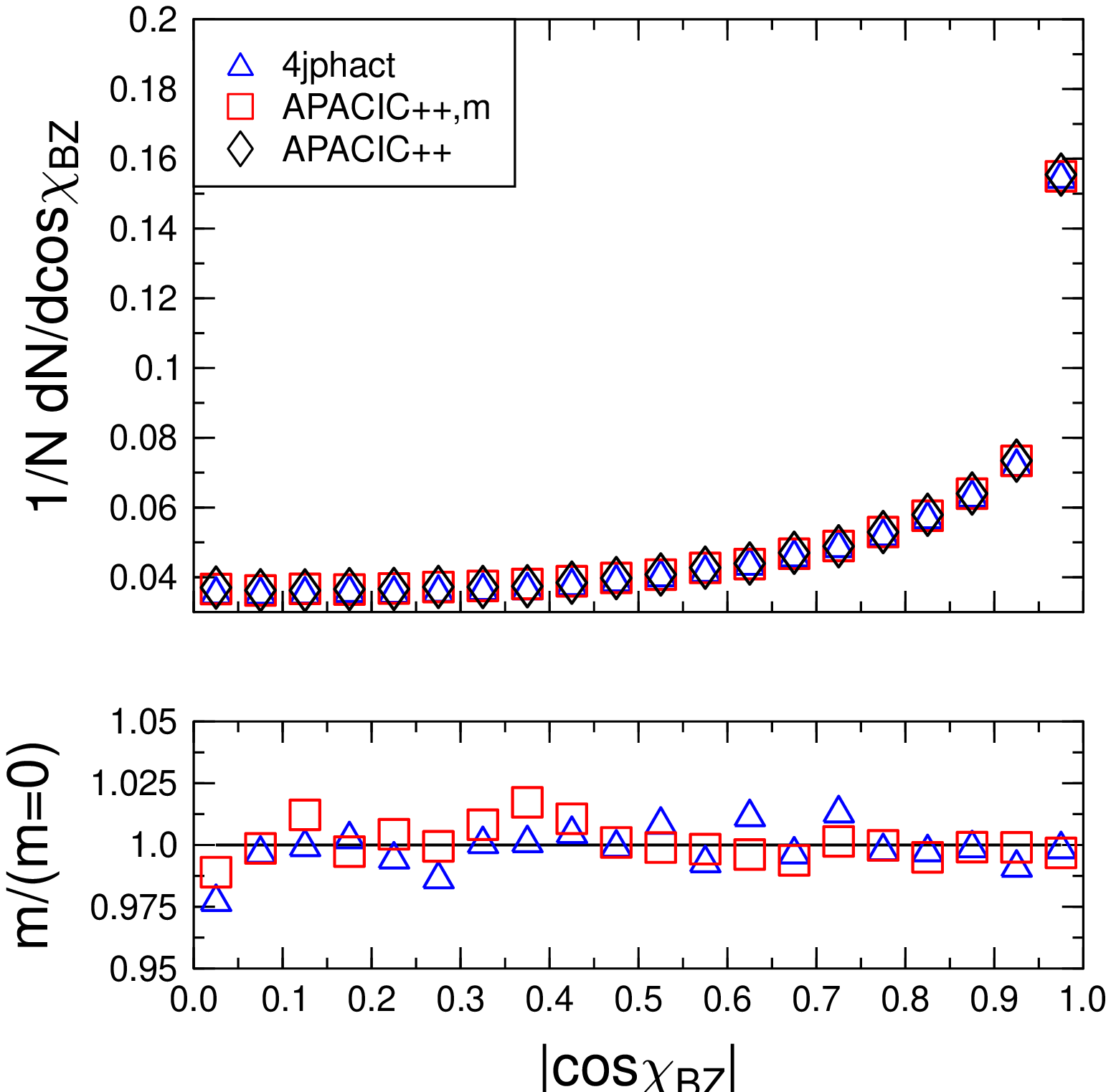} \\ 
\epsfxsize=8cm\epsffile{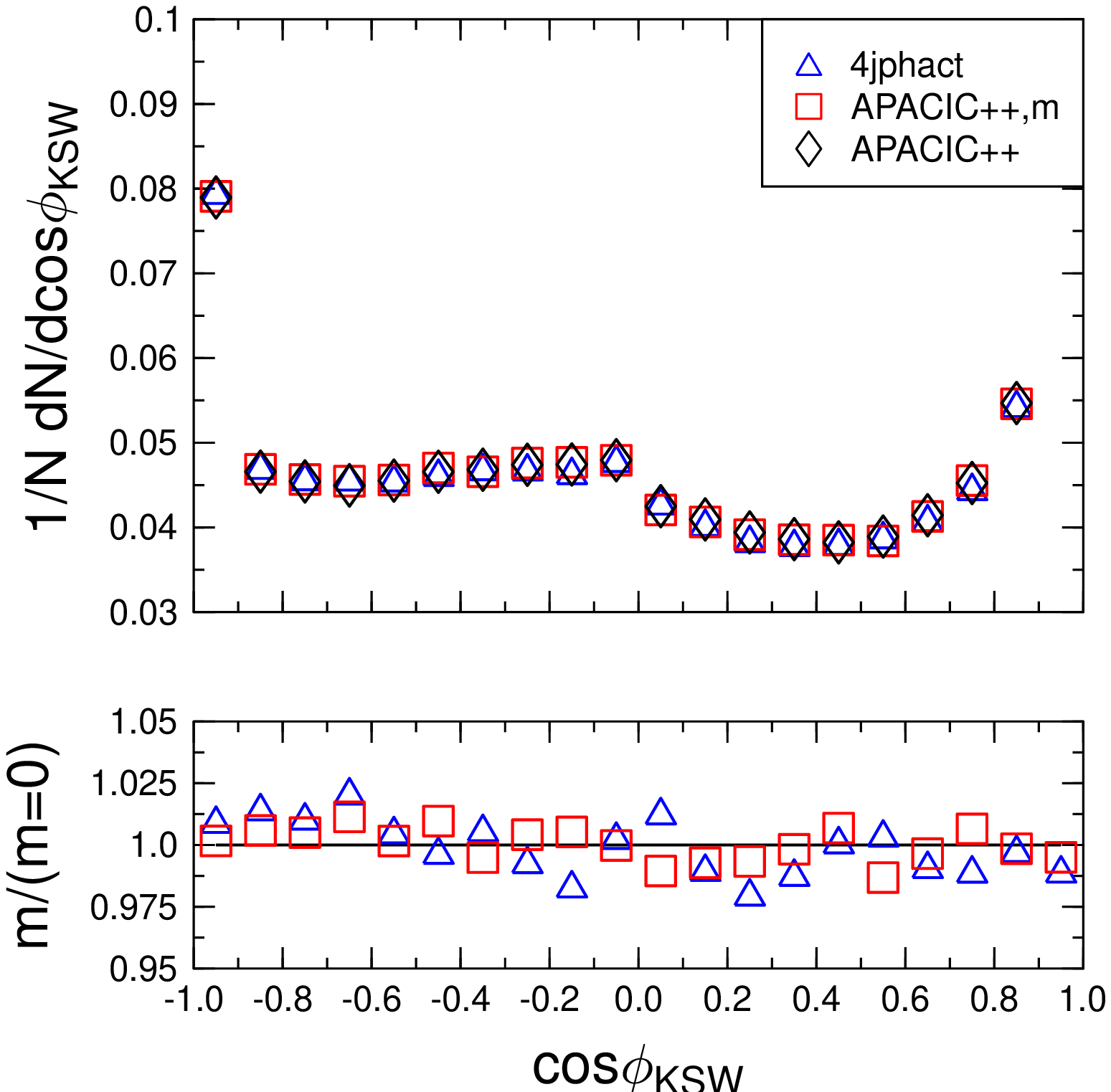} &
\epsfxsize=8cm\epsffile{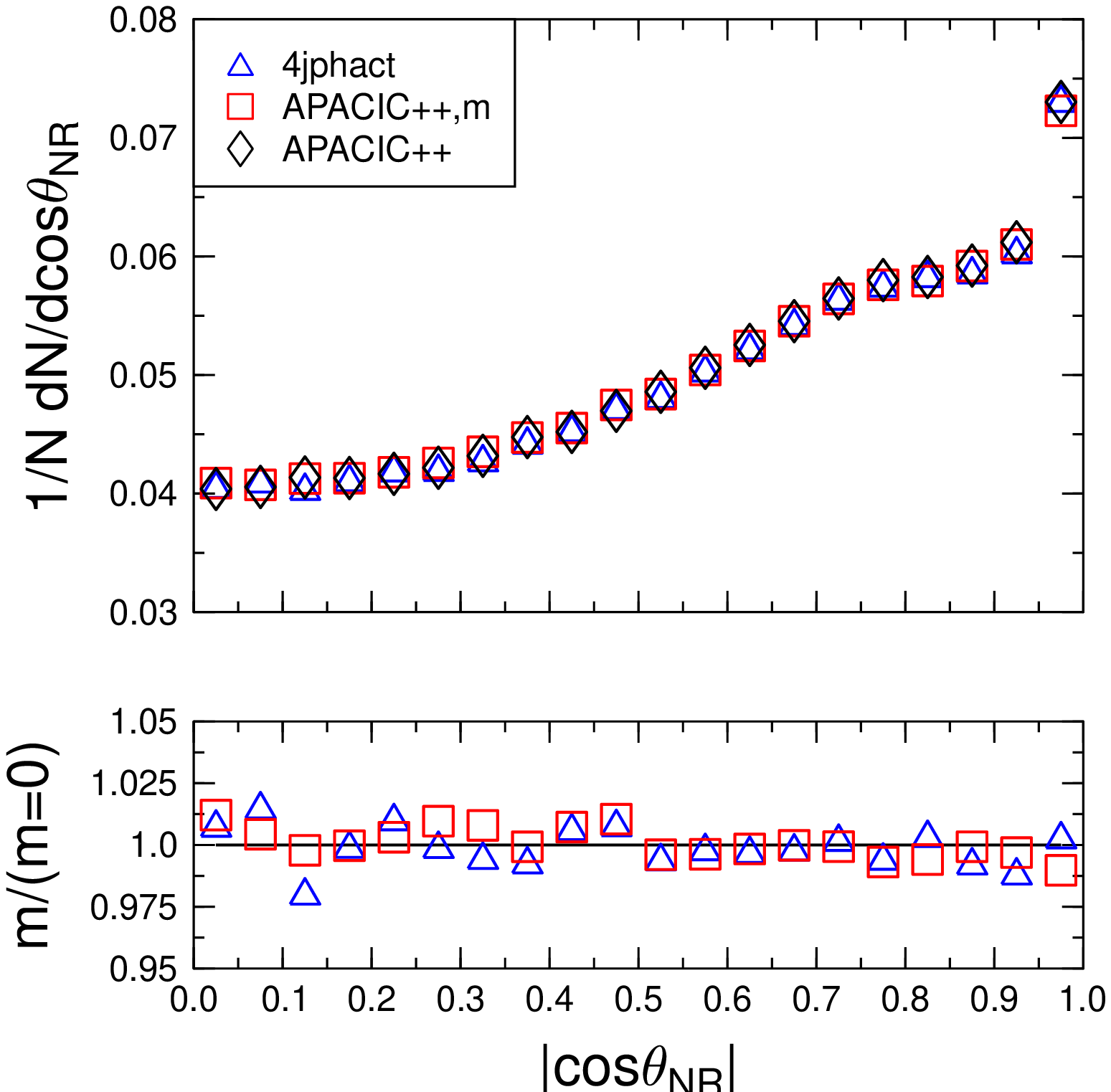} 
\end{tabular}
\caption{\label{angm} Comparison of massive and massless results for the four jet angles
on the level of LO--matrix elements.}
\end{figure}

In general, the inclusion of masses has only a minor effect on the angular distributions. 
This result can be read off \fig{angm}. In the upper plots the massive distributions 
of \fourjphact\  and \apacic\ are confronted with the massless 
result of \apacic,
and in the corresponding lower plots the ratios massive/massless are depicted. 
In most of the bins the effect of masses is of the order of $2\%$.
\begin{figure}[ht]
\begin{tabular}{cc}
\epsfxsize=8cm\epsffile{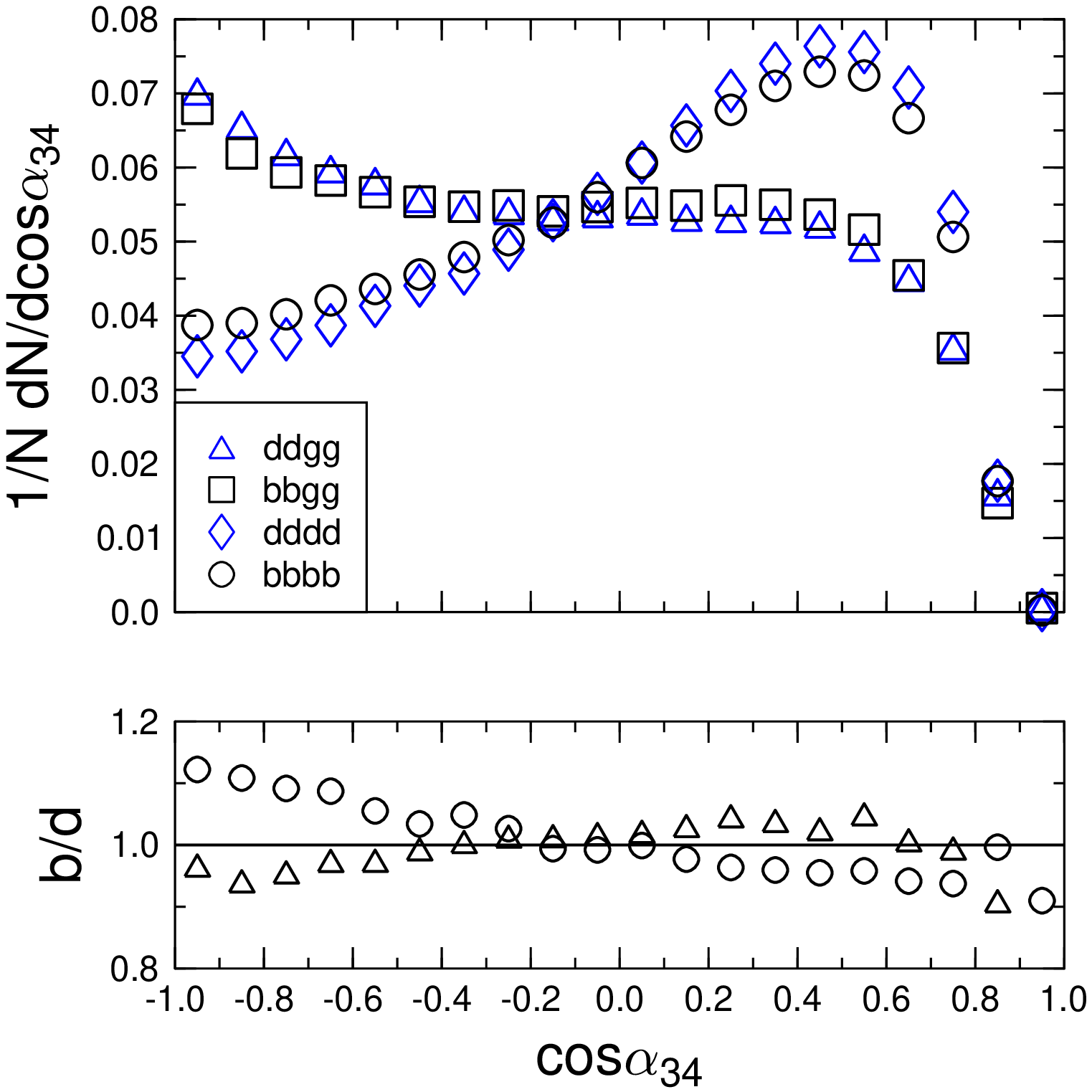} &
\epsfxsize=8cm\epsffile{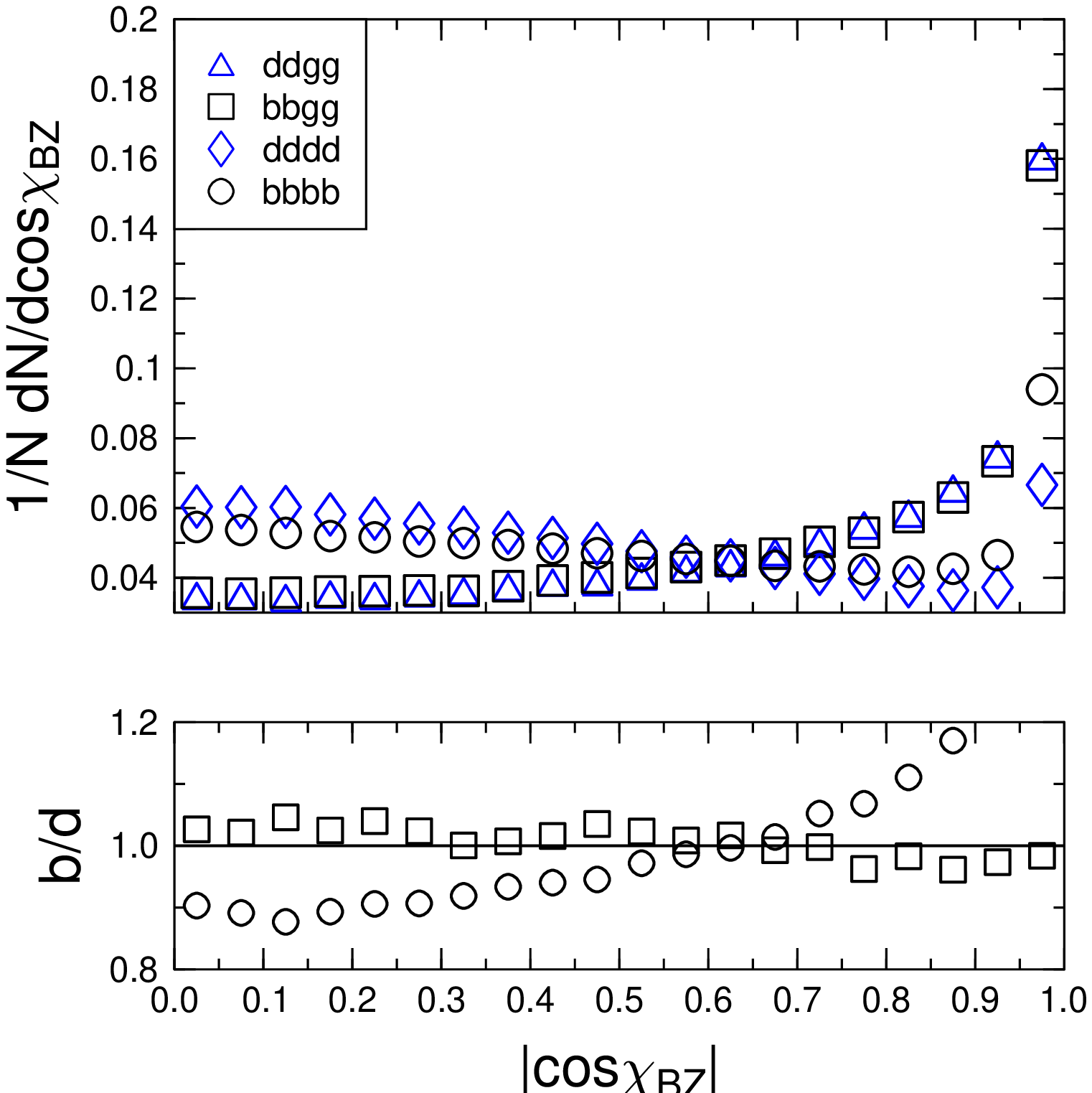} \\ 
\epsfxsize=8cm\epsffile{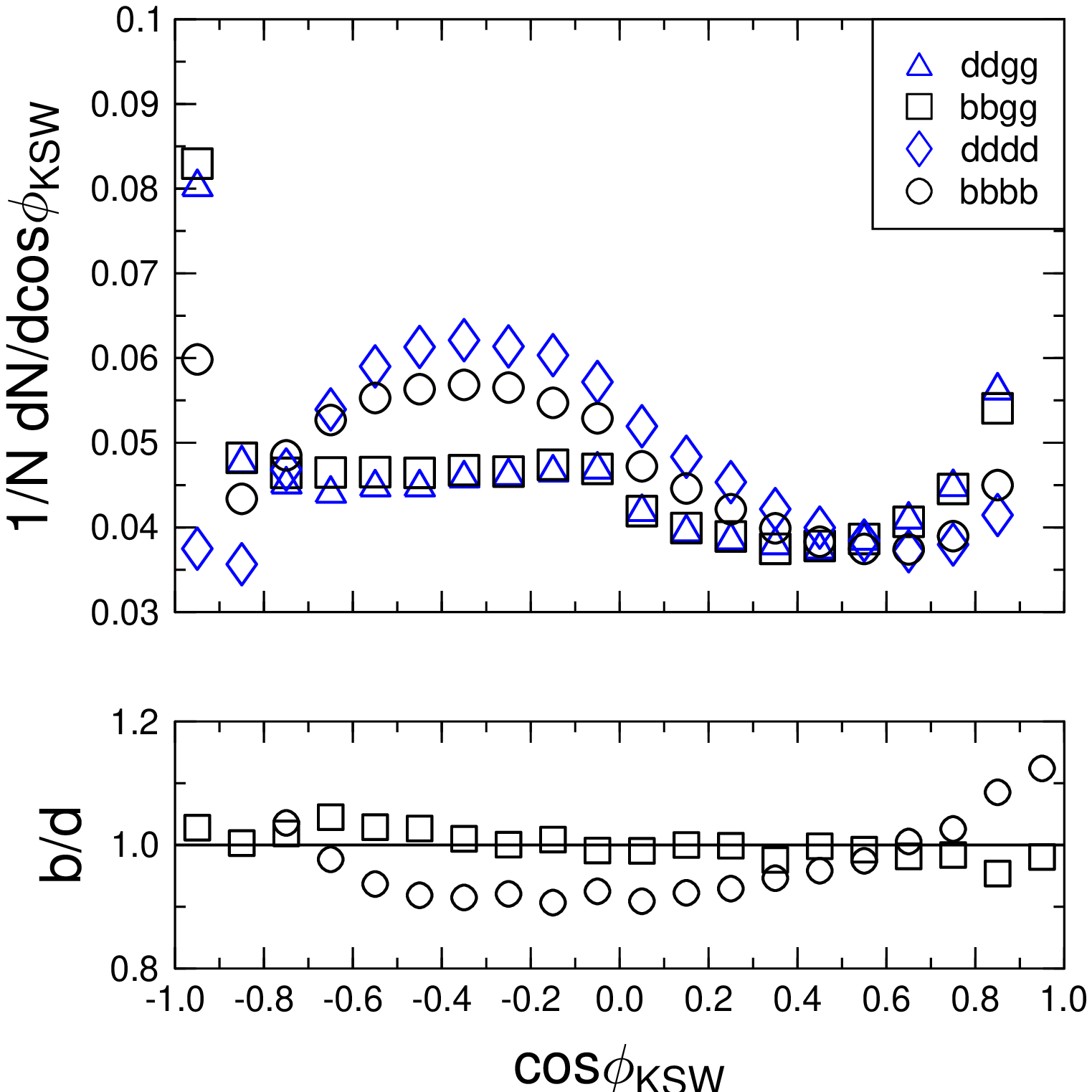} &
\epsfxsize=8cm\epsffile{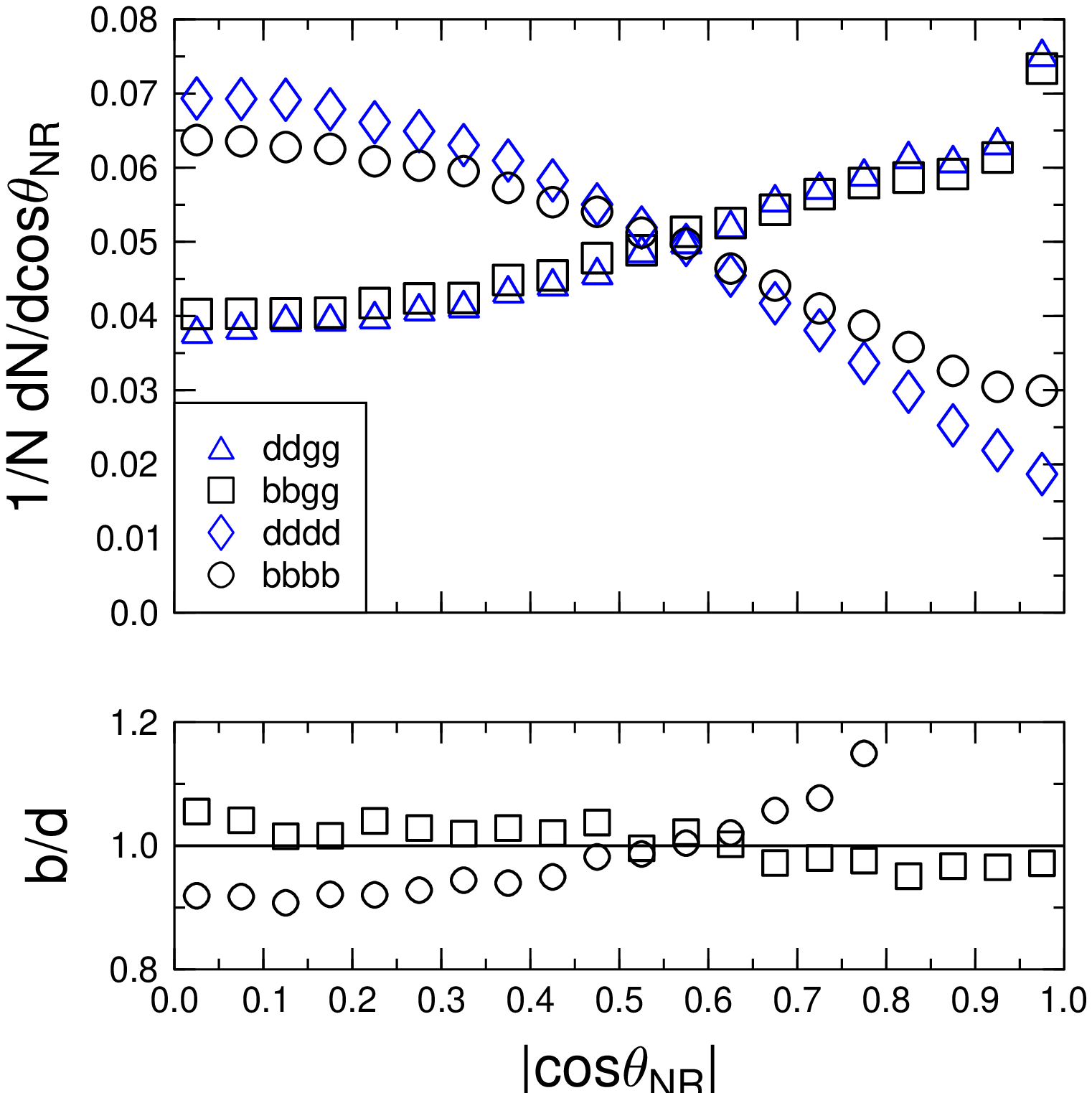} 
\end{tabular}
\caption{\label{angind} The four jet angles for some selected channels with and without masses.}
\end{figure}
This result is not too surprising, however, since only the $b$--quark mass induces any sizeable 
effect. In \fig{angind}, the results of different channels as given 
by \apacic\ are compared, 
namely $b\bar bb\bar b$, $b\bar bgg$, $d\bar dd\bar d$, and $d\bar dgg$. Again, in the upper plots 
the appropriately normalized number of events per bin is displayed for each channel, and the lower 
plots depict the corresponding ratios $(b\bar bb\bar b)/(d\bar dd\bar d)$ and 
$(b\bar bgg)/(d\bar dgg)$.
Closer inspection of this figure reveals that the by far dominant four 
jet $b$ channel, 
namely  $b\bar bgg$, is 
affected on the level well below $10\%$ in most of the phase space by mass effects. 
On the other hand, large effects, best seen in the comparison of $b\bar bb\bar b$ 
versus $d\bar dd\bar d$, are suppressed
by the small relative rates of the corresponding channels. 

\begin{figure}[ht]
\centerline{\epsfxsize=12cm\epsffile{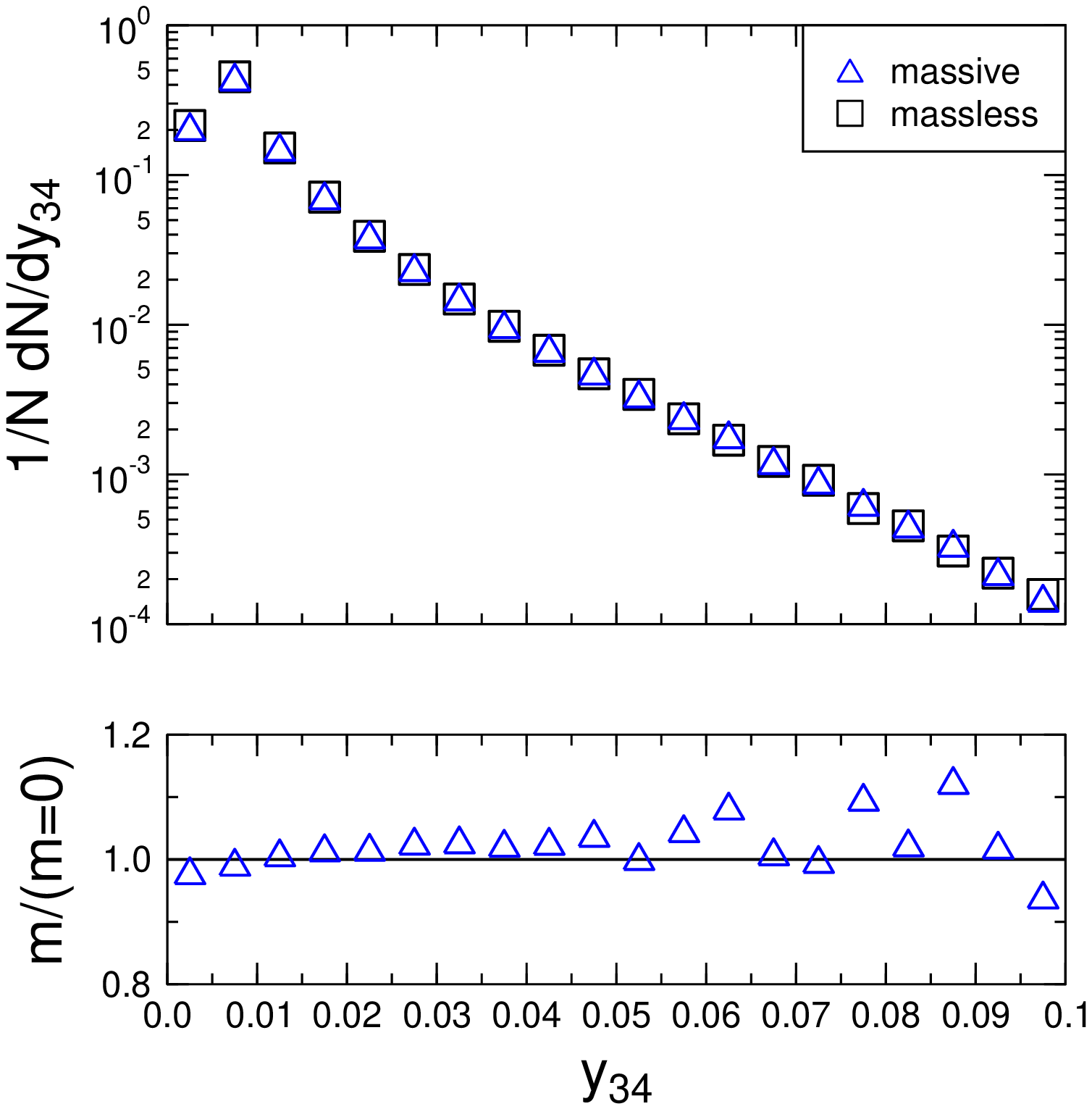}
} 
\caption{\label{ym} The $y_{34}^D$ distribution with and without the inclusion of masses.}
\end{figure}

In the $y_{34}$--distribution, mass effects are completely negligible, 
see \fig{ym}. 
In the upper  plot, the $y_{34}$ distribution as given by \apacic\ with and 
without the inclusion of masses 
are depicted. The lower plot shows the ratios of the massive and the massless 
distributions. 
As expected, the inclusion of masses results in a slight shift from 
relatively soft (small $y_{34}$) to 
comparably hard (large $y_{34}$) events, 
since masses shield the collinear regions of particle 
emission. However, this effect is only of minor size, 
and in most bins the ratios massive/massless 
are quite close to 1. Large deviations can be seen in bins with limited statistics only.
%

\subsubsection{Comparison of shower models}
  \label{subsec:GD_pscomp}

In this section we want to compare the predictions of various
parton shower models. When working with the new MC options which
allow for generating parton showers starting from a four-parton
configuration, care has to be taken for particular aspects of
these models. In order to avoid singularities in the four-parton
generation according to the LO matrix elements, some intrinsic
cut-off has to be applied for these programs, for example a \durham\
resolution criterion in case of \herwig.
Therefore the resolution criterion with which jets are selected
at the analysis level has to be larger than this intrinsic one.
In case of \herwig, the intrinsic cut-off chosen is 
$y^{intr}_{cut}=0.004$, whereas jets are selected with $y_{cut}=0.008$.
The analysis level can be the final state after the parton shower
or after the hadronization step.

Attention has also to be paid to the edges of certain phase space
regions, such as the end points in the angular distributions.
These regions could be sensitive to the details in the implementations
of the models, such as the handling of the masses and virtualities
of the partons from which the parton showers start. 

The hadronization parameters for the various models are taken
from the tuned parameter sets as used by \aleph. No study with
respect to variations in these parameters has been performed.

\begin{figure}[htbp]
\begin{center}
  \begin{tabular}{ll}
  \includegraphics[width=8cm]{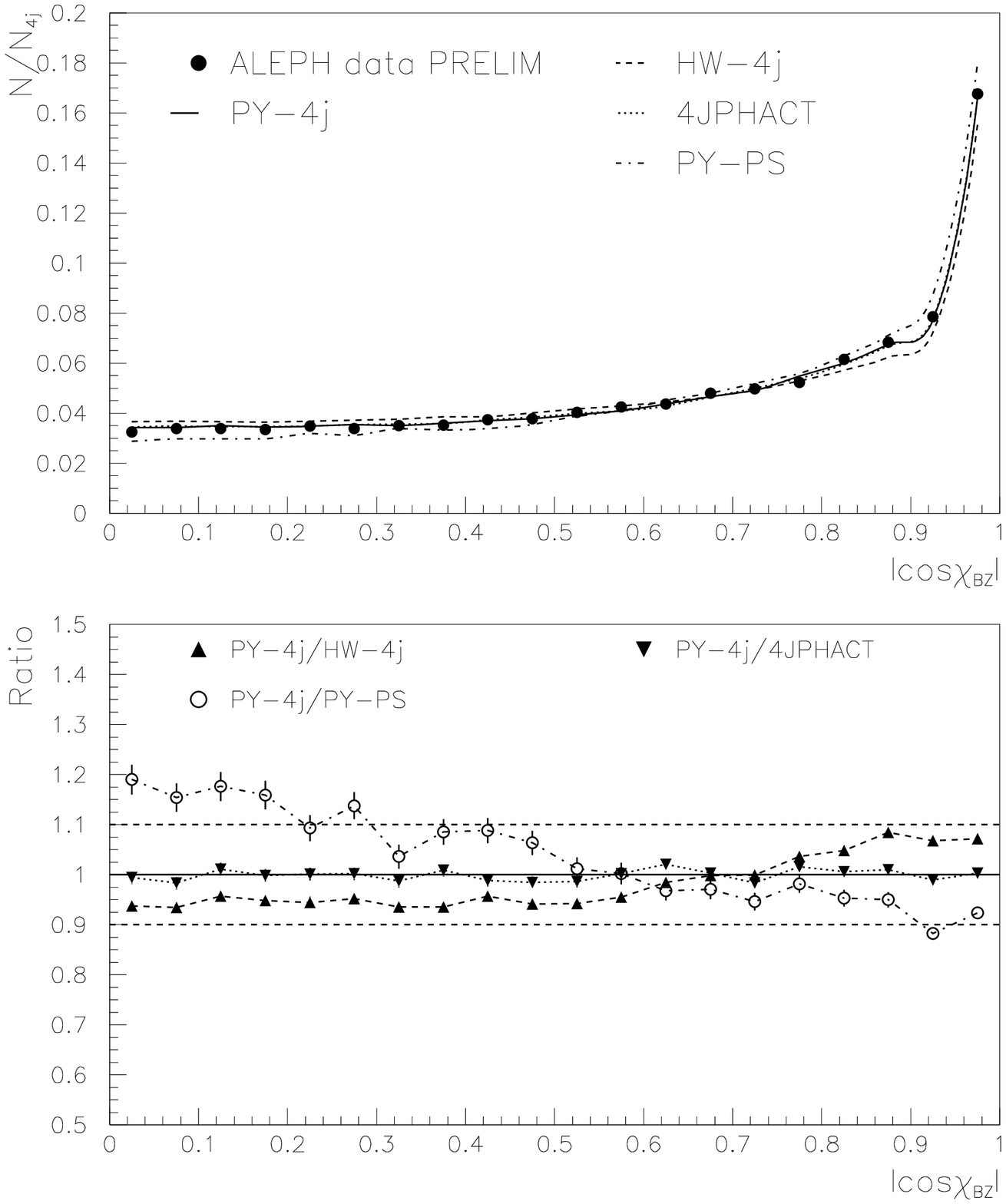} &
  \includegraphics[width=8cm]{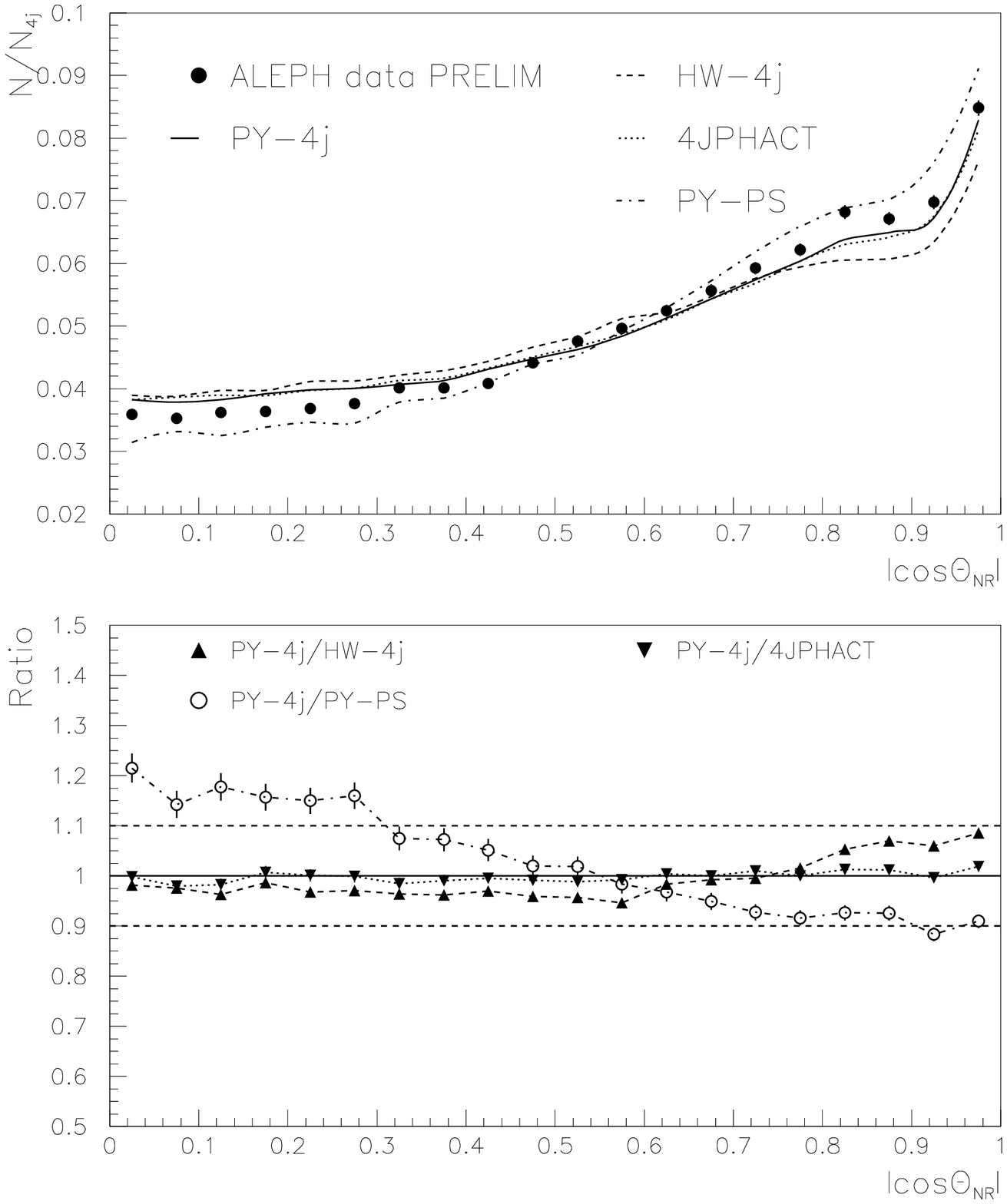} \\
  \includegraphics[width=8cm]{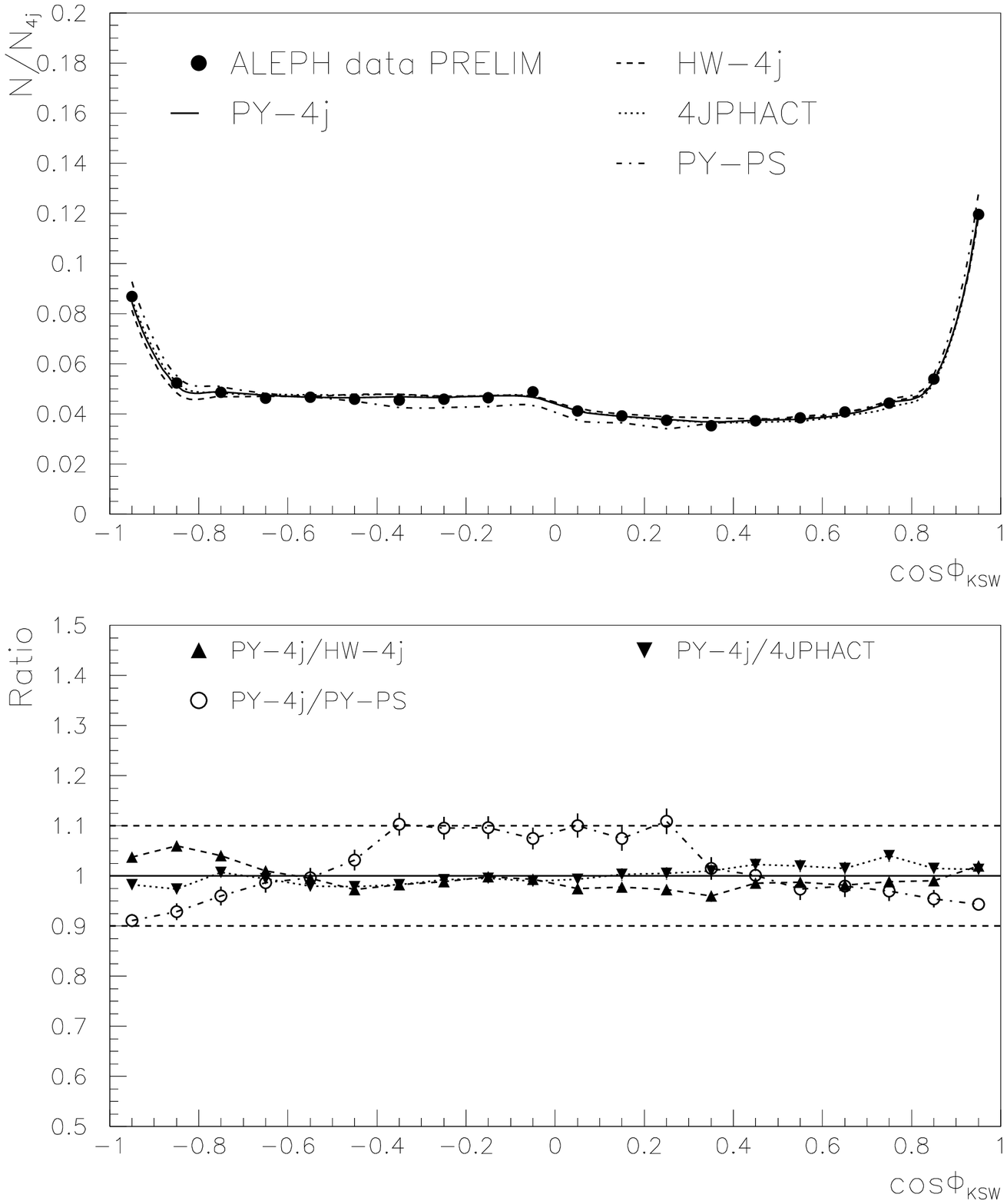} &
  \includegraphics[width=8cm]{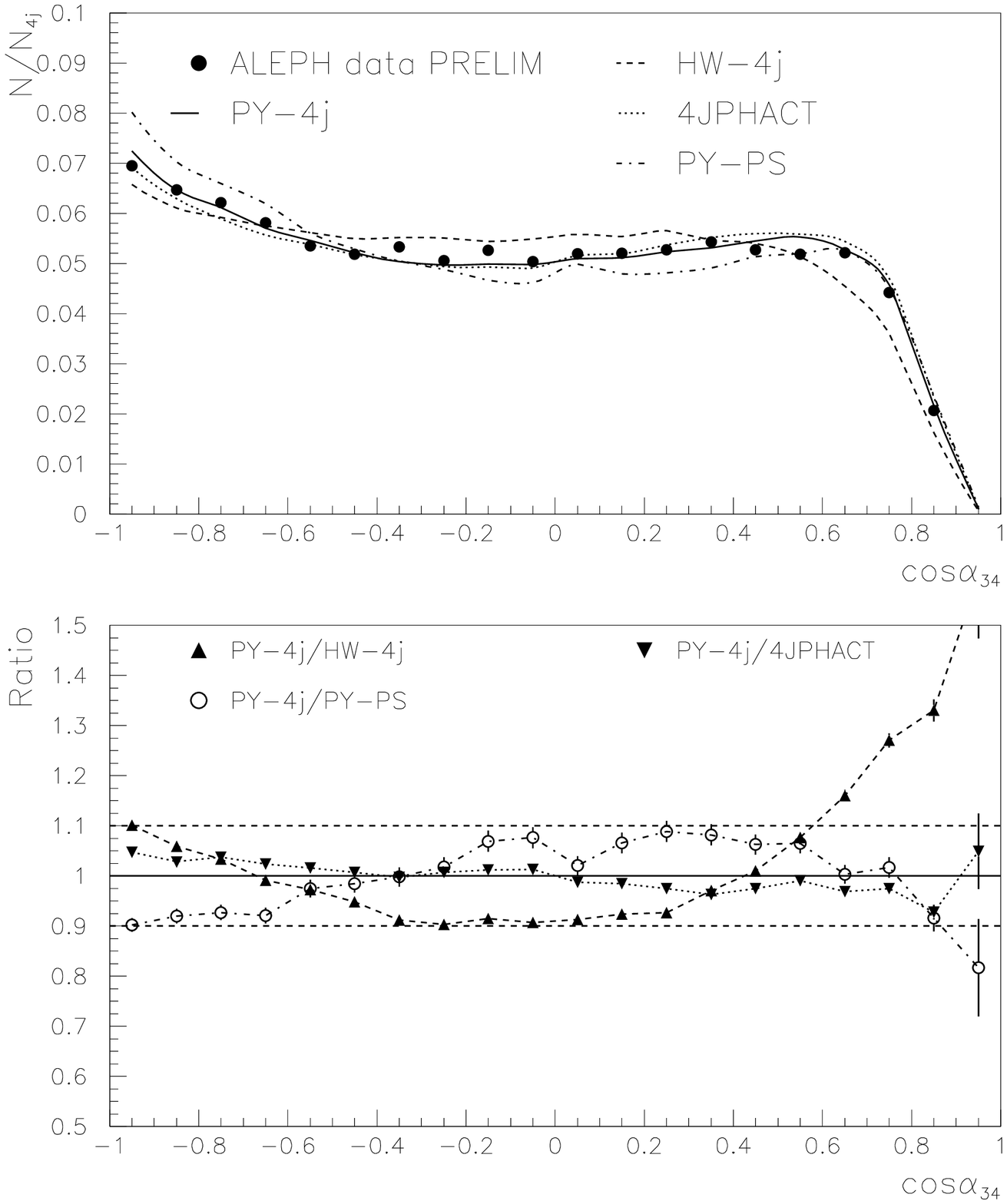}
\end{tabular}
\caption{Comparison of \aleph\ data (preliminary, statistical errors
only) to various MC model predictions for the four-jet angular 
distributions. Shown are also the ratios of the model predictions
at hadron level. PY-4j=\pythia\ 6.1, 4-parton generation plus parton 
shower; HW-4j=\herwig\ 6.1, 4-parton generation plus parton 
shower; \fourjphact=4-parton generation including mass effects, parton
shower via \pythia\ 6.1; PY-PS=\pythia\ 6.1, standard parton shower
starting from a $q\bar{q}$ pair.\label{fig:gd_4angles}} 
\end{center}
\end{figure}

A study of the changes of the shapes for the angular distributions
when going from parton to hadron level is outlined below in
\sect{sec:HW_hadronization} for the case of  \herwig. Here
we concentrate on the comparison of the model predictions at hadron
level, i.e., after parton showering and hadronization. Also shown
is a comparison to data by \aleph. These data are preliminary,
with statistical errors only, since the main purpose is to 
give a qualitative benchmark for the MC predictions.
The data have been corrected for detector acceptance and 
resolution effects by means of bin-to-bin correction factors,
as described in \sect{sec:GD_observ}. The Monte Carlo
program employed for this purpose is based on a standard
parton shower approach, starting from a  $q\bar{q}$ pair.

In case of the angular distributions, the normalization is
with respect to the total number of four-jet events found, in
order to concentrate on the shape. The event shape
distributions are normalized to the total number of hadronic
events generated.

From \fig{fig:gd_4angles} the following observations are made~:
The new option in \pythia\ 6.1, which interfaces a four-parton event
to a parton shower, gives generally a very good description of
the angular distributions, whereas the standard parton shower
option shows deviations. The two other four-jet MC programs differ
from the \pythia\ four-jet option by about 5-10\%, with larger discrepancies
seen only at the high end of the $\cos\alpha_{34}$ distribution, which
is sensitive to mass effects and details of the implementation
of the interface, since there two soft jets at close angles are probed.
Quark mass effects, which are implemented in \fourjphact, do not have 
a sizeable impact on the shape of the distributions for the
measured sample, which is a normal flavour mixture. However, when
studying event samples enriched in heavy quarks, the mass effects
should definitely be taken care of.

\begin{figure}[htbp]
\begin{center}
  \begin{tabular}{ll}
  \includegraphics[width=8cm]{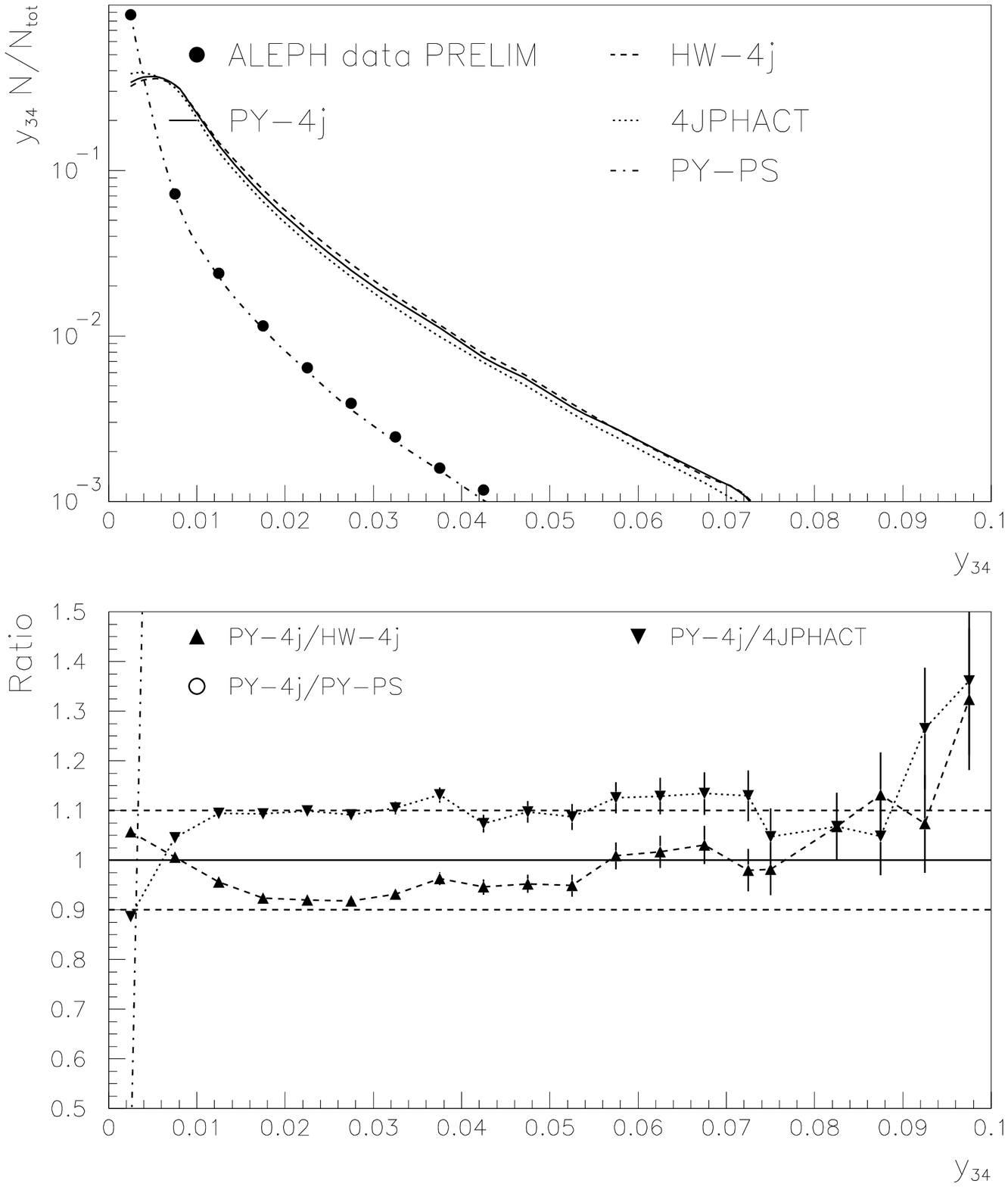} &
  \includegraphics[width=8cm]{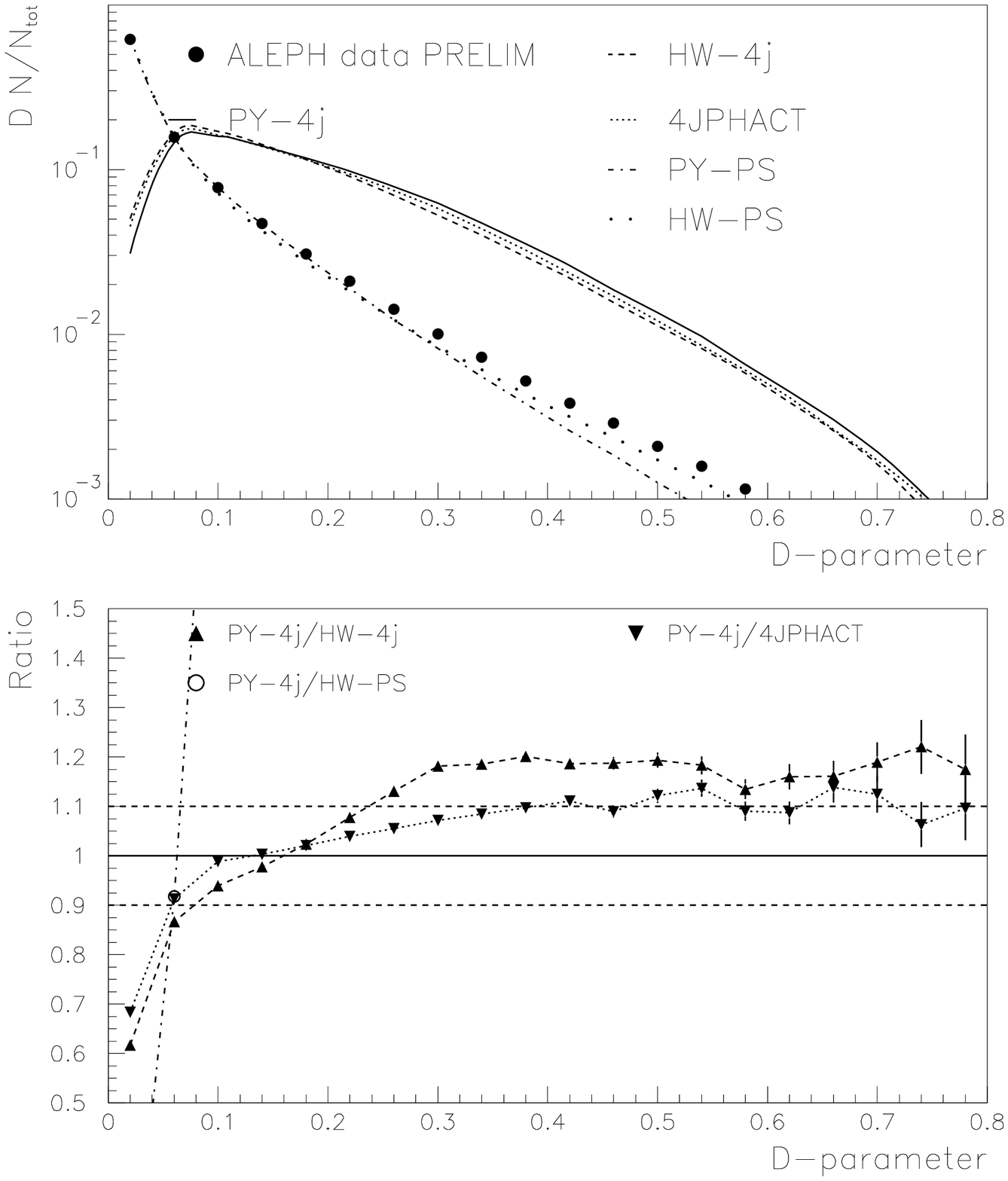} 
\end{tabular}
\caption{Comparison of \aleph\ data (preliminary, statistical errors
only) to various MC model predictions for the event shape 
distributions $y_{34}$ and D parameter. 
Shown are also the ratios of the model predictions
at hadron level. PY-4j=\pythia\ 6.1, 4-parton generation plus parton 
shower; HW-4j=\herwig\ 6.1, 4-parton generation plus parton 
shower; \fourjphact=4-parton generation including mass effects, parton
shower via \pythia\ 6.1; PY-PS=\pythia\ 6.1, 
HW-PS=\herwig\ 6.1, standard parton shower
starting from a $q\bar{q}$ pair.\label{fig:gd_evshapes}} 
\end{center}
\end{figure}

In \fig{fig:gd_evshapes} similar comparisons are shown
for the event shape distributions $y_{34}$ and D parameter.
Here a rather different picture arises. The four-jet MCs
definitely fail to describe these distributions, whereas
the standard parton shower approach gives good predictions
apart from the very tails. This can be understood from the
fact that the low end of the event shape distributions is 
rather sensitive to contributions from two- and three-jet
events, which vanish at leading order in perturbative QCD,
but can become sizeable after the parton shower simulation
and after hadronization. These contributions are not
implemented in the four-jet MC programs. Furthermore, the 
predictions for the low regions
of the event shapes depend strongly on 
the intrinsic four-jet resolution parameter.

Therefore from these observations we conclude that the new
four-jet MC options are well suited for the description of
the shape of angular distributions in four-jet events, with
remaining uncertainties of  the order of 5\%, if critical
phase space regions are avoided. However, they cannot be used
for observables which are sensitive to contributions from 
two- and three-jet events, which is the case for event shape
distributions. Also the relative jet rates cannot be
predicted correctly by these programs.

\subsection{Hadronization corrections}
  \label{sec:HW_hadronization}

A preliminary study of hadronization effects on the four-jet angular
distributions was made using the \herwig\ four-jet matrix element
+ parton shower option, described in \sect{sec_hw4jet}.

\begin{figure}
\begin{center}
\includegraphics[width=13cm]{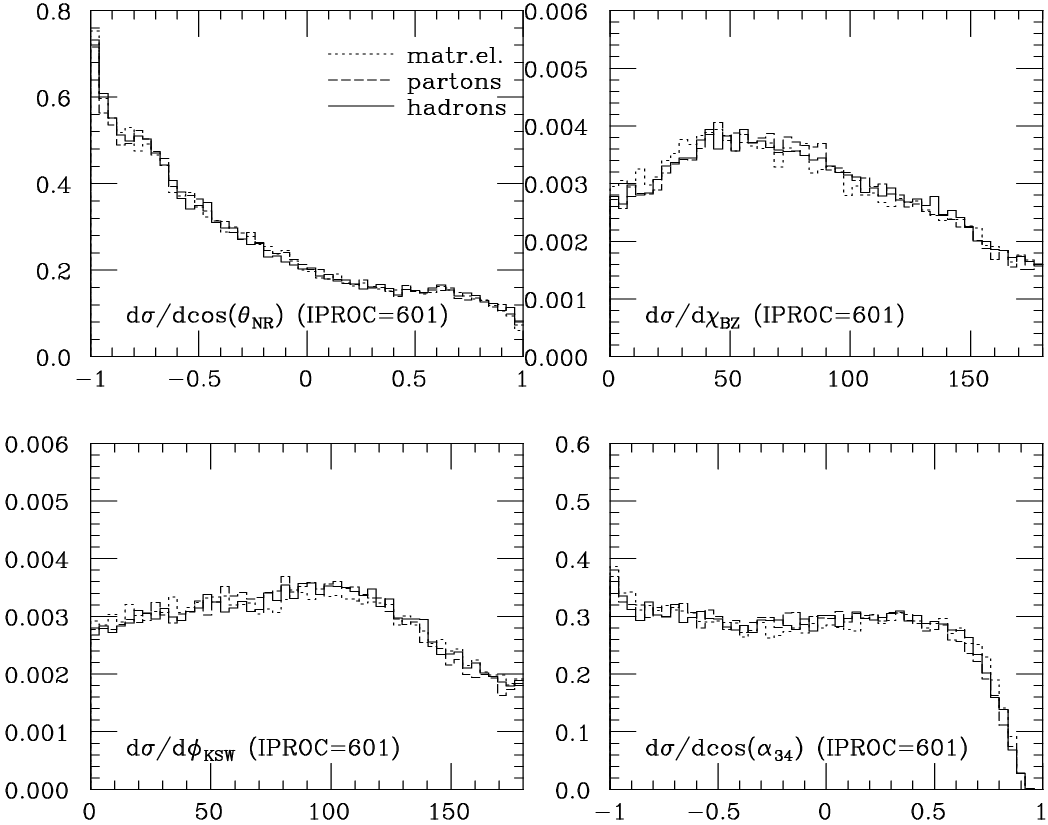}
\caption{Parton- and hadron-level 4-jet angular distributions obtained with
\herwig\ version 6.1 for light primary quarks, compared with matrix element.}
\label{fig_hw4jet}
\end{center}
\end{figure}
\begin{figure}
\begin{center}
\includegraphics[width=13cm]{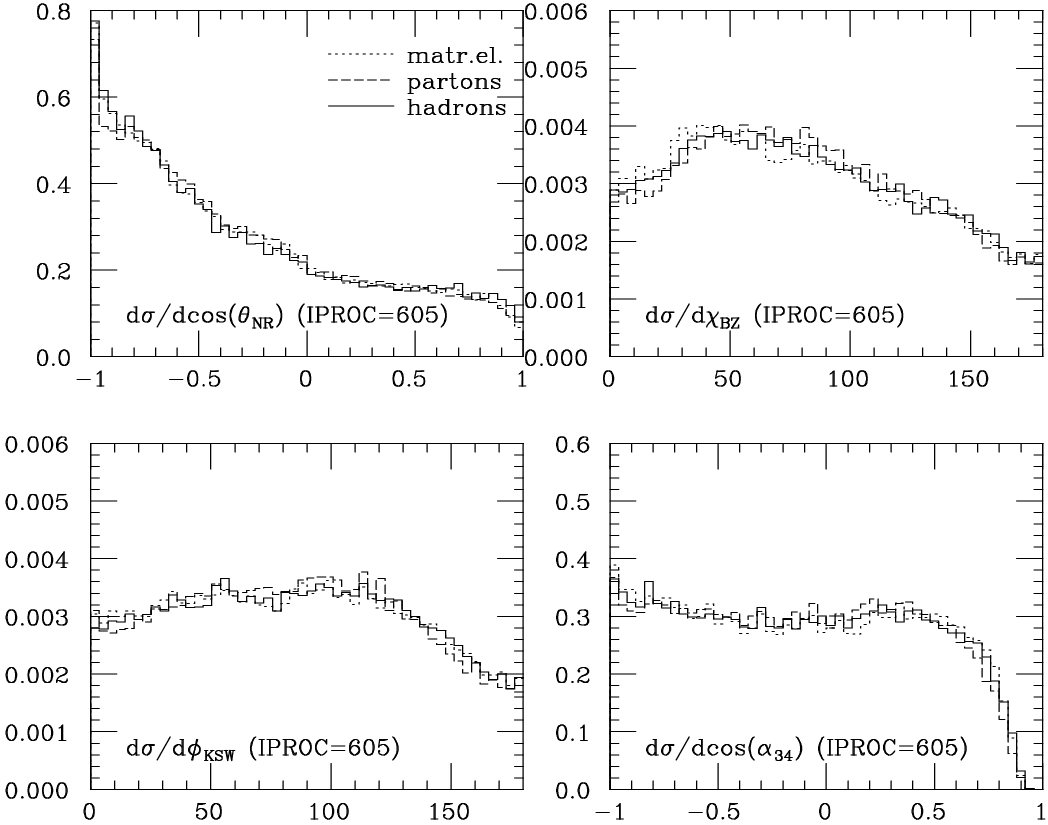}
\caption{Parton- and hadron-level 4-jet angular distributions obtained with
\herwig\ version 6.1 for primary $b$ quarks, compared with matrix element.}
\label{fig_hw4jetb}
\end{center}
\end{figure}

Figures~\ref{fig_hw4jet} and \ref{fig_hw4jetb} show the
distributions obtained at $s=\mz^2$ for light primary quarks
({\tt IPROC} = 601) and primary $b$-quarks ({\tt IPROC} = 605),
respectively.  The \durham\ jet metric was used with matrix-element
cutoff {\tt Y4JT} = $0.004$ and $y_{cut}=0.008$ for the actual resolution
of jets.  The dotted histograms show the matrix element distributions,
with the argument of $\as$ running as explained in \sect{sec_hw4jet},
while the dashed and solid ones are the reconstructed jet distributions
at the parton level (after showering) and hadron level (after decays),
respectively.  In each case the distributions are normalized to
the number of 4-jet events found (with $y_{cut}=0.008$) at the
relevant level.

One sees that hadronization effects on the shapes of the 4-jet
angular distributions are generally not large. The hadron/parton
level ratios are, within the limited statistics of the present study,
broadly similar to those obtained using the \herwig\ 2+3 jet ME+PS option
({\tt IPROC} = 101,105). There are, however, indications that hadronization
effects may be overestimated by the 2+3 jet ME+PS option at the 5-10\%
level in certain places, e.g.\ the central regions of $\theta_{NR}$ and
$\chi_{BZ}$.   Thus the use of correction factors obtained from the 2+3 jet 
ME+PS option needs to be treated with caution if precision better than 10\%
is required.

Close study of Figures~\ref{fig_hw4jet} and \ref{fig_hw4jetb} reveals
the rather surprising fact that the hadron-level results are often
closer than the parton-level ones to the (massless) matrix-element
distributions. This is particularly true for primary $b$-quarks,
suggesting that in \herwig\ the kinematic effects of the $b$-quark
mass during parton showering are largely cancelled by the effects
of hadronization and B-hadron decays.

Thus first results from the \herwig\ 4-jet ME+PS option suggest that
hadronization effects are similar to those obtained using the old 2+3
jet option at the 10\% level,
and that effects of parton showering and hadronization
tend to cancel in the 4-jet angular distributions.
This should be investigated further as a function of c.m.\ energy
and jet resolution. Further studies of 4-jet hadronization using the
\pythia\ generator as well as the new combined 2,3 and 4 jet ME+PS
\herwig\ option (see \sect{sec:hw234jet}) are also needed.

\subsection{Conclusions}
  \label{sec:GD_conclusions}

A study of the description of 4-jet final states by various
Monte Carlo models has been presented. Particular
emphasis has been put on the comparison of new Monte Carlo
generators, which produce 4-parton final states according
to the leading order QCD matrix elements, and then add a 
parton shower and hadronization. In general good agreement
between the different generators has been found. These new
models give a better description of 4-jet angular distributions
than the standard parton shower models, where the parton shower
starts from a quark-antiquark initial state, only.
However, jet rates as well as
event shape distributions sensitive to 4-jet production,
such as $y_{34}$, 
cannot be described by these models, since here also the
contributions from 2- and 3-jet events are important. These
observations should be taken into account when using the 4-jet MCs
for background studies.

Quark mass effects are small for the distributions under
consideration, for samples with normal flavour mixture. 
NLO contributions have minor impact on the shape
of 4-jet angular distributions, but they change considerably
the 4-jet rate and event shape distributions such as the
D parameter. Possible large non-perturbative power corrections
to observables such as the mean value of the D parameter
have not been studied here.

A preliminary study of hadronization effects on 4-jet angular distributions,
using the 4-jet ME+PS option of the \herwig\ generator only, showed with
limited statistics that such effects are broadly similar to those
seen with the 2+3 jet ME+PS option, although differences at the 5-10\%
level are apparent in certain configurations.

\section{B QUARK FRAGMENTATION FUNCTION}

Heavy quark fragmentation functions are a powerful tool in testing the
predictivity of perturbative QCD (pQCD), since effects of non-perturbative
origin are much more limited in size than in the light-flavour case.
At the origin of this behaviour lies the fact that the mass $m$ of the
heavy quark is much larger than the QCD scale $\Lambda$. 

Indeed, on one side
the large mass acts as an infrared cutoff for the mass singularities
which would appear in the perturbative calculation, ensuring a finite
result. The energy distribution of the $b$ quark prior
to hadronization can therefore be calculated perturbatively.
On the other side, hadronization effects have to be phenomenologically
modelled, but happen to be small: a heavy quark only loses a momentum
fraction of order $\Lambda/m$ when binding with a light one to 
form a heavy-light meson~\cite{bfrag-bj}.

\subsection{Experimental results}

Results for the normalized energy distribution of $B$
hadrons, i.e.
\begin{equation}
D(Q,x_E) \equiv {1\over\sigma} {d\sigma\over{d x_E}} \; ,
\label{bfrag-eq1}
\end{equation}
in $e^+e^-$ collisions are given by the \lep\ collaborations and by the 
\sld\ experiment at SLC, at $Q = \mz$. The scaling variable $x_E$ is given
by the ratio of the observed $B$ particle energy to the beam energy $E_{beam}$.

\begin{table}[ht]
\begin{center}
\begin{tabular}{||c|c|c||}
\hline\hline
$\langle x_E\rangle$                      &  $B$              &  Expt     \\
\hline
\hline
0.7394$\pm$0.0054(stat)$\pm$0.0057(syst) & L              & \aleph\ 2000
~\cite{bfrag-aleph2000}\\
\hline
0.7198$\pm$0.0045(stat)$\pm$0.0053(syst) & wd             & \aleph\ 2000
~\cite{bfrag-aleph2000}\\
0.714$\pm$0.005(stat)$\pm$0.007(syst)$\pm$0.002(mod) & wd & \sld\ 1999
~\cite{bfrag-sld1999}\\
\hline
0.702$\pm$0.008                                   & wd & \lep\ HFWG avg. 1996
~\cite{bfrag-lepavHF1996}\\
\hline
0.695$\pm$0.006(stat)$\pm$0.003(syst)$\pm$0.007(mod) & wd & \opal\ 1995
~\cite{bfrag-opal1995}\\
0.716$\pm$0.0006(stat)$\pm$0.007(syst)               & L & \delphi\ 1995
~\cite{bfrag-delphi1995}\\
\hline\hline
\end{tabular}
\end{center}
\caption{\label{bfrag-table1} Mean scaled energy of $B$ hadrons from various
$e^+e^-$ experiments at $Q = \mz$.}
\end{table}

A typical observable measured by experiments is the mean scaled energy
fraction $\langle x_E\rangle$. \tab{bfrag-table1} shows some of
the most recent determinations of this quantity.
In this table the second column identifies the kind of $B$ particle
observed in the
final state, be it the ``leading'' (also called ``primary'') (L) or the 
``weakly decaying'' one
(wd). Of course, the average energy of the latter is lower than that of
the former, since it has undergone further decaying processes. 
It should also be noted that the
precise details of what the observed final state actually is will at
least slightly vary from experiment to experiment. The numbers quoted in
the table under the same label ``wd'' are therefore not exactly 
comparable, though probably homogeneous enough within the experimental
uncertainties so that one can average them. Needless to say, it
would be useful if all analyses at some point finally agreed on a single 
definition for this final state.

The most recent analyses also report fairly accurate data for the full
fragmentation function eq.~(\ref{bfrag-eq1}), with $x_E$ ranging
from near zero to one. As expected, these distributions peak very close to
one, around $x_E\simeq$~0.8-0.9.

\subsection{Theoretical predictions}

The challenge for the theoretical calculations is of course to reproduce
not only the mean scaled energy but also, as far as possible, the full
fragmentation distribution. A certain degree of phenomenological
modelling will be necessary, as perturbative calculations cannot of
course describe the hadronization of the $b$ quark into $B$ mesons
and/or baryons. The full fragmentation function is therefore usually
described in terms of a convolution between a calculable perturbative
component and a phenomenological one:
\begin{equation}
D(Q,x;m,\Lambda;\epsilon_1,\ldots,\epsilon_n) = D^{pert}(Q,x;m,\Lambda) 
\otimes D^{np}(x;\epsilon_1,\ldots,\epsilon_n) \; .
\label{bfrag-convol}
\end{equation}
In this equation the perturbative component depends on the
centre-of-mass energy $Q$, the QCD coupling and the heavy quark pole mass $m$, 
while the
non-perturbative one is assumed to depend only on some given set of
phenomenological parameters $(\epsilon_1,\ldots,\epsilon_n)$, 
to be determined by fitting the experimental data.

The perturbative component can be either calculated by analytical means
or extracted from Monte Carlo simulations of the emission of radiation
by the fragmenting heavy quark. In the latter case the theoretical
accuracy will of course be lower.

As far as fixed order analytical calculations are concerned, today's state
of the art is the work of ref.~\cite{bfrag-no}. It
is accurate up to order $\as^2$ and also fully includes finite
mass terms of the form $(m/Q)^p$ with $p\ge 1$.

Fixed order results do however display two classes of large logarithmic
terms: collinear logs, of the form $\log(Q^2/m^2)$, and Sudakov logs,
$\log(1-x)$. These terms become large when the centre-of-mass energy is
much larger than the heavy quark mass, a fact certainly true at \lep\ and
\sld, and at the $x\simeq 1$ endpoint respectively. All-order
resummations for such terms, very important for producing a reliable
result,  have and are being considered \cite{bfrag-res}, and are now
available at next-to-leading log (NLL) accuracy for both classes of logarithms.
Ref.~\cite{bfrag-no} also provides a merging between the fixed
order calculation and the collinear-resummed one.

Once pQCD has produced a reliable prediction for the $b$
quark fragmentation, one has to ``dress'' it with some phenomenological
modelling in order to describe the observed $B$ hadrons distribution, as
discussed before. It is important to realize that, since  only the
convolution of the two factors in \eqn{bfrag-convol} has physical
meaning, the parameters fitted in the non-perturbative part will
strictly depend on the kind of description adopted for the perturbative
term. Different descriptions and/or different parameters in
$D^{pert}(Q,x;m,\Lambda)$ will lead to different values for the fitted
$(\epsilon_1,\ldots,\epsilon_n)$ set. {\sl Once fitted, such a set will
therefore  not be usable in conjunction with perturbative descriptions
other than the one it has been fitted with.} 

\begin{table}[ht]
\begin{center}
\begin{tabular}{||c|c|c|c||}
\hline\hline
$\langle x_E \rangle_{pert}$ 
& $\Lambda_5 = 100$ MeV  & $\Lambda_5 = 200$ MeV & $\Lambda_5 = 300$ MeV\\
\hline
$m_b = 4$ GeV   & 0.790 & 0.753 & 0.723\\
$m_b = 4.5$ GeV & 0.802 & 0.767 & 0.740\\
$m_b = 5$ GeV   & 0.813 & 0.780 & 0.755\\
\hline\hline
\end{tabular}
\end{center}
\caption{\label{bfrag-table2} Perturbative predictions for $\langle
x_E\rangle_{pert}$ at $Q = 91$ GeV, for different values of $m_b$ and 
$\Lambda_5$. } 
\end{table}

As an example, \tab{bfrag-table2}
shows the average scaled energy value of the $b$ quark after fragmentation,
$\langle x_E\rangle_{pert}$, as predicted by the perturbative term only, for
different values of $\Lambda$ and $m$. The calculation used in this
example resums to NLL accuracy both collinear and Sudakov logs,
but neglects the non-logarithmic finite mass terms. One can clearly see that, 
on one hand,
the purely perturbative predictions are too large to directly describe the
experimental results, unless probably unrealistic value for $\Lambda_5$
and $m_b$ are used. On the other hand, each different $(m_b,\Lambda_5)$
choice will of course imply a different fitted value for the non-perturbative 
parameters if the same measured $\langle x_E\rangle$ is to be correctly
described.

While the calculation of the perturbative component does of course
follow the rules dictated by pQCD, much more freedom is
instead available when choosing the form of the non-perturbative
contribution $D^{np}(x)$. A lot of different parametrizations have been
employed and can be found in the literature: some of them possibly more
physically motivated, some chosen only because of practical
advantages like an easy Mellin transform or enough parameters to ensure
that the shape of the data can be properly described. A list, probably
incomplete, of these
functional forms can be found in~\cite{bfrag-nonpert}, 
though it should be noted that some of them appear now to be disfavoured by 
comparisons with experimental data.

As an example of such a comparison, let us consider one of the
best-sellers of these non-perturbative forms, namely the Peterson et al.
fragmentation function. Such a function is meant to be one of the
physically motivated ones, and has the attractive feature of depending
on only one phenomenological parameter $\epsilon$, which can moreover be
roughly related to more fundamental quantities via the relation
$\epsilon \sim \Lambda^2/m^2$. This function reads
\begin{equation}
D^{np}(x;\epsilon) = N(\epsilon) {1\over x} \left( 1 - {1\over x} -
{\epsilon\over{1-x}}\right)^{-2}
\label{bfrag-peterson}
\end{equation}
where $N(\epsilon)$ is the normalization factor.

\begin{figure}
\begin{center}
\includegraphics[width=12.5cm]{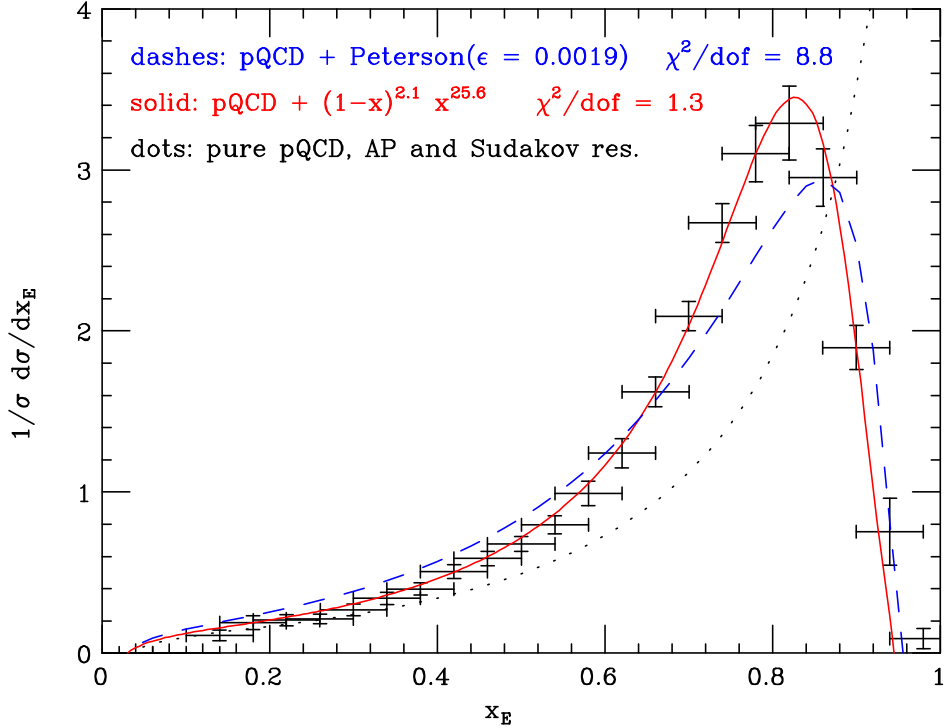} 
\caption{\label{bfrag-fig1} The data from
\sld\ \protect\cite{bfrag-sld1999} and two theoretical fits. The pQCD
prediction is in all cases collinear (or AP, for Altarelli-Parisi) and Sudakov
resummed.}
\end{center}
\end{figure}
The plot in \fig{bfrag-fig1} shows the result of attempting a
point-by-point fit of 
the Peterson form, convoluted with the same perturbative calculation
used in \tab{bfrag-table2}, to the latest data from \sld. It can be
clearly appreciated how the Peterson model, coupled to this perturbative
description, does not seem to offer a valid description of the data. The
same conclusion was reached by the \sld\ Collaboration by coupling 
this model to the \jetset\ Monte Carlo description of the perturbative
component.

It should however be noted that \eqn{bfrag-peterson} was derived
under the assumption of describing the hadronization of a heavy quark
into a heavy-light meson by picking up a light quark from the vacuum.
No attempt was made to include the description of the subsequent
decays transforming the leading $B$ particle into the weakly decaying
ones, which are the ones observed here. Such decays will modify the
shape of the fragmentation function, and might at least partially
explain the observed discrepancy.

\fig{bfrag-fig1} also shows a fit performed with a different
non-perturbative function, namely
\begin{equation}
D^{np}(x;\alpha,\beta) = N(\alpha,\beta) (1-x)^\alpha x^\beta\;.
\end{equation}
This particular form has no immediate physical origin, but it is often
used because it has a very simple Mellin transform and is flexible
enough to describe the data well. One can indeed see from the plot that it
allows for a very good fit of the experimental data.

\subsection{Monte Carlo predictions}
\begin{figure}
\begin{center}
\includegraphics[width=10.0cm]{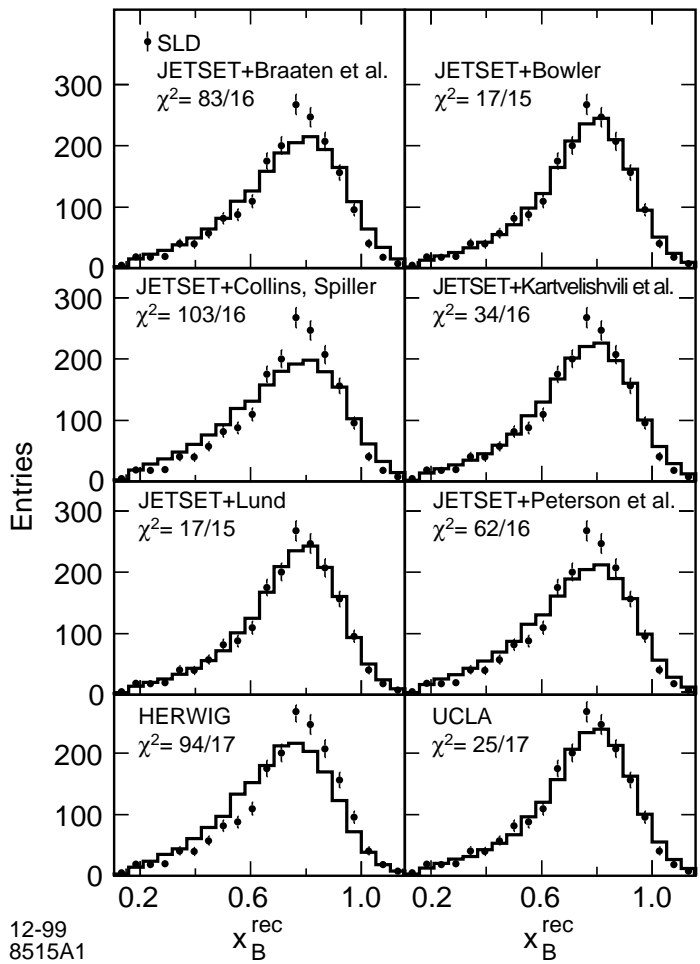} 
\caption{\label{bfrag-fig2} B-fragmentation data from
\sld\ \protect\cite{bfrag-sld1999} compared with various models.}
\end{center}
\end{figure}

An important issue that remains to be addressed is the performance of the
main Monte Carlo event generators in comparison with the latest data on
$b$-quark fragmentation.
\fig{bfrag-fig2} shows the results of combining \jetset\ (version 7.4)
parton showers with various models \cite{bfrag-nonpert} for $b$-quark
fragmentation into a weakly-decaying B-hadron, compared to the recent
\sld\ data \cite{bfrag-sld1999}. \jetset\ plus Lund
fragmentation gives a good description of the data whereas, as was the case
in the analytical calculations discussed above, the Peterson model does not.

The prediction from \herwig\ (version 5.7) is seen to be too soft in
comparison to the \sld\ data. As already remarked in \sect{sec:hwbfrag},
a harder B-hadron spectrum can be obtained in \herwig\ 6.1 by
varying the $b$-quark fragmentation parameters separately. However,
detailed tuning of version 6.1 to these data has not yet been attempted.

\subsection{Concluding remarks}

Accurate theoretical preditions exist for the perturbative part of the
heavy quark fragmentation function. Collinear and Sudakov logarithms can
be resummed to next-to-leading accuracy, and the finite mass terms are
known up to order $\as^2$. All the various contributions can be 
merged into a single result. 

On top of this perturbative result a non-perturbative contribution will
always have to be included, and precise experimental data can help
identifying the proper shape for such a function.

Predictions for heavy quark fragmentation from the latest versions of
Monte Carlo generators have yet to be compared and tuned to the
most recent data.



\section{GLUON SPLITTING INTO BOTTOM QUARKS}
\label{sec_gbbar}

\subsection{Experimental data}

The current experimental results on the quantity
$$
g_{bb} = \frac{\Gamma(Z^0\to q\bar q g, g\to b\bar b)}
{\Gamma(Z^0\to\mbox{hadrons})}
$$
are summarized in \tab{tab_gbbarex}. The first three results are
based on extracting a secondary  $b\bar b$ signal from a 4-jet sample,
whereas the latest \delphi\ value is obtained from a measurement of the
4$b$ rate $\Gamma(Z^0\to b\bar b b\bar b)$, which should be less
model-dependent and give better phase-space coverage.

\begin{table}[H]
\begin{center}
\begin{tabular}{||l|c|c||}
\hline\hline
Experiment & Ref. & $g_{bb}$ (\%)\\
\hline

\delphi\ &\cite{Abreu:1997nf}& $0.21\pm 0.11\pm 0.09$ \\

\aleph\ &\cite{Barate:1998vs}& $0.277\pm 0.042\pm 0.057$ \\

\sld\ &\cite{Abe:1999qg}& $0.307\pm 0.071\pm 0.066$ \\

\delphi\ &\cite{Abreu:1999qh}& $0.33\pm 0.10\pm 0.08$ \\
\hline\hline
\end{tabular}
\caption{Experimental data on gluon splitting to $b\bar b$.}
\label{tab_gbbarex}
\end{center}
\end{table}

\subsection{Analytical predictions}

References~\cite{Seymour:1994ca,Seymour:1995bz,Miller:1998ig}
give leading-order [${\cal O}(\as^2)$]
predictions as well as the results of resummation of
leading [$\as^n\log^{2n-1}(s/m_b^2)$] and next-to-leading
[$\as^n\log^{2n-2}(s/m_b^2)$] logarithms (NLL) to all orders in
$\as$, matched with leading order.
The results, for $\as=0.118$, $s=\mz^2$ and $m_b=5.0$ GeV,
are summarized in \tab{tab_gbbarth}.

\begin{table}[H]
\begin{center}
\begin{tabular}{||l|c||}
\hline\hline
Calculation & $g_{bb}$ (\%)\\
\hline
Leading order & 0.110 \\
LO + NLL \cite{Seymour:1995bz} & 0.207 \\ 
LO + NLL \cite{Miller:1998ig}  & 0.175 \\ \hline\hline
\end{tabular}
\caption{Calculations of gluon splitting to $b\bar b$.}
\label{tab_gbbarth}
\end{center}
\end{table}

We see that resummation of logarithms gives a substantial enhancement.
Bearing in mind that $\log(\mz^2/m_b^2)\simeq 6$ and
$\as\log^2(\mz^2/m_b^2)\simeq 4$, this is not
surprising. Still missing are higher-order terms of the form
$\as^n\log^{2n-m}(s/m_b^2)$ with $n,m>2$, which could well
become comparable with the leading term at high orders.
The difference between the predictions of \cite{Seymour:1995bz}
and \cite{Miller:1998ig}
is due to their different treatment of NNLL ($m=3$) terms.

In conclusion, the theoretical prediction for $g_{bb}$ must still
be regarded as quite uncertain, and not in serious disagreement with
the data.

Reference~\cite{Miller:1998ig} also contains predictions of secondary
heavy quark production as a function of an event shape variable
(the heavy jet mass). There are no data available yet on this,
so the authors compare with Monte Carlo results. Their predictions
are similar to those of the main event generators discussed below.

Monte Carlo predictions of $g_{bb}$ are in principle even more unreliable
than the theoretical results presented above, since they do not fully include
next-to-leading logarithms or matching to fixed order. Nevertheless they do
include some real effects absent from the analytical calculations, such as the
effects of phase-space limitations. Different options for the treatment
of subleading terms, such as the choice of argument for $\as$, can easily
be explored by providing suitable switches in the programs.
Also of course they provide a complete model of the final state,
which allows the effects of experimental cuts to be simulated.
Relevant developments in the main Monte Carlo programs are
described in the following three subsections.

\subsection{Monte Carlo developments: PYTHIA}
\label{ts:sec:gsplit}

\subsubsection{Strong coupling argument and kinematics}

The default behaviour in \pythia\ is to let $\as$ have $p_T^2$
as argument. Actually, since the exact kinematics has
not yet been reconstructed when $\as$ is needed,
the squared transverse momentum is represented by the
approximate expression $z(1-z)m^2$, where $z$ is the
longitudinal splitting variable and $m$ the mass of the
branching parton. Since $\as$ blows up when its
argument approaches $\Lambda_{QCD}$, this translates into
a requirement on $p_T^2$ or on $z$ and $m$, restricting allowed
emissions to $p_T > Q_0/2$, where $Q_0$ is the shower cutoff
scale. Also when full kinematics is reconstructed,
this is reflected in a suppression of branchings with
small $p_T$. Therefore, if the angular distribution of the
$g$ decay is plotted in its rest frame, the quarks do not
come out with the $1+\cos^2\theta$ angular distribution one 
might expect, but rather something peaked at $90\deg$
and dying out at $0\deg$ and $180\deg$.

For $g\to q\bar q$ branchings, the soft-gluon results that
lead to the choice of $p_T^2$ as scale are no longer
compelling, however. One could instead use some other
scale that does not depend on $z$ but only on $m$. A
reasonable, but not unique, choice is to use $m^2/4$, where
the factor 4 ensures continuity with $p_T^2$ for $z=1/2$.
This possibility has been added as new option {\tt MSTJ(44)}=3.
In order for this new option to be fully helpful, a few
details in the treatment of the kinematics have also been
changed for the $g\to q\bar q$ branchings. These changes
are not completely unimportant, but small on the scale 
of the other effects discussed here.

Actually, the change of $\as$ argument in itself
leads to a reduced $g\to q\bar q$ splitting rate, while 
the removal of the $p_T > Q_0/2$ requirement increases it. 
The net result is an essentially unchanged rate, actually
decreased by about 10\% for charm and maybe 20\% for bottom,
based on not overwhelming statistics. The kinematics of
the events is changed, so experimental consequences would
have to be better quantified. However, the changes are not 
as big as might have been expected -- see the following.

\subsubsection{Coherence}

In the above subsection, it appears as if the $1+\cos^2\theta$
distribution would be recovered in the new option
{\tt MSTJ(44)}=3. However, this neglects the coherence
condition, which is imposed as a requirement in the
shower that successive opening angles in branchings become
smaller. Such a condition actually disfavours branchings
with $z$ close to 0 or 1, since the opening angle becomes
large in this limit. It should be noted that the opening
angle discussed here is not the true one, but the one
based on approximate kinematics, including neglect of
masses. One may question whether the coherence arguments
are really watertight for these branchings, especially if
one considers $g\to q\bar q$ close to threshold, where the
actual kinematics is quite different from the one assumed
in the massless limit used in the normal coherence
derivation.

As a means to exploring consequences, two new coherence level 
options {\tt MSTJ(42)}=3 and 4 have thus been introduced. In the
first, the $p_T^2$ of a $g\to q\bar q$ branching is reduced by 
the correct mass-dependent term, $1-4m_q^2/m_g^2$, while the 
massless approximation is kept for the longitudinal momentum. 
This is fully within the uncertainty of the game, and no less
reasonable than the default {\tt MSTJ(42)}=2. 
In the second, no angular ordering at all is imposed on 
$g\to q\bar q$ branchings. This is certainly an extreme scenario, 
and should be used with caution. However, it is still 
interesting to see what it leads to.

It turns out that the decay angle distribution of the gluon is 
much more distorted by the coherence than by the $\as$ 
and kinematics considerations described earlier. Both 
modifications are required if one would like to have a 
$1+\cos^2\theta$ shape, however. Also other distributions, like 
gluon mass and energy, are affected by the choice of options.

The most dramatic effect appears in the total gluon branching
rate, however. Already the introduction of the mass-dependent  
factor in the angular ordering requirement can boost the 
$g\to b\bar b$ rate by about a factor of two. The effects are
even bigger without any angular ordering constraints at all.  
It is difficult to know what to make of these big effects.
The options described here would not have been explored
had it not been for the \lep\ data that seem 
to indicate a very high secondary charm and bottom production rate. 
Experimental information on the angular distribution of secondary 
$c\bar c$ pairs might help understand what is going on better, 
but probably that is not possible experimentally.

\subsubsection{Summary}

In order to study uncertainties in the $g\to  b\bar b$ rate,
some new \pythia\ options have been introduced, {\tt MSTJ(44)}=3 and
{\tt MSTJ(42)}=3 and 4, none of them as default (yet). Taken together,
they can raise the $g\to  b\bar b$ rate by a significant factor,
as summarized in \tab{tab_gbbarpy}. A study of the effects of these
options on the 3-jet rate ratio $R_4^{b\ell}$ is described in
\sect{sec:gbbimpact}.

\begin{table}[H]
\begin{center}
\begin{tabular}{||c|c|c|c|c||}
\hline\hline
{\tt MSTJ(44)} &  {\tt MSTJ(42)} & $g_{uu+dd+ss}$ (\%) & 
$g_{cc}$ (\%) & $g_{bb}$ (\%)\\
($\as$) & (coherence) &  &  & \\
\hline
    2   &  2   &   14.3   &    1.26   &    0.16 \\
    2   &  3   &   20.9   &    1.93   &    0.26 \\
    2   &  4   &   38.5   &    3.07   &    0.32 \\
    3   &  2   &   12.9   &    1.16   &    0.15 \\
    3   &  3   &   19.9   &    1.77   &    0.28 \\
    3   &  4   &   42.9   &    3.48   &    0.46 \\
\hline\hline
\end{tabular}
\caption{\pythia\ options for gluon splitting to $q\bar q$.
Rates at 91.2 GeV for the normal flavour mixture.}
\label{tab_gbbarpy}
\end{center}
\end{table}

The {\tt MSTJ(42)}=4 option is clearly extreme, and to be used with
caution, whereas the others are within the (considerable) 
range of uncertainty.

The corrections and new options are available starting with
\pythia\ 6.130, obtainable from\\
{\tt www.thep.lu.se/}$\sim${\tt torbjorn/Pythia.html}.

\subsection{Monte Carlo developments: HERWIG}\label{sec:hwgbbar}

\subsubsection{Angular distribution in $g\to q\bar q$}
In \herwig, the angular-ordering constraint, which is derived for soft
gluon emission, is applied to all parton shower vertices, including
$g\to q\bar q$.  In versions before 6.1, this resulted in a severe
suppression (an absence in fact) of configurations in which the gluon
energy is very unevenly shared between the quarks.  For light quarks
this is irrelevant, because in this region one is dominated by gluon
emissions, which are correctly treated.  However for
heavy quarks, this energy sharing (or equivalently the quarks' angular
distribution in their rest frame) is a directly measurable quantity, and
was badly described.

Related to this was an inconsistency in the calculation of the Sudakov
form factor for $g\to q\bar q$. This was calculated using the entire allowed
kinematic range (with massless kinematics) for the energy fraction $x$,
$0\le x\le 1$, while the $x$
distribution generated was actually confined to the angular-ordered
region, $x,1\!-\!x \ge m/\sqrt{E^2\xi}$ (see \sect{sec_hwshower}).

In \herwig\ version 6.1, these defects are corrected as follows.
We generate the $E^2\xi$ and $x$ values for the shower as before.
We then apply an {\it a posteriori\/} adjustment to the
kinematics of the $g\to q\bar q$ vertex during the kinematic
reconstruction.  At this stage, the masses of the $q$ and $\bar q$ showers
are known.   We can therefore guarantee
to stay within the kinematically allowed region.  In fact, the
adjustment we perform is purely of the angular distribution of the $q$
and $\bar q$ showers in the $g$ rest frame, preserving all the masses
and the gluon four-momentum.  Therefore we do not disturb the kinematics
of the rest of the shower at all.

Although this cures the inconsistency above, it actually introduces a
new one: the upper limit for subsequent emission is calculated from the
generated $E^2\xi$ and $x$ values, rather than from the finally-used
kinematics.  This correlation is of NNL importance, so we can
formally neglect it.  It would be manifested in an incorrect correlation
between the masses and directions of the produced $q$ and $\bar q$ jets.
This is, in principle, physically measurable, but it seems
less important than getting the angular distribution itself right.  In
fact the solution we propose maps the old angular distribution smoothly
onto the new, so the sign of the correlation will still be preserved,
even if the magnitude is wrong.

Even with this modification, the \herwig\  kinematic reconstruction can
only cope with particles that are emitted into the forward hemisphere in
the showering frame.  Thus one cannot populate the whole of
kinematically-allowed phase space.  Nevertheless, we find that this is
usually a rather weak condition, and that most of phase space is
actually populated.

Using this procedure, we find that the predicted angular distribution for
secondary $b$ quarks at \lep\ energies is well-behaved, i.e.\ it looks
reasonably similar to the leading-order result ($1+\cos^2\theta^*$),
and has relatively small hadronization corrections.

\subsubsection{Predictions for $g_{bb}$}
Reference~\cite{Seymour:1995bz} contains comparisons between
analytical predictions for $g_{bb}$ and those of \herwig. One result of the
analytical calculation is that, to NLL accuracy, one can use the massless
formula for the splitting $g\to b\bar b$, provided one also sets a cutoff on
the virtual gluon mass of $m_g>e^{5/6}m_b = 2.3 m_b$ instead of the kinematic
cutoff  $m_g>2m_b$. Somewhat fortuitously, this is similar to the \herwig\
method, which uses the massless formula with a cutoff
$m_g >2(m_b+Q_0)$ with $Q_0\simeq 0.5$ GeV. The  comparisons in
show that the resummed and \herwig\
predictions are quite similar at \lepone\ and \leptwo\ energies.

\herwig\ results for $Z^0$ decay are summarized in \tab{tab_gbbarhw},
for the version used in the original comparisons (5.7) and the latest
version, 6.1 \cite{Corcella:1999qn}.
The main difference between the two versions in this
context is a change in the default $b$-quark mass from $m_b=5.2$ GeV to
4.95 GeV, which is justified by the approximate relation $m_B=m_b+m_l$
where $m_l$ is the light quark mass. We see that the \herwig\ results
are somewhat higher than the resummed predictions in \tab{tab_gbbarth}
and in better agreement with the data in \tab{tab_gbbarex}.

\begin{table}[H]
\begin{center}
\begin{tabular}{||c|c||}
\hline\hline
Version & $g_{bb}$ (\%)\\
\hline
5.7 & 0.23 \\
6.1 & 0.25 \\ 
\hline\hline
\end{tabular}
\caption{\herwig\ predictions for gluon splitting to $b\bar b$.}
\label{tab_gbbarhw}
\end{center}
\end{table}

\subsection{Monte Carlo developments: ARIADNE}
\label{ll:gbbplit}

The splitting of gluons into a $q\bar{q}$ pair does not fit into the
dipole picture in an obvious way, since this splitting is related
directly to a single gluon rather than to any dipole between two
partons. Also, all gluons in emitted in the cascade are massless, and
to be able to split into massive quarks, energy has to be required
from somewhere. The way the process is included in \ariadne\ is
described in ref.\ \cite{arigqq} The splitting probability of a gluon
is simply divided in two equal parts, each of which is associated to
each of the two connecting dipoles. The splitting process can then
again be treated as a two-to-three process, where a spectator parton
is used to conserve energy and momentum. It can be shown that this is
equivalent to standard parton shower approaches in the limit of
strongly ordered emssions. But the differences when extrapolating away
from that limit can become large, and \ariadne\ typically gives twice
as many secondary b$\bar{\mbox{b}}$ pairs as compared to eg.\ \jetset.

But this treatment of secondary heavy quarks may lead to rather
strange situations (as noted in ref.\ \cite{Mikeg2qq}).  Since
transverse momentum of the $q\bar{q}$ splitting can become small even
for heavy quarks, it is possible to split a gluon so that the mass
$m_{q\bar{q}}^2$ is larger than the transverse momentum scale --
$p_{\perp g}^2$ -- at which the gluon was emitted -- although the
ordering of the emissions, $p_{\perp q\bar{q}}^2<p_{\perp g}^2$, is
still respected. To avoid such situations there is an option in
\ariadne\footnote{\texttt{MSTA(28)$\ne$0} in the \texttt{/ARDAT1/}
  common block.} which introduces an extra limit,
$m_{q\bar{q}}^2<p_{\perp g}^2$, on gluon splitting.

As discussed already in \sect{ts:sec:gsplit}, it is not quite clear if or
how the ordering of emissions should be enforced in the case of gluon
splitting into massive quarks and, for that reason, \ariadne\ also
includes an option where these splittings are allowed to be
non-ordered, ie.\ $p_{\perp q\bar{q}}$ is allowed to be larger than
the transverse momentum of the preceeding
emission.\footnote{\texttt{MSTA(28)$<$0} in the \texttt{/ARDAT1/}
  common block. The current version (4.10) contains a bug for this
  option.  A bug fix can be obtained on request to leif@thep.lu.se.}
The corresponding rates of gluon splittings are given in table
\ref{tab:arisplit}

\begin{table}[htbp]
  \begin{center}
    \begin{tabular}{|c|c|c|c|}
      \hline
      \texttt{MSTA(28)} & $g_{u\bar{u}+d\bar{d}+s\bar{s}}$
      & $g_{c\bar{c}}$ & $g_{b\bar{b}}$\\
      \hline
     0 & 25.9 & 2.18 & 0.34\\
     1 & 17.7 & 1.09 & 0.13\\
   --1 & 28.8 & 1.88 & 0.16\\
      \hline
    \end{tabular}
    \caption{\ariadne\ options for gluon splitting into $q\bar{q}$.
      Rates at 91.2 GeV for the normal flavour mixture.}
    \label{tab:arisplit}
  \end{center}
\end{table}

\section{OVERALL CONCLUSIONS AND RECOMMENDATIONS}
Here we summarise the main results from each Section above and
the recommendations that follow from them.
The term `jet rates' always refers to the \durham\ algorithm unless
it is explicitly stated otherwise.

\subsection{Monte-Carlo developments}
The relevant features of the main event generators, \pythia, \herwig\ and
\ariadne, were reviewed with emphasis on relevant new developments.  In
many cases, important modifications and new options were introduced as a
result of discussions in the working group.

In \pythia, improved treatment of quark masses in the parton shower permits
a better description of the overall 3-jet rate, but there remains a
problem of underestimation of the 4-jet rate. Modification of the way in
which matrix element corrections are applied has little effect on this.

In \herwig, parameters for the cluster hadronization of $b$-quarks have been
separated from those for ligher quarks, so that improved tuning to 
$b$-quark fragmentation data will be possible.

In both  \pythia\ and \herwig\ there are now options to interface parton
showers to the massless 4-parton matrix elements. In addition,
\fourjphact\ interfaces the {\em massive}  4-parton matrix elements to
\pythia\ parton showers.  These options provide an improved description
of 4-jet final states, but are not suitable for describing
features that receive important 2- and 3-jet contributions.
A very recent development in \herwig\ is an option to combine
2,3 and (massless) 4-parton matrix elements together with parton showers
in a way that aims to avoid the worst aspects of double counting.
Comparison and tuning of this option to \lepone\ data is in progress.

In \ariadne, options exist for switching on and off the dead-cone effect in
QCD radiation from heavy quarks. A massive leading-order matrix element
correction has also been introduced, as exists in \pythia\ and \herwig.
Overall, \ariadne\ gives a better description of jet rates than either
\pythia\ or \herwig. However, the description of mass effects in jet rates
in \ariadne\ is not so satisfactory, and in fact turning off the
treatment of quark masses altogether appears to provide a better description.

A major new development is the introduction of the new event generator
\apacic, which for the first time interfaces $n$-parton matrix elements
and parton showers for $n=2,\ldots,5$. A number of options are available for
choosing the relative jet rates, and for initialising and evolving the
parton showers. As in  \pythia\ and \ariadne, the \jetset\ string
hadronisation model is used. A first attempt at tuning to \lepone\ data
gives encouraging results, with fits to most event shape and single-particle
distributions of a quality similar to the established generators. Differential
jet rates show some features which may be associated with merging the
different parton multiplicities. The tuned shower cutoff is high, so that
the parton showers have little phase space for evolution and final-state
structure is mostly determined by matrix elements and hadronization.

\subsection{Jet rates (inclusive)}
Both \pythia\ and \herwig\ have problems with fitting the 3- and 4-jet
rates simultaneously as functions of the jet resolution $y_c$. For a given
tuning, one can describe e.g.\ the 3-jet rate well, but then the rates for
higher jet multiplicities are overestimated by \herwig\ and underestimated by
\pythia. None of the modifications tried was able to eliminate this problem.
For analyses at \leptwo\ energies using  \pythia\ or \herwig, we recommend
tuning to the relevant jet rate at \lepone\ in order to minimize the
associated systematic error, which then results only from the
change in that jet rate from \lepone\ to \leptwo. If this is done,
then a systematic error of 2\% in the 4-jet rate at \leptwo\ could be
achieved. On the other hand, if only a general tuning at \lepone\ is
performed, the systematic error could be as large as 5\%.

\ariadne\ gives the best overall description of jet rates and should therefore
be considered as the generator of choice for estimating multi-jet backgrounds,
e.g.\ in hadronic WW decay studies. Using \ariadne\ could lead to a further
reduction of systematic errors.

\subsection{Jet rates (mass effects)}
Full NLO massive matrix element calculations of the 3-jet rate are now
available. They were used to study the effect of the $b$-quark mass on
the ratio of $b$-quark to light-quark rates,  $R_3^{b\ell}(y_c)$. This
is an observable in which many systematic uncertainties tend to cancel.
The difference between the running-mass and pole-mass schemes was used
as an estimate of higher-order contributions. This difference was indeed
reduced relative to the LO calculation, with the predicted NLO band lying
within the LO one.

In the case of the 4-jet rate ratio, $R_4^{b\ell}(y_c)$, only LO massive
predictions are available. Therefore, to be cautious, theoretical uncertainties
on both  $R_3^{b\ell}$ and  $R_4^{b\ell}$ were estimated at $\pm$ half the
difference between the LO pole-mass and running-mass predictions. The
\delphi\ \lepone\ data do fall within this band over the range measured
($0.01\leq y_c\leq 0.06$). For detailed numerical estimates of the
uncertainties, see \sect{sec:R34conc}.

The performance of the Monte Carlo event generators was judged against
the theoretical predictions with the estimated uncertainties. Generally
speaking, the generators tend to overestimate mass effects, i.e.\ they
underestimate $R_3^{b\ell}$ and  $R_4^{b\ell}$. Overall, \herwig\ gave
the best agreement at $\sqrt{s}=\mz$, although the new mass treatment in
\pythia\ describes $R_3^{b\ell}$ better. \ariadne\ underestimates more
severely, with dead-cone effects and the new massive matrix element
correction tending to worsen agreement.

A full NLO massive matrix element calculations of the 4-jet rate would
undoubtedly be helpful in reducing the systematic uncertainty on
$b$-quark mass effects in jet rates, and in testing and improving
the performance of event generators. 

A study of hadronization effects showed that use of the \cambridge\ jet
algorithm can considerably reduce hadronization corrections to
$R_3^{b\ell}$, which should be helpful in determinations of the
$b$-quark mass. Comparison of the latest versions of \pythia\ and
\herwig\ showed that they give closer estimates of hadronization
corrections following improvements in \herwig,
and that decay effects are small for sufficiently large values of $y_c$.

\subsection{Four-jet observables}
The Monte Carlo generators with specific 4-jet options (\pythia, \herwig\ and
\apacic) were compared with each other and with matrix element calculations,
for the standard set of 4-jet angular distributions as well as the
differential 4-jet rate and the D-parameter distribution. No significant
differences were found at the matrix element level, except for the
differential jet rate in \pythia, which was thought to be due to an
intrinsic \jade\ (mass) cut on the 4-parton configurations generated
by that program. Good agreement between the programs was found after
parton showering and hadronization. Quark mass effects were found
to be small (2\% level) for the \lep\ flavour mixture.

NLO corrections to the 4-jet angular distributions are small but can
have a significant effect on the extracted colour factors, owing to
their different functional dependence on these quantities. On
the other hand NLO effects are very large (70-130\%) in the
4-jet rate and D-parameter distributions. This indicates
that resummation of large higher-order corrections is required.
The 4-jet options of the event generators are not able to
describe these distributions owing to the lack of 2- and 3-jet
contributions.  The default 2+3 jet + parton shower options in
\pythia\ and \herwig\ are more successful here.  
The new combined multijet + shower options in \herwig\ and
\apacic, developed during the workshop, may provide
a better simultaneous description of these distributions and of the
4-jet angular distributions, provided successful 
tuning to the \lepone\ data can be achieved.
  
\subsection{B fragmentation}
The data, theory and models for $b$-quark fragmentation into B-hadrons
were reviewed. New theoretical calculations with resummation of
large higher-order terms suggest that non-perturbative effects are
small but significantly different from the conventional Peterson
parametrization.

No new work could be undertaken on the important topic of comparing
the performance of Monte Carlo generators with the data and with
theoretical calculations in this area. Comparisons with new data
presented recently by the
\sld\ collaboration suggest that the \pythia/\jetset\ description
{\em with the original Lund parametrization} of fragmentation
is satisfactory at $\sqrt{s}=\mz$. 

\subsection{$g \rightarrow b {\bar b}$ splitting}
The experimental results on the rate of gluon splitting into $b$-quark
pairs (around 3\permil\ at $\sqrt{s}=\mz$) are somewhat higher than the
best theoretical estimates (around 2\permil). However the theoretical
uncertainties due to unknown sub-leading logarithmic corrections
easily cover the discrepancy.  This point is emphasised by the
sensitivity of Monte Carlo generator predictions to the treatment of
subleading and kinematic effects.  In \pythia\ a number of new options
have been introduced to vary the treatment of such effects, and
particular choices can readily bring the $g \rightarrow b {\bar b}$
rate up to the observed value. We provisionally recommend the setting
{\tt MSTJ(42)}={\tt MSTJ(44)}=3. In \herwig\ and \ariadne, the default
settings in the latest versions already give adequate agreement with
the data. However, an estimated uncertainty as large as 30\% remains
appropriate.


\end{document}